\documentclass[12pt,a4paper]{report}

\usepackage{epsfig}
\usepackage{amssymb}
\usepackage{array}
\usepackage{amsmath}

\newcommand{\ncmd}{\newcommand}

\newtheorem{defi}{Definition}
\newtheorem{theo}{Theorem}
\newtheorem{prop}{Proposition}
\newtheorem{lem}{Lemma}
\newtheorem{cor}{Corollary}

\newtheorem{rem}{Remark}

\newcommand{\GL}[1]{\mbox{$\mathbf{GL}(#1)$}}
\newcommand{\SU}[1]{\mbox{$\mathbf{SU}(#1)$}}
\newcommand{\U}[1]{\mbox{$\mathbf{U}(#1)$}}
\newcommand{\SP}[1]{\mbox{$\mathbf{Sp}(#1)$}}
\newcommand{\SO}[1]{\mbox{$\mathbf{SO}(#1)$}} 

\newcommand{\OG}[1]{\mbox{$\mathbf{O}(#1)$}}

\ncmd{\btheo}{\begin{theo}$\!\!\!$. -- }
\ncmd{\etheo}{\end{theo}}
\ncmd{\bpro}{\begin{prop}$\!\!\!$. -- }
\ncmd{\epro}{\end{prop}}
\ncmd{\preuve}{{\sc Preuve --}\ }
\ncmd{\bdefi}{\begin{defi} $\!\!\!$. -- } 
\ncmd{\edefi}{\end{defi}}
\ncmd{\bco}{\begin{cor}$\!\!\!$. -- }
\ncmd{\eco}{\end{cor}}
\ncmd{\ble}{\begin{lem}$\!\!\!$. -- }
\ncmd{\ele}{\end{lem}}
\ncmd{\bno}{\begin{notation}$\!\!\!\!\!$. -- }
\ncmd{\eno}{\end{notation}} 
\ncmd{\bre}{\begin{rem}$\!\!\!$. --  \begin{em}}
\ncmd{\ere}{\end{em} \end{rem}}
\ncmd{\beq}{\begin{equation}}
\ncmd{\eeq}{\end{equation}}
\ncmd{\ben}{\begin{enumerate}}
\ncmd{\een}{\end{enumerate}}
\ncmd{\bit}{\begin{itemize}}
\ncmd{\eit}{\end{itemize}}
\ncmd{\refp}[1]{(\ref{#1})}
\ncmd{\A}{\mathcal{A}}
\ncmd{\pl}{\langle\langle}
\ncmd{\pr}{\rangle\rangle}
\ncmd{\Fi}{\mathbb{F}}
\ncmd{\Oc}{\mathbb{O}} 
\ncmd{\Ha}{\mathbb{H}}
\ncmd{\R}{\mathbb{R}}
\ncmd{\C}{\mathbb{C}}
\ncmd{\Z}{\mathbb{Z}}
\ncmd{\N}{\mathbb{N}}
\ncmd{\Sph}{\mathbb{S}}
\ncmd{\T}{\mathbb{T}} 
\ncmd{\D}{\mathbb{D}}
\ncmd{\Lp}{\mathfrak||\Pi||} 
\ncmd{\Lg}{\mathfrak{g}}
\ncmd{\La}{\mathfrak{a}}
\ncmd{\Lk}{\mathfrak{k}}
\ncmd{\Lm}{\mathfrak{m}}
\ncmd{\Lh}{\mathfrak{h}}
\ncmd{\End}{\mbox{End}}
\ncmd{\Aut}{\mbox{Aut}}
\ncmd{\ad}{\mbox{ad}} 
\ncmd{\Ad}{\mbox{Ad}}
\ncmd{\HH}{\mbox{H}}
\ncmd{\V}{\mbox{V}}
\ncmd{\w}{\widetilde\omega}
\ncmd{\tr}{\mbox{tr}}
\ncmd{\Ker}{\mbox{Ker}}
\ncmd{\I}{\mbox{I}}
\ncmd{\CC}{\mathcal{C}}
\ncmd{\Exp}{\mbox{Exp}}
\ncmd{\di}{\displaystyle} 
\ncmd{\bs}{\backslash}
\ncmd{\ov}{\overline}
\ncmd{\no}{\noindent}
\ncmd{\ra}{\rightarrow}
\ncmd{\lra}{\longrightarrow}
\ncmd{\eps}{\epsilon}
\ncmd{\Lra}{\Longrightarrow}
 
\ncmd{\scalar}[2]{\mbox{$\mathcal{h} #1,#2 \mathcal{i}$}}
\ncmd{\ichap}{\^{\i}}

\title{Geometrical Methods in Gauge Theory}
\author{Henrique de Andrade Gomes\footnote{Universidade de S\~ao Paulo, Instituto de Matematica e Estat\'istica, gomes.ha@gmail.com 
}} 

\begin{document}
\maketitle
\tableofcontents

\pagebreak

\section{Preface}
\begin{quote} {\it Na matem\'atica, um manipulador formal frequentemente experiencia a desconfort\'avel sensa\c c\~ao de que seu l\'apis \'e mais inteligente do que ele.} - Howard Eves\end{quote}

Esta disserta\c c\~ao \'e baseada em \cite{Palais} - notas de um curso ministrado por Richard Palais em 1981 entitulado {\it Geometriza\c c\~ao da F\'isica}. Este curso, dado no curto per\'iodo de seis semanas, est\'a em forma bastante resumida, com  demonstra\c c\~oes muitas vezes omitidas ou brevemente ``indicadas".  

Algumas das demonstra\c c\~oes contidas no presente trabalho s\~ao  mais ``sujas" do que as encontradas nos livros textos usuais, por terem sido elaboradas independentemente,  com a ajuda de meu orientador, ou ainda com os norteamentos gerais das notas do Palais. Pe\c co paci\^encia aos leitores se lhes parecerem \'obvias demonstra\c c\~oes alternativas. 

  Em todo momento a exposi\c c\~ao \'e acompanhada por um subtexto interpretativo, numa tentativa de compreender em um n\'ivel mais intuitivo e geom\'etrico os resultados a que chegamos. Tomei esta atitude unicamente em benef\'icio pr\'oprio, com a  esperan\c ca de permanecer ``mais inteligente do que meu l\'apis". Nessa empreitada sigo o esp\'irito de  \cite{Baez}, e \cite{Feynman}. Preferi essa atitude a destinar minha disserta\c c\~ao unicamente aos meus examinadores, que por seu conhecimento profundo do assunto t\^em provavelmente pouco a ganhar com uma exposi\c c\~ao menos seca. 

O objetivo deste estudo foi o de me introduzir ao vasto arcabou\c co matem\'atico necess\'ario ao f\'isico te\'orico moderno, um horizonte um pouco mais difuso do que a disserta\c c\~ao de mestrado usual (baseada em um teorema ou artigo), e respons\'avel pela aparente desconex\~ao de alguns t\'opicos, como o Teorema de Hodge, que, apesar de sua enorme import\^ancia,  n\~ao \'e reutilizado no decorrer da exposi\c c\~ao.  

Primeiramente, gostaria de agradecer encarecidamente ao meu orientador, Professor Claudio Gorodski, pelas ricas discuss\~oes e por ensinar-me  tanto sobre como fazer matem\'atica. Seu modo de pensar ser\'a sempre uma refer\^encia para mim.  

Dedico esse trabalho aos meus pais, que sempre aturaram com paci\^encia meus h\'abitos anormais, estimularam a busca pelo conhecimento, instigando em mim uma admira\c c\~ao infind\'avel pela nobreza dessa empreitada, e por vezes at\'e fingiram interesse em meus estudos.  
 
Gostaria de agradecer tamb\'em \`a minha companheira Helena, pelo seu est\'imulo aos meus h\'abitos anormais, paci\^encia com minha busca pelo conhecimento, e por, vez em quando, tamb\'em  fingir interesse.

\section{Introduction}
\begin{quote}{{\it{A Filosofia \'e escrita neste grande livro - o Universo - que permanece sempre aberto ao nosso mirar, mas que n\~ao pode ser compreendido a n\~ao ser que primeiro se aprenda a compreender a linguagem e interpretar os caract\'eres em que \'e escrito. Ele \'e escrito na linguagem da matem\'atica, e seus caract\'eres s\~ao tri\^angulos, c\'irculos, e outras figuras geom\'etricas, sem as quais \'e humanamente imposs\'ivel compreender dele uma s\'o palavra.}}- Galileo Galilei}\end{quote} 
A rela\c c\~ao \'intima entre geometria e f\'isica, apesar de bem explicitada pela cita\c c\~ao acima, n\~ao come\c cou com Galileo e certamente n\~ao se extinguiu desde ent\~ao. Deveras, desde Pit\'agoras, Euclides e Arquimedes se busca refletir a irregular realidade no claro espelho da geometria. 

Ao tornarmos o olhar para o desenvolvimento cient\'ifico atual, podemos dizer com seguran\c ca que a teoria da Relatividade Geral foi um dos maiores desenvolvimentos da ci\^encia no s\'eculo XX, um cap\'itulo important\'issimo no grande livro de Galileo, um espelho geom\'etrico l\'impido  que inspirou in\'umeras ramifica\c c\~oes na f\'isica e na matem\'atica. 
Na mec\^anica Newtoniana, o espa\c co e o tempo eram encarados como palcos fixos, onde tinha lugar a din\^amica do Universo conhecido. Estes palcos no entanto, permaneciam alheios, indiferentes  ao que neles se desenrolava. Matematicamente, o espa\c co nesta teoria seria representado por uma c\'opia de $\R^3$ e o tempo por uma c\'opia de $\R$, i.e.: $\R^3\times\R$, ambos fixos e imisc\'iveis, . 

Com o advento da teoria eletromagn\'etica da Maxwell, foi necess\'aria uma revis\~ao destes conceitos. Maxwell mostrou que duas teorias aparentemente distintas, a el\'etrica e a magn\'etica, eram simplesmente dois aspectos do ``campo eletromagn\'etico''. Explicou assim a luz como um fen\^omeno decorrente de dist\'urbios no campo el\'etrico gerando dist\'urbios no campo magn\'etico e vice-versa, criando um mecanismo retroativo, propagando-se no espa\c co e no tempo. `Acidentalmente', essa forma de propaga\c c\~ao deveria ter uma velocidade fixa, o mecanismo deveria ``girar'' sempre \`a mesma velocidade, o que significava que a luz emitida por um corpo seria percebida com a mesma velocidade por todos os observadores, irrespectivamente de seu movimento em rela\c c\~ao \`a fonte.  

Isso levou Lorentz, Poincar\'e e especialmente Einstein, a perceberem que para acomodar a teoria de Maxwell, o espa\c co e o tempo n\~ao poderiam permanecer imisc\'iveis. Assim originou-se a revolu\c c\~ao da relatividade restrita, que transformou a nossa vis\~ao do espa\c co e do tempo, fundindo os dois em um s\'o, o espa\c co-tempo. De  $\R^3\times\R$, passamos a enxergar o palco da realidade como $\Ha^4$, o espa\c co de Minkowski quadri-dimensional\footnote{O espa\c co plano de assinatura $+++-$}. 

Einstein ao tentar expandir essa teoria para que englobasse tamb\'em a gravidade, naturalmente tentou seguir o modelo de Maxwell, por\'em foi for\c cado a construir uma teoria em que o espa\c co-tempo n\~ao poderiam mais assistir ao desenrolar dos eventos placidamente, mas deveriam ser afetados e afetar a din\^amica dos corpos de uma maneira fundamental. Grosseiramente, as equa\c c\~oes de Einstein\footnote{$$R_{\mu\nu}-\frac{1}{2}g_{\mu\nu}R=\kappa{T}_{\mu\nu}$$ onde $R_{\mu\nu}$ \'e o tensor de Ricci,$R$ \'e o escalar de curvatura e $g_{\mu\nu}$ \'e a m\'etrica do espa\c co -tempo, $\kappa$ \'e uma constante, todos os termos ser\~ao definidos apropriadamente mais adiante.} dizem que energia e momento afetam a forma como medimos dist\^ancia e dura\c c\~ao no espa\c co-tempo assim como cargas e correntes afetam o campo eletromagn\'etico. O espa\c co-tempo deixou de ser globalmente identificado a $\Ha^4$ e passou a ser localmente identificado a peda\c cos de $\Ha^4$, a ser encarado como uma variedade Lorentziana quadri-dimensional, onde as trajet\'orias naturais de part\'iculas\footnote{Part\'iculas sem carga, cor, etc.} seriam representadas por geod\'esicas. Isso alimentou as esperan\c cas de que talvez grande parte ou at\'e mesmo toda a f\'isica  tivesse car\'ater puramente geom\'etrico, e fertilizou o solo para a inven\c c\~ao das teorias de gauge.

Em f\'isica, as chamadas teorias de gauge (ou calibre) s\~ao baseadas na id\'eia de que certas transforma\c c\~oes de simetria de um sistema podem ser efetuadas tanto local, como (\`as vezes) globalmente, sem afetar os resultados da teoria, i.e.: que a teoria \'e invariante por certas transforma\c c\~oes. Estas simetrias localmente sempre refletem uma redund\^ancia na descri\c c\~ao de um sistema, o que \'e encarada como uma transforma\c c\~ao ``passiva", ou de coordenadas, enquanto que globalmente tamb\'em podem ser relacionadas com transforma\c c\~oes `` de fato", ou ativas.   Como veremos, a teoria eletromagnetica de Maxwell pode ser considerada o caso mais simples das chamadades teorias de gauge, seu exemplo ``par excellence". Antecipando um pouco nossa exposi\c c\~ao, podemos perceber que a defini\c c\~ao de ``terra'' do potencial el\'etrico \'e um exemplo de simetria de gauge: o valor absoluto do potencial \'e imaterial, o que importa em qualquer sistema el\'etrico \'e a diferen\c ca entre potenciais. Esse \'e um caso de uma transforma\c c\~ao global, se considerarmos tamb\'em mudan\c cas do potencial magn\'etico,  \'e possivel fazer mudan\c cas locais sem afetar qualquer resultado.  

A import\^ancia desta simetria no entanto permaneceu despercebida at\'e uma tentativa de Hermann Weyl \cite{We} de unificar eletromagnetismo e relatividade geral.  Inspirado na simetria conforme das teorias de Maxwell, Weyl procurou interpretar o eletromagnetismo como uma distor\c c\~ao de comprimentos relativ\'isticos produzidos pelo deslocamento sobre uma curva fechada. Conjecturou que invari\^ancia por mudan\c cas em escala (ou calibre) poderiam tamb\'em ser uma simetria local da relatividade geral. Superficialmente, seu racioc\'inio foi de que escala local n\~ao deveria ser observ\'avel, j\'a que um aumento das dimens\~oes de todos os objetos ao nosso redor n\~ao  poderia ser detect\'avel\footnote{Matematicamente, ele postulou um transporte paralelo alternativo, que n\~ao preservava a norma.}. Contudo, formalmente chegou ao resultado de que o espectro de radia\c c\~ao dos \'atomos dependeria de sua hist\'oria, um resultado n\~ao encontrado na Natureza \cite{Ga}, o que foi apontado por Einstein. 

No entanto, ap\'os o desenvolvimento da mec\^anica qu\^antica e suas fun\c c\~oes complexas, tornou-se claro que fase, e n\~ao escala, era o que deveria construir a ponte com eltromagnetismo\footnote{ Ou em linguagem moderna, o grupo de Gauge deveria ser modelado em  $S^1$ ao inv\'es do grupo multipilcativo $\R$.}.  Weyl, Vladimir Fock e Fritz London reutilizaram a id\'eia inicial, substituindo o fator de escala por uma vari\'avel complexa, transformando a mudan\c ca de escala em uma mudan\c ca de fase (uma simetria de gauge $U(1)$, como veremos mais tarde). Alcan\c caram assim uma bela explica\c c\~ao para o efeito do campo eletromagn\'etico sobre a descri\c c\~ao qu\^antica de uma part\'icula carregada \footnote{Assunto que n\~ao abordaremos nessa disserta\c c\~ao, j\'a que nos ateremos \`as formula\c c\~oes cl\'assicas ( i.e.: n\~ao qu\^anticas) da teoria.}. Infelizmente, nessa nova encarna\c c\~ao n\~ao havia mais uma maneira de incorpor\'a-la na relatividade geral, e ela tinha de ser superposta como estrutura adicional sobre o espa\c co-tempo. Nascia assim a teoria de gauge.

Em 1954, na tentativa de resolver problemas na teoria de part\'iculas elementares, Chen Ning Yang e Robert Mills introduziram teorias de gauge com grupos de simetria n\~ao abelianos como modelos para a intera\c c\~ao forte; a cola que permite a coes\~ao dos n\'ucleos at\^omicos. No entanto sua liga\c c\~ao com a teoria de fibrados permaneceu largamente ignorada ou considerada irrelevante at\'e os anos 70, quando aspectos n\~ao perturbativos relacionados \`as solu\c c\~oes cl\'assicas das equa\c c\~oes de Yang-Mills (instantons) vieram \`a tona, incorporando quest\~oes globais da teoria de fibrados. 

Paradoxalmente, o formalismo matem\'atico das teorias de gauge proveram uma estrutura para a unifica\c c\~ao das teorias qu\^anticas de campos, notavelmente incompat\'iveis com a relatividade geral, um dos conceitos progenitores da teoria de gauge. O Modelo Padr\~ao descreve com alto grau de precis\~ao as intera\c c\~oes fraca, forte e eletromagn\'etica, atrav\'es de um grupo de simetrias n\~ao-abeliano $SU(3)\times{SU(2)}\times{U(1)}$.  
   
Apesar da motiva\c c\~ao inicial destas teorias terem sido de ordem f\'isica, assim como em praticamente todas as outras constru\c c\~oes relevantes do campo, a intera\c c\~ao e benesse  rec\'iproca entre f\'isica e matem\'atica provinda das teorias de gauge foi extremamente f\'ertil. Aos anos 70, Michael Atiyah, estudou as solu\c c\~oes das equa\c c\~oes cl\'assicas de Yang-Mills, e em 1983, Donaldson, aluno de Atiyah, utilizando este trabalho revolucionou o estudo de variedades de dimens\~ao 4. Michael Freedman, continuando esse estudo, conseguiu exibir estruturas diferenci\'aveis ``fake'' do $\R^4$ ( i.e.: diferentes da can\^onica). Isso levou a um grande interesse em teorias de gauge por seus resultados puramente matem\'aticos. Em 1994, Edward Witten e Nathan Seiberg, inventaram  m\'etodos de calcular invariantes topol\'ogicos baseados em teorias de gauge. 

Obviamente, nosso \^ambito nessa disserta\c c\~ao \'e bem mais modesto. Pretendemos fornecer somente um apanhado inicial, e pessoal, da teoria de gauge cl\'assica, com \^enfase no aspecto geom\'etrico. O presente trabalho \'e uma tentativa de apresentar as ferramentas matem\'aticas necess\'arias a essa empreitada; \`a geometriza\c c\~ao da f\'isica. As chamadas teorias de gauge (ou de calibre) conseguem uma representa\c c\~ao puramente geom\'etrica das intera\c c\~oes n\~ao gravitacionais, utilizando intensamente o maquin\'ario matem\'atico da teoria de fibrados. Vamos a ela.      

\section{Basic No(ta)tions}

\begin{quote}{{\it{Por essa raz\~ao \'e seu nome Babel; porque o Senhor ali confundiu a linguagem de toda a Terra.}}- Genesis 11;9}\end{quote}

 Chamaremos normalmente a variedade espa\c co-tempo de  $M$, mas faremos poucas refer\^encias ao seu car\'ater Lorentziano. O tratamento de teorias f\'isicas modelando o espa\c co-tempo atrav\'es de variedades diferenci\'aveis automaticamente implementa um dos maiores insights provindos da relatividade geral: a invari\^ancia das leis f\'isicas por mudan\c cas de coordenadas\footnote{ E sua invari\^ancia por difeomorfismos globais, que sucitou algumas discuss\~oes interessantes em rela\c c\~ao ao `` argumento do buraco" (hole argument) \cite{Wald}.}.

 Para facilitar a nota\c c\~ao, salvo aviso, utilzaremos a conven\c c\~ao de soma de Einstein, pela qual se soma \'indices repetidos em cima e em baixo, e.g.: $$A^iB_i=\sum_i A^iB_i$$ Aproveitamos o ensejo para apontar que se $A,B$ s\~ao matrizes $k\times{k}$ n\'os temos que $[AB]^i_j=A^i_kB^k_j=B^k_jA^i_k\neq[BA]^i_j=B^i_kA^k_j$ i.e.: para passarmos da nota\c c\~ao escalar para a matricial precisamos arranjar a ordem correta, pr\'e estabelecida dos termos.

Ainda no t\'opico ``\'indices", utilizaremos para bases de $p$-formas muitas vezes a nota\c c\~ao de multi-\'indices, normalmente denotados por $I$ e $J$, que {\it n\~ao} devem ser confundidos com os \'indices latinos ma\'iusculos $A$ e $B$ que utilizaremos na se\c c\~ao sobre a teoria de Kaluza-Klein, onde denotam \'indices normais que percorrem os \'indices usuais sobre $M$ {\it e} os \'indices do espa\c co interno. 

Buscando satisfazer o f\'isico que h\'a em todos n\'os, bem como ilustrar um m\'etodo bastante \'util em c\'alculos expl\'icitos, 
 tentaremos alternar demonstra\c c\~oes usando bases locais com demonstra\c c\~oes mais intr\'insecas, sem \'indices.

Assumiremos um conhecimento b\'asico de variedades diferenci\'aveis,  campos vetoriais e grupos de Lie, o referente a um subconjunto denso dos primeiros tr\^es cap\'itulos de \cite{Warner}. Em especial mencionamos os dois seguintes teoremas:

\begin{theo}\label{transversalteo}
Sejam $M, N$ variedades suaves de dimens\~oes $m$ e $n$ respectivamente e $f:M\ra N$ suave. Se $Q$ \'e subvariedade mergulhada  em $N$, ent\~ao $S=f^{-1}(Q)$ \'e subvariedade mergulhada de $M$ se e somente se para todo $p\in S$ n\'os tivermos $T_{f(p)}N=\mbox{Im}(df_p)+T_{f(p)}Q$. \end{theo}
H\'a uma bela e simples demonstra\c c\~ao deste teorema em \cite{Guillemin}. 

Um outro teorema que nos servir\'a em diversas discuss\~oes \'e o de Frobenius.

\begin{theo}\label{frobeniusteo} 
 Uma $k$-distribui\c c\~ao suave involutiva  em uma variedade $M^m$, $m\geq k$, \'e integr\'avel.\end{theo}

Onde uma $k$-distribui\c c\~ao em uma variedade $M$ \'e um mapa $\mathfrak{D}:M\ra TM$ que assinala a cada ponto $m\in M$ um subespa\c co $k$-dimensional de $T_mM$,  $\mathfrak{D}_m$.  A $k$-distribui\c c\~ao \'e dita suave se  cada ponto $m\in M$ tem vizinhan\c ca aberta $U$ onde $\mathfrak{D}$ \'e gerado por $k$ campos suaves em $U$ linearmente independentes. Uma variedade integral de $\mathfrak{D}$ \'e uma subvariedade $N^k$ de $M$ tal que $T_mN=\mathfrak{D}_m$ para todo $m\in N$. $\mathfrak{D}$ \'e dita involutiva se para todos os campos suaves $X,Y\in \mathfrak{D}$ , n\'os tivermos $[X,Y]\in\mathfrak{D}$. As demonstra\c c\~oes geom\'etricas s\~ao padr\~ao, procedendo por indu\c c\~ao  a partir do Teorema do Fluxo Tubular, ver \cite{Boothby}.

\chapter{Vector Bundles}

\section{Foundations}

\begin{quote}{{\it{As equa\c c\~oes de Maxwell e os princ\'ipios da mec\^anica qu\^antica levaram \`a id\'eia de invari\^ancia de gauge. Tentativas de generalizar essa id\'eia, motivada pelos conceitos f\'isicos de fases, simetrias e leis de conserva\c c\~ao, levaram \`a teoria de gauge de campos n\~ao abelianos. Que campos de gauge n\~ao abelianos s\~ao conceitualmente id\^enticos a id\'eias na bela teoria de fibrados, desenvolvida por matem\'aticos sem refer\^encia ao mundo f\'isico, deixava-me estupefato. Em 1975, discuti minhas considera\c c\~oes con Chern, e disse`` isso \'e tanto emocionante quanto desconcertante, j\'a que voc\^es matem\'aticos inventaram esses conceitos do nada.'' Imediatamente ele protestou: `` N\~ao, n\~ao. Estes conceitos n\~ao foram sonhados. Eles foram naturais e reais."}}- C. N. Yang} \end{quote} 

\subsubsection{Fiber Bundles}

Ao medirmos qualquer quantidade f\'isica, a estamos medindo localmente, i.e.: sobre um aberto de $M$. Podemos dizer que essa quantidade mora em um espa\c co `` interno" $E_x$ sobre cada ponto $x$ do espa\c co-tempo, e sua medi\c c\~ao implica uma proje\c c\~ao sobre $M$. A forma mais simples de ``campo" seria assim uma fun\c c\~ao $f:M\rightarrow{E}$, supondo que o campo sobre cada ponto `` mora " no mesmo espa\c co, $E\simeq{E_x}$. No entanto, tal asser\c c\~ao implica que podemos comparar valores do campo sobre toda variedade $M$, por outro lado, n\~ao necessariamente possu\'imos um sistema de coordenadas global, no qual descrever\'iamos qualquer quantidade f\'isica, ent\~ao parece precipitado instituir a priori que  podemos descrever o campo globalmente. A priori, n\'os temos apenas uma uni\~ao disjunta dos espa\c cos internos sobre cada ponto, um ``feixe" de espa\c cos internos\footnote{ Essa seria a tradu\c c\~ao mais adequada para ``bundle", infelizmente j\'a \'e utilizada em outra \'area da matem\'atica.}.

\begin{defi} {Um {{fibrado}} consiste de variedades $E$ (chamado de espa\c co total), e $M$ (espa\c co base) e um mapa diferenci\'avel sobrejetor $\pi:E\rightarrow{M}$ (proje\c c\~ao). $E_x=\pi^{-1}(x)$ \'e chamada de {\bf{fibra}} sobre $x$.} \end{defi} 
Como a proje\c c\~ao \'e sobrejetora, \'e claro que $\bigcup_{x\in{M}}E_x=E$. Em geral, as fibras $E_x$ n\~ao precisariam  ser isomorfas mas esse \'e o caso interessante para n\'os, por isso assumiremos essa condi\c c\~ao, i.e.: $E_x\simeq{E_{x'}}$ para todos $x,x'\in{M}$. N\'os estamos ent\~ao tomando uma \'unica variedade para a descri\c c\~ao das fibras e n\~ao uma cole\c c\~ao de variedades, uma sobre cada ponto.  Chamamos um fibrado de localmente trivial se tivermos uma cobertura de $M$, \{$U_\alpha, \alpha\in\Lambda$\}, onde $\bigcup_{\alpha\in\Lambda}U_\alpha=M$, e para cada $U_\alpha$ exista $\psi_\alpha$ difeomorfismo tal que: 
\beq\label{triv}\psi_\alpha:\pi^{-1}(U_\alpha)\longrightarrow{U_\alpha}\times{F}\eeq
onde $F$ \'e uma variedade diferenci\'avel fixada para todo $\alpha$, chamada de fibra t\'ipica de $E$. Agora, n\'os queremos tamb\'em que
\beq\label{phi}\begin{array}{ll} 
\psi_{\alpha}: & \pi^{-1}(x)\rightarrow{\{x\}\times{F}} \\  
~ & u\longmapsto(x,\psi_{\alpha, x}(u))\end{array}\eeq
 onde, como $\psi$ \'e difeomorfismo, 
$\psi_{\alpha,x}:\pi^{-1}(x)\rightarrow{F}$ \'e tamb\'em difeomorfismo. \'E importante notar que a equa\c c\~ao (\ref{triv}) por si s\'o n\~ao implica em (\ref{phi}), j\'a que se $\tilde{U}_\alpha\subset{U_\alpha}$, n\~ao \'e  necessariamente verdade que a  restri\c c\~ao de $\psi$ seja da seguinte forma: 
\beq\label{triv2}\psi_\alpha:\pi^{-1}(\tilde{U}_\alpha)\longrightarrow{\tilde{U}_\alpha}\times{F}\eeq 

Essa condi\c c\~ao, i.e.: (\ref{triv2})  para todo aberto $\tilde{U}_\alpha\subset{U_\alpha}$, equivale a (\ref{triv}). Essa caracteriza\c c\~ao sobre a trivializa\c c\~ao \'e imprescind\'ivel para que ela mantenha o car\'ater de fibras sobre os pontos do fibrado. Em outras palavras, impomos na defini\c c\~ao que $pr_1\circ\psi=\pi$, onde $pr_1$ \'e a proje\c c\~ao na primeira coordenada. 

Ou seja, um fibrado localmente trivial (ao qual nos referiremos mais adiante simplesmente como fibrado) \'e simplesmente uma variedade que localmente \'e uma variedade produto\footnote{Veremos daqui a pouco porque variedades localmente produtos s\~ao mais interessantes para n\'os.}. 

{\bf{EXEMPLOS:}}

\begin{itemize}
 \item{{\bf{O fibrado trivial}}

$E=M\times{N}$. Esse \'e o caso mais simples de fibrado, chamado de trivial. A proje\c c\~ao \'e simplesmente a proje\c c\~ao na primeira coordenada, i.e.: 
\begin{eqnarray*}
\pi:E\rightarrow{M} \\ 
(x,v)\mapsto{x}\end{eqnarray*} 
Este fibrado \'e claramente localmente trivial, pela restri\c c\~ao da primeira coordenada.} 
 
 \item{{\bf{Faixa de Moebius}}

Seja $\sigma:\{I\times{\R\}}\rightarrow\{I\times{\R\}}/_\sim=E$ onde a rela\c c\~ao de equival\^encia \'e dada por $(0,t)\sim(1,-t)$.
Ent\~ao  $\pi:{I\times{\R}}\rightarrow{I}$  induz um mapa $\tilde{\pi}:E\rightarrow{S^{1}}$ e a fibra aqui \'e $\R$, e E seria homeomorfo \`a faixa de M\"obius sem as fronteiras. Com essa fibra\c c\~ao, a faixa de Moebius \'e um exemplo de um fibrado localmente trivial (basta vermos que $\sigma:]0,1[\times\R\rightarrow{E}$ \'e uma trivializa\c c\~ao local), por\'em n\~ao trivial, i.e.: n\~ao \'e uma variedade produto. Isso \'e facilmente observ\'avel j\'a que $S^1\times\R$ \'e um cilindro, logo orient\'avel e portanto n\~ao pode ser difeomorfo \`a faixa de Moebius.} 

 \item{{\bf{O fibrado tangente TM}}

$TM= \bigcup_{x\in{M}}T_xM=\{(x,v), x\in{M}, v\in{T_xM}\}$ ent\~ao \begin{eqnarray*}
\pi:TM\rightarrow{M}\\
(x,v)\mapsto{x}
\end{eqnarray*}
  i.e.: se $w\in{T}_xM$ ent\~ao $\pi(w)=x$, dada um atlas para M, $\{U_\alpha, \psi_\alpha\}$, n\'os temos que $TM$ \'e variedade diferenci\'avel com atlas dado por 
\begin{eqnarray*}
\left(\psi_\alpha\circ\pi,{d}\psi_\alpha\right):\pi^{-1}(U_\alpha)\rightarrow{W}\times{\R^n}\subset\R^n\times\R^n\\
(x,v)\mapsto\left(\psi_\alpha(x),(d\psi_\alpha)_x(v)\right)
\end{eqnarray*} 
 Aqui a fibra t\'ipica \'e isomorfa \`a $\R^n$ e uma trivializa\c c\~ao local pode ser dada como acima, substituindo $\psi_\alpha\circ\pi\rightarrow{\pi}$. \'E ainda f\'acil vermos que como as cartas s\~ao compat\'iveis, tamb\'em os ser\~ao as trivializa\c c\~oes locais.  } 

\end{itemize}

 Uma primeira quest\~ao que pode surgir naturalmente \'e se a condi\c c\~ao de trivialidade local n\~ao \'e um corlo\'ario das outras propriedades de fibrados. I.e.: se $E$ \'e diferenci\'avel, e as fibras s\~ao isomorfas, n\~ao seri razo\'avel que $E$ fosse localmente uma variedade produto? A resposta \'e negativa, mostremos um contra-exemplo baseado na faixa de Moebius ``torta". 

Como vimos, tomada como fibrado sobre a base $S^1$ com fibra t\'ipica dada por um segmento de reta, a faixa de Moebius \'e um exemplo de um fibrado n\~ao trivial. Mas partindo da constru\c c\~ao acima com $\{I\times\R\}/\sim$ , 
podemos escolher tomar como variedade base, ao inv\'es de $S^1$ (ou melhor, $I\times\{0\}/\sim$),  uma semi-reta perpendicular a $S^1$, por exemplo $$\{0\}\times[0,\infty[=\{0\}\times\R/\sim$$
 as fibras ent\~ao ser\~ao as proje\c c\~oes: $$\pi^{-1}(a)=\sigma(I\times\{a\}), a\in[0,\infty[$$
 ou seja, variedades compactas unidimensionais. Logo, a fibra t\'ipica ser\'a dada por $S^1$. Mas agora, trivializa\c c\~oes locais deveriam ser difeomorfos a $J\times{S^1}$, onde  $J$ \'e um intervalo de $[0,\infty[$. Ou seja, deveriam ser segmentos de cilindros. Mas \'e f\'acil ver que para qualquer intervalo aberto $W\in\R$ que contenha a origem, $\sigma(I\times{W})$ \'e uma faixa de Moebius, portanto n\~ao orient\'avel, portanto n\~ao difeomorfa a um cilindro.

\begin{defi}{{{Um fibrado vetorial}} \'e um fibrado localmente trivial, cujas fibras s\~ao espa\c cos vetoriais, e tais que, para todo $x\in{M}$, $\phi_x$ definido em (\ref{phi}) \'e isomorfismo linear.}\end{defi} 

Chamamos de $n$-fibrado vetorial (real) se a dimens\~ao da fibra t\'ipica \'e $n$. Ou seja, n\'os temos que    
$$\psi:\pi^{-1}(x)\rightarrow{\{x\}\times\R^n}$$ \'e um isomorfismo linear para todo $x\in{M}$.
Poder\'iamos tomar da mesma forma $\C^n$ ao inv\'es de $\R^n$. 

Definimos um isomorfismo entre fibrados vetoriais sobre uma mesma base $M$ (ou seja, entre as triplas previamente definidas $(E_{1},\pi_{1},M)$ , $(E_{2},\pi_{2},M)$), como um difeomorfismo $f:E_1\rightarrow{E_2}$ tal que $\pi\circ{f}=\pi$, onde $f$ 
leva $\pi^{-1}_{1}(x)\rightarrow{\pi^{-1}_{2}(x)}$ por um isomorfismo linear. Analogamente, um morfismo entre $(E_{1},\pi_{1},M)$ , $(E_{2},\pi_{2},M)$) \'e uma fun\c c\~ao suave $f:E_1\rightarrow{E_2}$ tal que $\pi\circ{f}=\pi$, onde $f$ 
leva $\pi^{-1}_{1}(x)\rightarrow{\pi^{-1}_{2}(x)}$ linearmente. 

\begin{defi} {{{Uma se\c c\~ao}} de um fibrado vetorial $(E,\pi,M)$ \'e simplesmente uma fun\c c\~ao 
$s:{M\rightarrow{E}}$, tal que para cada $x\in{M}$ $s(x)\in\pi^{-1}(x)$.}\end{defi}

Assim uma se\c c\~ao suave sobre o dom\'inio de uma trivializa\c c\~ao local $\theta$, pode ser identificado com uma fun\c c\~ao suave sobre a fibra t\'ipica, i.e.:, para todo $x\in\theta$ $s:x\mapsto (x,f_s(x))$. 

Isso nos permite vizualizar concretamente o motivo principal para que tomemos o fibrado como sendo localmente trivial: \'e nele poss\'ivel tomar se\c c\~oes suaves como sendo localmente expressas por fun\c c\~oes {\it suaves} a valores nas fibras t\'ipicas. Para se convencer de que isso n\~ao \'e poss\'ivel em fibrados que n\~ao s\~ao localmente triviais, construa uma se\c c\~ao na faixa de Moebius ``torta" e uma fun\c c\~ao $f:[0,b]\ra S^1$ que a represente. Verifique que $f$ \'e descont\'inua na origem.

Denotaremos por $\Gamma(E_{|\theta})$ o espa\c co de todas as se\c c\~oes sobre $\theta\subset{M}$. 
$s=s_1,\cdots,s_k\in\Gamma(E_{|\theta})$  \'e chamado de base local de se\c c\~oes de $E$ sobre $\theta$  se a parametriza\c c\~ao:\begin{eqnarray*} 
\psi_{\theta}^{-1}:{\theta\times{\R^k}}\rightarrow{\pi^{-1}(\theta)=E_{|\theta}}\\ 
(p,\alpha_{1},\cdots,\alpha_k)\longmapsto{\alpha_{1}e_{1}(p)+\cdots+\alpha_{k}e_{k}(p)}
\end{eqnarray*}
 \'e um difeomorfismo, isto \'e: se as se\c c\~oes $e_1, \cdots, e_k$ s\~ao l.i. em cada ponto. \'E claro que $e_i(p)=\psi_{\theta}^{-1}(p,\tilde{e}_{i})$, onde $\tilde{e}_i$ denota o i-\'esimo vetor vetor da base can\^onica de $\R^k$. 
Ent\~ao, localmente, dada uma base, uma se\c c\~ao pode ser expressa de forma un\'ivoca por combina\c c\~ao $C^\infty(M)$-linearmente em termos dessa base.

Os mapas $F^\theta=\psi_{\theta}^{-1}$ s\~ao chamados de gauges locais de E sobre $\theta$.
Suponhamos que temos dois gauges:
$F^\theta=\psi_{\theta}^{-1}$,$F^\beta=\psi_{\beta}^{-1}$, ent\~ao j\'a que 
\begin{eqnarray*} 
\psi_{\beta}^{-1}:x\times\{\R^n\}\rightarrow\pi^{-1}(x)\\
\psi_{\theta}:\pi^{-1}(x)\rightarrow{x}\times\{\R^n\}\end{eqnarray*} s\~ao ambos isomorfismos lineares, $\psi_{\theta}\circ\psi_{\beta}^{-1}:x\times\{\R^n\}\rightarrow{x\times\{\R^n\}}$ \'e isomorfismo linear. Logo n\'os temos 

\beq\label{g}\begin{array}{ll}
\psi_\theta\circ{\psi_\beta}^{-1}:\theta\cap\beta\times\R^{k}\longrightarrow\theta\cap\beta\times\R^{k}\\
(x,v)\longmapsto(x,g_{\theta\beta}(x)v)\end{array}\eeq

 Onde, para cada $x\in\theta\cap\beta$, n\'os temos $g_{\theta\beta}(x)={(F_x^{\theta})^{-1}\circ{F_x^\beta}:\R^k\rightarrow\R^k}$ \'e isomorfismo linear, isto \'e: $g_{\theta\beta}:\theta\cap\beta\rightarrow{GL(k)}$ \'e chamado de mapa de transi\c c\~ao de gauge. \'E importante notar que apesar de uma fibra sobre um dado ponto ser isomorfa \`a $\R^k$, n\~ao existe isomorfismo can\^onico, qualquer trivializa\c c\~ao local deve ser t\~ao boa quanto outra, qualquer gauge \'e igualmente apropriado.

Por consist\^encia devemos ter $g_{\alpha\alpha}=Id$.
Agora, aplicando $\psi_{\beta}\circ{\psi_{\theta}^{-1}}$ de ambos os lados de (\ref{g}), n\'os temos  \beq{(x,v)=\psi_{\beta}\circ{\psi_{\theta}^{-1}}(x,g_{\theta\beta}(x)v)=(x,g_{\beta\theta}g_{\theta\beta}(x)v)}\eeq  

Ent\~ao 
\beq{g_{\beta\theta}g_{\theta\beta}=1}\eeq 
 Como essa equa\c c\~ao \'e v\'alida pra todo $x\in M$ e $g_{\theta\beta}(x)\in{G}L(k)$, n\'os temos que
\beq{g_{\theta\beta}^{-1}=g_{\beta\theta}}\eeq
Al\'em disso \'e claro que estamos identificando os pontos em $E$
\beq\label{1}\psi_\beta^{-1}(x,v)=\psi_\theta^{-1}(x,g_{\theta\beta}(x)v)\eeq
 Logo \beq{\psi_\gamma\circ\psi_\beta^{-1}(x,v)=(x,g_{\gamma\beta}(x)v)=\psi_\gamma\circ{\psi_\theta^{-1}}(x,g_{\theta\beta}(x)v)=(x,g_{\gamma\theta}g_{\theta\beta}(x)v)}\eeq 
Ent\~ao $$g_{\gamma\beta}(x)=g_{\gamma\theta}g_{\theta\beta}(x), \forall{x\in{U_\theta\cap{U_\beta}\cap{U_\gamma}}}$$ de onde , utilizando a equa\c c\~ao (\ref{1}), tiramos a condi\c c\~ao de cociclo: \beq{g_{\gamma\theta}g_{\theta\beta}g_{\beta\gamma}=1}\eeq 

Se, para todo ponto $x\in{M}$, e quaisquer trivializa\c c\~oes $\theta$ e $\beta$, n\'os tivermos que a imagem de $g_{\theta\beta}$ est\'a contida em um subgrupo $G\subset\GL k$, dizemos que $E$ \'e um $G$-fibrado vetorial sobre $M$. Generalizando, se tivermos um automorfismo local $T\in\Gamma({\Aut(E|_\beta)})$, tal que para uma  trivializa\c c\~ao qualquer $\psi_\beta$ sobre $\beta$, n\'os tivermos 
\beq\label{triv3}\psi_\beta\circ T(\psi_\beta^{-1}(x,v))=(x,g_T(x)v)\eeq 
onde a imagem de $g_T$ esteja contida em $G\subset\GL k$, chamaremos  $T$ de transforma\c c\~ao de gauge local. \'E trivial ver que se (\ref{triv3}) vale para uma trivializa\c c\~ao local  sobre $\beta$, valer\'a tamb\'em para qualquer outra $G$-trivializa\c c\~ao sobre $\beta$: 
$$T(\psi_\beta^{-1}(x,v))=T(\psi_\beta'^{-1}(x,g_{\beta'\beta}(x)v))=\psi_\beta^{-1}(x,g_T(x)v)=\psi_\beta'^{-1}(x,g_{\beta'\beta}(x)g_T(x)v)$$ logo fazendo $v\ra(g_{\beta'\beta})^{-1}(v)$ vemos que  $g_T'(x)=g_{\beta'\beta}(x)g_T(x)(g_{\beta'\beta})^{-1}\in G\subset\GL k$.

O princ\'ipio da teoria de gauge \'e de que campos sejam se\c c\~oes de $G$-fibrados, e de que as leis da f\'isica sejam equa\c c\~oes diferenciais, tais que se $s$ for uma se\c c\~ao  solu\c c\~ao dessas equa\c c\~oes, ent\~ao $gs$ tamb\'em o deve ser, para toda transforma\c c\~ao de gauge $g$. 

POdemos, tomando outra atitude em rela\c c\~ao a fibrados vetoriais focada em sua trivialidade local, construir$E$ colando os fibrados triviais $U_\alpha\times{F}$ atrav\'es dos $g_{\alpha\beta}$ com as propriedades acima. A proje\c c\~ao $\pi$ \'e definida por 
\begin{eqnarray*}
{\pi:E\rightarrow{V}} \\ 
{[x,v]}_\alpha\mapsto{x}\end{eqnarray*} 
As propriedades sobre $g_{\alpha\beta}$ garantem que, se $x\in{U_\alpha}\cap{U_\beta}$, ent\~ao $(x,v)\in{U_\alpha}\times{F}$. Logo, se $(x,w)\in{U_\beta\times{F}}$, temos a seguinte rela\c c\~ao de equival\^encia:$$(x,v)\sim(q,w)\Leftrightarrow{x=g}~; w=g_{\alpha\beta}v$$   Isto \'e:

\begin{enumerate}
\item[(i)]{$(x,v)\sim(x,v)\Rightarrow{g_{\alpha\alpha}=1}$}
\item[(ii)]{$(x,v)\sim(x,w)\Rightarrow(x,w)\sim(x,v)\Rightarrow{g_{\alpha\beta}g_{\beta\alpha}=1}$} E finalmente
\item[(iii)]{$(x,v)\sim(x,w)\sim(x,u)\Rightarrow(x,v)\sim(x,u)\Rightarrow{g_{\alpha\gamma}=g_{\alpha\beta}g_{\beta\gamma}}\Rightarrow[x,v]_\alpha=[q,w]_\beta\Leftrightarrow{x=q},v=g_{\alpha\beta}w$} 
\end{enumerate}

E tomamos a estrutura vetorial em $\pi^{-1}(x)$ como: 
$[x,g_{\alpha\beta}U]_\beta+[x,g_{\alpha\beta}w]_\beta=[x,u]_\alpha+[x,w]_\alpha=[x,u+w]_\alpha=[x,g_{\alpha\beta}(u+w)]_\beta$ j\'a que $g_{\alpha\beta}\in\rho(G)\subset{GL(F)}$ 

Fazemos $E=(\underset{\alpha\in\Lambda}{\bigcup}U_\alpha\times{F})/_\sim$ e temos que, sobre cada $U_\alpha$,$E$ \'e trivial, isto \'e: temos o isomorfismo 
\begin{eqnarray}
\Phi_\alpha:\pi^{-1}(U_\alpha)\rightarrow{U_\alpha\times{F}}\\ 
{[x,u]}_\alpha\mapsto(x,u)
\end{eqnarray}

Dessa forma, temos se $x\in{U_\alpha}\cap{U_\beta}$, ent\~ao $\Phi_\alpha\circ\Phi_\beta^{-1}(x,u)=(x,g_{\alpha\beta}u)$ e completamos a volta.

 Se $S$ \'e alguma estrutura em $\R^k$ invariante por $G$, ent\~ao podemos passar suavemente $S$ para cada $E_x$ pelos isomorfismos $$F^i_x:\R^k\rightarrow{E_x}$$

Por exemplo
uma estrutura riemanniana em $\R^k$\'e invariante pelo grupo $O(k)$, portanto se $E$ for um $O(k)$-fibrado; podemos induzir suavemente uma estrutura riemanniana em $E$ por qualquer gauge (j\'a que a transi\c c\~ao de gauge est\'a no grupo). 

Um atlas de um $G$-fibrado \'e uma cobertura aberta $\{\theta_{\alpha}\}_{\alpha\in{A}}$ de M em conjunto com os mapas $g_{\alpha\beta}:\theta_\alpha\cap\theta_p\rightarrow{G}$ satisfazendo a condi\c c\~ao de cociclos.

Dado um $G$-fibrado $E$, ent\~ao um $G$-referencial para $E$ em $x$ \'e um isomorfismo linear dado por um gauge $F_x^{\theta}:\R^k\rightarrow{E_x}$. Dado um tal $G$-referencial $f_0$ n\'os temos que ${f=f_0\circ{g}}$ \'e tamb\'em um $G$-referencial para todo $g\in{G}$ e o mapa $g\rightarrow{f_0\circ{g}}$ \'e uma bije\c c\~ao de $G$ com o conjunto de todos os $G$-referenciais de $E$ em $x$. 

$\mbox{Aut}(E)$ \'e o grupo de automorfismos (isto \'e: isomorfismos de fibrados vetoriais entre $E$ e $E$) e se $E$ \'e um $G$-fibrado vetorial, ent\~ao $\mbox{Aut}_G(E)$ denota o sub-grupo de automorfismos de $E$ como $G$-fibrado vetorial, i.e.: tal que para $\psi\in\Gamma(\Aut(E))$, $\psi(x)=\rho(g)$, onde $g\in G$ e $\rho:G\ra \Aut(E_x)$ .

Se $E_1$ e $E_2$ forem fibrados sobre $M$, os elementos de $\Gamma(E_1\otimes{E_2})$ s\~ao gerados pelos elementos da forma $s^1\otimes{s}^2$ onde $s^1\in\Gamma(E_1)$ e $s^2\in\Gamma(E_2)$. Vejamos porque: dados dois espa\c cos vetoriais $V,~W$, definimos $V\otimes W$ como a soma bilinear de elementos  da forma\footnote{ I.e.: identificando em $V\times W$ os elementos, para  $v_1,v_2\in V$, $w_1,w_2\in W$:
\begin{itemize}
\item { $(v,w_1+w_2)= (v,w_1)+(v,w_2)$}
\item {$(v_1+v_2,w)=(v_1,w)+(v_2,w)$}
\end{itemize}}
 $(v,w)$, $v\in V~;~w\in W$. \'E f\'acil ver que dadas bases $\{e_k\}_{k=1}^{\mbox{dim}(V)}$ e  $\{b_i\}_{i=1}^{\mbox{dim}(W)}$ de $V$ e $W$,  tanto $v$ quanto $w$ tem  representa\c c\~oes \'unicas em termos destas bases (fato elementar de \'algebra linear), logo escrevendo $v$ e $w$ por extenso e utilizando a bilinearidade, cada elemento $(v,w)$ da soma se decomp\~oe em uma combina\c c\~ao linear de elementos da base $(b_i,e_k)$. Juntando os coeficientes de cada $i,k$ obtemos uma combina\c c\~ao linear de elementos $\{(b_i,e_k)\}$.  
 Portanto voltando ao caso dos fibrados vetoriais $E_1$ e $E_2$, se $\{e_i\}_{i=1}^k$ e $\{b_i\}_{i=1}^l$ s\~ao bases locais de se\c c\~oes (referenciais) de, respectivamente, $E_1$ e $E_2$  sobre $\theta$,  elas geram  univocamente as  se\c c\~oes locais sobre o fibrado produto. Obviamente n\'os temos que para $\psi\in\Aut_G(E_1\otimes E_2)$, ent\~ao, para $x\in M$, $\psi(x)=\rho(g)$ onde $\rho$ \'e uma representa\c c\~ao linear $$\rho:G\ra L\Big((E_1|_x\otimes E_2|_x)~;~(E_1|_x\otimes E_2|_x)\Big)$$ 
   
Como exemplo, vamos destrinchar o caso do fibrado $\End(E)$, que como mostraremos corresponde a $E_1=E, E_2=E^*$. Primeiramente, alguns resultados elementares de \'algebra linear :
Suponhamos  $T:V\ra W$ isomorfismo linear, como $T^*:W^*\ra V^*$, o isomorfismo linear induzido por $T$ que leva $V^*$ em $W^*$ \'e justamente $(T^{-1})^*$. Portanto, para $\lambda\in V^*$, $w\in W$, temos que $T$ age sobre $\lambda$ da  forma natural (generalizado pelo pull-back): 
$$(T^{-1})^*\lambda(w)=\lambda T^{-1}(w)\Rightarrow (T^{-1})^*\lambda=\lambda T^{-1}$$
Portanto se $\Omega=\lambda\otimes v\in V\otimes V^*$ a a\c c\~ao natural de um isomorfismo linear $T\in \Aut(V)$ \'e pela aplica\c c\~ao adjunta: $\Omega\ra T\Omega T^{-1}$. Portanto se $\rho$ for uma representa\c c\~ao de $G$ em $V$, a representa\c c\~ao  correspondente em $V\otimes V^*$ \'e pela representa\c c\~ao adjunta \beq\label{represad}\widetilde\rho(g):\Omega\ra\rho(g)\Omega(\rho(g))^{-1}\eeq 
N\'os temos ainda que existe um isomorfismo natural entre $L(V;V)$ e $V\otimes V^*$.  Afirmamos que existe um isomorfismo can\^onico $L(V;V)\simeq V\otimes V^*$, dado pela matriz resultante de aplica\c c\~ao de $A\in L(V;V)$ em uma base de $V$, i.e.: tomando base $e=\{e_i\}_{i=1}^n$ de $V$, sua dual $\{e^i\}_{i=1}^n=e^*$, base de $V^*$, escrevemos 
\beq\label{transmatriz} A=A^i_je_i\otimes e^j \mbox{~~onde~~}A^i_j=e^i(A(e_j))\eeq
 Para $T\in \Aut(V)$, tomando as bases induzidas naturalmente por $T$,   $\tilde{e}=\{\tilde{e}_i\}_{i=1}^n=\{Te_i\}_{i=1}^n$ e  $\tilde{e^*}=\{\tilde{e}^i\}_{i=1}^n=\{e^iT^{-1}\}_{i=1}^n$, em $V$ e $V^*$ respectivamente, obtemos que o isomorfismo independe de base simplesmente aplicando a defini\c c\~ao (\ref{transmatriz}) para as duas bases e utilizando (\ref{represad}) . 

Portanto induzimos um isomorfismo natural  independente de base $\End(V)\simeq V\otimes V^*$ (significando  que a a\c c\~ao de uma transforma\c c\~ao linear independe da base em que \'e representada), ou seja, $\End(E)\simeq{E\otimes{E^*}}$. O que significa que uma se\c c\~ao $s\in\Gamma(\End(E))$ sob um isomorfismo de fibrados $g:M\ra\Aut(E)$ sofre a seguinte transforma\c c\~ao: $s(x)\ra g(x)s(x)g^{-1}(x)\in\End(E_x)$. Ou seja, nesse caso as transforma\c c\~oes de gauge agem pela a\c c\~ao adjunta dos automorfismos, na nota\c c\~ao mais completa, se $\rho_x:G\ra\Aut(E_x)$ \'e a representa\c c\~ao usual do grupo sobre $E_x$, n\'os temos a nova representa\c c\~ao correspondente: $\widetilde\rho_x(g(x))=\Ad(\rho(g(x)))$ .

\section{Linear Differential Operators}

Definimos $\alpha=(\alpha_1,\cdot,\alpha_n)\in(\Z^+)^n$ e $|\alpha|=\alpha_1+\cdots+\alpha_n$. Definimos ainda:
$$D^\alpha:=\partial_\alpha=\frac{\partial^{|\alpha|}}{\partial{x}_1^{\alpha_1}\cdots\partial{x}_n^{\alpha_n}}:C^\infty(\R^n;\R^k)\rightarrow{C}^\infty(\R^n;\R^k)$$ 

Uma aplica\c c\~ao linear $L:C^\infty(\R^n;\R^k)\rightarrow{C^\infty(\R^n;\R^l)}$ \'e chamada de operador diferencial de ordem menor ou igual a $r$ se \'e da forma:

\beq\label{dif}{(Lf)(x)=\sum_{|\alpha|\leq{r}}a_\alpha(x)(D^\alpha{f})(x)}\eeq onde $a_\alpha\in{C}^\infty(\R^n;L(\R^k,\R^l))$, $f\in{C}^\infty(\R^n;\R^k)$ e  $D^\alpha{f}(x)\in\R^k$. Se $L$ tem ordem menor ou igual a $r$ e n\~ao tem ordem menor ou igual a $r-1$, $L$ \'e dito pertencer a Diff$~^r(\R^n\times\R^k;\R^n\times\R^l)$.

Sejam ent\~ao $(E,\pi_E,M)$ e $(F,\pi_F,M)$ fibrados vetoriais sobre M. Definimos que $L:\Gamma(E)\rightarrow\Gamma(F)$ \'e um operador diferencial linear entre $E$ e $F$ de ordem no m\'aximo igual a $r$ se \'e um morfismo entre $E$ e $F$ que  pode localmente ( i.e.: no do dom\'inio de gauges)  ser representado na forma de (\ref{dif}). N\'os precisamos mostrar que essa defini\c c\~ao independe das trivializa\c c\~oes. Em primeiro lugar, precisamos escrever o que significa um operador diferencial ser representado localmente na forma de (\ref{dif}). 
 
Sejam ent\~ao $\theta$ aberto em $M$, tal que $\pi_{_E}^{-1}(\theta)$ e  $\pi_{_F}^{-1}(\theta)$ s\~ao dom\'inios de  trivializa\c c\~oes de gauge $\phi:\pi_{_E}^{-1}(\theta)\rightarrow\theta\times\R^k$ e $\psi:\pi_{_F}^{-1}(\theta)\rightarrow\theta\times\R^l$. Chamemos de $pr_2^\phi:\theta\times\R^k\rightarrow\R^k$ e $pr_2^\psi:\theta\times\R^l\rightarrow\R^l$ as proje\c c\~oes can\^onicas nas fibras t\'ipicas e $a_\alpha\in{C}^\infty(\theta;L(\R^k,\R^l))$.Ent\~ao $L:\Gamma(E_1)\rightarrow\Gamma(E_2)$ \'e localmente representado na forma de (\ref{dif}) se, para toda se\c c\~ao $s\in\Gamma(E_1)$ e todo $m\in\theta$: 

\beq\label{diff}{ pr_2^\psi\circ\psi\circ{L(s(m))}=\sum_{|\alpha|\leq{r}}a_\alpha(m)D^\alpha(pr_2^\phi\circ\phi(s))(m)}\eeq  
Como para $m\in\theta$, n\'os temos $\psi_m:=pr_2^\psi\circ\psi_{|\pi_{_F}^{-1}(m)}:\pi_{_F}^{-1}(m)\ra\R^l$ isomorfismo linear (e $\phi_m$ definido analogamente), 
 
\beq\label{diff2}L(s)(m)=\psi_m^{-1}\left(\sum_{|\alpha|\leq{r}}a_\alpha(m)D^\alpha(pr_2^\phi\circ\phi(s))(m)\right)\eeq
Sejam agora, definidos da mesma forma, trivializa\c c\~oes locais $\widetilde\psi$ e $\widetilde\phi$. Temos ent\~ao 
 
\beq\label{diff3}L(s)(m)=\widetilde\psi_m^{-1}\left(\sum_{|\alpha|\leq{r}}(\widetilde\psi_m\circ\psi_m^{-1})\circ{a}_\alpha(m)D^\alpha(pr_2^\phi\circ\phi(s))(m)\right)\eeq 
Mas por (\ref{g}), $(\widetilde\psi_m\circ\psi_m^{-1})={g_{\widetilde\psi\psi}(m)}$ 
Agora, chamando de $g_{\phi\widetilde{\phi}}:\theta\ra{GL(k)}$ o mapa de transi\c c\~ao entre as trivializa\c c\~oes $\phi$ e $\widetilde\phi$,  n\'os temos, para qualquer $m\in\theta$:

\begin{eqnarray*}
pr_2^\phi\circ\phi(s(m))&=&(pr_2^\phi\circ\phi)(\widetilde\phi^{-1}\circ\widetilde\phi)(s(m))\\
~&=&pr_2^\phi\left(\phi\circ\widetilde\phi^{-1}\right)(m,pr_2^{\widetilde\phi}\circ\widetilde\phi(s(m)))\\
&=& pr_2^\phi\left(m,g_{\phi\widetilde{\phi}}pr_2^{\widetilde\phi}\circ\widetilde\phi(s(m))\right)\\
&=&g_{\phi\widetilde{\phi}}(m)pr_2^{\widetilde\phi}\circ\widetilde\phi(s(m)) 
\end{eqnarray*}
 podemos reescrever (\ref{diff3}):
 
\beq\label{diff4}L(s)(m)=\widetilde\psi_m^{-1}\left(\sum_{|\alpha|\leq{r}}{g_{\widetilde\psi\psi}(m)}\circ{a}_\alpha(m)D^\alpha\left(g_{\phi\widetilde{\phi}}(pr_2^{\widetilde\phi}\circ\widetilde\phi(s))\right)(m)\right)\eeq 
Utilizando regra da cadeia : 

$$L(s)(m)=\widetilde\psi_m^{-1}\left(\sum_{|\alpha|\leq{r}}{g_{\widetilde\psi\psi}(m)}\circ{a}_\alpha(m)\sum_{|\gamma|\leq |\alpha|}\binom{|\alpha|}{|\gamma|}D^{\alpha-\gamma}(g_{\phi\widetilde{\phi}})(m)D^\gamma(pr_2^{\widetilde\phi}\circ\widetilde\phi(s))(m)\right)$$ onde a nota\c c\~ao simplificada subsume que  $\gamma$ \'e tal que $\gamma^i\leq\alpha^i$.
Mas $g_{\phi\widetilde{\phi}}\in{C}^\infty(\theta;L(\R^k,\R^k))\simeq{C}^\infty(\theta;\R^{k^2})$, portanto $D^{\alpha-\gamma}(g_{\phi\widetilde{\phi}})\in{C}^\infty(\theta;L(\R^k,\R^k))$ . Finalmente

$$L(s)(m)=\widetilde\psi_m^{-1}\left(\sum_{|\gamma|\leq{r}}\left(\sum_{r\geq|\alpha|\geq|\gamma|}{g_{\widetilde\psi\psi}(m)}\circ{a}_\alpha(m)\binom{|\alpha|}{|\gamma|}D^{\alpha-\gamma}(g_{\phi\widetilde{\phi}})(m)\right)D^\gamma(pr_2^{\widetilde\phi}\circ\widetilde\phi(s))(m)\right)$$
 como ${g_{\widetilde\psi\psi}}\in{C}^\infty(\theta;L(\R^l,\R^l))$ n\'os temos que    $$\left(\sum_{r\geq|\alpha|\geq|\gamma|}{g_{\widetilde\psi\psi}(m)}\circ{a}_\alpha(m)\binom{|\alpha|}{|\gamma|}D^{\alpha-\gamma}(g_{\phi\widetilde{\phi}})(m)\right)=:\widetilde{a}_\gamma\in{C}^\infty(\theta;L(\R^k,\R^l))$$ logo $\widetilde{a}_\gamma\in{C}^\infty(\theta;L(\R^k,\R^l))$ e portanto verificamos que vale 
 
\beq{L}(s)(m)=\widetilde\psi_m^{-1}\left(\sum_{|\gamma|\leq{r}}\widetilde{a}_\gamma(m)D^\gamma(pr_2^{\widetilde\phi}\circ\widetilde\phi(s))(m)\right)\eeq 

\subsubsection{ Formal Adjoints for Differential Operators}

Sejam $E,F$ fibrados vetoriais riemannianos sobre $M$.
O subespa\c co de $\Gamma(E)$ composto por se\c c\~oes de suporte compacto ser\'a denotado por $\Gamma_C(E)$. Se $s_1,s_2\in\Gamma_C(E)$, ent\~ao \'e claro que $x\rightarrow\langle{s_1(x),s_2(x)}\rangle$ tem suporte compacto. Logo, utilizando o produto interno pontual em $E$  $\langle\cdot,\cdot\rangle$,  definimos o produto interno em $\Gamma_C(E)$, denotado $\langle\langle\cdot,\cdot{\rangle\rangle}$, por 
$$\langle\langle{s}_1,s_2{\rangle\rangle}=\int_M \langle{s_1,s_2}\rangle\mu$$ 
\'E trivial mostrar que  \'e bilinear e positivo definido, j\'a que $\langle\cdot,\cdot\rangle$ o \'e, e pela suavidade das se\c c\~oes elas n\~ao podem ser n\~ao nulas em um conjunto de medida zero. 
 Agora, se $L:\Gamma(E)\ra\Gamma(F)$ \'e tal que $L\in\mbox{Diff}^r(E,F)$, ent\~ao chamamos de adjunto formal de $L$, o operador diferencial linear $L^*:\Gamma(F)\ra\Gamma(E)$, tal que valha, para todos $s_1\in\Gamma_C(E)$ e $s_2\in\Gamma_C(F)$: 
$$\langle\langle{L}(s_1),s_2\rangle\rangle=\int_M\langle{L(s_1),s_2}\rangle_{_F}\mu=\int_M\langle{s_1,L^*(s_2)}\rangle_{_E}\mu={\pl}s_1,L^*(s_2)\rangle\rangle$$ \'E claro que se tal operador adjunto existir, pela n\~ao degeneresc\^encia do produto interno acima, ele ser\'a \'unico. 

Como vimos, para toda se\c c\~ao $s_1\in\Gamma(E)$ e para $m\in\theta$, por (\ref{diff2}) $L:\Gamma(E)\rightarrow\Gamma(F)$ pode ser escrito como : 

$$L(s_1)(m)=\psi_m^{-1}\left(\sum_{|\alpha|\leq{r}}a_\alpha(m)D^\alpha(pr_2^\phi\circ\phi(s_1))(m)\right)$$
Como $\psi_m$ \'e isometria ,para $u,v\in\R^l$:
 $$\langle{u},v\rangle_{\R^l}=\langle{\psi^{-1}_m(u)},\psi^{-1}_m(v)\rangle_{_F}$$Agora, como \'e isometria, o adjunto de $\psi^{-1}_m$ \'e igual a   $\psi_m$ e n\'os obtemos: 

\begin{eqnarray*}\langle{L(s_1)(m),s_2(m)}\rangle_{_F}&=&\langle{\psi_m^{-1}\sum_{|\alpha|\leq{r}}a_\alpha(m)D^\alpha(pr_2^\phi\circ\phi(s_1))(m),s_2(m)})\rangle_{_F}\\
&=&\langle\sum_{|\alpha|\leq{r}}a_\alpha(m)D^\alpha(pr_2^\phi\circ\phi(s_1))(m),(\psi_m\circ{s}_2(m)\rangle_{\R^l} 
\end{eqnarray*}
Como $a_\alpha(m)\in{L(\R^k;\R^l)}$, podemos tomar tamb\'em o seu adjunto $a_\alpha(m)^*$, utilizando ent\~ao a bilinearidade da m\'etrica obtemos:

$$\langle{L(s_1)(m),s_2(m)}\rangle_{_F}=\sum_{|\alpha|\leq{r}}\langle{D}^\alpha(pr_2^\phi\circ\phi(s_1))(m),{a}_\alpha(m)^*\circ\psi_m(s_2(m))\rangle_{\R^k}$$
Integrando sobre $\theta$ e utilizando integra\c c\~ao por partes sucessivamente obtemos: 

$$\int_\theta\langle{L(s_1),s_2}\rangle_{_F}\mu=\sum_{|\alpha|\leq{r}}\int_\theta(-1)^{|\alpha|}\langle\phi_m(s_1(m)),{D}^\alpha\left({a}_\alpha^*({pr}_2^\psi\circ\psi)\circ{s_2}\right)(m)\rangle_{\R^k}\mu$$
 E finalmente: 

$$\int_\theta\langle{L(s_1),s_2}\rangle_{_F}\mu=\int_\theta\langle{s_1(m)},\sum_{|\alpha|\leq{r}}(-1)^{|\alpha|}\phi_m^{-1}{D}^\alpha\left({a}_\alpha^*({pr}_2^\psi\circ\psi)\circ{s_2}\right)(m)\rangle_{\R^k}\mu$$
Onde o operador ao lado direito \'e claramente um morfismo entre $F$ e $E$. Portanto provamos que localmente existe um adjunto formal. Se tivermos dois adjuntos formais sobre $\theta$ , $L^*_{|\theta}~ ,~ \widetilde{L}^*_{|\theta}$ ent\~ao claramente 
$$\frac{1}{2}\left(L^*_{|\theta}+\widetilde{L}^*_{|\theta}\right)$$
\'e tamb\'em um adjunto formal, ou seja, combina\c c\~oes lineares convexas de adjuntos formais locais s\~ao adjuntos formais locais. Portanto, como $L(s)_{|U}=L(s_{|U})$, tomando uma parti\c c\~ao da unidade subordinada a uma cobertura de $M$ por abertos que sejam dom\'inios de trivializa\c c\~oes locais obtemos um adjunto formal global, que como mencionamos \'e \'unico. Calcularemos explicitamente alguns adjuntos formais ao longo da exposi\c c\~ao.
 
\subsubsection{Vector Bundle Valued Differential Forms}
Se $V$ \'e um espa\c co vetorial denotamos por $\Lambda^p(V)$, todas as aplica\c c\~oes anti-sim\'etricas p-lineares de $V$ em $\R$. Se $W$ \'e um espa\c co vetorial, ent\~ao $\Lambda^p(V)\otimes{W}$ denota o espa\c co das formas a valores em $W$, e \'e gerado linearmente por elementos da forma $\eta\otimes{w}$ onde $\eta\in\Lambda^p(V)$ e $w\in{W}$. Sejam  
ent\~ao $v_1,\cdots,v_p\in{V}$, ent\~ao n\'os temos que 
$$\eta\otimes{w}(v_1,\cdots,v_p):=\eta(v_1,\cdots,v_p)w$$
que \'e alternante p-linear. 

Se $E$ \'e um fibrado sobre $M$ chamamos $\Lambda^p(TM^*)\otimes{E}$ de fibrado de p-formas em $M$ a valores em $E$. Notemos que se $\lambda\in\Gamma(\Lambda^p(TM^*)\otimes{E})$, ent\~ao para todo $x\in{M}$, $\lambda_x\in\Lambda^p(T_xM)\otimes{E_x}$  \'e um mapa alternante p-linear de $T_xM$ em $E_x$. Agora, $\Gamma(\Lambda^p(TM^*)\otimes{E})$ tamb\'em \'e gerado $C^\infty(M,\R)$-linearmente por elementos da forma $\eta\otimes{s}$, onde $\eta\in\Gamma\Big(\Lambda^p(TM^*)\Big)$ 
e $s\in\Gamma(E)$, logo se $X_1,\cdots,X_p\in\Gamma(TM)$, n\'os temos que 
 
\beq\eta\otimes{s}(X_1,\cdots,X_p)=\eta(X_1,\cdots,X_p)s\eeq
e como $$\eta(X_1,\cdots,X_p)\in{C^\infty(M)}\Rightarrow\eta\otimes{s}(X_1,\cdots,X_p)\in\Gamma(E)$$ 
Isto \'e, se $\lambda\in\Gamma(\Lambda^p(TM^*)\otimes{E})$, a aplica\c c\~ao:

\begin{eqnarray}
\lambda(X_1,\cdots,X_p):M\rightarrow{E}\\
x\mapsto\lambda_x\Big((X_1)_x,\cdots,(X_p)_x\Big)\end{eqnarray}
\'e uma se\c c\~ao de $E$, p-linear e anti-sim\'etrica nos $X_1,\cdots,X_p$ (j\'a que a \'e em cada ponto). 

Agora sejam $\eta^1\in\Gamma(\Lambda^{p_1}(TM^*)\otimes{E_1})$ e $\lambda^2\in\Gamma(\Lambda^{p_2}(TM^*)\otimes{E_2})$,
definimos 

$$\lambda^1\tilde\wedge\lambda^2\in\Gamma(\Lambda^{p_1+p_2}(TM^*)\otimes{E_1\otimes}E_2)$$
 por 

\beq\label{ext3}\lambda^1\tilde\wedge\lambda^2(X_1,\cdots,X_{p_1+p_2})=\frac{p_1!p_2!}{(p_1+p_2)!}\sum_{\sigma\in{P(p_1+p_2)}}\tau(\sigma)\lambda^1(X_{\sigma(1)},\cdots,X_{\sigma(p_1)})\otimes\lambda^2(X_{\sigma(p_1+1)},\cdots,X_{\sigma(p_1+p_2)})\eeq 

Onde $P(p_1+p_2)$ \'e o grupo de permuta\c c\~oes de $p_1+p_2$ elementos, e $\tau(\sigma)$ \'e a paridade da permuta\c c\~ao $\sigma$.
 Em outras palavras, o operador $\tilde\wedge$ age como produto externo s\'o na parte de formas da se\c c\~ao. Se chamarmos o produto externo usual de 
$$\wedge_\R=\Gamma\Big(\Lambda^{p_1}(TM^*)\Big)\times\Gamma\Big(\Lambda^{p_2}(TM^*)\Big)\rightarrow\Gamma(\Lambda^{p_1+p_2}(TM^*)\Big)$$ temos, para  elementos da forma $\eta^{1}\otimes{s^{1}}$ onde $\eta^{1}\in\Gamma(\Lambda^{p_1}(TM^*))$ e $s^{1}\in\Gamma(E_1)$ que a equa\c c\~ao (\ref{ext3}) fica: 
\beq\label{wedge10}(\eta^{1}\otimes{s^{1}})\tilde\wedge(\eta^{2}\otimes{s^{2}})=(\eta^{1}\wedge_\R\eta^{2})\otimes({s^{1}}\otimes{s^{2}})\eeq

Mais rigorosamente, seja $\{\eta^i\}_{i=1}^n$ base local de $\Gamma(\Lambda^1(TM^*))$ e $\{e_j\}_{i=1}^k$  base local de $\Gamma({E_1}{|_\theta})$. Lembramos que ${E_1}{|_\theta}:=\pi_1^{-1}(\theta)$ onde $\pi_1:E_1\rightarrow{M}$. N\'os temos que $\{\eta^I\}$ \'e base local de $\Gamma\Big(\Lambda^{p_1}(TM^*{|_\theta})\Big)$ , onde o superscrito mai\'usculo $I$ \'e a chamada nota\c c\~ao de multi-\'indices, que denota uma combina\c c\~ao de $p_1$ elementos da forma $\eta^i$. I.e.: se $I=(i_1,\cdots,i_{p_1})$ com $1\leq{i_1}<\cdots<i_{p_1}\leq{n}$ ent\~ao $\eta^I=\eta^{i_1}\wedge\cdots\wedge{\eta}^{i_{p_1}}$.

Logo $\Gamma(\Lambda^{p_1}(TM^*)\otimes{E_1})$  \'e localmente gerado $C^\infty(M,\R)$-linearmente pela base $\{\eta^I\otimes{e_j}\}$ de forma \'unica.  Seja  $J=(j_1,\cdots,j_{p_2})$, e $\{b_j\}_{i=1}^l$ base local de $\Gamma({E_2}{|_\theta})$, temos ent\~ao, para $\lambda^1\in\Gamma(\Lambda^{p_1}(TM^*{|_\theta})\otimes{{E_1}{|_\theta}})$ e $\lambda^2\in\Gamma(\Lambda^{p_2}(TM^*{|_\theta})\otimes{E_2{{|_\theta}}})$, $\lambda^1=f_I^i\eta^I\otimes{e_i}$, $\lambda^2=g_J^j\eta^J\otimes{b_j}$ , $f_I^i,g_J^j:\theta\rightarrow\R$. Portanto, obtemos :   
\beq{\lambda^1\tilde\wedge\lambda^2=f_I^ig_J^j(\eta^I\wedge_\R\eta^J)\otimes({e_i}\otimes{b_j})}\eeq 

 No caso de $p_1=p_2=1$ ent\~ao 

\beq\label{ext}\lambda^1\tilde\wedge\lambda^2(X_1,X_2)=\frac{1}{2}\Big(\lambda^1(X_1)\otimes\lambda^2(X_2)-\lambda^1(X_2)\otimes\lambda^2(X_1)\Big)\in\Gamma(E_1)\otimes\Gamma(E_2)\eeq

Onde $X_1,X_2$ s\~ao se\c c\~oes de $TM$. Se tivermos no entanto uma aplica\c c\~ao bilinear $\mu:E_1\otimes{E_2}\rightarrow{E_3}$, ao inv\'es de $\tilde\wedge$, substituindo o produto tensorial em (\ref{ext}) por $\mu$, podemos definir um produto externo $\wedge$, que vai de formas a valores em $E_1$ e $E_2$, respectivamente, em $\Gamma(\Lambda^{p_1+p_2}(TM^*)\otimes{E_3})$. I.e.:
$$
\wedge: \Gamma(\Lambda^{p_1}(TM^*)\otimes{E_1})\otimes\Gamma(\Lambda^{p_2}(TM^*)\otimes{E_2})\rightarrow\Gamma(\Lambda^{p_1+p_2}(TM^*)\otimes{E_3})$$
E para o caso de 1-formas: \beq\label{formaspi}\lambda^1\wedge\lambda^2(X_1,X_2)=\frac{1}{2}\Bigg(\mu\Big(\lambda^1(X_1),\lambda^2(X_2)\Big)-\mu\Big(\lambda^1(X_2),\lambda^2(X_1)\Big)\Bigg) 
\eeq

Por exemplo, podemos ter $E_1=E$ e $E_2=E^*$ com $\mu=C$ sendo o operador de contra\c c\~ao e nesse caso $E_3=M\times\R$. Ou ainda, $E_1=E_2=E$ e $\mu=g$ uma m\'etrica sobre $E$ (j\'a que cada fibra \'e linear e portanto comporta produto interno), novamente com $E_3=M\times\R$. Outro exemplo bastante \'util \'e se $E_1=M\times\R$, ou seja, $p$-formas a valores reais, neste caso $\mu$ \'e simplesmente a multiplica\c c\~ao por fun\c c\~oes reais. 

Mas o caso mais importante para n\'os \'e se $E_1=E_2=\End(E)$, o grupo dos endomorfismos de $E$.  Nesse caso temos uma aplica\c c\~ao natural de composi\c c\~ao de endomorfismos $\End(E)\times\End(E)\rightarrow{\End(E)}$.

Em termos de  uma base $\{s^1_k\}=\{s^2_k\}=\{e^i\otimes{e_j}\}$ onde  $\{e_i\}$ \'e base local de $\Gamma({E}{|_\theta})$ e $\{e^i\}$ \'e sua base dual, base de $\Gamma({E^*}{|_\theta})$,  para $\lambda^1\in\Gamma\Big(\Lambda^p(TM^*{|_\theta})\otimes\End({E}){|_\theta}\Big)$,  
$\lambda^1={\lambda^1}_j^i\otimes{e_i\otimes}e^j$ onde ${\lambda^1}_j^i\in\Gamma\Big(\Lambda^p(TM^*{|_\theta})\Big)$ podemos ilustrar a opera\c c\~ao acima como uma simples contra\c c\~ao:

\beq\label{contrac}\lambda^1\wedge\lambda^2={\lambda^1}_j^i\wedge_\R{\lambda^2}_k^le_i\otimes\Big(e^j(e_l)\Big)\otimes{e^k}={\lambda^1}_j^i\wedge_\R{\lambda^2}_k^je_i\otimes{e^k}\eeq
Fazemos aqui a importante observa\c c\~ao que, salvo aviso, tomaremos sempre este produto exterior entre formas a valores em $\End(E)$.

Agora, quando $E_1=E_2=E$, se o fibrado for um fibrado de \'algebras, existe uma aplica\c c\~ao $E\otimes{E}\rightarrow{E}$.
 Com essa aplica\c c\~ao podemos novamente definir um produto externo a valores em $E$   e (\ref{ext}) pode tomar a forma: 

\beq\label{alg}\lambda^1\tilde\wedge\lambda^2(X_1,X_2)=\frac{1}{2}\Big(\lambda^1(X_1)\lambda^2(X_2)-\lambda^1(X_2)\lambda^2(X_1)\Big)\eeq
  Pela forma de (\ref{alg}), a nota\c c\~ao para tal aplica\c c\~ao no fibrado se sugere como: 
\beq\lambda^1\tilde\wedge\lambda^2(X_1,X_2)=[\lambda^1(X_1),\lambda^2(X_2)]=[\lambda^1,\lambda^2](X_1,X_2)\eeq
Em particular, se for um fibrado de \`algebras anti-comutativas: 
$$\lambda(X_1)\lambda(X_2)=-\lambda(X_2)\lambda(X_1)\Longrightarrow\lambda\wedge\lambda=\lambda\otimes\lambda$$ 
Portanto fica aqui claro que n\~ao temos, como no caso da formas a valores reais (que \'e um fibrado de \`algebras comutativas), que 
$\lambda\wedge\lambda=0$. Em geral, temos localmente, em termos de uma base $\{e_i\}$ de $\Lg$, $\lambda=\lambda^ie_i$, denotando o produto da \'algebra por $[\cdot, \cdot]$: 
\begin{eqnarray}
\label{omex}
\lambda\wedge\lambda(X_1,X_2)&=&\frac{1}{2}\left(\lambda(X_1)\lambda(X_2)-\lambda(X_2)\lambda(X_1)\right)\\
~&=&\lambda^i(X_1)\lambda^j(X_2)e_ie_j-\lambda^j(X_2)\lambda^i(X_1)e_je_i\\
~&=&\lambda^i(X_1)\lambda^j(X_2)[e_i,e_j] 
\end{eqnarray}

\subsubsection{The Exterior Derivative}

Seja $E=M\times{V}$, neste caso (j\'a que podemos manter se\c c\~oes de $\Gamma(E)$ ``constantes''), teremos um operador diferencial de primeira ordem:
\begin{eqnarray*}
d:\Gamma(\Lambda^p(M)\otimes{E})\rightarrow\Gamma(\Lambda^{p+1}\otimes{E})\\ 
\lambda\longmapsto{d\lambda}\phantom{aaaaaaaaaa}\end{eqnarray*} Onde para, $X_1,\cdots,X_{p+1}\in\Gamma(TM)$,
\begin{multline}\label{derivext}
d\lambda(X_1,\cdots,X_{p+1})=\sum_{i=1}^{p+1}(-1)^{i+1}X_i\lambda(X_1,\cdots,\hat{X}_i,\cdots,X_{p+1})+\\ 
\sum_{1\leq{i\le}j\leq{p+1}}(-1)^{i+j}\lambda([X_i,X_j],X_1,\cdots,\hat{X}_i,\cdots,\hat{X}_j,\cdots,X_{p+1})
\end{multline} 
Que \'e exatamente an\'alogo \`a defini\c c\~ao da derivada exterior de  $p$-formas a valores reais, e generaliza $$d\lambda(X,Y)=X[\lambda(Y)]-Y[\lambda(X)]-\lambda([X,Y])$$ 
Mostremos que $d\lambda$ \'e $C^\infty(M)$-multilinear, como a defini\c c\~ao \'e obviamente anti-sim\'etrica, basta que provemos em uma entrada.  Seja ent\~ao $f\in{C}^\infty(M)$:
\begin{eqnarray*}
d\lambda(fX_1,\cdots,X_{p+1})&=&fX_1[\lambda(X_2,\cdots,X_{p+1})]\\
~&+&\sum_{i=2}^{p+1}(-1)^{i+1}X_i[\lambda(fX_1,\cdots,\hat{X}_i,\cdots,X_{p+1})]\\ 
~&+&\sum_{i=2}^{p+1}(-1)^{i+1}\lambda([fX_1,X_i],\cdots,\hat{X}_i,\cdots,X_{p+1})\\
~&+&\sum_{2\leq{i\le}j\leq{p+1}}(-1)^{i+j}\lambda([X_i,X_j],fX_1,\cdots,\hat{X}_i,\cdots,\hat{X}_j,\cdots,X_{p+1})
\end{eqnarray*}
Mas $\lambda$ \'e multilinear, ent\~ao 
\begin{eqnarray*}
\sum_{i=2}^{p+1}(-1)^{i+1}X_i[\lambda(fX_1,\cdots,\hat{X}_i,\cdots,X_{p+1})]&=&
\sum_{i=2}^{p+1}(-1)^{i+1}X_i[f]\lambda(X_1,\cdots,\hat{X}_i,\cdots,X_{p+1})\\
~&+&
f\sum_{i=2}^{p+1}(-1)^{i+1}X_i[\lambda(X_1,\cdots,\hat{X}_i,\cdots,X_{p+1})] 
\end{eqnarray*}
Al\'em disso, como $[fX_1,X_i]=f[X_1,X_i]-X_i[f]X_1$,
\begin{gather*}
\sum_{i=2}^{p+1}(-1)^{i+1}\lambda([fX_1,X_i],\cdots,\hat{X}_i,\cdots,X_{p+1})\\
=f\sum_{i=2}^{p+1}(-1)^{i+1}\lambda([X_1,X_i],\cdots,\hat{X}_i,\cdots,X_{p+1})- 
\sum_{i=2}^{p+1}(-1)^{i+1}X_i[f]\lambda(X_1,\cdots,\hat{X}_i,\cdots,X_{p+1})
\end{gather*}
 Juntando todos os termos obtemos 
\beq{d\lambda(fX_1,\cdots,X_{p+1})=fd\lambda(X_1,\cdots,X_{p+1})}\eeq
Logo $d\lambda$ \'e tensorial, s\'o depende dos valores dos campos nos pontos calculados.Temos as seguintes propriedades, id\^enticas \`aquelas v\'alidas para formas a valores reais: 
\begin{enumerate}
\item[(i)]{$d$ \'e linear}
\item[(ii)]{$d(\lambda^1\wedge\lambda^2)=d\lambda^1\wedge\lambda^2+(-1)^{p_1}\lambda^1\wedge{d\lambda^2}$ ,para $\lambda^i\in\Gamma(\Lambda^{p_i}(TM^*)\otimes{E}))$.}
\item[(iii)]{$d^2=0$} 
\item[(iv)]{Se $\lambda$ tem suporte compacto $C$ ent\~ao $$\int_Cd\lambda=\int_{\partial{C}}\lambda$$ (Teorema de Stokes).}
\end{enumerate}

Bem coloquialmente,  o que \'e  a derivada exterior? \'E a potencialidade de varia\c c\~ao. A potencialidade de varia\c c\~ao de uma fun\c c\~ao 
por exemplo, se toma ao longo de dire\c c\~oes, e a de uma fun\c c\~ao sobre dire\c c\~oes (1-formas) se toma ao longo de elementos de \'area direcionados\footnote{ As caracter\'isticas Grassmanianas de 2-formas v\^em de serem relacionadas n\~ao a um elemento de \'area qualquer, e sim a um elemento de \'area direcionado. } \ Agora, porque a potencialidade de varia\c c\~ao {\it{da}} potencialidade de varia\c c\~ao \'e nula? Porque a potencialidade de varia\c c\~ao n\~ao \'e uma quantia escalar, e sua dire\c c\~ao em pontos (ou elementos de \'area,  etc.) vizinhos \'e oposta \cite{Feynman}. Pense em uma distribui\c c\~ao de temperatura, se voc\^e vai de $A$ para $B$ a temperatura aumenta tanto quanto diminui se voc\^e toma a dire\c c\~ao oposta, de $B$ pra $A$. Portanto a soma das potencialidades se cancela, deixando somente o valor na fronteira da regi\~ao tomada (Teorema de Stokes). Como a fronteira de uma fronteira \'e nula, a integral $$\int_Sd^2\eta=0$$ para qualquer $p$-forma $\eta$ e qualquer variedade $S$, logo $d^2=0$.

\section{Hodge Decomposition Theorem}
\subsubsection{The Hodge $*$ Operator}

Suponhamos que $M^n$ tenha estrutura Riemanniana $(M,\langle\cdot,\cdot\rangle)$, h\'a uma maneira natural de induzir um isomorfismo entre $T_xM$ e $T_xM^*$ dado pela m\'etrica, a saber, dado $\lambda\in{T_xM^*}$ e $u,v\in{T_xM}$ definimos  ${u}^\sharp\in{T_xM^*}$ por $\langle{u},\cdot\rangle$: 
$${u}^\sharp(v)=\langle{u,v\rangle}$$ assumindo que a m\'etrica \'e n\~ao degenerada, e que o espa\c co dual tem a mesma dimens\~ao, \'e f\'acil verificar que $\sharp:{T_xM}\ra{T_xM^*}$ \'e isomorfismo: se $u,v\in{T_xM}$ tal que $\langle{u},w\rangle=\langle{v},w\rangle$ para todo $w\in{T_xM}$, ent\~ao 
$$\langle{u-v},w\rangle=0\Longrightarrow{u-v}=0$$
 Denotamos o inverso de $\sharp$ por $\flat:{T_xM^*}\ra{T_xM}$.  Podemos vizualizar este isomorfismo da seguinte maneira: dado um vetor $v\in{T_xM}$ , $v^\sharp$ seria representado por uma ``pilha'' de hiperplanos em $T_xM$ ortogonais a $v$, de forma que $v^\sharp(w)$ fosse a velocidade com que o vetor $w$ atravessa os hiperplanos. Dessa maneira, se $w$ \'e ortogonal a $v$, $v^\sharp(w)=0$. Claramente precisamos da m\'etrica para nos dizer o que \'e ortogonalidade. 

O produto exterior de duas 1-formas, seguindo esse racioc\'inio, seria uma fam\'ilia de vetores (elementos de linha), dadas pela intersec\c c\~oes dos seus respectivos hiperplanos, e sua aplica\c c\~ao feita em elementos de \'area (correspondentes a dois vetores), seria a velocidade com a qual esses elementos de linha  atravessam esses elementos de \'area \cite{Penrose}.
Em dimens\~oes mais altas a vizualiza\c c\~ao se torna totalmente abstrata, mas, seguindo esse racioc\'inio, uma $p$-forma em um ponto $x\in{M}$  seria equivalente a uma pilha de subespa\c cos $n-p$ dimensionais em $T_xM$, e sua aplica\c c\~ao a $p$ vetores ordenados seria a velocidade com que a pilha \'e ``atravessada" pelos subespa\c cos orientados $p$ dimensionais formados por esses vetores.  

O produto interno em $T_xM$ induz um produto interno em  $\Lambda^{p}(T_xM)$. Como o produto interno deve ser uma aplica\c c\~ao bilinear, basta definirmos tal aplica\c c\~ao em elementos formados pelo produto externo de $p$ 1-formas. Seja ent\~ao $\lambda_i, \theta_i\in\Lambda^1(T_xM)$, denotaremos o produto interno em $\Lambda^{p}(T_xM)$ por $\leqslant\cdot,\dots,\cdot\geqslant$: 

\begin{gather}
\label{priformas}\leqslant{\lambda_{1}}\wedge{\cdots}\wedge{\lambda}_p,\theta_{1}\wedge\cdots\wedge\theta_{p}\geqslant:=\sum_{\sigma\in{S_n}}\tau(\sigma)\langle\lambda_{\sigma(1)},\theta_1\rangle\cdots\langle\lambda_{\sigma(p)},\theta_p\rangle =\mbox{det}(\langle\lambda_i,\theta_j\rangle)
\end{gather}
onde definimos  $\langle\lambda,\theta\rangle:=\langle\lambda^\flat,\theta^\flat\rangle$ , $S_p$ denota o grupo de permuta\c c\~ao de $p$ elementos, $\tau(\sigma)$ \'e a paridade da permuta\c c\~ao $\sigma$.  Se $\{e^i\}_{i=1}^n$ \'e uma base ortonormal para $T^*_xM$, ent\~ao os $\binom{n}{p}$ elementos formados por $e^I$, onde o superscrito mai\'usculo $I$ \'e um multi-\'indice de $p$ elementos\footnote{ I.e.: se $I=(i_1,\cdots,i_{p})$ com $1\leq{i_1}<\cdots<i_{p}\leq{n}$ ent\~ao $\lambda^I=\lambda^{i_1}\wedge\cdots\wedge{\lambda}^{i_{p}}$.}, forma uma base ortonormal para $\Lambda^p(T_xM)$.

Logo $\Lambda^n(T_xM)$ \'e 1-dimensional e tem dois elementos de norma 1. Se pudermos escolher $\mu\in\Gamma(\Lambda^n(TM^*))$ tal que $||\mu_x||=1$, ent\~ao $M$ \'e orient\'avel e uma escolha \'e chamada de orienta\c c\~ao de $M$, $\mu$ \'e o  elemento de volume Riemanniano. 

Consideremos o mapa bilinear:
\begin{gather*}
{B_p}:\Gamma(\Lambda^p(TM^*))\times\Gamma(\Lambda^{n-p}(TM^*))\rightarrow{C^\infty(M,\R)}\\
\phantom{aaaaaaaaaaaB_r:}(\lambda,\nu)\longmapsto{B}_p(\lambda,\nu)\mu=\lambda\wedge\nu 
\end{gather*} 

\begin{prop}$B_p$ \'e n\~ao-degenerado e portanto determina unicamente um isomorfismo $*$: $\Gamma(\Lambda^p(TM^*))\rightarrow\Gamma(\Lambda^{n-p}(TM^*))$ tal que: 
\beq\label{star}\lambda\wedge*\nu=\leqslant\lambda,\nu\geqslant\mu\eeq 
 onde $\lambda,\nu\in\Lambda^p(M),\mu\in\Lambda^n(M)$.
\end{prop}
{\bf{Dem:}}
 Seja $\{e_i\}_{i=1}^n$ base ortonormal de $T_xM$, e $I=(i_1,\cdots,i_p)$ com $1\leq{i_1}<\cdots<i_p\leq{n}$, chamamos de $I^C$ o complementar de $I$ em $(1,2,\cdots,n)$ em ordem tamb\'em crescente. Chamaremos novamente de $\tau(I)$ a paridade da permuta\c c\~ao levando $(1,2,\cdots,n)$ em $(i_1,i_2,\cdots,i_n)$: 
$$\binom{1~ 2~ \cdots~ p~ p+1~ \cdots~ n}
{i_1~ i_2~ \cdots~ i_p~ j_1~ \cdots~ j_{n-p}}$$ obviamente
$e_I\wedge{e_{I^c}}=\tau(I)\mu$.
 N\'os temos ainda que para  qualquer $J$ subconjunto crescente de $n-p$ elementos de $\{1,\cdots,n\}$ tal que $J\neq{I^c}$,   necessariamente ${e_I\wedge{e_J}}=0$. Logo se $I\neq{J}$ ent\~ao $e_{I^C}\neq{e}_{J^C}$ e portanto $\leqslant{e_{I^C},e_{J^C}}\geqslant=0$.  

Chamando de ~$\CC$~ a cole\c c\~ao de todos os subconjuntos crescentes de $p$-elementos  de $\{1,\cdots,n\}$, n\'os temos ent\~ao que   $\{e_I\}_{I\in{C}}$ e $\{\tau(I)e_{I^c}\}_{I\in{C}}$ s\~ao bases ortonormais de $\Lambda^p(T_xM)$ e $\Lambda^{n-p}(T_xM)$ (j\'a que t\^em a mesma dimens\~ao). 
Agora, como $*$ deve satisfazer: $$e_I\wedge*e_I=\leqslant{e_I,e_I}\geqslant\mu=\mu=\tau(I)\tau(I)\mu=e_I\wedge\tau(I)e_{I^C}$$ podemos definir o operador linear $*$ como agindo em uma base da seguinte forma $$*e_I:=\tau(I)e_{I^C}$$  
Como leva base em base, o operador $*$ \'e isomorfismo linear, que por constru\c c\~ao obedece (\ref{star}),  como $B_p$ \'e n\~ao degenerado, o operador est\'abem definido e \'e \'unico. Claramente se $e_I=\mu$, ent\~ao $*e_I=1$.
$~~~~~~\blacksquare$ \medskip 
 
Aqui nossa maneira de visualizar formas como elementos de \'area direcionados vem a calhar. Todo subespa\c co de um espa\c co vetorial tem um subespa\c co ortogonal, mas somente subespa\c cos direcionados (com orienta\c c\~ao) t\^em subespa\c cos ortogonais direcionados, \'e da\'i que vem todas as caracater\'isticas do operador de Hodge. Similarmente ao operador $\sharp$ podemos encarar o operador $*$ de Hodge como levando cada elemento de \'area $p$-dimensional orientado ao elemento de \'area $(n-p)$-dimensional ortogonal de orienta\c c\~ao compat\'ivel. Por exemplo, em $\R^3$ com a m\'etrica can\^onica: 
$$*dx=dy\wedge{dz}~,~~*dy=dz\wedge{dx}~,~~*dz=dx\wedge{dy}$$

Continuando, $*e_I=\tau(I)e_{I^c}$ ent\~ao
$$*(*e_I)=\tau(I)\tau(I^c)e_I=\tau(I\cdot{I^c})e_I$$ que \'e a permuta\c c\~ao:
$$\binom{i_1~ i_2~ \cdots~ i_p~ j_1~ \cdots~ j_{n-p}}{j_1~\cdots~j_{n-p}~i_1~\cdots~i_p}$$ cuja paridade \'e $(-1)^{p(n-p)}$. 
Em particular se $p=\frac{n}{2}$ e $n=4m\Rightarrow{p(n-p)}=4m^2$ ent\~ao $(-1)^{p(n-p)}=1$.

 Portanto, $(*)^2=Id$, e como $*:\Lambda^p(T_xM):\ra\Lambda^p(T_xM)$ logo temos, para $\lambda\in\Lambda^p(T_xM)$,
\begin{eqnarray}
\lambda&=&\frac{1}{2}\left(\lambda_++\lambda_-\right)\\
~&=&\frac{1}{2}\left((\lambda+*\lambda)+(\lambda-*\lambda)\right)\\ 
\label{autodecomp}\therefore~~~*\lambda&=&\frac{1}{2}\left(*\lambda_++*\lambda_-\right)\\
~&=&\frac{1}{2}\left((*\lambda+\lambda)+(*\lambda-\lambda)\right)\\
~&=&\frac{1}{2}\left(\lambda_+-\lambda_-\right)\end{eqnarray}
Ent\~ao denotando o autoespa\c co do autovalor $k$ do operador linear $*$ no espa\c co em quest\~ao como $A(k)$, n\'os obtemos $\Lambda^p(T_xM)=A(-1)\oplus{A}(1)$, decomposi\c c\~ao que \'e importante no estudo das equa\c c\~oes de Yang-Mills.  

Agora, para $E$  fibrado vetorial Riemanniano, 
  utilizando o produto externo nestes espa\c cos que incorpora o produto interno riemanniano nas fibras (que denotaremos nesse caso por $g$), como explicitado na equa\c c\~ao (\ref{formaspi}), se $\lambda\in\Lambda^p(T_xM)\otimes{E_x}$ e $\nu\in\Lambda^{n-p}(T_xM)\otimes{E_x}$ pela equa\c c\~ao (\ref{ext3}) definimos este produto externo como: 
\begin{gather*}\lambda\wedge\nu(X_1,\cdots,X_{n})
:=\frac{p!(n-p)!}{n!}\sum_{\sigma\in{P(n)}}\tau(\sigma)g\Big(\lambda(X_{\sigma(1)},\cdots,X_{\sigma(p)}),\nu(X_{\sigma(p+1)},\cdots,X_{\sigma(n)})\Big)\end{gather*} 

Ent\~ao assim como para formas a valores reais, incorporando o produto interno riemanniano obtemos uma forma bilinear n\~ao degenerada de $(\Lambda^p(T_xM)\otimes{E_x})\otimes(\Lambda^{n-p}(T_xM)\otimes{E_x})\ra\R$:
$$\lambda\wedge\nu=B_p(\lambda,\nu)\mu\in\Lambda^{n}(T_xM)$$
Seguindo a demonstra\c c\~ao do lema anterior, obtemos um \'unico isomorfismo $*:\Lambda^p(M)\otimes{E}\rightarrow\Lambda^{n-p}(M)\otimes{E}$ caracterizado por 
$$\lambda\wedge*\nu=\langle\lambda,\nu\rangle{\mu}$$ a saber, o isomorfismo levando base ortonormal de $\Lambda^p(T_xM)\otimes{E_x}$ em base ortonormal de $\Lambda^{n-p}(T_xM)\otimes{E_x}$: 
\beq\label{hodgeprod}*(e_I\otimes{b}_j)=\tau(I)e_{I^c}\otimes{b}_j\eeq
onde $\lambda\in\Lambda^p(M)\otimes{E}$ e o produto interno, para o qual utilizamos a mesma nota\c c\~ao do produto interno de 1-formas, aqui incorpora tanto o produto interno pontual para $p$-formas, $\leqslant\cdot,\cdot\geqslant$, quanto o produto interno pontual riemanniano.

\subsubsection{ Exterior Co-derivative}
Pela defini\c c\~ao, se $\lambda,\nu\in\Gamma_C(\Lambda^p(M)\otimes{E})$
$$\langle\langle\lambda,{\nu}\rangle\rangle=\int_M\langle\lambda,\nu\rangle\mu=\int_M\lambda\wedge*\nu$$ 

Vamos calcular explicitamente o adjunto formal da derivada exterior $d:\Gamma(\Lambda^p(TM^*)\otimes{E})\ra\Gamma(\Lambda^{p+1}(TM^*)\otimes{E})$, que chamaremos de $\delta:\Gamma(\Lambda^{p+1}(TM^*)\otimes{E})\ra\Gamma(\Lambda^p(TM^*)\otimes{E})$, a coderivada exterior. Lembramos antes de mais nada que s\'o existe um conceito natural de derivada exterior sobre {\it{fibrados produto}} $E=M\times{V}$ (ou localmente para a trivializa\c c\~ao $\theta\times{V}$), j\'a que a\'i h\'a uma maneira natural de manter um campo``fixo". 

Seja $\lambda\in\Gamma_C(\Lambda^p(TM^*)\otimes{E})$ e $\nu\in\Gamma_C(\Lambda^{p+1}(TM^*)\otimes{E})$. Lembramos que $*\nu\in\Gamma_C(\Lambda^{n-p-1}(TM^*)\otimes{E})$ e portanto $\lambda\wedge*\nu\in\Gamma_C(\Lambda^{n-1}(TM^*))$ e $d(\lambda\wedge*\nu)\in\Gamma_C(\Lambda^{n}(TM^*))$. Portanto podemos utilizar  Stokes: 
\begin{eqnarray}
\int_Md(\lambda\wedge*\nu)&=&\int_Md\lambda\wedge*\nu+\int_M(-1)^p\lambda\wedge{d(*\nu)}\\
~&=&\int_{\partial{M}}\lambda\wedge*\nu=0\\
\label{integral}\therefore~~~~\int_Md\lambda\wedge*\nu&=&\int_M(-1)^{(p+1)}\lambda\wedge{d(*\nu)}\end{eqnarray} 
 Mas pela defini\c c\~ao de $*$, sobre cada ponto de $M$, $d\lambda\wedge*\nu=\langle{d\lambda,\nu\rangle}\mu$, e por outro lado, como $**=(-1)^{p(n-p)}$ ent\~ao $(-1)^{p(n-p)}**=1$, e n\'os temos 
\begin{gather*}(-1)^{p+1}\lambda\wedge{d(*\nu)}=(-1)^{p+1}\lambda\wedge((-1)^{p(n-p)}**{d(*\nu)})=(-1)^{p+p(n-p)+1}\lambda\wedge*(*{d(*\nu)})\\ 
\therefore~~~~~~~~~~(-1)^{p+1}\lambda\wedge{d(*\nu)}=(-1)^{p+p(n-p)+1}\langle\lambda,*{d(*\nu)}\rangle\mu\end{gather*}
Subsituindo em (\ref{integral}), obtemos
$$\int_M\langle{d\lambda,\nu\rangle}\mu=(-1)^{p+p(n-p)+1}\int_M\langle\lambda,*{d(*\nu)}\rangle\mu$$ 
Ent\~ao $${\langle\langle}d\lambda,\nu{\rangle\rangle}=(-1)^{p+p(n-p)+1}{\langle\langle}\lambda,*d*\nu{\rangle\rangle}$$ e portanto 
$$\delta_{p+1}=(-1)^{np+1}*_{n-p}d_{n-(p+1)}*_{p+1}$$ onde utilizamos que $-1^{p^2-p}=1$ e os subscritos denotam o grau das formas a que os operadores est\~ao sendo aplicados; a coderivada exterior est\'a sendo aplicada em $p+1$-formas a valores no fibrado e as levando para $p$ formas a valores no fibrado. Ent\~ao finalmente obtemos para $\delta:\Gamma(\Lambda^{p}(TM^*)\otimes{E})\ra\Gamma(\Lambda^{p-1}(TM^*)\otimes{E})$ 
\beq\delta_{p}=(-1)^{n(p+1)+1}*_{n-p+1}d_{n-p}*_p\eeq
Temos as seguintes propriedades para a coderivada exterior, facilmente verific\'aveis (aplicaremos em formas de grau $p$):
\begin{enumerate}
\item[(i)]{$\delta\circ\delta=d*\delta=\delta*d=0$ 

Como $d$ e $*$ s\~ao lineares, $\delta\circ\delta=\pm{*d\circ{d}*}=\pm{d*}\delta=\pm\delta*d=0$.}
\item[(ii)]{$*d\delta=\delta{d}*$~,~$*\delta{d}=d\delta*$

Simplesmente escrevendo por extenso os dois lados das equa\c c\~oes obtemos os resultados.}
\item[(iii)]{$\delta_{n-p}*_p=(-1)^{p+1}*d$

N\'os temos $\delta_{n-p}*_p=(-1)^{n(n-p+1)+1+p(n-p)}*_{p+1}d_p$,  e fazendo as contas obtemos \\~~~~~~~~~~~~$(-1)^{n(n-p+1)+1+p(n-p)}=(-1)^{p+1}$ }

\item[(iv)]{$*_{p-1}\delta_p=(-1)^pd*$

J\'a que $(-1)^{(p-1)(n-p+1)+n(p+1)+1}=(-1)^p$}
\end{enumerate}

\subsubsection{The Laplacian}
Em matem\'atica e f\'isica, o  Laplaciano, denotado por $\Delta$, \'e um operador diferencial  de suma import\^ancia, sendo utilizado na modelagem de propaga\c c\~ao de ondas e fluxo de calor. \'E ainda central na teoria eletromagn\'etica e na mec\^anica qu\^antica, onde representa o operador de energia cin\'etica. Definido como o divergente do gradiente, em coordenadas cartesianas de $\R^3$ o operador assume a bem conhecida f\'ormula: 
$$\widetilde\Delta=\frac{\partial^2}{\partial{x}^2}+\frac{\partial^2}{\partial{y}^2}+\frac{\partial^2}{\partial{z}^2}=\sum_{i=1}^3\partial_i^2$$
\'E poss\'ivel provar que em uma variedade Riemmaniana $M$ qualquer, podemos escrever o operador de Laplace acima, a menos de um sinal negativo, como 
\footnote{Na nota\c c\~ao mais comum entre os f\'isicos, o negativo do divergente do gradiente em uma variedade riemanniana $M$ aplicado em $f\in{C}^\infty(M,\R)$~ \'e escrito como:
 $$ \Delta(f)=-\widetilde\Delta(f)=-\frac{1}{\sqrt{|g|}}\, \partial_i(\sqrt{|g|}\,\partial^i f)$$ 
  utilizindo a nota\c c\~ao f\'isica:~$|g|=\mbox{det}(g_{ij})$~ e  $g^{ij}g_{jk}=\delta^i_k$ \'e a matriz inversa da m\'etrica, e $\partial^i=g^{ij}\partial_j$ \'e o levantamento do campo $\partial_i$ pela m\'etrica, i.e.: $\partial_i^\sharp$. Utilizando novamente a nota\c c\~ao de multi-\'indices para letras mai\'usculas, $\mu$ como a forma volume  temos: 
  \begin{gather*}\Delta f={d}\delta f + \delta\,{d}f = 
\delta\, {d}f =\delta \, \partial_i f \, dx^i =- *{d}{*\partial_i f \, {d}x^i} = - *{d}(\tau({i J})  \sqrt{|g|}\partial^i f \, {d}x^J)\\ 
 =- *\tau({i J}) \, \partial_j(\sqrt{|g|}\partial^i f)\, {d} x^j \, {d}x^J =- * \frac{1}{\sqrt{|g|}} \, \partial_i (\sqrt{|g|}\,\partial^i f) \mu = -\frac{1}{\sqrt{|g|}}\, \partial_i (\sqrt{|g|}\,\partial^i f)\end{gather*} } 
 
\beq\label{Laplace}\Delta_p=d_{p-1}\delta_p+\delta_{p+1}d_p\eeq
ou em nota\c c\~ao mais compacta: $\Delta=d\delta+\delta{d}$, chamaremos este operador de Laplaciano, ao inv\'es de $\widetilde\Delta$, prefer\^encia justificada por ser assim um operador positivo definido, como mostramos a seguir. \'E trivial perceber que adjunto do Laplaciano, $\Delta^*=\delta d+d\delta=\Delta$, ou seja, \'e um operador auto-adjunto. O n\'ucleo de $\Delta$ \'e chamado de espa\c co de $p$-formas   harm\^onicas a valores em $V$. 
\begin{prop}Temos as seguintes propriedades do Laplaciano:
\begin{enumerate}
\item[(i)]{O laplaciano \'e auto-adjunto.}
\item[(ii)]{ Para $\lambda\in\Gamma_C(\Lambda^p(T^*M)\otimes{E})$, $\lambda$ \'e harm\^onico se e somente se tivermos ambas as condi\c c\~oes: $d\lambda=0$ {\it{e}} $\delta\lambda=0$.} 
\item[(iii)]{$*\Delta=\Delta*$. Logo se $\lambda$ \'e harm\^onica ent\~ao $*\lambda$ tamb\'em \'e.}
\end{enumerate}
\end{prop}
{\bf{Dem:}}
 J\'a comentamos o primeiro item. Para o segundo, supondo que $\lambda$ \'e harm\^onica temos: 
\begin{eqnarray*}
d(\delta\lambda)+\delta(d\lambda)=0&\Longrightarrow&{\langle\langle}d(\delta\lambda)+\delta(d\lambda),\lambda\rangle\rangle=0\\
~&=&\pl{d(\delta\lambda),\lambda\pr}+\pl\delta(d\lambda),\lambda\pr\\
~&=&\pl\delta\lambda,\delta\lambda\pr+\pl{d\lambda},d\lambda\pr\\ 
~&\therefore&~~~~~~ \delta\lambda=0~~\mbox{e}~~d\lambda=0
\end{eqnarray*}
onde utilizamos na \'ultima passagem que o produto interno $\pl\cdot,\cdot\pr$ \'e definido positivo. A afirma\c c\~ao de (ii) \'e \'obvia. Fica claro tamb\'em dessa demonstra\c c\~ao que o Laplaciano \'e assim positivo definido. 

Para o terceiro item, basta observar o item (ii) das propriedades da coderivada exterior acima. ~~~~$\blacksquare$\medskip

De agora em diante, chamaremos  $\Gamma_C(\Lambda^p(T^*M)\otimes{E})$ de $\mathcal{A}^p(E)$ ou abreviando ainda mais, $\A^p$,  e o espa\c co de $p$-formas harm\^onicas em $\mathcal{A}^p$ de $\mathcal{H}^p$. Um corol\'ario trivial desta \'ultima proposi\c c\~ao \'e o Teorema de Liouville, que diz que se $M$ \'e compacto, orientado e conexo, ent\~ao qualquer fun\c c\~ao harm\^onica, i.e.: tal que $\Delta f=0$, \'e constante (j\'a que $df=0$). Temos ainda que se $\lambda$ for uma $n$-forma harm\^onica, ent\~ao $\lambda$ \'e um m\'ultiplo constante da forma volume, j\'a que $\lambda=f\mu$ e portanto $*\Delta\lambda=\Delta*\lambda=\Delta f=df=0$. 
 
\begin{prop} $\mathcal{H}^p~,~\mbox{Im}(d_{p-1})~\mbox{e}~\mbox{Im}(\delta_{p+1})$ s\~ao mutuamente ortogonais em $\mathcal{A}^p$. \end{prop} 
{\bf{Dem:}} Seja $\lambda\in\mathcal{A}^{p-1}$ , $\nu\in\mathcal{A}^{p+1}$ e $\eta\in\mbox{H}^p$. 
\begin{itemize}
\item{$\mbox{Im}(d_{p-1})\perp \mbox{Im}(\delta_{p+1})$: 
$$\pl d\lambda,\delta\nu\pr=\pl\lambda,\delta(\delta\nu)\pr=0$$}
\item{$\mbox{Im}(d_{p-1})\perp\mbox{H}^p$ :
$$\pl d\lambda,\eta\pr=\pl\lambda,\delta\eta\pr=0$$} 
\item{$\mbox{Im}(\delta_{p+1})\perp\mbox{H}^p$ :
$$\pl\eta,\delta\nu\pr=\pl d\eta,\nu\pr=0$$}
\end{itemize}
~~~~~~~~~~~$\blacksquare$\medskip
 
\subsubsection{Hodge's Decomposition Theorem}

\begin{theo}\label{hodge5}Para $M$ compacto,  e $E=M\times V$ fibrado riemanniano sobre $M$. Ent\~ao $\mathcal{A}^{p}=\mathcal{H}^p\oplus\mbox{Im}(d_{p-1})\oplus\mbox{Im}(\delta_{p+1})$ \end{theo}

N\'os temos que $\mathcal{H}^p\oplus\mbox{Im}(d_{p-1})\oplus\mbox{Im}(\delta_{p+1})$ \'e uma soma n\~ao s\'o direta, mas perpendicular, contida em $\mathcal{A}^{p}$. Infelizmente $\A^p$ \'e de dimens\~ao infinita, e uma prova desse teorema involve uma incurs\~ao em an\'alise funcional e cohomologia de Rham que n\~ao farermos aqui (ver \cite{Warner}, \cite{Morita}). Se $\A^p$ fosse de dimens\~ao finita seria suficiente provar que um elemento de  $\mathcal{A}^{p}$ ortogonal a $\mathcal{H}^p\oplus\mbox{Im}(d_{p-1})\oplus\mbox{Im}(\delta_{p+1})$ \'e obrigatoriamente nulo. Isto \'e: 
se $\lambda\in\mathcal{A}^{p}$ \'e ortogonal a $\mbox{Im}(d_{p-1})$, n\'os temos para todo $\nu\in\mathcal{A}^{p-1}$,
$$\pl d\nu,\lambda\pr=\pl\nu,\delta\lambda\pr=0$$ portanto, como tomamos $\nu$ qualquer, $\delta\lambda=0$. Da mesma forma obtemos que se $\lambda$ \'e ortogonal a $\mbox{Im}(\delta_{p+1})$ ent\~ao $d\lambda=0$. Agora, claramente se $\lambda$ for ortogonal a $\mbox{Im}(d_{p-1})\oplus\mbox{Im}(\delta_{p+1})$ ent\~ao $\lambda\in\Ker(d_{p-1})\oplus\Ker(\delta_{p+1})=\mathcal{H}^p$, e claramente se for ortogonal aos tr\^es, $\lambda=0$.  
 Agora  assumiremos  que para todo $\lambda$ ortogonal a $\mathcal{H}^p$ existe  $\nu\in\mathcal{A}^{p}$  que satisfaz  a equa\c c\~ao: $$\Delta\nu=\lambda$$ um fato advindo da teoria de equa\c c\~oes diferenciais parciais el\'ipticas \cite{Warner}. 

Como o n\'ucleo de qualquer operador linear \'e um subespa\c co fechado, e intersec\c c\~oes arbitr\'arias de conjuntos fechados \'e fechada, temos que $\mathcal{H}^p$ \'e fechado em um espa\c co  normado, $\A^p$ . Logo dado um elemento $\nu\in\A^p$,   temos que existem muitos $h\in\mathcal{H}^p$ e $h^\perp\in\A^p-\mathcal{H}^p$ tais que $h^\perp=\nu-h$, mas \'unicos tais que a norma de $h^\perp=\nu-h$ \'e m\'inima, ou seja temos uma decomposi\c c\~ao ortogonal $\A^p=(\mathcal{H}^p)^\perp\oplus\mathcal{H}^p$. Chamaremos a proje\c c\~ao em $\mathcal{H}$ de $\widehat{\mathcal{H}}$.   

Para todo $\eta\in\A^p$ n\'os teremos que $\eta-\widehat{\mathcal{H}}(\eta)\in(\mathcal{H}^p)^\perp$ e portanto por hip\'otese existe $\nu\in\A^p$ tal que 
$$\Delta\nu=d(\delta\nu)+\delta(d\nu)=\eta-\widehat{\mathcal{H}}(\eta)$$  e finalmente 
$\eta=d(\delta\nu)+\delta(d\nu)+\widehat{\mathcal{H}}(\eta)$ o que nos fornece a decomposi\c c\~ao de Hodge. 
\begin{prop}
Se $\lambda\in\A^p$ \'e fechada, i.e.: $d\lambda=0$, ent\~ao existe um \'unico $h\in\mathcal{H}^p$ e um \'unico  $\nu\in\A^{p-1}$ tal que $\lambda=h+d\nu$. \end{prop} 
{\bf{Dem:}} Pela decomposi\c c\~ao de Hodge n\'os temos
$$\lambda=h+d\nu+\delta\eta$$
Mas como $d\lambda=0$, obtemos $d\delta\eta=0$. Logo 
$$\pl d\delta\eta,\eta\pr=\pl\delta\eta,\delta\eta\pr=0\Longrightarrow \delta\eta=0$$ 
e obtemos $\lambda=h+d\nu  $ , temos que cada classe de cohomologia cont\'em um \'unico representante harm\^onico. ~~~~~~~$\blacksquare$\medskip

Na verdade a rec\'iproca tamb\'em vale; o resultado do {\bf Teo.\ref{hodge5}} fornece uma resposta \`a seguinte pergunta: dada uma $p$-forma  $\lambda\in\mathcal{A}^{p}$ em um fibrado trivial Riemanniano sobre uma variedade compacta  $M$, 
sob quais condi\c c\~oes existe  $\nu\in\mathcal{A}^{p}$  que satisfaz  a equa\c c\~ao:
$$\Delta\nu=\lambda~~?$$
A resposta \'e que a condi\c c\~ao necess\'aria e suficiente \'e $\lambda$ ser ortogonal a $\mathcal{H}^p$. Que \'e necess\'aria  \'e facilmente demonstr\'avel: suponha que $\Delta\nu=\lambda$, ent\~ao para todo $\eta\in\mathcal{H}^p$ temos: 
$$\pl\eta,\lambda\pr=\pl\eta,\Delta\nu\pr=\pl\Delta\eta,\nu\pr=0$$
A sufici\^encia deriva do Teorema de Decomposi\c c\~ao de Hodge:
Assumindo que $\lambda\in(\mathcal{H}^p)^\perp$ n\'os temos que $\lambda=d\nu+\delta\eta$. Agora afirmamos que existem $\theta, \alpha\in\A^p$ tal que $\Delta\theta=d\nu$ e $\Delta\alpha=\delta\eta$. De fato, sucessivamente aplicando o teorema de Hodge obtemos: 
$$\nu=d\nu_1+\delta\eta_1+\gamma_1~\Longrightarrow d\nu=d\delta\eta_1$$
$$\eta_1=d\nu_2+\delta\eta_2+\gamma_2~\Longrightarrow \delta\eta_1=\delta d\nu_2$$
portanto substituindo uma na outra: 
$d\nu=d\delta(d\nu_2)=\Delta(d\nu_2)=\Delta\theta$, onde o $\theta$ que procur\'avamos \'e dado por $d\nu_2$.  E portanto, fazendo a mesma conta para $\eta$ obtemos, para algum $\alpha$, $\delta\eta=\Delta\alpha$ e finalmente $\lambda=\Delta(\alpha+\theta)$.

\section{Connections in Vector Bundles}
N\'os vimos que no caso de fibrados localmente triviais, ao redor de qualquer ponto da base h\'a uma vizinhan\c ca sobre cujas fibras existe o conceito de  uma  se\c c\~ao se manter"constante". Poder\'iamos  escolher comparar vetores segundo essa trivializa\c c\~ao, n\'os ter\'iamos um `` pano de fundo" local em cada fibra segundo o qual  poder\'iamos dizer se um campo variou ou permaneceu constante. Ainda assim, esse `` pano de fundo'' depende da trivializa\c c\~ao. Uma se\c c\~ao de um fibrado designa para cada ponto da {\it{base}} um elemento da fibra sobre aquele ponto, e n\~ao existe forma can\^onica de compara\c c\~ao entre elementos de fibras diferentes, h\'a muitas formas distintas de se fazer isto. Para comparar elementos de diferentes fibras  n\'os precisamos de um isomorfismo entre estas fibras, precisamos {\it{escolher}} um pano de fundo, uma forma de compara\c c\~ao. No caso de uma trivializa\c c\~ao, isto equivale a se utilizar da ``estrutura produto" local e o isomorfismo natural das coordenadas do espa\c co produto para estabelecer-se uma equival\^encia entre as fibras. A escolha de uma forma de compara\c c\~ao de valores entre diferentes espa\c cos internos chama-se conex\~ao. 

\begin{defi} {Uma conex\~ao em um fibrado vetorial $E$ sobre $M$ \'e um mapa linear 

$$\nabla: \Gamma(E)\rightarrow\Gamma(T^*M\otimes{E})$$ 

tal que se $f\in{C^\infty}(M,\R)$ e se $s\in\Gamma(E)$ ent\~ao 

\beq\label{nabla}{\nabla(fs)=f\nabla{s}+ df\otimes{s}}\eeq}\end{defi}

\begin{theo}{Qualquer que seja ${E}$ fibrado vetorial sobre M, existe uma conex\~ao em $E$.}\end{theo}
{\bf{Dem:}} Dividiremos nossa demonstra\c c\~ao em quatro partes:

\begin{enumerate}

\item[(i)]{Se $\phi:E_1\rightarrow{E_2}$ \'e um isomorfismo de fibrados vetoriais, 
i.e.: $\pi_1\circ\phi=\pi_2$ onde $\pi_1:E_1\rightarrow{M}~,~\pi_2:E_2\rightarrow{M}$ e 

$$\phi_{|\pi_1^{-1}(x)}:\pi_1^{-1}(x)\rightarrow\pi_2^{-1}(x)$$ 
\'e isomorfismo linear. Ent\~ao seja 

$$_1\nabla:\Gamma(E_1)\rightarrow\Gamma(T^*M\otimes{E}_1)$$ 
uma conex\~ao em $E_1$, e seja $$\tilde\phi:T^*M\otimes{E_1}\rightarrow{T}^*M\otimes{E_1}$$ dada por $\tilde\phi=Id\otimes\phi$ 
(claramente $C^\infty$ bilinear). 

Definimos ent\~ao 
\begin{eqnarray*}
{_2\nabla}:\Gamma(E_2)\rightarrow\Gamma(T^*M\otimes{E}_2)\\
s\mapsto\tilde\phi\circ\left(_1\nabla(\phi^{-1}(s))\right)\end{eqnarray*}

Agora seja $f\in{C}^\infty(M,\R)$ e $s\in\Gamma(E_2)$ ent\~ao temos 

$$_2\nabla(fs)=\tilde\phi\circ_1\nabla(\phi^{-1}(fs))=\tilde\phi(_1\nabla{f}\phi^{-1}(s))$$ 
$$=\tilde\phi\left((df\otimes\phi^{-1}(s))+f(_1\nabla(\phi^{-1}(s))\right)=df\otimes{s}+f\tilde\phi\circ_1\nabla(\phi^{-1}(s))$$ 
$$=df\otimes{s}+f(_2\nabla{s})$$

Logo $_2\nabla$ \'e uma conex\~ao em $\Gamma(E_2)$ induzida por $\phi$.}

\item[(ii)]{Agora se $E=M\times{V}\Rightarrow\Gamma(E)=C^\infty(M,V)$ e se $s\in\Gamma(E)\Rightarrow{s=s^ie_i}$ onde $s^i\in{C}^\infty(M,\R)$
e n\'os temos que se $f\in{C}^\infty(M,\R)\Rightarrow{d}(s^ie_i)=ds^i\otimes{e}_i$ \'e a conex\~ao flat de $E:d(fs^ie_i)=df\otimes{s}^ie_i+fds^i\otimes{e_i}$ 

\item[(iii)]{Pelos itens (i) e (ii) n\'os temos que um gauge $F:\theta\times\R^k\rightarrow{E_{|\theta}}$ define uma conex\~ao $\nabla^F$ para $E_{|\theta}$.}

\item[(iv)]{Agora seja $\{\theta_\alpha\}$ 
uma cobertura de M localmente finita (e pequena suficiente para que cada $\theta_\alpha$ seja um dom\'inio de uma carta de gauge) e $\{\phi_\alpha\}$ uma parti\c c\~ao de unidade a ela subordinada. 
Chamamos de $\nabla^\alpha$ a conex\~ao flat induzida pelo gauge em $E{|\theta_\alpha}$. 

Seja $\nabla:\Gamma(E)\rightarrow\Gamma(T^*M\otimes{E})$ dada por $\nabla=\sum_\alpha\phi_\alpha\nabla^\alpha$, ent\~ao, para $f\in{C}^\infty(M,\R)$ e $s\in\Gamma(E)$
n\'os temos 
\begin{eqnarray*}
\nabla(fs)=\sum_\alpha\phi_\alpha\nabla^\alpha(fs)=\sum_\alpha\phi_\alpha(df\otimes{s}+f\nabla^\alpha{s})\\ 
=df\otimes{s}+f\sum_\alpha\phi_\alpha\nabla^\alpha{s}=df\otimes{s}+f\nabla{s}\\
~~~~~~~~~~~~~~~~~~~~~~~~\blacksquare\end{eqnarray*}

}}

\end{enumerate}

\smallskip
Agora definimos 
$$\nabla_X:\Gamma(E)\longrightarrow\Gamma(E)$$ 
$$\phantom{\nabla_XX:}s\longmapsto\nabla{s}(X)$$ 
Como $\nabla$ \'e linear, $\nabla_X$ tamb\'em o \'e. Podemos ver isso facilmente localmente, j\'a que 
$$\nabla{s}=a_i^je^i\otimes{s_j}\Rightarrow\nabla{s}(fX)=f\nabla{s}(X)$$ 
Resumindo n\'os temos as seguintes propriedades 
\begin{itemize}
\item{$\nabla_X\in{D}iff^1(E,E)$ e \'e linear}
\item{$X\longrightarrow\nabla_X$ \'e linear} 
\item{Por (\ref{nabla}) e a defini\c c\~ao de $\nabla_X$, temos $\nabla_Xfs=X[f]s+f\nabla_Xs$} 
\end{itemize}

\subsubsection{Curvature of a Connection}

Seja $E=\theta\times{V}$ ,$\nabla$ a conex\~ao trivial vinda deste gauge, i.e.: $\nabla^\phi=d$. Se n\'os n\~ao conhecermos a trivializa\c c\~ao espec\'ifica de antem\~ao, existe alguma forma de detectarmos que existe um gauge para o qual $\nabla$ tem a forma acima? Seja $f\in\Gamma(E)\simeq{C}^\infty(M,V)$ e $X,Y\in\Gamma(TM)$ 
$$\Rightarrow\nabla_Xf=X[f]\Rightarrow[\nabla_X,\nabla_Y]f=[X,Y]f=\nabla_{[X,Y]}f$$ 
$$\therefore[\nabla_X,\nabla_Y]=\nabla_{[X,Y]}$$

I.e.: $X\longrightarrow\nabla_X$ \'e um homomorfismo de \'algebras de Lie entre $\Gamma(TM)$ em $\mbox{Diff}(E,E)$, que \'e a condi\c c\~ao pela qual definiremos uma conex\~ao flat, ou plana. Em geral, este n\~ao ser\'a o caso, o que sugere que estudemos o mapa: 
\begin{eqnarray}
\Omega:\Gamma(TM)\times\Gamma(TM)\rightarrow{\mbox{Diff}}^0(E,E)\\
\label{curv}(X,Y)\longmapsto[\nabla_X,\nabla_Y]-\nabla_{[X,Y]}\end{eqnarray} 
que mede o quanto o mapa $X\longrightarrow\nabla_X$ falha em ser homomorfismo de \'algebras de Lie. Por defini\c c\~ao, o comutador de elementos de uma \'algebra de Lie, pertence a pr\'opria \'algebra de Lie. A falha do comutador de uma distribui\c c\~ao em pertencer a distribui\c c\~ao \'e uma medida da sua falta de integrabilidade. Como veremos, uma falha da aplica\c c\~ao acima em ser um homomorfismo de \'algebras de Lie em um dado ponto representa a falta de  integrabilidade de qualquer referencial local paralelo (ou ainda, a impossibilidade de escolhermos um referencial que n\~ao observe os efeitos da curvatura).

Em termos mais pedestres, podemos dizer que  o primeiro termo do lado direito de (\ref{curv}), o comutador, mede a diferen\c ca entre derivar covariantemente primeiro em uma dire\c c\~ao e depois na outra, e tomar as derivadas na ordem inversa. Mas isso n\~ao diz muita coisa, j\'a que mesmo em um fibrado trivial, com a conex\~ao trivial, as derivadas covariantes podem n\~ao comutar simplesmente porque os campos vetoriais $X,Y$ podem ter o seu colchete de Lie n\~ao nulo. I.e.: para uma se\c c\~ao do fibrado trivial $f:M\rightarrow\R^k$, e para a derivada usual do $\R^k$, os primeiros termos equivalem \`a $(XY-YX)f$ , mas n\~ao \'e necesariamente verdade que $(XY-YX)f=0$. O segundo termo corrige este efeito. Mostremos pois algumas caracter\'isticas b\'asicas do tensor de curvatura.  

\begin{theo}{Para todo $x\in{M}$ existe um mapa linear $\Omega(X,Y)_x:E_p\rightarrow{E_p}$ tal que se $s\in\Gamma(E)$ ent\~ao 
$$(\Omega(X,Y)s)(x)=\Omega(X,Y)_xs(x)$$}\end{theo}

O enunciado do teorema equivale a dizer que 
$\Omega(X,Y)\in{\mbox{Diff}}^0(E,E)$, ou seja, que $\Omega(X,Y)$, tem car\'ater tensorial, 
j\'a que o resultado de sua aplica\c c\~ao s\'o depende do valor dos campos no ponto de aplica\c c\~ao. 
 
{\bf{Dem:}} N\'os temos que 
\beq
([\nabla_X,\nabla_Y]-\nabla_{[X,Y]})(fs)=f([\nabla_X,\nabla_Y]-\nabla_{[X,Y]})s\eeq

Portanto se $f(x)=0$ n\'os temos: 
$(\Omega(X,Y)fs)(x)=0$
Provamos ent\~ao  que $\Omega(X,Y)$ \'e $C^\infty(M)$ linear, i.e.: tem car\'ater tensorial, ou ainda, pertence a Diff$^0(E,E)$. Para provar que isso implica que s\'o depende do valor no ponto basta 
tomarmos uma fun\c c\~ao que se anule em uma vizinhan\c ca arbitr\'aria do dado ponto. $\blacksquare$

\begin{theo}
Existe uma 2-forma $\Omega$ em M com valores em $\End(E)$ (ou seja, $\Omega\in\Gamma(\Lambda^2(M)\otimes{\End(E)}$ ) tal que para todos $X,Y\in\Gamma(TM)$,
$$\Omega_x(X_x,Y_x)=\Omega(X,Y)_x$$\end{theo} 

{\bf{Dem:}} $\Omega(X,Y)$ \'e claramente anti-sim\'etrica, vimos tamb\'em que $\Omega(X,Y)\in{\End(E)}$, agora nos resta ver que  $\Omega$
\'e $C^\infty$-linear na primeira entrada (i.e.: tem car\'ater tensorial e portanto s\'o depende de seus valores no ponto). Calculando $\Omega(fX,Y)$ obtemos, depois de um pouco de \'algebra, 
$\Omega(fX,Y)=f\Omega(X,Y)$. 

\subsubsection{Structure of the Space of Connections on $E$}

Chamamos de $\mathcal{C}(E)$ o espa\c co de todas as conex\~oes em $E$ e seja 
$$\Delta(E)=\Gamma\left(T^*M\otimes\End(E)\right)$$

\begin{defi}{ Se $\omega\in\Delta(E)$ e $s\in\Gamma(E)$, definimos $\omega(s)\in\Gamma(T^*M\otimes{E})$ por 
\beq{\omega(s)(X)=\omega\left(X,s(p)\right)\in{E_x}}\eeq 
onde $X\in{T_pM}$.}\end{defi}
Na verdade estamos mudando o enfoque sobre $\omega$ de 
\beq{\Delta(E)=\Gamma\left(T^*M\otimes\End(E)\right)\rightarrow{\Gamma\left(E^*\otimes(T^*M\otimes{E})\right)=\mbox{Diff}^0(E,T^*M\otimes{E})}}\eeq
Ou seja, existe um isomorfismo trivial entre os dois espa\c cos que s\'o muda a ordem de opera\c c\~ao de seus elementos. 

\begin{theo}{Se $\nabla^0\in\mathcal{C}(E)$ e para todos $\omega\in\Delta(E)$ n\'os definirmos 
\begin{eqnarray}
\nabla^\omega:\Gamma(E)\rightarrow\Gamma(T^*M\otimes{E})\\
s\mapsto\nabla^0s+\omega(s)\end{eqnarray} 
 Ent\~ao $\nabla^\omega\in\mathcal{C}(E)$ e o mapa $\omega\mapsto\nabla^\omega$ \'e bijetor.}\end{theo} 

{\bf{Dem:}} $\nabla^\omega$ \'e $\R$-linear j\'a que ambos $\nabla^0$ e $\omega$ o s\~ao. Se $f\in{C}^\infty(M,\R)$ e $s\in\Gamma(E)$, ent\~ao temos 
\beq{\nabla^\omega{fs}=f\nabla^0s+df\otimes{s}+f\omega(s)=f\nabla^\omega{s}+df\otimes{s}}\eeq 

Agora provemos que $\omega\mapsto\nabla^\omega$ \'e bijetor.
Que \'e injetora \'e trivial. Seja ent\~ao $\nabla^\alpha\in\mathcal{C}$, n\'os temos que:
\begin{eqnarray}
\label{nabla1}\nabla^\alpha(fs)=df\otimes{s}+f\nabla^\alpha{s}\\ 
\label{nabla2}\nabla^0(fs)=df\otimes{s}+f\nabla^0s
\end{eqnarray}
De (\ref{nabla1}) e (\ref{nabla2}) n\'os temos:

 $$(\nabla^\alpha-\nabla^0)(fs)=f(\nabla^\alpha-\nabla^0)s$$
 ou seja $\nabla^\alpha-\nabla^0$ \'e linear. I.e.: 
$$\nabla^\alpha-\nabla^0\in{\mbox{Diff}}^0(E,T^*M\otimes{E})\therefore\nabla^\alpha-\nabla^0=\omega$$ para algum $\omega\in\Delta(E)$, ent\~ao $\nabla^\alpha=\nabla^0+\omega$ ~~~~~~~~~~~$\blacksquare$\medskip~

Logo $\Delta(E)$ \'e um subespa\c co afim de $\mbox{Diff}^1(E, T^*M\otimes{E})$ e n\'os temos $$\mathcal{C}(E)\simeq\nabla^0+\mbox{Diff}^0(E, T^*M\otimes{E})$$
N\'os chamaremos de $\Delta(E)=\Gamma\left(T^*M\otimes{\End(E)}\right)$ o espa\c co das formas de conex\~ao. 

J\'a que a curvatura tem car\'ater tensorial, \'e  interessante notar que faz sentido atribuir-la um valor zero em um dado ponto, independente do gauge,  o que n\~ao podemos fazer com formas de conex\~ao, que s\~ao afins. 
Uma forma de conex\~ao 
n\~ao define por si s\'o uma conex\~ao $\nabla^\omega$, mas somente relativamente a outra conex\~ao $\nabla^0$, que pode ser considerado a origem segundo a qual uma conex\~ao \'e nula.  Logo $\Delta(E)$ \'e o espa\c co das diferen\c cas de conex\~ao. Pela propriedade da trivializa\c c\~ao local, podemos trabalhar localmente como se o fibrado fosse o produto (ou trivial) ent\~ao \'e v\'alido que estudemos conex\~oes no fibrado trivial para depois globalizarmos algumas de suas propriedades. 

\subsubsection{Connections on a Trivial Bundle}

Seja $E$ o fibrado trivial em $M\times\R^k$, ent\~ao $\Gamma(E)=\C^\infty(M,\R^k)$ e $\Gamma(\Lambda^1(M)\otimes{E})$ \'e o espa\c co das formas a valores em $\R^k$.
Como vimos, a escolha natural para $\nabla^0$ \'e $d$, a diferencial usual de uma fun\c c\~ao a valores vetoriais. Temos: 

$$\Delta(E)=\Gamma\left(T^*M\otimes\End(E)\right)=\Gamma\left(T^*M\otimes{L}(\R^k,\R^k)\right)$$ ent\~ao $\omega\rightarrow\nabla^\omega=d +\omega$ nos d\'a uma bije\c c\~ao entre $\mathcal{C}(E)$ e formas a valores em $M(k\times{k})$. 
Escrevamos ent\~ao $\omega$ em termos de uma base de $\R^k$ , $\mathcal{B}=\{e_i\}_{i=1}^k$ e tomamos a base dual  $\mathcal{B}^*=\{e^i\}_{i=1}^k$, ent\~ao \'e claro que 
$\{e_i\otimes{e^j}\}$ para $\{{1\leq{i}\leq{k},1\leq{j}\leq{k}}\}$ \'e a base associada de $L(\R^k,\R^k)$. Ent\~ao, escrevendo $\omega$ nessa base, temos: 
$$\omega=\omega_\alpha^\beta\otimes{e^\alpha\otimes}e_\beta$$
onde $\omega_\alpha^\beta\in\Gamma(T^*M)$ . Agora seja $s=s^ie_i\in\Gamma(E)=\C^\infty(M,\R^k)$ ent\~ao
\beq{\nabla^\omega{s}=d(s^ie_i)+\omega(s^ie_i)=ds^j\otimes{e_j}+s^i\omega_i^j\otimes{e}_j}\eeq Logo,
$$(\nabla^\omega{s})^j=ds^j+s^i\omega_i^j$$ Logo se $X\in(T_xM)$, 
\beq{\nabla^\omega{s}(X)=(\nabla^\omega{s})^j(X)e_j\Rightarrow(\nabla^\omega{s})^j(X)=X[s^j]+s^i(x)\omega^j_i (X)}\eeq
Tomando $s=e_k$ temos 
\beq\nabla^\omega{s}=\omega(e_k)=\omega_k^\beta\otimes{e_\beta}\Rightarrow (\nabla^\omega{e_k})^\beta=\omega^\beta_k\eeq E finalmente obtemos : \beq{\nabla}^\omega_X{e_k}=\omega^\beta_k(X)e_\beta\eeq 
N\'os acabamos de provar que:
\begin{theo}\label{corresp2}{Existe uma bije\c c\~ao $\omega\rightarrow\nabla^\omega$ entre matrizes $k\times{k}$ de 1-formas $\omega$ sobre $M$ e conex\~oes $\nabla^\omega$ no fibrado produto $E=M\times\R^k$. $\nabla^\omega$ \'e determinado por $\omega$ pela rela\c c\~ao 
$$(\nabla^\omega_vs(x))=v[s]+\omega(v)s(x)$$ para $v\in{T}_xM$ e $s\in\Gamma(E)=C^\infty(M,\R)$. Rec\'iprocamente, $\nabla\in\mathcal{C}(E)$ determina $\omega$ tomando uma base de se\c c\~oes $\{e^i\}_{i=1}^k$ e expandindo a a\c c\~ao de $\nabla_X$ nessa base, para $X\in\Gamma(E)$: 
\beq\label{con}\nabla_Xe_i=\omega^j_i(X)e_j=\omega(X)(e_i)\eeq}\end{theo}
 
Ent\~ao $\Omega\in\Gamma(\Lambda^2(M)\otimes{L}(\R^k,\R^k)$ est\'a relacionada a uma conex\~ao $\omega$. Vejamos como.

\begin{eqnarray}
\Omega:\Gamma(TM\times{TM})\rightarrow{L(}\R^k,\R^k)\\
(X,Y)\mapsto\Omega_l^j(X,Y)e^l\otimes{e}_j\end{eqnarray}

Onde $\Omega_l^j(X,Y)$ \'e uma matriz $k\times{k}$ de 2-formas em $M$. Logo, aplicando a  $e_i$ ambos os lados da \'ultima equa\c c\~ao:  $$\Omega_i^j(X,Y)e_j=\Omega(X,Y)(e_i)=([\nabla_X,\nabla_Y]-\nabla_{[X,Y]})e_i$$ onde lembramos o leitor que \'indices repetidos indicam uma somat\'oria de termos.  
Mas 
\beq\nabla_Xe_i=\omega^j_i(X)e_j\Rightarrow\nabla_Y(\nabla_Xe_i)=\Big(Y[\omega_i^j(X)]+\omega^k_i(X)\omega^j_k(Y)\Big)e_j\eeq
Para calcularmos $\nabla_X\nabla_Ye_i$ basta invertermos $X\leftrightarrow{Y}$. Por \'ultimo resta $$\nabla_{[X,Y]}e_i=\omega^j_i([X,Y])e_j$$ 
Juntando todos os termos obtemos:
\begin{eqnarray*}
\Omega_i^j(X,Y)e_j&=&\Big({X}[\omega^j_i(Y)]+\omega^k_i(X)\omega^j_k(Y)-Y[\omega^j_i(X)]-\omega_i^k(Y)\omega_k^j(X)-\omega^j_i([X,Y])\Big)e_j\\
~&=&d\omega^j_i(X,Y)+\omega^k_i\wedge\omega_k^j(X,Y)\end{eqnarray*}
Portanto obtemos
 
\beq\label{Oomega}\Omega_i^j=d\omega^j_i+(\omega\wedge\omega)^j_i\eeq
Onde $(\omega\wedge\omega)$ \'e a matriz de 2-formas resultante da multiplica\c c\~ao das matrizes de 1-forma $\omega$, onde a multiplica\c c\~ao de cada termo se d\'a com o produto exterior. 
Portanto, definindo $d\omega:=d\omega^j_i\otimes{e^i\otimes}e_j$ chegamos a 
\beq\Omega=d\omega+\omega\wedge\omega\eeq
Ou em palavras:
\begin{theo}{Se $\omega$ \'e uma matriz $k\times{k}$ de 1-formas em $M$ e $\nabla=\nabla^\omega=d+\omega$ \'e a conex\~ao correspondente no fibrado produto $E=M\times\R^k$, ent\~ao a forma de curvatura $\Omega^\omega$ relacionada a $\nabla$ \'e a matriz  $k\times{k}$ de 2-formas em $M$ dada por $\Omega^\omega=d{\omega}+\omega\wedge\omega$.}\end{theo} 

Esta forma de expressar $\Omega$ facilita em muito a deriva\c c\~ao da Identidade de Bianchi:
\beq\label{Bianchi}
{d\Omega^\omega+\omega\wedge\Omega^\omega-\Omega^\omega\wedge\omega=0}\eeq

{\bf{Dem:}}
\begin{eqnarray*}
d\Omega&=&d(d\omega)+d(\omega\wedge\omega)\\
~&=& d\omega\wedge\omega-\omega\wedge{d}\omega\\
~&=&(\Omega-\omega\wedge\omega)\wedge\omega-\omega\wedge(\Omega-\omega\wedge\omega)\\
&=&\Omega\wedge\omega-\omega\wedge\Omega+\omega\wedge\omega\wedge\omega-\omega\wedge\omega\wedge\omega\\
&=&\Omega\wedge\omega-(-1)^2\Omega\wedge\omega
\end{eqnarray*}
A identidade de Bianchi, \'e uma rela\c c\~ao geom\'etrica que, como veremos, representa leis de conserva\c c\~ao.  
Claramente estes resultados se estendem naturalmente para uma dada trivializa\c c\~ao local. Mas quanto dependem estes resultados das nossas escolhas de trivializa\c c\~ao? 

\subsubsection{ Gauge Transformation}

Sejam $\{e_i\}_{i=1}^k$ e $\{\tilde{e}_i\}_{i=1}^k$  duas bases locais de $\Gamma(E_{|\theta})$, relativas \`as trivializa\c c\~oes $F=\psi^{-1}:\theta\times{\R^k}\rightarrow\pi^{-1}(\theta)$ e $G=\phi^{-1}:\theta\times{\R^k}\rightarrow\pi^{-1}(\theta)$ respectivamente. E seja $g:\theta\rightarrow{\Gamma(\Aut(E|_\theta))}$ o mapa de transi\c c\~ao de um ao outro. Ent\~ao temos $\tilde{e}_i=g(e_i)$. Pela equa\c c\~ao (\ref{con}),  $\omega(X)\in{\End(E_x)}$, e poder\'iamos nos perguntar se $\omega$ como se\c c\~ao de transforma\c c\~oes lineares  escrita em outra se\c c\~ao de bases seria similar (no contexto de transforma\c c\~oes lineares) \`a pr\'opria $\omega$. A resposta \'e negativa, pois aqui a pr\'opria transforma\c c\~ao de bases varia sobre as fibras, e temos de levar esse efeito em conta  \footnote{Lembrando que em termos de bases locais o mapa de transi\c c\~ao pode ser encarado como uma aplica\c c\~ao $g:\theta\ra{GL(k)}$, poder\'iamos considerar o mapa de transi\c c\~ao constante se um mesmo elemento de $GL(k)$ ligasse as bases sobre todos os pontos de $\theta$, mas sob outra trivializa\c c\~ao isso n\~ao seria necessariamente verdade. }. 

\begin{eqnarray*}
\widetilde\omega(X)\tilde{e}_i&=&\nabla_X\tilde{e}_i=\nabla_X\Big(g(e_i)\Big)=dg(X)(e_i)+g(\nabla_Xe_i)={d}g(X)(e_i)+g\omega(X)(e_i)\\
~&=&\Big(dg(X)+g\omega(X)\Big)e_i=\Big(dg(X)+g\omega(X)\Big)g^{-1}(\tilde{e}_i)\end{eqnarray*}
Portanto \beq\label{mudanca}{\widetilde\omega=(dg)g^{-1}+g\omega{g}^{-1}} \eeq

Calculemos ent\~ao a mudan\c ca na forma de curvatura $$\widetilde\Omega=d\widetilde\omega+\widetilde\omega\wedge\widetilde\omega$$
N\'os temos que 
\beq\label{2}d\widetilde\omega=d\Big((dg)g^{-1}\Big)+d(g\omega{g}^{-1})=-(dg\wedge{g}^{-1}dg{g}^{-1})+dg\wedge\omega{g}^{-1}+gd\omega{g}^{-1}-g\omega\wedge{g}^{-1}dg{g}^{-1}\eeq e por outro lado 
\beq\label{3}\widetilde\omega\wedge\widetilde\omega=dg{g}^{-1}\wedge dg{g}^{-1}+dg\wedge\omega{g}^{-1}+g\omega{g}^{-1}\wedge dg{g}^{-1}+g\omega\wedge\omega{g}^{-1}\eeq
Somando (\ref{2}) e (\ref{3}), obtemos
\beq\widetilde\Omega=gd\omega{g}^{-1}+g\omega\wedge\omega{g}^{-1}=g\Omega{g}^{-1}\eeq 
Logo se $\Omega(X,Y)$ for nula, ela ser\'a nula em todos os gauges, representando bem o seu car\'ater tensorial. Enquanto que para formas de conex\~ao, mesmo que $\omega(X)=0$, n\'os temos de levar em conta o termo $dgg^{-1}$, que representa o car\'ater afim da forma de conex\~ao (sem origem). Notemos que n\~ao derivamos o resultado usual de $$\widetilde\omega=gdg^{-1}+g\omega g^{-1}$$ 
isso ocorre porque ao contr\'ario da maioria das abordagens, aqui o grupo age sobre uma base \`a esquerda e n\~ao \`a direita. \'E f\'acil ver que utilizando a a\c c\~ao \`a direita recupera-se o resultado usual. 
    
\subsubsection{ Alternative Approach}
Esta se\c c\~ao pode ser ignorada sem preju\'izo para a continuidade da exposi\c c\~ao. 

Em \'algebra linear, n\'os sabemos que a a\c c\~ao de mudan\c cas de base sobre transforma\c c\~oes lineares \'e efetuada por conjuga\c c\~ao por automorfismos, o que  confere a esta conjuga\c c\~ao tamb\'em uma ``interpreta\c c\~ao passiva" (de significar a mesma transforma\c c\~ao linear sob diferentes bases). 
A distin\c c\~ao entre tais conjuga\c c\~oes ``ativas" e ``passivas" fica borrada no caso de transforma\c c\~oes lineares  porque n\~ao utilizamos em espa\c cos vetoriais mudan\c cas de base locais, os automorfismos s\~ao globais e portanto sua a\c c\~ao pode ser considerada uma nova transforma\c c\~ao linear. No caso de variedades suaves a distin\c c\~ao entre transi\c c\~ao de cartas e difeomorfismos globais \'e \'obvia.

Tomando o ponto de vista ativo, ou global,  aqui tamb\'em podemos ter conex\~oes ``equivalentes", e assim como em transforma\c c\~oes lineares, julgaremos duas conex\~oes equivalentes se forem relacionadas pela conguga\c c\~ao de automorfismos    ( aqui um difeomorfismo $f:E\rightarrow{E}$ tal que $\pi\circ{f}=\pi$, onde $f$ 
leva $\pi^{-1}(x)\rightarrow{\pi^{-1}(x)}$ por um isomorfismo linear), i.e.:  devemos tamb\'em ter o mapa comutativo (para todo $X\in\Gamma(TM)$):
\begin{eqnarray*}
\Gamma(E) & \xrightarrow{\nabla_X\phantom{1}} & \Gamma(E)\\ 
g\downarrow & \phantom{\xrightarrow{\nabla_X\phantom{1}}} & \downarrow{g}\\ 
\Gamma(E) & \xrightarrow[\widetilde\nabla_X]{\phantom{\nabla_X{1}}} & \Gamma(E)\\
\end{eqnarray*} 
Exatamente como ocorre com a representa\c c\~ao de transforma\c c\~oes lineares sob isomorfismos lineares. 
I.e.: n\'os podemos representar a conex\~ao $\nabla$ sob um isomorfismo de fibrados, e obteremos uma nova conex\~ao 
\beq\label{conex}\widetilde\nabla_X(s)=g\nabla_X({g^{-1}s})\Rightarrow\widetilde\nabla_X=g\nabla_X{g^{-1}}\eeq 
ou ainda, $g\nabla_X=\widetilde\nabla_Xg$. Chequemos pois que $\widetilde\nabla$ \'e realmente uma conex\~ao:
\begin{enumerate}
\item[(i)]{Se $s\in\Gamma(E)$, $f\in{C^\infty(M,\R)}$ e $X\in\Gamma(TM)$, ent\~ao
\beq{\widetilde\nabla_X(fs)=g\nabla_X({g^{-1}fs})=g\nabla_X(f{g^{-1}s})=gX[f]g^{-1}s+fg\nabla_X(g^{-1}s)=X[f]s+f\widetilde\nabla_X(s)}\eeq} 
\item[(ii)]{O mapa $X\rightarrow\nabla_X$ \'e $C^\infty(M,\R)$-linear: 
\beq{\widetilde\nabla_{fX}=g\nabla_{fX}g^{-1}=fg\nabla_{X}g^{-1}=f\widetilde\nabla_{X}}\eeq}

\end{enumerate}

Agora podemos nos perguntar, como provamos que existe bije\c c\~ao entre o espa\c co das conex\~oes e o espa\c co das formas, se $\nabla$ estiver relacionado a $\omega$ e $\widetilde\nabla$ estiver relacionado a $\widetilde\omega$ ent\~ao qual \'e a rela\c c\~ao entre $\omega$ e $\widetilde\omega$? Sobre $\pi^{-1}(\theta)$, o dom\'inio de uma trivializa\c c\~ao podemos colocar $\widetilde\nabla^0=d$, fazendo $g\nabla_X=\widetilde\nabla_Xg$: 

\begin{eqnarray*}
g\nabla_Xe_i&=&g\omega^k_i(X)e_k=\omega^k_i(X)g^j_ke_j=(\omega^l_i(X)g^k_l)e_k\\
\widetilde\nabla_X(ge_i)&=&\widetilde\nabla_X(g^j_ie_j)=(dg^k_i+g^j_i\widetilde\omega^k_j(X))e_k\\
~&\therefore&~~~dg^k_i+g^j_i\widetilde\omega^k_j(X)=\omega^l_i(X)g^k_l
\end{eqnarray*}

V\'alido para todo $X\in\Gamma(TM_|\theta)$, obtemos ent\~ao:
\beq\widetilde\omega=(dg)g^{-1}+g\omega{g}^{-1}\eeq

Para calcularmos a curvatura sob uma transforma\c c\~ao de gauge, notemos que, a partir de (\ref{conex}), obtemos, para todos $X,Y\in\Gamma(TM)$: 
$$\widetilde\nabla_X\widetilde\nabla_Y= g\nabla_X{g^{-1}}g\nabla_Y{g^{-1}}=g\nabla_X\nabla_Y{g^{-1}}$$ e que 
$$\widetilde\nabla_{[X,Y]}=g\nabla_{[X,Y]}{g^{-1}}$$ portanto obtemos:

\beq\widetilde\Omega=g\Omega{g^{-1}}\eeq

E realmente, o efeito local  de automorfismos globais \'e uma transforma\c c\~ao de gauge. Vejamos pois outras formas de construir novas conex\~oes a partir de antigas, constru\c c\~oes necess\'arias para a introdu\c c\~ao dos importantes conceitos de paralelismo e holonomia. 
\section{Parallel Transport and Holonomy} 

\subsubsection{Building New Connections}
\begin{prop} {Seja $\{\theta_i\}_{i\in{I}}$ uma cobertura de $M$ e $\nabla^i$ a conex\~ao em $\pi^{-1}(\theta_i)$  de forma que $\nabla^i$ e $\nabla^j$ concordam em $E_{|\theta_i\cap\theta_j}$, para todos $i, j\in I$. Ent\~ao existe uma \'unica conex\~ao $\nabla$ em $E$ tal que $\nabla_{|{\pi^{-1}\theta_i}}=\nabla^i$}\end{prop} 

{\bf{Dem:}} Que existe uma \'e f\'acil demonstrar usando parti\c c\~oes da unidade associadas \`a cobertura $\{\theta_i\}$. Chamando essa parti\c c\~ao de $\{\sigma_i\}$ e  fazendo $\nabla=\sum_{i\in{I}}\sigma_i\nabla^i$ podemos facilmente verificar que obtemos a conex\~ao desejada. Por outro lado, se houvesse duas diferentes, elas teriam de diferir em pelo menos um aberto (j\'a que s\~ao lisas), podemos supor sem perda de generalidade que seriam ent\~ao diferentes em um aberto contido em $E_{|\theta_i}$, logo n\~ao podem ambas ter restri\c c\~ao igual em $E_{|\theta_i}$. 

\begin{theo} Se $\nabla$ \'e qualquer conex\~ao em $E$ existe uma conex\~ao \'unica $\nabla^*$ no fibrado dual $E^*$ tal que se $\sigma\in\Gamma(E^*)$ e $s\in\Gamma(E)$, ent\~ao temos, para todo $X\in\Gamma(E)$:
\beq\label{dual}{X[\sigma(s)]=\nabla^*_X\sigma(s)+\sigma\nabla_Xs}\eeq\end{theo} 

{\bf{Dem:}} A forma mais f\'acil de demonstrar esse fato \'e simplesmente tomando 
\beq{\nabla^*_X\sigma(s)=X[\sigma(s)]-\sigma\nabla_Xs}\eeq
e mostrando que essa defini\c c\~ao preenche os requisitos de uma conex\~ao. 
Ser\'a \'util no entanto achar a forma de conex\~ao em termos de uma trivializa\c c\~ao local, em termos de bases locais, $\{e_i\}$ e $\{e^i\}$ de $\Gamma(E_{|\theta_i})$ e $\Gamma(E^*_{|\theta_i})$ respectivamente. 
Ent\~ao, chamando $\nabla^*_Xe^j=\omega^*(X)e^j$, usando $\nabla_Xe_i=\omega(X)e_i$ e a propriedade que a conex\~ao comuta com a contra\c c\~ao, i.e.: (\ref{dual}), obtemos:
$$X[e^j(e_i)]=(\omega^*(X)e^j)(e_i)+e^j\omega(X)e_i=0\Rightarrow(\omega^*(X)e^j)(e_i)=-e^j(\omega(X)e_i)$$ 
Mas lembremos que se $S\in{L(\R^k,\R^k)}$, se $\lambda\in\Lambda^1(\R^k)$ e $v$ \'e um vetor em $\R^k$, ent\~ao
$\lambda(Sv)=S^T\lambda(v)$, i.e.: $\lambda S=S^T\lambda$, 
Onde o sobrescrito $T$ denota transposi\c c\~ao da matriz. Ent\~ao, utilizando o caso geral: 

\beq\label{abcdef}{-e^j(\omega(X)e_i)=(-\omega(X)^Te^j)(e_i)=(\omega^*(X)e^j)(e_i)}\eeq
 Logo, como isso \'e v\'alido para todos $i, j, X$ obtemos, $\omega^*=-\omega^T$ .
 Ou ainda,   sendo $\{\lambda_i\}$ uma base de campos para $\Gamma(E^*_{|\theta_i})$ dual a $\{e_i\}$, obtemos facilmente 
$$\nabla_X^*\lambda_i(e_j)+\lambda_i(\nabla_Xe_j)=\omega^{*k}_i(X)\lambda_k(e_j)+\lambda_i\omega^k_j(X)e_k=0$$ 
$$\therefore\omega^{*j}_i=-\omega^i_j$$
Demos a primeira demonstra\c c\~ao porque a segunda contraria um pouco a nota\c c\~ao e pr\'atica dos f\'isicos, de n\~ao considerar formas duais como campos vetoriais, e de denot\'a-las por \'indices superiores ao inv\'es de inferiores. Considerando formas simplesmente como se\c c\~oes de um fibrado vetorial (e n\~ao somente como duais de se\c c\~oes) a demonstra\c c\~ao se torna trivial.

\begin{theo}\label{tensorderiv} Se $\nabla^i$ \'e conex\~ao em um fibrado $E^i$ sobre $M$, i=1,2,
ent\~ao existe uma \'unica $\nabla=\nabla^1\otimes{1}+1\otimes\nabla^2$ em
$E^1\otimes{E^2}$, tal qual se $s^i\in\Gamma(E^i)$ e $X\in\Gamma(TM)$ 
Ent\~ao
\beq\label{1}\nabla_X(s^1\otimes{s^2})=\nabla^1_X{s^1}\otimes{s^2}+s^1\otimes\nabla^2_X{s^2}\eeq
\end {theo}

{\bf{Dem:}}  Novamente tomamos duas bases $\{e_i\}^k_{i=1}$,
$\{b_i\}^l_{i=1}$ sobre $E^1|_\theta$ e $E^2|_\theta$ respectivamente.
Ent\~ao temos, se $\nabla$ satisfaz (\ref{1}):
$$\nabla_X(e_i\otimes{b_j})=\nabla^{1}_X{e_i}\otimes{b_j}+e_i\otimes\nabla^{2}_X{b_j}=$$ 
$$\sum_{m=1}^k\omega^{(1)}(e_i)\otimes{b_j}+\sum_{p=1}^le_i\otimes\omega^{(2)}({b_j})$$

Logo como $\{e_m\otimes{b_j}\}_{i=1, j=1}^{i=k, j=l}$ \'e base de $(E^1\otimes{E^2}){|_\theta}$, para que $\nabla$ obede\c ca (\ref{1}), e j\'a que $\omega^1$ e $\omega^2$ s\~ao \'unicos para $\nabla^1$ e $\nabla^2$, $\omega$ \'e \'unica e tem a forma 
\beq\label{abcde}\omega=\omega^1\otimes{Id}+Id\otimes\omega^2\eeq 
$\phantom{aaaaaaaaaaaaaaaaaaaaaaaaaaaaaaaaaaaaaaaaaaaaaaaaaaa}\blacksquare$\medskip

 Agora , do fibrado $(E,\pi,M)$, e dado  um mapa $\phi:N\rightarrow{M}$, n\'os obtemos o fibrado do pull-back $\phi^*(E)$ sobre $N$ tomando como fibra sobre $x\in{N}$, 
$$\big(\phi^*(E)\big)_x\simeq(E_{\phi(x)})$$ i.e.: n\'os puxamos as fibras juntamente com os pontos de $\phi(N)$. Logo n\'os obtemos o mapa linear
$$\begin{array}{llll}
\phi^*: & \Gamma(E) & \longrightarrow & \Gamma\Big(\phi^*(E)\Big)\\
~ & s & \longmapsto & {s}\circ\phi
\end{array} $$

\begin{theo} Dado $\nabla\in\mathcal{C}(E)$, existe uma \'unica conex\~ao $\nabla^\phi$ para $\phi^*(E)$ tal que se $s\in\Gamma(E)$, $Y\in\Gamma(TN)$ e $X=d\phi(Y)$ ent\~ao 
\beq\nabla^\phi_Y(\phi^*s)=\phi^*(\nabla_Xs)\eeq\end{theo} 

{\bf{Dem}:} Novamente escolhemos uma base $\{e_i\}^k_{i=1}$ em $E{|_\theta}$, e obtemos
\beq{\phi^*(\nabla_Xe_i)=\phi^*(\omega^j_i(X)e_j)=\omega_i^j(d\phi(Y))\phi^*e_j)}\eeq
\'E claro que $\{\phi^*(e_i)\}^k_{i=1}=\{\tilde{e_i}\}^k_{i=1}$ \'e base de $\Gamma(\phi^*(E))$ ent\~ao 
\beq\phi^*(\nabla_Xe_i)=\omega_i^j(d\phi(Y))\tilde{e_j}\Rightarrow\tilde\omega_i^j=\phi^*\omega_i^j\eeq
Portanto temos
\beq\label{phi*}(\nabla^\phi\phi^*e_i)^j=\phi^*\omega_i^j\eeq
$\phantom{aaaaaaaaaaaaaaaaaaaaaaaaaaaaaaaaaaaaaaaaaaaaaaaaaaaaaaaaaaaaaaa}\blacksquare$ 

{\bf{Exemplos}}
\begin{itemize}

\item{Se $N$ for subvariedade de $M$ e $\imath:{N}\hookrightarrow{M}$ ent\~ao $\imath^*(E)=E_{|N}$. \'E claro que se $\theta\subset\imath(N)$ \'e dom\'inio de um gauge, $X\in\Gamma(TM_{|\theta})$ e $s^1,s^2\in\Gamma(E)$ tal que se $s^1_{|\theta}=s^2_{|\theta}$, ent\~ao $\nabla_Xs^1=\nabla_Xs^2$} 
\item{Se $N=I=[a,b]$ e $\phi=\sigma:I\longrightarrow{M}$, temos 
$\sigma^*(s)_t=s(\sigma(t))$
e escrevemos para $\nabla^\sigma\in\mathcal{C}(\sigma^*(E))$ e $Y\in\Gamma(TI)$,
\beq{\nabla^\sigma_Y\sigma^*s=\nabla^\sigma_\frac{\partial}{\partial{t}}\sigma^*s=\frac{D}{dt}(s\circ\sigma)=\nabla_{\sigma'}s}\eeq 
chamada de derivada covariante ao longo de $\sigma$. Novamente tomando $s=e_i$, por (\ref{phi*}) n\'os temos 
\beq\Big(\nabla^\sigma_\frac{\partial}{\partial{t}}\sigma^*(e_i)\Big)^j=\omega_i^j\Big(d\sigma\left(\frac{\partial}{\partial{t}}\right)\Big)=\omega_i^j(\sigma')=(\nabla_{\sigma'}e_i)^j\eeq 
Portanto, como vimos:
$(\nabla{s})^j=ds^j+s^i\omega_i^j$ e obtemos
\beq(\nabla^\sigma_\frac{\partial}{\partial{t}}\sigma^*s)^j=ds^j\left(\frac{\partial}{\partial{t}}\right)+s^i\omega_i^j(\sigma')=\frac{ds^j}{d{t}}+s^i\omega_i^j(\sigma')\eeq 
se $\{x^i\}_{i=1}^n$ s\~ao coordenadas locais de $M$ escrevemos $\omega_i^j=\Gamma^j_{\phantom{j}ik}dx^k$ e portanto
\beq\label{parallel}\frac{Ds^j}{d{t}}=\frac{ds^j}{d{t}}+\Gamma^j_{\phantom{j}ik}\frac{d\sigma^k}{d{t}}s^i\eeq 

Note-se que essa equa\c c\~ao \'e uma equa\c c\~ao diferencial ordin\'aria linear de primeira ordem, com coeficientes lisos na fun\c c\~ao vetorial $(s^i(t),\cdots,s^k(t))\in\R^k$.} 

\end{itemize}

\subsubsection{Parallel Transport}

Como mencionamos $\nabla$ \'e chamada de flat ou plana, se e somente se $\Omega^\nabla=0$. Como $\nabla$ \'e uma 2-forma, se dim$M=1$, $\Omega\equiv{0}$. Logo, se $\sigma:I=[0,1]\rightarrow{M}$ e $\nabla\in\mathcal{C}(E)$, ent\~ao $\nabla^\sigma$ \'e flat. Logo para cada tal $\nabla$ existe uma forma can\^onica de comparar vetores ao longo de $\sigma$. 
\begin{defi}
Seja $$\frac{D}{dt}=\nabla^\sigma_{\sigma'}:\Gamma(\sigma^*(E))\rightarrow\Gamma(\sigma^*(E))$$ 
que \'e linear. Chamamos $P(\sigma):=Ker\nabla^\sigma_{\sigma'}$, o espa\c co de campos paralelos ao longo de $\sigma$.\end{defi} 
\begin{theo} O mapa 
\begin{eqnarray*}
\sigma_{t}:P(\sigma)\rightarrow{E}_{\sigma(t)}\\
\phantom{\sigma}s\longmapsto{s(t)}
\end{eqnarray*} \'e um isomorfismo linear.\end{theo}

{\bf{Dem:}} Pelo  teorema de exist\^encia e unicidade de solu\c c\~oes, dado $v\in{E}_{|\sigma(t_0)}$ qualquer, existe uma \'unica solu\c c\~ao para (\ref{parallel}) (i.e.: existe uma \'unica se\c c\~ao $s\in\Gamma(\sigma^*E)$) que satisfaz  $s(t_0)=v$ e   
$\nabla^\sigma_{\sigma'}s(t)=0$ para todo $t\in[0,1]$, ou seja,  existe um \'unico elemento associado de $P(\sigma)$. Al\'em disso, a equa\c c\~ao diferencial \'e linear, portanto depende linearmente de seus par\^ametros iniciais. Portanto o mapa  $\sigma_{t}$ que leva o espa\c co vetorial $P(\sigma)$ no espa\c co vetorial ${E}_{\sigma(t)}$ \'e uma bije\c c\~ao linear. $\blacksquare$ 

\begin{defi} Para $t_1,t_2\in{I}$ definimos o isomorfismo linear (chamado de transporte paralelo):
\begin{eqnarray*}
P_\sigma(t_1,t_2):E_{\sigma(t_1)}\rightarrow{E_{\sigma(t_2)}}\\
\phantom{P_\sigma(t_1,t_2)}v\longmapsto{s(t_2)} 
\end{eqnarray*} onde $s\in{P}(\sigma)\subset\Gamma(\sigma^*(E))$ \'e dado por $s=(\sigma_{t_1})^{-1}(v)$.\end{defi}
Valem as seguintes propriedades:
\begin{enumerate}
\item[(i)]{$P_\sigma(t,t)=Id_{E_{\sigma(t)}}$} 
\item[(ii)]{$P_\sigma(t_3,t_2)P_\sigma(t_2,t_1)=P_\sigma(t_3,t_1)$ j\'a que se $v=s(t_1)$, como a solu\c c\~ao \'e \'unica: 
$P_\sigma(t_3,t_2)P_\sigma(t_2,t_1):s(t_1)\mapsto{s(t_2)\mapsto}s(t_3)=P_\sigma(t_3,t_1)s(t_1)$} 
\item[(iii)]{Dos itens (i) e (ii) obtemos facilmente que $P_\sigma(t_1,t_2)=P_\sigma(t_2,t_1)^{-1}$.}
\end{enumerate}
Como dissemos, dado $\nabla$ podemos comparar  vetores de forma can\^onica ao longo de qualquer curva, e o transporte paralelo fornece o m\'etodo geom\'etrico de manter campos ``constantes'' sob essa compara\c c\~ao. Notemos  que a mesma constru\c c\~ao n\~ao vale para uma imers\~ao $\phi:N\rightarrow{M}$ se $\nabla^\phi$ n\~ao for plana, j\'a que se fosse poss\'ivel poder\'iamos achar uma trivializa\c c\~ao local de $\phi^*(E)$ onde $\nabla^\phi=d$ e portanto $\Omega^\phi=0$, o que \'e contradit\'orio. Veremos mais adiante que isto est\'a relacionado com a integrabilidade de bases de campos paralelos ( a pr\'opria equa\c c\~ao (\ref{curv}) aponta para o Teorema de Frobenius em termos de formas.)

\begin{prop} A conex\~ao $\nabla$ pode ser recuperada do transporte paralelo.\end{prop}
{\bf{Dem:}}
Dado $s\in\Gamma(E)$ e $X\in\Gamma(TM)$, tomamos $\sigma:[0,1]\rightarrow{M}$ qualquer tal que $\sigma'(0)=X_{\sigma(0)}$. 
N\'os temos que, seja $\{e_i(0)\}_{i=1}^k$ base de $E_{\sigma(0)}$, tomamos, para todo $i=1\cdots{k}$ 
$$e_i(t)=P_\sigma(t,0)e_i(0)$$ Como $P_\sigma(t,0)$ \'e isomorfismo linear, $\{e_i(t)\}_{i=1}^k$ \'e base de $E_{\sigma(t)}$. Logo podemos escrever $$s(\sigma(t))=s^i(\sigma(t))e_i(t)$$ 
e usando a regra de Leibnitz:
\beq{\nabla_{\sigma'(0)}s=\frac{D}{dt}(s(\sigma(t))_{|t=0}=\frac{d}{dt}_{|t=0}\Big(s^i(\sigma(t))\Big)e_i(0)}\eeq
Definimos ent\~ao a curva em $E_{\sigma(0)}$:
\begin{eqnarray}
\tilde{s}:I\rightarrow{E_{\sigma(0)}}\phantom{\tilde{s}}\\
t\longmapsto{P_\sigma(0,t)s(\sigma(t))}
\end{eqnarray}
Mas $$P_\sigma(0,t)s(\sigma(t))=P_\sigma(0,t)\Big(s^i(\sigma(t))e_i(t)\Big)=s^i(\sigma(t))e_i(0)$$ 
Portanto chegamos a 
\beq\label{deriv}(\nabla_Xs)_{\sigma(0)}=\nabla_{\sigma'(0)}s=\frac{d}{dt}_{|t=0}\tilde{s}(t)\phantom{aaaaaaaaaa}\blacksquare\eeq \medskip

Portanto podemos considerar a derivada covariante como a forma infinitesimal do transporte paralelo, isto \'e como o mecanismo operacional do conceito geom\'etrico de transporte paralelo levado ao limite.

Sejam $\{e_i\}_{i=1}^k$ e $\{\tilde{e_i}\}_{i=1}^k$  bases locais sobre $\sigma^*(E)$ e $\omega$ e $\widetilde\omega$ as respectivas formas de conex\~ao. Se $g:[0,1]\rightarrow{GL(k)}$ \'e fun\c c\~ao de transi\c c\~ao de gauge, n\'os sabemos que $\omega=g\widetilde\omega g^{-1}+dg^{-1}g$ , agora se a base local $\{\tilde{e_i}\}_{i=1}^k$ \'e tal que 
$$\tilde{e_i}(t)=P_\sigma(t,0)\tilde{e_i}(0)\Rightarrow\nabla^\sigma\tilde{e_i}=0\Rightarrow\widetilde\omega=0$$
Logo obtemos $\nabla^\sigma=d$ na base $\{\tilde{e_i}\}_{i=1}^k$ e $\nabla^\sigma=dg^{-1}g$ na base $\{{e_i}\}_{i=1}^k$. 
Note-se que n\~ao mudamos a base $\{{e_i}\}_{i=1}^k$ em si, mas as novas fun\c c\~oes de transi\c c\~ao absorvem o efeito do antigo $\widetilde\omega$. 

\subsubsection{Holonomy}

Sejam $\alpha$ e $\beta$ dois caminhos cont\'inuos tal que $\alpha(0)=p$ e $\alpha(1)=x=\beta(0)$ e $\beta(1)=q$. Definimos os caminhos $\alpha^{-1}$ e $\beta\circ\alpha$ por
$\alpha^{-1}(t)=\alpha(1-t)$ e por
\[\qquad 
\beta\circ\alpha(t)=\left\{\begin{array}{ll}
    \alpha(2t)&
\mbox{se $0\leq{t}\leq1/2$,}\\
    \beta(2t-1)&\mbox{se $1/2\leq{t}\leq1$,} \end{array}\right. \]

Logo, pela propriedade (2) do transporte paralelo:
\beq\label{P1}{P_{\beta\alpha}(1,0)={P_{\beta\alpha}}\left(1,\frac{1}{2}\right)\circ{P_{\beta\alpha}}\left(\frac{1}{2},0\right)=P_\beta(1,0)\circ{P_\alpha(1.0)}}\eeq 
\beq\label{P2}{P_{\alpha^{-1}}(1,0)\circ{P}_\alpha(1,0)=Id\Rightarrow{P}_\alpha(1,0)^{-1}={P_{\alpha^{-1}}(1,0)}}\eeq
Agora seja $$\Lambda_x:=\{\sigma:[0,1]\rightarrow{M}| \sigma(0)=x=\sigma(1)\}$$  N\'os temos que, abreviando a nota\c c\~ao, $P_\sigma(1,0):=P_\sigma:E_x\rightarrow{E_x}$ \'e isomorfismo linear, logo $P_\sigma\in{GL(E_x)}$. 
Definimos ent\~ao o grupo de holonomia de $\nabla$ no ponto : 
$$\mbox{Hol}_x(\nabla):=\{P_\sigma : \sigma\in\Lambda_x\}\subset{GL(E_x)}$$ 
$\mbox{Hol}_x(\nabla)$  \'e realmente um subgrupo de $GL(E_x)$ j\'a que por (\ref{P1}), se $\alpha, \beta\in\Lambda_x$ ent\~ao para $P_\alpha ,P_\beta\in \mbox{Hol}_x(\nabla)$ n\'os temos que $P_\alpha\circ{P}_\beta=P_{\alpha\beta}\in \mbox{Hol}_x(\nabla)$ e os termos cont\'em inversa por  (\ref{P2}), j\'a que se $\alpha\in\Lambda_x\Rightarrow\alpha^{-1}\in\Lambda_x$. Em outras palavras, o mapa $\sigma\rightarrow{P_\sigma}$ \'e um homomorfismo de grupos entre $\Lambda_x$ e um subgrupo de $GL(E_x)$. 
 Supondo que $M$ seja conexo, dados $x,y\in{M}$ ,  existe curva $\gamma:[0,1]\rightarrow{M}$ tal que $\gamma(0)=x$ e $\gamma(1)=y$, e $P_\gamma(1,0):E_x\rightarrow{E_y}$ \'e isomorfismo linear. Logo, como se $\alpha\in\Lambda_x$ ent\~ao $\gamma\circ\alpha\circ\gamma^{-1}\in\Lambda_y$ e $P_{\gamma\alpha\gamma^{-1}}=P_\gamma\circ{P}_\alpha\circ{P}_\gamma^{-1}$ , ent\~ao existe uma conjuga\c c\~ao entre $\mbox{Hol}_x(\nabla)$ e $\mbox{Hol}_y(\nabla)$. I.e.:
$$\mbox{Hol}_y(\nabla)=P_\gamma\circ{\mbox{Hol}_x(\nabla)}\circ{P}_\gamma^{-1}$$
Isso mostra que o grupo de holonomia $\mbox{Hol}_x(\nabla)$ independe do ponto base no seguinte sentido: suponha que $E$ tenha fibra isomorfa a $\R^k$, ent\~ao qualquer identifica\c c\~ao $E_x\simeq\R^k$ induz um isomorfismo entre o grupo das transforma\c c\~oes lineares invers\'iveis de $E_x$ com o grupo das matrizes invers\'iveis $GL(E_x)\simeq{G}L(k,\R)$. Ent\~ao identificamos $\mbox{Hol}_x(\nabla)=H_x(\nabla)<GL(k,\R)$. Mas se escolhermos outra base para $E_x$, i.e.: outra identifica\c c\~ao, n\'os temos de aplicar uma transforma\c c\~ao de semelhan\c ca em todos os elementos de $H_x(\nabla)$, ou seja, obtemos o subgrupo $aH_x(\nabla)a^{-1}$ 
onde $a\in{GL}(k,\R)$ \'e a  transforma\c c\~ao entre as bases. Obtemos que o  grupo de holonomia est\'a bem definido como subgrupo de $\GL k,\R$ a menos de conjuga\c c\~ao.  

\begin{prop} 
Seja $(E,\pi,M)$ fibrado vetorial com fibra t\'ipica $\R^k$, $\nabla\in\mathcal{C}(E)$ e  $M$ simplesmente conexo.Ent\~ao $\mbox{Hol}(\nabla)$ \'e um subgrupo de Lie de $GL(k,\R)$ conexo.\end{prop}
{\bf{Dem:}} 
Seja $\gamma\in\Lambda_x$ e $F(s,t)$ uma homotopia entre $\gamma$ e o la\c co constante. Ent\~ao $s\mapsto{P_{F_s}}(1,0)$ \'e um mapa cont\'inuo\footnote{Os teroremas de solu\c c\~oes de E.D.O.s garantem que as solu\c c\~oes dependentes de par\^ametros, dependem continuamente de seus par\^ametros, bem como de suas condi\c c\~oes iniciais.}  entre $[0,1]$ e $\mbox{Hol}(\nabla)$. Como $P_{F_0}(1,0)=Id$ e $P_{F_1}(1,0)=P_\gamma(1,0)$, cada $P_\gamma(1,0)$ pode ser ligado a identidade por um caminho cont\'inuoem $\mbox{Hol}(\nabla)$. Por um teorema de Yamabe \cite{Ya} todo subgrupo conexo por caminhos de um grupo de Lie \'e um subgrupo de Lie conexo.   ~~~~~~$\blacksquare$

\begin{defi} $\mbox{Hol}^0(\nabla):=\{P_\gamma : \gamma$ \'e homot\'opica a um ponto$\}$. \end{defi}

\begin{theo}\label{variation} Se $\sigma$ \'e homot\'opica a um ponto, ent\~ao  a conex\~ao $\nabla$ \'e flat se e somente se  $P_\sigma=Id$\end{theo}

{\bf{Dem:}}

Sejam $\gamma_0$ e $\gamma_1$ homot\'opicas, ou seja, 
existe fun\c c\~ao cont\'inua:
\begin{eqnarray*}
F:[0,1]\times[0,1]\rightarrow{M}\\
(s,t)\longmapsto{F(s,t)}
\end{eqnarray*}
tal que, para $s$ fixo: $F(s,t)=F_s(t)=\gamma_s(t)$ fam\'ilia cont\'inua de curvas interpolando $\gamma_1$ e $\gamma_0$. Al\'em disso $\gamma_s(0)=x$ e $\gamma_s(1)=y$. Agora sejam 
\beq\frac{\partial}{\partial{s}}_{\vert(s_0,t_0)}:=F_*\frac{d}{d{s}}_{\vert(s_0,t_0)}~~~;~~~~ \frac{\partial}{\partial{t}}_{\vert(s_0,t_0)}:=F_*\frac{d}{d{t}}_{\vert(s_0,t_0)}\eeq
E seja $v\in{E}_x$. Definimos $X\in\Gamma(F^*E)$ por $X_{(F(s,t))}:=P_{\gamma_s}(t,0)v$  . Obviamente, para todo $s\in[0,1]$,  $X_{(F(s,0))}=v$, portanto 
\beq\label{hom2}\left(\nabla_{\frac{\partial}{\partial{s}}}X\right)_{\vert(s_0,0)}=0\eeq para todo $s_0\in[0,1]$. Agora, como a curvatura \'e nula:
\beq\label{hom}{R(\frac{\partial}{\partial{s}},\frac{\partial}{\partial{t}})X=\nabla_\frac{\partial}{\partial{t}}\nabla_\frac{\partial}{\partial{s}}X-\nabla_\frac{\partial}{\partial{s}}\nabla_\frac{\partial}{\partial{t}}X=0}\eeq 
J\'a que, como $(s,t)$ s\~ao coordenadas da parametriza\c c\~ao $F(s,t)$:
$$\left[\frac{\partial}{\partial{s}},\frac{\partial}{\partial{t}}\right]=F_*\left[\frac{d}{d{s}},\frac{d}{d{t}}\right]=0$$

Agora, pela defini\c c\~ao de $X$, n\'os temos $\left(\nabla_{\frac{\partial}{\partial{t}}}X\right)_{\vert(s_0,t_0)}=0$ , para todos $s_0$ , $t_0$. Ent\~ao, usando (\ref{hom})
\beq\nabla_\frac{\partial}{\partial{t}}\nabla_\frac{\partial}{\partial{s}}X=0\eeq 
Ora, mas isso quer dizer que para cada $s_0\in[0,1]$ o campo $\left(\nabla_{\frac{\partial}{\partial{s}}}X\right)$ \'e transportado paralelamente ao longo de $\gamma_s$, i.e.:
\beq\label{hom3}\left(\nabla_{\frac{\partial}{\partial{s}}}X\right)_{\vert(s_0,t_0)}=P_{\gamma_s}(t_0,0)\Bigg(\left(\nabla_{\frac{\partial}{\partial{t}}}X\right)_{\vert(s_0,0)}\Bigg)\eeq 
Ent\~ao  n\'os temos por (\ref{hom2}) que (\ref{hom3}) ser\'a o transporte paralelo de um vetor nulo, logo obtemos que $$\left(\nabla_{\frac{\partial}{\partial{s}}}X\right)_{\vert(s_0,t_0)}=0$$
Portanto , em particular, n\'os que $\left(\nabla_{\frac{\partial}{\partial{s}}}X\right)_{\vert(s_0,1)}=0$ . Ou seja, a curva sobre a fibra $E_y$ , $X(s,1)\in{E_y}$ \'e uma curva constante, independe de $s\in[0,1]$.  Logo o transporte paralelo de conex\~oes flat \'e invariante por curvas homot\'opicas. Isso significa que em um dom\'inio simplesmente conexo, se tivermos uma conex\~ao flat, podemos estabelecer uma base de vetores paralelos, j\'a que em cada ponto $y$ a base paralelamente transportada desde o ponto $x$  n\~ao depender\'a da curva que liga os dois pontos. ~~~~~~~~~~~~~~$\blacksquare$\medskip 

 Provemos pois alguns teoremas \'uteis sobre grupos e \'algebras de holonomia.
\begin{prop}{$\mbox{Hol}^0(\nabla)$ \'e um subgrupo de Lie conexo de $GL(k,\R)$, \'e a componente conexa de $\mbox{Hol}(\nabla)$ que cont\'em a identidade e \'e um subgrupo normal de $\mbox{Hol}(\nabla)$. Al\'em disso, existe um homeomorfismo de grupos sobrejetivo natural $$\Phi:\pi_1(M)\rightarrow{\mbox{Hol}(\nabla)/(\mbox{Hol}^0(\nabla))}$$} 
\end{prop}
Demonstra\c c\~ao: Pela proposi\c c\~ao anterior, $\mbox{Hol}^0(\nabla)$ \'e um subgrupo de Lie conexo. Sejam $\alpha, \gamma\in\Lambda_{p},\gamma$ homot\'opico a $p$. Ent\~ao $\alpha\circ\gamma\circ\alpha^{-1}$ tamb\'em \'e homot\'opico a $p$. Portanto, $P_{\alpha\gamma\alpha^{-1}}=P_\alpha\circ{P_\gamma}\circ{P_\alpha^{-1}}\in{\mbox{Hol}_p^0(\nabla)}$, sendo que $P_\alpha\in{\mbox{Hol}_p(\nabla)}$, logo $\mbox{Hol}_p^0(\nabla)$ \'e normal. 
Seja $$\Phi:\pi_1(M)\rightarrow{\mbox{Hol}_x}(\nabla)/\mbox{Hol}_x^0(\nabla)$$ $$[\gamma]\mapsto[P_\gamma]$$
Mostramos que \'e bem definido, isto \'e, que $\alpha,\gamma\in[\gamma]$, isto \'e:
$$\alpha\simeq\gamma\Rightarrow[P_\alpha]=[P_\gamma]$$ ou seja, que existe $a\in{\mbox{Hol}_x^0}(\nabla)$, tal que $P_\alpha=P_\gamma\circ{a}$. 
J\'a que $\alpha\simeq\gamma$, ent\~ao $\gamma^{-1}\alpha\simeq{Id}$. Portanto, $P_{\gamma^{-1}\alpha}=P_{\gamma^{-1}}\circ{P_\alpha}=a'\in{\mbox{Hol}_x^0}\Rightarrow{P_\alpha=P_\gamma\circ{a}}$, \'e claro por sua defini\c c\~ao que $\Phi$ \'e um homo sobrejetor (epimorfismo). 
Como $\pi_1(M)$ \'e enumer\'avel, o grupo quociente $\mbox{Hol}_x(\nabla)/\mbox{Hol}_x^0(\nabla)$ tamb\'em o \'e, logo $\mbox{Hol}_x(\nabla)$ \'e um grupo de Lie e  $\mbox{Hol}_x^0(\nabla)$ \'e a componente conexa de $\mbox{Hol}_x(\nabla)$, que cont\'em a identidade.

Definimos a \'algebra de holonomia $\Lh\mbox{ol}(\nabla)$ como sendo a \'algebra de $\mbox{Hol}^0(\nabla)$. \'E uma sub\'algebra de $\Lg\mbox{l}(k,\R)$, definido a menos de  conjuga\c c\~ao (a a\c c\~ao adjunta de $\GL{k,\R}$), j\'a que $\mbox{Hol}^0(\nabla)$ \'e um subgrupo de $\GL {k,\R}$ definido a menos de conjuga\c c\~ao. 
Da mesma forma $\Lh\mbox{ol}_x(\nabla)$ \'e a \'algebra de $\mbox{Hol}_x^0(\nabla)$ (que \'e um subgrupo de $\GL{E_x}$). Logo, \'e uma sub\'algebra de $\End(E_x)$.
Mesmo sendo $\mbox{Hol}^0(\nabla)$ um subgrupo de Lie de $\GL {k,\R}$, n\~ao \'e necessariamente um subgrupo fechado de $\GL {k,\R}$, e mesmo que o seja, $\mbox{Hol}(\nabla)$ n\~ao o \'e necessariamente.

\section{Admissible Connections and Exterior Covariant Derivative}

\subsubsection{Admissible Connections in A $G$-vector bundle}

Estudaremos agora conex\~oes que de uma forma ou de outra s\~ao compat\'iveis com estruturas em $E$, significando que o transporte paralelo manter\'a as propriedades que caracterizam essas estruturas.

Seja $(E,\pi,M)$ um $G$-fibrado vetorial sobre $M$.  
Dada uma trivializa\c c\~ao local, chamaremos de referencial admiss\'ivel \'aquele induzido pela base can\^onica de $\R^k$ atrav\'es desta trivializa\c c\~ao. N\'os sabemos que as fun\c c\~oes de transi\c c\~ao das trivializa\c c\~oes est\~ao, sobre todo ponto $x\in{M}$, dentro do grupo $G$, portanto, referenciais admiss\'iveis para $E$ s\~ao ligados por representa\c c\~oes de $G$, i.e.: sobre cada $x\in{M}$ as bases s\~ao ligadas por um \'unico $g\in{G}$. 

Por outro lado, dado um mapa de transi\c c\~ao de gauge qualquer  $\sigma:\theta\rightarrow{\rho(G)\subset{GL(k)}}$ e uma trivializa\c c\~ao local $\psi:\pi^{-1}\theta\rightarrow\theta\times\R^k$, n\'os vimos que $\tilde\psi$ dado por 
$$\psi\circ\tilde\psi^{-1}(x,u)=(x,\sigma(x)u)$$
 tamb\'em \'e trivializa\c c\~ao local do $G$-fibrado. Logo se  $\{e\}$ \'e  referencial admiss\'ivel  sobre $\theta$ dado por $\psi$, existe um \'unico referencial admiss\'ivel $\{\tilde{e}\}$ dado por $\tilde\psi$.   Logo dadas duas bases admiss\'iveis $\{{e_i(x)}\}_{i=1}^k$ e $\{{\tilde{e}_i(x)}\}_{i=1}^k$   sobre $x$, existe um \'unico $\sigma(x)=g\in{G}$ tal que\footnote{ Estamos identificando a representa\c c\~ao  de $g$, com $g$, i. e.: $\rho(g)=g$, ou seja, considerando a inclus\~ao $G\subset\GL k$.}  $\{{e_i}\}_{i=1}^k=\{{g\tilde{e}_i}\}_{i=1}^k$.  Logo, uma vez escolhida uma base inicial em uma fibra, existe uma bije\c c\~ao (que claramente n\~ao \'e can\^onica) entre as bases admiss\'iveis e o grupo $G$. 
 
Por exemplo, dado uma m\'etrica sobre $M$, o fibrado das bases ortonormais sobre $M$ corresponde ao $O(n)$-fibrado vetorial. Esta constru\c c\~ao ser\'a utilizada quando introduzirmos fibrados principais e os relacionarmos a fibrados vetoriais. 
    
Um mapa linear $T\in{L}(E_x, E_y) $ \'e chamado de $G$-mapa se leva base admiss\'ivel em base admiss\'ivel, ou seja, se dadas base admiss\'iveis em $x$ e $y$, a matriz de $T$ est\'a em $G$. Denotaremos o conjunto de tais mapas por $L_G(E_x,E_y)$. Se $x=y$, n\'os temos que $L_G(E_x,E_x)$ \'e o espa\c co de $G$-automorfismos, denotado por Aut$_G(E_x)=GL(E_x)$ e \'e um subgrupo de $\End(E_x)$, isomorfo \`a $G$. Logo n\'os temos que a \'algebra de Lie $\mathfrak{g}$ de $G$ \'e um subespa\c co de  End$(E_x)$ chamado de End$_\mathfrak{g}(E_x)$. 
\begin{defi}
$\nabla\in\mathcal{C}(E)$ \'e admiss\'ivel se para toda curva $\sigma:[0,1]\rightarrow{M}$, o transporte paralelo $P_\sigma$
for um $G$-mapa entre $E_{\sigma(0)}$ e $E_{\sigma(0)}$. \end{defi}

No exemplo anterior, do fibrado ortonormal sobre $M$, conex\~oes compat\'iveis com a m\'etrica, i.e.: tais que $\nabla\langle\cdot~,~\cdot\rangle=\langle\nabla(\cdot)~,~\cdot\rangle+\langle\cdot~,~\nabla(\cdot)\rangle$ claramente levar\~ao bases ortonormais em bases ortonomais por transporte paralelo. 

\begin{theo}\label{admiss}
 $\nabla$ \'e admiss\'ivel se e somente se, para todo $x\in M$ e todo $v\in T_xM$, em termos de bases admiss\'iveis, $\omega(v)\in\Lg$ no sentido que existe uma representa\c c\~ao $d\rho:\Lg\ra \End(E_x)$ tal que $\omega(v)=d\rho(h)$ para algum $h\in\Lg$. \end{theo}

{\bf{Dem:}}
Se  $\nabla $ \'e admiss\'ivel, ent\~ao para todo $v\in{T_xM}$ tal que $\sigma'(0)=v$, $\sigma:[0,1]\rightarrow{M}$, $P_\sigma(0,t)$ \'e um $G$-mapa, ou seja, leva bases admiss\'iveis em bases admiss\'iveis. Seja $E$ um $G$-fibrado sobre$M$,  
 e $\{{e_i}\}_{i=1}^k=e$ referencial sobre $\theta$. Designaremos uma base para $\sigma^*(E_{|\theta})$ dada por $\{{e_i}|_{(\sigma(t)}\}_{i=1}^k=\{\sigma^*{e_i}\}_{i=1}^k$. 
 Simplesmente utilizando (\ref{deriv}) para se\c c\~oes de campos do referencial, obtemos: 
\beq\label{prinvet}\omega(v)e_i(\sigma(0))=\frac{d}{d{t}}_{|t=0}\Big(P_\sigma(0,t)e_i(\sigma(t))\Big)\eeq
Mas $\Big((P_\sigma(0,t)e_1(\sigma(t)), \cdots , (P_\sigma(0,t)e_k(\sigma(t))\Big)=P_\sigma(0,t)e(\sigma(t))$ \'e uma curva lisade referenciais admiss\'iveis {\bf\it{em $\sigma(0)$}}. Tendo escolhido $e(\sigma(0))$ como referencial admiss\'ivel em 
$E_{\sigma(0)}$ h\'a uma bije\c c\~ao entre os referenciais admiss\'iveis e $G$, portanto existe curva lisa $g:[0,1]\rightarrow{G}$ para a qual  $$P_\sigma(0,t)e(\sigma(t))=\rho(g(t))e(0)$$ Claramente $g(0)=Id$, portanto 
\beq\frac{d}{d{t}}_{|t=0}P_\sigma(0,t)e(\sigma(t))=d\rho(g'(0))e(0)\eeq 
e n\'os obtemos que $g'(0)\in\Lg$. Portanto $$\omega(v)e(0)=d\rho(g'(0))e(0)$$ onde $e(0)$ \'e base em x, logo, em rela\c c\~ao a bases admiss\'iveis, $\omega(v)\in\Lg$ no sentido explicitado acima (omitiremos no restante dessa se\c c\~ao a representa\c c\~ao para facilitar a nota\c c\~ao).  

Por outro lado suponhamos  que $\omega(v)\in\Lg$ para todo $v$. Seja $e(\sigma(t)$ curva de referenciais admiss\'iveis sobre $\sigma$. Definimos  $\tilde{e}(\sigma(t)):=P_\sigma(t,0)e(\sigma(0))$. N\'os temos que existe $h:[0,1]\rightarrow{GL(k)}$  ~ tal que para todo $t\in[0,1]$ ~,~ $e(\sigma(t))=h(t)\tilde{e}(\sigma(t))$ . Claramente $h(0)=Id$. Agora 
\begin{gather*}
\omega(\sigma'(t))h(t)\tilde{e}(\sigma(t))=\omega(\sigma'(t))e(\sigma(t))=\nabla_{\sigma'(t)}e(\sigma(t))\\
=\nabla_{\sigma'(t)}h(t)\tilde{e}(\sigma(t))=h'(t)\tilde{e}(\sigma(t))+h(t)\nabla_{\sigma'(t)}\tilde{e}(\sigma(t))=h'(t)\tilde{e}(\sigma(t))\\ 
\therefore~~~\omega(\sigma'(t))h(t)=h'(t)\end{gather*}

Agora, usamos o seguinte lema provindo da teoria de grupos de Lie (para uma demonstra\c c\~ao ver \cite{Lie}):
\begin{lem}\label{CurvaG}
Dada uma curva lisa na \'algebra de Lie $\xi:[0,1]\rightarrow\Lg$ ~ , ~ existe uma \'unica curva lisa no grupo $G$ , ~$g:[0,1]\rightarrow{G}$ tal que $g(0)=Id$ e \beq\label{curvG}{g^{-1}(t)g'(t)=\xi(t)}\eeq \end{lem} 
N\'os sabemos que $G\subset{GL(k)}$ , $\Lg\subset\Lg{l(k)}$ e $h$ \'e ent\~ao a \'unica curva lisa em $GL(k)$ que satisfaz (\ref{curvG}). Mas pelo lema existe uma \'unica curva {\it{em}} $G$ que satisfaz (\ref{curvG}), logo $h:[0,1]\rightarrow{G}$. E finalmente temos $$e(\sigma(t))=h(t)P_\sigma(t,0)e(\sigma(0))\Longrightarrow{P}_\sigma(t,0)e(\sigma(0))=h^{-1}(t)e(\sigma(t))$$ 
e portanto o transporte paralelo leva base admiss\'ivel em base admiss\'ivel.  

\begin{prop}Se $\nabla$ \'e admiss\'ivel, ent\~ao $\Omega$ toma valores em $L_\Lg(E,E)\subset{L(}E,E)$\end{prop}
{\bf{Dem:}} Lembrando que podemos escrever a curvatura com $\Omega=d\omega+\omega\wedge\omega$ a demonstra\c c\~ao torna-se trivial, j\'a que $\Lg$ \'e subespa\c co vetorial fechado por comuta\c c\~ao. 

\subsubsection{Quasi-Canonical Gauge}

Seja $(\phi,\theta)$ um sistema de coordenadas convexo em $M$,  centrado em $x_0\in\theta$ , i.e.: tal que $\phi_{x_0}=0\in\R^n$. Como em $\R^n$ existe uma escolha privilegiada de curva para ligar um dado ponto $\phi(y)$ e a origem, i.e.: $t\phi(y)$, podemos transportar esta estrutura para $M$ e com a ajuda do transporte paralelo escolher um referencial privilegiado relacionado \`a carta $\phi$. I.e.: dada uma base $\{{v_i}\}_{i=1}^k$ de $E_{x_0}$ a estendemos a um referencial local $\{{e_i}\}_{i=1}^k:=e$ tal que $e(y)=P_\sigma(t,0)v$ onde definimos o ``raio'' ligando $x$ a $y$ como sendo $\sigma(t)=\phi^{-1}(t\phi(y))$ i.e.: $\phi(\sigma(t))=t\phi(p)$. 
Se $\nabla$ \'e uma conex\~ao admiss\'ivel e $v$ \'e uma base admiss\'ivel em $x$, ent\~ao $e(x)$ \'e uma base admiss\'ivel em $\theta$ e chamada de gauge quase-can\^onico para $E$ sobre $\theta$. 

Como j\'a mencionamos, equa\c c\~oes diferenciais ordin\'arias dependentes de par\^ametros, t\^em depend\^encia diferenci\'avel n\~ao s\'o nas condi\c c\~oes iniciais mas tamb\'em nos par\^ametros. Logo $e(x)$ \'e um referencial liso. Al\'em disso se $\omega$ \'e a forma de conex\~ao relativa \`a $e(x)$, $\omega(x_0)\equiv0$. 

{\bf{Dem:}}
Seja $u\in{T}_{x_0}$ e seja $\gamma:[-\epsilon,\epsilon]\rightarrow{M}$ tal que $\phi(\gamma(t))=td\phi(u)$, ent\~ao, como 
$$\frac{d}{d{t}}_{|t=0}\Big(\phi(\gamma(t))\Big)=d\phi(\gamma'(0))=d\phi(u)\therefore \gamma'(0)=u$$ 
N\'os temos que a reparametriza\c c\~ao $\tilde\gamma(t)=\gamma(\epsilon{t}):[0,1]\rightarrow{M}$ \'e o raio que liga $x$ a $\gamma(\epsilon)=y$. Ou seja, 
\beq\phi(\gamma(\epsilon{t}))=\epsilon{t}d\phi(u)=t\phi(\gamma(\epsilon))\Rightarrow\phi(\tilde\gamma({t}))=t\phi(y)\eeq 
Agora 
\beq\omega(u)(v_i)=\frac{d}{d{t}}_{|t=0}\Big(P_{\tilde\gamma}(0,t)e_i(\tilde\gamma(t)\Big)=\frac{d}{d{t}}_{|t=0}\Big(P_{\tilde\gamma}(0,t)P_{\tilde\gamma}(t,0)v_i\Big)=0\eeq
$\blacksquare$

\subsubsection{The  Exterior Covariant Derivative}
A derivada exterior que conhecemos at\'e agora, heuristicamente falando, mesmo quando aplicada em elementos de $\Gamma(\Lambda^p(TM^*)\otimes{E})$ s\'o opera nas formas a valores reais, e s\'o \'e v\'alida para formas a valores em fibrados triviais (j\'a que l\'a temos uma no\c c\~ao intr\'inseca de ``deixar campos constantes''). Queremos uma derivada exterior que n\~ao seja assim limitada, que derive tamb\'em a parte de se\c c\~oes, que leve em conta  tanto as caracter\'isticas usuais quanto a conex\~ao em $E$. A essa derivada chamaremos de derivada exterior de gauge e denotaremos por $D^\nabla_p$. Dado que $\Gamma(\Lambda^p(TM^*)\otimes{E})$ \'e gerado por elementos da forma $\lambda\otimes{s}$, onde $\lambda\in\Gamma(\Lambda^p(M))$ e $s\in\Gamma(E)$, observando o crit\'erio acima, a forma natural de fazer isso seria: 
 \beq\label{D}\widetilde{D}^\nabla_p(s\otimes{\lambda})=s\otimes{d\lambda}+\nabla{s}\tilde\wedge\lambda\eeq

J\'a que $\nabla{s}\in\Gamma(\Lambda(TM^*)\otimes{E})$ e como vimos na se\c c\~ao {\bf{1.1.3}}, existe um produto exterior natural $$\Gamma(\Lambda^{p_1}(TM^*)\otimes{E})\times{\Gamma}(\Lambda^{p_2}(TM^*))\rightarrow\Gamma(\Lambda(TM^*)^{p_1+p_2}\otimes{E})$$ 
que chamaremos genericamente a partir agora tamb\'em de $\wedge$ para quaisquer $p_1,p_2$, j\'a que implicitamente fazemos a identifica\c c\~ao do fibrado produto $\R\otimes{E}\simeq{E}$. Fa\c camos um r\'apido interl\'udio. Note o leitor que  inadvertidamente invertemos a ordem $\lambda\otimes{s}\rightarrow{s}\otimes\lambda$. Porque o fizemos? Notemos que j\'a para a derivada exterior da multiplica\c c\~ao de uma fun\c c\~ao $g:M\rightarrow\R$ por uma forma fechada $d\lambda$, n\'os temos que se usarmos a regra de Leibniz com uma ordem obtemos um resultado diferente do que com outra: 
$d(df\otimes{g})=df\wedge_\R{dg}=-dg\wedge_\R{df}=-d(g\otimes{df})$. Mas n\'os definimos a derivada exterior do produto de uma fun\c c\~ao por uma forma como sendo $d(g\otimes{df})$. Como estamos perseguindo uma analogia entre a derivada exterior e a derivada exterior covariante, escolhemos aqui a mesma ordena\c c\~ao na defini\c c\~ao. Como mostraremos essa condi\c c\~ao ser\'a tamb\'em necess\'aria para a compatibilidade das duas em um sentido que veremos adiante. No entanto, como estamos usando um conceito de ``formas a valores em fibrados vetoriais'' e portanto estamos ordenando os elementos como $\lambda\otimes{s}$, definimos: 
\beq\label{Dcerto}\widetilde{D}^\nabla_p(\lambda\otimes{s})=d\lambda\otimes{s}+(-1)^p\lambda\tilde\wedge\nabla{s}\eeq  

 A forma de (\ref{Dcerto}) sugere fortemente que a transi\c c\~ao equivale a fazer uma substitui\c c\~ao das derivadas normais de fun\c c\~oes a valores reais, para derivadas covariantes de se\c c\~oes de fibrados.  

Explicitando: como mencionamos anteriormente, localmente sempre podemos expandir $\eta\in\Gamma(\Lambda^p(TM^*)\otimes{E})$ de maneira un\'ivoca em termos de uma base $dx^I\otimes{e_i}$, onde $\{e_i\}_{i=1}^k$ \'e base local de $\Gamma(E)$ , $\{dx^i\}_{i=1}^n$ s\~ao coordenadas locais em $M$ e o superscrito mai\'usculo $I$ \'e a nota\c c\~ao de multi-\'indices (com ordem $p$). 

 Portanto localmente escrevemos $\eta=f_{\phantom{i}I}^idx^I\otimes{e_i}=dx^I\otimes{f_{\phantom{i}I}^i}e_i$, ent\~ao por (\ref{Dcerto}) temos, para $f_{\phantom{i}I}^i:\theta\subset{M}\rightarrow\R$,
$$\widetilde{D}^\nabla_p\eta=\widetilde{D}^\nabla_p(dx^I\otimes{f_{\phantom{i}I}^i}e_i)=(-1)^pdx^I\wedge\nabla(f_{\phantom{i}I}^ie_i)$$ 
Ou seja, ao generalizar formas a valores reais para formas a valores em fibrados vetoriais, devemos tamb\'em generalizar a equa\c c\~ao $d(f_I\otimes{d}x^I)=d(f_Idx^I)=df_I\wedge_\R{dx^I}=(-1)^p{dx^I}\wedge_\R{df_I}$, relamente substituindo a derivada exterior usual pela derivada covariante! 

Finalmente, dado $\nabla\in\mathcal{C}(E)$, devidamente motivados, simplesmente substitu\'imos em (\ref{derivext}) a deriva\c c\~ao por $X$ pela deriva\c c\~ao $\nabla_X$ , ou seja, $d\rightarrow\nabla$. Definimos ent\~ao os mapas lineares: 
$$D^\nabla_p=D_p:\Gamma(\Lambda^p(M)\otimes{E})\rightarrow\Gamma(\Lambda^{p+1}\otimes{E})$$ 
que, para $X_1,\cdots,{X}_{p+1}\in\Gamma(TM)$, \'e dado por 
\begin{multline}\label{Derivext}(D_p\eta)(X_1,\cdots,{X}_{p+1})=\sum_{i=1}^{p+1}(-1)^{i+1}\nabla_{X_i}\eta(X_1,\cdots,\hat{X}_i,\cdots{X}_{p+1})+\\ 
\sum_{i=1}^{p+1}(-1)^{i+j}\eta([X_i,X_j],X_1,\cdots,\hat{X}_i,\cdots,\hat{X}_j,\cdots,X_{p+1})\end{multline} 

Para mostrar que $D_p\eta\in\Gamma(\Lambda^{p+1}\otimes{E})$ , basta notar que por defini\c c\~ao $D_p\eta$ \'e totalmente anti-sim\'etrica e que \'e $C^\infty(M)$ linear em cada entrada, j\'a que $X[f]=\nabla_Xf$, e portanto a demonstra\c c\~ao \'e exatamente a mesma que para (\ref{derivext}). Com essa defini\c c\~ao n\~ao estamos mais restritos a tomar elementos da forma $\lambda\otimes{s}$, o que \'e \'util j\'a que s\'o podemos decompor um elemento de $\Gamma(\Lambda^p(TM^*)\otimes{E})$  de forma \'unica para cada base, i.e.: localmente.

\begin{theo} Se $ \nabla^i$ \'e uma conex\~ao em $E_i$, $i=1,2$ e $\nabla$ \'e a conex\~ao correspondente em $E_1\otimes{E_2}$, ent\~ao, para $\eta^1\in\Gamma(\Lambda^{p_1}(TM^*)\otimes{E_1})$ e $\eta^2\in\Gamma(\Lambda^{p_2}(TM^*)\otimes{E_2})$ temos 
\beq{D^\nabla_{p_1+p_2}(\eta^1\tilde\wedge\eta^2)=D^{\nabla_1}_{p_1}\eta^1\tilde\wedge\eta^2+(-1)^{p_1}\eta^1\tilde\wedge{D}^{\nabla_2}_{p_2}\eta^2}\eeq\end{theo}

{\bf{Dem:}} Se escolhermos em $x$ um gauge quase can\^onico, ent\~ao as formas de conex\~ao relativas a $\nabla^i$ s\~ao nulas, i.e.: $\omega^i(x)=0$, logo $\nabla^i=d$ e portanto $\nabla=d\otimes{Id}+Id\otimes{d}$ que \'e simplesmente a regra da cadeia. Portanto a equa\c c\~ao (\ref{Derivext}) volta \`a forma de (\ref{derivext}), para a qual j\'a demonstramos que vale o teorema.  

Gauges quase can\^onicos s\~ao instrumentos poderosos na teoria de fibrados. Para ilustrarmos o quanto eles podem facilitar nossa vida, bem como para demonstrarmos  como bases locais nos podem ser \'uteis, daremos outra demonstra\c c\~ao do teorema, mais trabalhosa, mas tamb\'em elucidativa. 

 Expressemos $\eta^1$ e $\eta^2$ localmente por uma base como acima, i.e.: usando multi-\'indices $I$ e $J$ de ordem $p_1$ e $p_2$ respectivamente, $\{e_i\}$, $\{b_i\}$ bases de $\Gamma(E_1)$ e  $\Gamma(E_2)$, respectivamente, e  $f_{\phantom{i}I}^i,g_{\phantom{b_j}J}^j:M\rightarrow\R$, temos: $\eta^1=f_{\phantom{i}I}^idx^I\otimes{e_i}$, $\eta^2=g_{\phantom{i}J}^jdx^J\otimes{b_j}$. Utilizando (\ref{Dcerto}) e $\nabla=\nabla^1\otimes{Id}+Id\otimes\nabla^2$: 
\begin{eqnarray*}
\eta^1\tilde\wedge\eta^2&=&(f_{\phantom{i}I}^ie_i\otimes{g}_{\phantom{i}J}^j{b_j})\otimes(dx^I\wedge_\R{dx}^J)\\
D^\nabla_{p_1+p_2}(\eta^1\tilde\wedge\eta^2)&=&\nabla(f_{\phantom{i}I}^ie_i\otimes{g_{\phantom{i}J}^j}{b_j})\otimes(dx^I\wedge_\R{dx^J})\\ 
~&=&\Big(\nabla^1(f_{\phantom{i}I}^ie_i)\otimes{g_{\phantom{i}J}^j}{b_j}+f_{\phantom{i}I}^ie_i\otimes\nabla^2({g_{\phantom{i}J}^j}{b_j})\Big)(dx^I\wedge_\R{dx}^J)\\
~&=&D^{\nabla^1}_{p_1}({f}_{\phantom{i}I}^ie_i\otimes{d}x^I)\tilde\wedge({g}_{\phantom{i}J}^jb_j\otimes{dx}^J)+(-1)^{p_1}({f}_{\phantom{i}I}^ie_i\otimes{d}x^I)\tilde\wedge{D}^{\nabla^2}_{p_2}({g}_{\phantom{i}J}^jb_j\otimes{dx}^J)\\ 
~&=&D^{\nabla^1}_{p_1}\eta^1\tilde\wedge\eta^2+(-1)^{p_1}\eta^1\tilde\wedge{D}^{\nabla^2}_{p_2}\eta^2
\end{eqnarray*}
{{$\blacksquare$}}\medskip
 
 A partir de agora abreviaremos $D^{\nabla}_{p}$ por $D$, deixando os \'indices subentendidos no contexto.  
 Por esse resultado, tiramos imediatamente dois corol\'arios \'uteis:
\begin{prop} $\lambda^1$ for uma $p$-forma e $\lambda^2$  uma $p'$-forma a valores em $E$, n\'os temos: 
$$D(\lambda^1\wedge\lambda^2)=d\lambda^1\wedge\lambda^2+(-1)^p\lambda^1\wedge{D\lambda^2}$$\end{prop} 
\begin{prop} Se $\lambda\in\Gamma(\Lambda^p(TM^*))$ e $s\in\Gamma(E)$, ent\~ao $${D}^\nabla_p(\lambda\otimes{s})=d\lambda\otimes{s}+(-1)^p\lambda\wedge\nabla{s}$$\end{prop}
Ou seja, a opera\c c\~ao de $D$ sobre elementos dessa forma \'e a apropriada ($D$ estende $\widetilde{D}$).  

De onde tiramos que, ao localmente escrevermos para $\eta\in\Gamma(\Lambda^p(TM^*_{|\theta})\otimes{E_{|\theta}})$ em rela\c c\~ao a uma base $\{e_i\}$ de $\Gamma(E)$, i.e.: $\eta=\eta^i\otimes{e_i}$ para $\eta^i\in\Gamma(\Lambda^p(TM^*_{|\theta}))$ obtemos: 
$$D\eta=D(\eta^i\otimes{e}_i)=d(\eta^i)\otimes{e}_i+(-1)^p\eta^i\wedge(\omega(e_i))=d(\eta^i)\otimes{e_i}+\omega\wedge(\eta^i\otimes{e_i})$$
Logo \beq\label{Dlocal}D\eta=d\eta+\omega\wedge\eta\eeq
Onde utilizamos a derivada exterior sobre formas em fibrados triviais, $d$ e em $\omega\wedge\eta$ o produto exterior que incorpora a a\c c\~ao bilinear $\End(E)\otimes{E}\rightarrow{E}$, resultante da a\c c\~ao da forma de conex\~ao sobre as se\c c\~oes de $E$.

N\'os sabemos que  $d^2=d\circ{d}=0$ expressa justamente o fato de que as derivada ordin\'arias usuais comutam. Isto \'e, se $f:M\rightarrow\R$, $X,Y\in\Gamma(TM)$, calculamos
\beq\label{123456}d(df)(X,Y)=X[df(Y)]-Y[df(X)]-df([X,Y])=(XY-YX-[X,Y])f=0\eeq 
Novamente, fazendo a substitui\c c\~ao $d\rightarrow{D}$, i.e.: $X\rightarrow\nabla_X$ em (\ref{123456}) obtemos 
$$\nabla_X\nabla_Y-\nabla_Y\nabla_X-\nabla_{[X,Y]}=\Omega(X,Y)$$
Logo devemos ter $D_1\circ{D_0}=i_\Omega$ (produto interior por $\Omega$). De fato, se $\eta$ \'e uma p-forma a valores em $E$, utilizando (\ref{Dlocal}) temos: 
\begin{eqnarray*}
D(\eta)&=&d\eta+\omega\wedge\eta\\
{D}(D(\eta))&=&d\omega\wedge\eta-\omega\wedge{d\eta}+\omega\wedge{d\eta}+\omega\wedge\omega\wedge\eta\\
~&=&\Omega\wedge\eta
\end{eqnarray*} 

\begin{theo}
Dada conex\~ao $\nabla$ em $E$, seja $\widetilde\nabla$ a conex\~ao induzida em $\End(E)\simeq{E}^*\otimes{E}$. Se $\omega$ \'e a matriz de 1-formas de conex\~ao para $\nabla$ relativa \`a base $\{e_i\}$ e $\alpha$ \'e uma p-forma em $M$ a valores em $\End(E)$ , utilizando o produto externo natural entre formas a valores em $\End(E)$ que descrevemos na se\c c\~ao {\bf{ 1.1.3}} temos:
\beq\label{Dtio}D^{\widetilde\nabla}\alpha=d\alpha+\omega\wedge\alpha-(-1)^p\alpha\wedge\omega\eeq
\end{theo} 

{\bf{Dem:}}
Novamente, utilizando (\ref{Dlocal}) para formas a valores no fibrado vetorial $\End(E)$, temos:
$$D^{\widetilde\nabla}\alpha=d\alpha+\widetilde\omega\wedge\alpha$$
 onde assim como em  (\ref{Dlocal}), temos de deixar a a\c c\~ao da forma  sobre as se\c c\~oes do fibrado impl\'icitas no produto exterior, o que aqui, ao inv\'es de induzir a identifica\c c\~ao $\End(E)\otimes{E}\rightarrow{E}$,  induz $L(\End(E),\End(E))\otimes\End(E)\rightarrow{\End(E)}$. 

Se $\widetilde\omega$ \'e a conex\~ao de $\widetilde\nabla=\nabla\otimes\mbox{Id}+\mbox{Id}\otimes\nabla^*$,
lembremos que $\widetilde\omega=\omega\otimes\mbox{Id}+\mbox{Id}\otimes\omega^*$.
Mas como vimos, na identifica\c c\~ao $\End(E)\simeq{E}\otimes{E}^*$, temos  $\mbox{Id}\otimes\omega^*=-\mbox{Id}\otimes\omega^T$. Utilizaremos ainda que $\omega^T(e^j)=e^j\omega$, fato j\'a comentado. Portanto escrevendo localmente $\alpha=\alpha^i_j\otimes{e_i\otimes}e^j$ obtemos:
\begin{eqnarray*}
D\alpha&=&D(\alpha^i_j\otimes{e}_i\otimes{e}^j)\\
~&=&d(\alpha^i_j)\otimes({e}_i\otimes{e_j})+(-1)^p\alpha_j^i\wedge\left(\omega(e_i)\otimes{e}^j- 
e_i\otimes\omega^T({e}^j)\right)\\
~&=&d(\alpha_j^i)\otimes{e_i}\otimes{e^j}+\omega\wedge(\alpha_j^i\otimes{e_i}\otimes{e^j})-(-1)^p\alpha_j^i\wedge{e}_i\otimes{e}^j\omega\\
~&=&d\alpha+\omega\wedge\alpha-(-1)^p\alpha\wedge\omega 
\end{eqnarray*}
J\'a que a multiplica\c c\~ao exterior de matrizes de formas \'e simplesmente a multiplica\c c\~ao de matrizes usando o produto externo em cada termo. 
{{$\blacksquare$}}

\begin{prop} Deste teorema emergem os seguintes corol\'arios: 

\begin{itemize}
\item[(i)] {A identidade de Bianchi (\ref{Bianchi}) \'e equivalente \`a $D\Omega=0$.}
\item[(ii)] {$D\omega=\Omega$}
\end{itemize}
\end{prop}
{\bf{Dem:}} N\'os temos  $d\Omega=\omega\wedge\Omega-\Omega\wedge\omega$, portanto tomando no \'ultimo teorema $\alpha=\Omega$ obtemos $$D\Omega=d\omega+\omega\wedge\Omega-\Omega\wedge\omega$$ O segundo item se verifica similarmente tomando $\alpha=\omega$ em (\ref{Dlocal}).  
{{$\blacksquare$}}\medskip

Agora, dada conex\~ao $\nabla$ em $E$ e $\kappa\in\Gamma(\Lambda^1(TM^*)\otimes\End(E))$, definimos $\nabla^\kappa=\nabla+\kappa$. Se tivermos uma base local isto \'e equivalente a tomar, para a forma de conex\~ao de $\nabla^\kappa$, $\omega^\kappa=\omega+\kappa$ ent\~ao obtemos: 
\begin{gather*}d(\omega+\kappa)+(\omega+\kappa)\wedge(\omega+\kappa)=d\omega+d\kappa+\omega\wedge\omega+\kappa\wedge\omega+\omega\wedge\kappa+\kappa\wedge\kappa=\\
\Omega+(d\kappa+\omega\wedge\kappa-(-1)^1\kappa\wedge\omega)+\kappa\wedge\kappa\end{gather*} Logo obtemos a partir de (\ref{Dtio}): 
\beq\label{kappa}\Omega^\kappa=\Omega+D\kappa+\kappa\wedge\kappa\eeq

N\'os voltaremos \`a identidade de Bianchi quando discutirmos as equa\c c\~oes de Yang-Mills. N\'os veremos por exemplo que em eletromagnetismo ela representa conseva\c c\~ao de carga, e \'e verdade que em relatividade geral, pela equa\c c\~ao de Einstein  ela \'e equivalente \`a conserva\c c\~ao local de energia e momento.

\chapter{Principal Fiber Bundles}
\begin{quote} {\it{O Universo \'e embasado em um plano, um plano cuja profunda simetria est\'a de alguma forma presente na estrutura interna do nosso intelecto.}} - Paul Valery \medskip

{\it{Einstein em seu tempo trabalhou incessantemente para construir ``um sistema completo de f\'isica te\'orica". Ele procurou os `` conceitos e princ\'ipios fundamentais'' que permitiriam uma grande s\'intese da estrutura do mundo real. Centrais a essa s\'intese est\~ao as for\c cas, ou intera\c c\~oes que mant\^em unida a mat\'eria, que produzem a pletora de rea\c c\~oes de que consistem os fen\^omenos naturais.
Eu acredito que ainda estamos hoje muito longe desta grand s\'intese com a qual sonhou Einstein. Mas n\'os temos um de seus elementos  chave: o princ\'ipio que intera\c c\~oes s\~ao regidas por simetrias, utilizado primeiramente pelo pr\'oprio Einstein.}} - C.N. Yang
\end{quote}  

\section{Foundations}

Em muitos sentidos, a Teoria de Gauge \'e uma teoria que procura a natureza intr\'inseca das intera\c c\~oes, aquilo que independe das formas como elas s\~ao representadas. Como um objeto  ao meio de uma roda de observadores, cuja descri\c c\~ao \'e feita a partir de diferentes \^angulos (`` cada uma delas correta''), a Teoria de Gauge tenta desemaranhar propriedades inerentes de propriedades descritivas. Assim como para o objeto, as descri\c c\~oes s\~ao relacionadas por transforma\c c\~oes de simetria, por um grupo de simetria. Estudaremos agora uma forma de estudarmos as rela\c c\~oes entre as descri\c c\~oes feitas por estes diferentes observadores, ou referenciais.

Ao introduzirmos bases admiss\'iveis na se\c c\~ao {\bf{1.4}}, chamamos $\{{e_i}\}_{i=1}^k$ e $\{{\tilde{e}_i}\}_{i=1}^k$ de bases admiss\'iveis  de $E_x$ se existisse $g\in{G}$ tal que $\{{e_i}\}_{i=1}^k=\{{g\tilde{e}_i}\}_{i=1}^k$. Este $g$ \'e \'unico j\'a que existe um \'unico $g\in{GL(k)}\supset{G}$ que leva uma na outra, ent\~ao conclu\'imos que uma vez escolhida uma base inicial (uma origem), existe uma bije\c c\~ao entre as bases admiss\'iveis e o grupo $G$. Notemos ainda que n\~ao h\'a uma base que se destaca das outras, n\~ao h\'a uma base que possa ser considerada canonicamente como a identidade. Como veremos ao final deste cap\'itulo, o fibrado dos referenciais admiss\'iveis  de $E$, constitui uma ponte entre as no\c c\~oes de fibrado vetorial e principal. 

Motivados por essas constru\c c\~oes, definimos:
\begin{defi}
Uma variedade diferenci\'avel $P$ \'e chamada de fibrado principal com grupo associado $G$, se $G$ age livre e diferenci\'avelmente sobre $P$. Ou seja, se existe a\c c\~ao de $G$ em $P$ : $G\times{P}\rightarrow{P}$ lisa e tal que para cada $p\in{P}$  o grupo de isotropia de $p$ \'e a identidade, i.e.:
$$G_p=\{y\in{G} ~|~ gp=p\}=\{e\}$$ 
\end{defi}
Naturalmente constru\'imos uma proje\c c\~ao em $P$ ,  $\pi:P\rightarrow{M}$, dada por $p\simeq{q}\Leftrightarrow{p=g\cdot{q}}$ para algum $g\in{G}$ . Definimos ent\~ao o espa\c co base $M$ como sendo o espa\c co das \'orbitas de $P$ , $M=P/G$, com a topologia quociente, i.e.: caracterizada por $\pi$ ser aberta e cont\'inua. Pela defini\c c\~ao, $G$ age transitivamente sobre cada fibra.

\subsubsection{ Lie Group Actions}

Antes de come\c car o tratamento de fibrados principais em si, precisamos de algumas ferramentas da teoria de grupos de Lie: 
Seja, $G$ um grupo Lie compacto e $\Lg$ sua \'algebra de Lie, ou seja,
$\Lg\simeq{T_e{G}}$. 

Definimos a a\c c\~ao de conjuga\c c\~ao do grupo  como sendo:
\begin{flalign*}
\zeta(g):G\rightarrow{G}\\ 
a\mapsto{g}ag^{-1}\end{flalign*}
Cuja derivada em $e\in{G}$ denotaremos por $\Ad(g)=d(\zeta(g))_e$. \'E f\'acil ver que $\zeta(g)\zeta(h)=\zeta(gh)$,  ent\~ao pela regra da cadeia, $\Ad(gh)=d(\zeta(gh))_e=d(\zeta(g))_ed(\zeta(h))_e=\Ad(g)\Ad(h)$, logo $\Ad:G\ra \Aut(\Lg)$ \'e uma representa\c c\~ao do grupo sobre a \'algebra.  
Definimos ${\exp}tX$ como sendo o \'unico subgrupo a 1-par\^ametro tangente a $X$ em $e$.

\begin{prop}{$\zeta(g)({\exp}tX)={\exp}(t\Ad{(g)}X)$}
\end{prop}

{\bf{Dem:}} N\'os temos que $\zeta(g)e=e$,~ e ~$\zeta(g)(ab)=\zeta(g)a\zeta(g)b$~,~ isto \'e: ~ $\zeta(g)$ ~ \'e automorfismo de $G$, logo leva subgrupo a 1-par\^ametro em subgrupo \`a 1-par\^ametro.
Logo $\zeta(g){\exp}tX$ \'e subgrupo a 1-par\^ametro que passa pela origem, com tangente $$\frac{d}{dt}{|_{t=o}}\zeta(g){\exp}tX=\Ad(g)X$$ 
 Por outro lado, ${\exp}(t\Ad(g)X)$ \'e o \'unico subgrupo a 1-par\^ametro que passa pela origem com tangente $\Ad(g)X$.
{{$~~~~~\blacksquare$}}\medskip

Seja $P$ uma variedade onde $G$ age como grupo de difeomorfismos (a\c c\~ao que denotaremos por `$\cdot$'. Ent\~ao, para todo $p\in{P}$, $X\in\Lg$ e $t\in\R$, n\'os temos uma a\c c\~ao 
\begin{align*}
{\exp}(tX): & P  \longrightarrow  {P}\\ 
\phantom{{\exp}t} & p  \longmapsto  {\exp}(tX)\cdot{p}
\end{align*}
Logo definimos
\begin{align*}
\mbox{I}_p: & \Lg\longrightarrow{T_pP}\\
\phantom{Ip} & X\longmapsto{\frac{d}{dt}{|_{t=o}}}({\exp}(tX)\cdot{p}) 
\end{align*}
Portanto,
\begin{align*}
\mbox{I}(X): & P\rightarrow\Gamma(TP)\\
\phantom{a} & p\mapsto{\frac{d}{dt}{|_{t=o}}}({\exp}(tX)\cdot{p})\end{align*}
\'E claro que ${\exp}(tX)\cdot{p}=p\Leftrightarrow{\mbox{I}_p}(X)=0$ e, se a a\c c\~ao do grupo for livre, $\I_p$ \'e injetora. Al\'em disso n\'os temos que: 
 $$Im\mbox{I}_p=\{\tilde{X}\in{T_pP}~|~\tilde{X}=\frac{d}{dt}{|_{t=o}}({\exp}(tX)\cdot{p})~ \mbox{para algum}~ X\in\Lg\}$$
Por outro lado, ${\exp}$ \'e difeomorfismo local e, portanto, ${\exp}(tX)\cdot{p}$ gera $G\cdot p$ em uma vizinhan\c ca ao redor de $p$. Logo, $T_pG\cdot p=Im\mbox{I}_p$. Denotando a aplica\c c\~ao `$\cdot$'$:G\times{P}\ra{P}$ por $\mu$, fica f\'acil ver que $\I_p$ \'e linear: 
$$\I_p(X)=\frac{d}{dt}{|_{t=o}}\left(\Phi({\exp}(tX),p)\right)=d\mu_p(X,0)$$ 

Al\'em disso n\'os temos, pela \'ultima proposi\c c\~ao que 
\begin{eqnarray*}
\I_p\left(\Ad(g^{-1})X\right)=\frac{d}{dt}{|_{t=o}}({\exp}t\left(\Ad(g^{-1})X\right)\cdot{p})=\frac{d}{dt}{|_{t=o}}
(g^{-1}{\exp}(tX)g\cdot{p})\\ 
~~\therefore~~{d}g(\I_p\left(\Ad(g^{-1})X\right))=\frac{d}{dt}{|_{t=o}}({\exp}(tX)g\cdot{p})=\I_{g\cdot{p}}(X)\end{eqnarray*} 
Obtemos ent\~ao as duas propriedades
\begin{itemize}
\item{$Im(\I_p)=$ espa\c co tangente \`a \'orbita $G\cdot p$ em $p$. } 
\item{A a\c c\~ao de I em um ponto $q$ transladado do ponto $p$ por um elemento $g$ de $G$ \'e relacionada \`aquela no ponto $p$ da seguinte forma:
\beq\label{Ip}\I_{g\cdot{p}}(X)=dg(\I_p\left(\Ad(g^{-1})X\right))\eeq}\end{itemize} 

\subsubsection{ $G$-invariant M\'etrics}
Na teoria de fibrados principais, aten\c c\~ao especial deve ser dada \`a estruturas $G$-ivariantes, isto \'e, \`as estruturas em $P$ que n\~ao se alteram sob a a\c c\~ao do grupo. Este em si \'e um t\'opico rico e profundo, que n\~ao abordaremos em sua generalidade, mas somonte no tocante \`a uma estrutura: a m\'etrica. Para uma abordagem mais profunda veja \cite{Tausk}. 
\begin{defi}Seja $\gamma$ uma m\'etrica sobre $P$, dizemos que $\gamma$ \'e $G$-invariante se $G$ est\'a contido no grupo de isometrias de $\gamma$. I.e.: se para todo $g\in{G}$ , $g^*\gamma=\gamma$.\end{defi}

Uma quest\~ao que naturalmente surge \'e sobre a generalidade da exist\^encia de tais m\'etricas. Mostraremos que sempre existem para o caso de $G$ compacto e conexo. 

\begin{defi} A m\'edia de $\gamma$ por $G$ (compacto, conexo e de dimens\~ao$=m$) \'e dada ponto a ponto por:
        $$\tilde\gamma=\int_{g\in{G}}g^*(\gamma)\nu$$
onde $\nu$ \'e uma m-forma volume bi-invariante. (que sempre existe se $G$ \'e compacto e conexo).\end{defi} 
\begin{theo}\label{metricGinv} Seja $\gamma$ uma m\'etrica em uma variedade riemanniana $P$ onde $G$ age como grupo de difeomorfismos, ent\~ao a m\'edia de $\gamma$ \'e $G$-invariante.\end{theo}
{\bf{Dem:}} Sejam $u,v\in{T_pP}$, $h\in{G}$, definimos $f:G\rightarrow\R$ por $f(g)=g^*\gamma(u,v)$, ent\~ao 
\begin{eqnarray}
\tilde\gamma(u,v)=\int_{g\in{G}}\gamma(g_*(u),g_*(v))\nu=\int_{g\in{G}}f(g)\nu~~~\therefore\\
h_*\tilde\gamma(u,v)=\tilde\gamma(h_*u,h_*v)=\int_{g\in{G}}\gamma\Big(g_*(h_*(u)),g_*(h_*(v))\Big)\nu=\\ 
\int_{g\in{G}}\gamma\Big(g_*h_*(u),g_*h_*(v))\Big)\mbox{R}^*_h(\nu)=\int_{g\in{G}}(\mbox{R}^*_hf)(g)\mbox{R}^*_h(\nu)\end{eqnarray}
Mas a transla\c c\~ao \`a direita R$_h:G\rightarrow{G}$ \'e um difeomorfismo que preserva a orienta\c c\~ao, logo, $h_*\tilde\gamma=\tilde\gamma$. Dado que toda variedade diferenci\'avel admite m\'etrica riemanniana, o teorema est\'a provado.   ~~~~~$~~~~\blacksquare$ 

\subsubsection{Sections on a Principal Bundle}
 Tentaremos agora, assim como assumimos em fibrados vetoriais, demonstrar que existe uma estrutura local de variedade produto tamb\'em para os fibrados principais, onde a fibra t\'ipica coincide com o grupo associado, $G$. 
\begin{defi} Seja $\theta$ um aberto em $M$, definimos uma se\c c\~ao local de $P$ sobre $\theta$ como uma subvariedade $\Sigma$ de $P$ tal que $\Sigma$ \'e transversal \`as \'orbitas,~ $T_p\Sigma+\I_p(\Lg)=T_pP$,~ e $\Sigma$ intersecta \'orbitas em um \'unico ponto, i.e.: se $p\in\Sigma$ ent\~ao $G\cdot{p}\cap\Sigma=\{p\}$. \end{defi}

\begin{theo} Dado $p\in{P}$, existe uma se\c c\~ao local $\Sigma$ de $P$ contendo $p$. \end{theo}
 
{\bf{Dem:}} A id\'eia da prova \'e, usando uma m\'etrica invariante em $P$, exponenciar os vetores ortogonais a $G\cdot{p}=:N$~, de comprimento $\delta$ , obtendo uma outra subvariedade, $\Sigma$, transversal a $N$. Devemos tomar o cuidado necess\'ario para que  $\delta$ seja suficientemente pequeno de modo que haja uma s\'o intersec\c c\~ao entre 
$\Sigma$ e $N$. Provaremos um caso mais geral e depois mostraremos que nosso caso se encaixa. Provamos antes de mais nada que $G\cdot p$ \'e subvariedade mergulhada:

 Para todo $p\in P$, definimos o mapa suave $\theta^{(p)}:G\ra P$ por $\theta^{(p)}(g)=g\cdot p$, ou seja a \'orbita de $p$. Como o grupo age livremente \'e trivial ver que $\theta^{(p)}$ \'e injetora. Al\'em disso como existe identifica\c c\~ao can\^onica 
 $d\theta^{(p)}\simeq\I_p$, por (\ref{Ip}) o posto de $\theta^{(p)}$ \'e constante sobre $G$. Agora pelo teorema do posto, existem cartas apropriadas de $G$ e $P$ tal que a representa\c c\~ao local de $\theta^{(p)}$, que chamamos de $\widetilde\theta^{(p)}$, pode ser escrita como $\widetilde\theta^{(p)}(x^1,\dots,x^n)=(x^1,\dots, x^j,0,\dots,0)$ onde $n=\mbox{dim}{G}$. Mas como \'e injetora, $j=n$, i.e.: \'e uma imers\~ao. Agora, como $G$ \'e compacto, n\'os temos uma imers\~ao injetora de um compacto, que \'e portanto mergulho.  

Seja ent\~ao $N$ subvariedade compacta  de $P$, definimos o fibrado normal de N:
$$\nu{N}=\{(x,v) | x\in{N}~,~v\in{\nu_xN}\}$$
\'E claro  que $\nu{N}$ \'e fibrado vetorial, j\'a que $TN$ \'e distribui\c c\~ao $C^\infty$ em $TP$ e portanto o seu ortogonal tamb\'em \'e uma distribui\c c\~ao lisa, o que transforma $\nu{N}$ em sub-fibrado de $TP$. Achemos ent\~ao uma trivializa\c c\~ao local de $\nu{N}$. 

 Tomamos primeiramente uma carta de $P$ adaptada \`a $N$, $\psi:U\rightarrow \widetilde{U}\subset\R^m$ . Agora, pela propriedade de carta adaptada \`a subvariedade, temos que a proje\c c\~ao nas $n$ primeiras coordenadas, ${pr}^n\circ\psi:N\cap{U}\rightarrow\R^n\cap\widetilde{U}$ \'e carta de $N\cap{U}$. Seja $k=m-n$, e ${pr}^k$ a proje\c c\~ao nas \'ultimas $k$ coordenadas. Claramente $f:={pr}^k\circ\psi$ \'e submers\~ao e $U\cap{N}=f^{-1}(0)$, al\'em disso $\mbox{Ker}df_p=T_pN$. Agora\footnote{ A forma mais \'obvia de exibirmos uma subvariedade transversal a $N$ em $p$ seria simplesmente tomando a subvariedade dada por $\Sigma=\psi^{-1}(\R^k\cap\widetilde{U})$, ou seja, os pontos $\psi^{-1}(0,\dots,0,x^{n+1},\dots,x^m)$ onde $(0,\dots,0,x^{n+1},\dots,x^m)\in\widetilde{U}$. No entanto, a propaga\c c\~ao desta constru\c c\~ao ao longo da fibra de forma $G$-invariante se torna mais complicada do que o que faremos aqui utilizando a exist\^encia de uma m\'etrica $G$-invariante.} 
, induzindo uma m\'etrica em $R^m$ por $\psi$ (e portanto em $\R^k$ por $f$), denotamos o adjunto de $df_p$ por $df^*_p$, que \'e definido por
dados $u\in{T_pP}$ e $w\in\R^k=\mbox{Im}df_p$:
 $$\langle{df_p(u)~,~w}\rangle_{\R^k}=\langle{u~,~df^*_p(w)}\rangle_{T_pP}$$ Claramente,  como $df_p$ \'e isometria, $df^*_p=df_p^{-1}$. Agora se $u\in{T_pN}$ , qualquer que seja $w\in\R^k$, $\langle{u}~,~df_p^{-1}(w)\rangle_{T_pP}=0$. Isto \'e, $\mbox{Im}df_p^{-1}\subset{T}_pN^\perp$, mas ambos t\^em dimens\~ao $k$, logo $df_p^{-1}:\R^k\rightarrow{\nu_pN}$ \'e isomorfismo. Logo constru\'imos uma trivializa\c c\~ao local para $\nu{N}$ dada por \begin{align*} 
\phi:~ & \nu{N}_{|N\cap{U}}\longrightarrow({U}\cap{N})\times\R^k\\
\phantom{\phi:~} & (p,u)\longmapsto(p,df_p(u))\end{align*}

Utilizando a constru\c c\~ao de fibrado tangente (que vimos nos exemplos da {\bf{Se\c c\~ao 1}}) \'e f\'acil vermos que trivializa\c c\~oes dadas por cartas adaptadas $\psi$, $\widetilde\psi$,  compat\'iveis, ser\~ao tamb\'em compat\'iveis. Chamaremos a proje\c c\~ao suave deste fibrado vetorial de $\pi_N:\nu{N}\lra{N}$.   Estudemos ent\~ao a aplica\c c\~ao $\Exp: \nu{N}\rightarrow{P}$, que \'e simplesmente a restri\c c\~ao da aplica\c c\~ao exponencial usual em $TP$ \`a distribui\c c\~ao normal a $N$. Todo cuidado \'e pouco ao estudarmos fibrados tangentes de fibrados vetoriais, por isso para facilitar a vizualiza\c c\~ao utilizamos que $(U\cap{N})\times\R^k$ \'e fibrado trivial, e, para $p\in{N\cap}U$, qualquer vetor $\xi\in{T_{(p,0)}}\nu{N}$ \'e dado por $d\phi_{(p,0)}^{-1}\tilde\xi$ para algum $$\tilde\xi=(w,u)\in{T_{(p,0)}}((U\cap{N})\times\R^k)=T_p(U\cap{N})\times{T_0\R^k}$$ 
que por sua vez  \'e tangente a uma curva $(\gamma(t),tu)$, onde $\gamma:[0,1]\rightarrow{N\cap}U$. Ou seja, fazendo $(p,0)=q$: 
\begin{gather*}
\xi=d\phi_q^{-1}(w,u)=d\phi_q^{-1}(w,0)+d\phi^{-1}_q(0,u)\\
=d\phi_q^{-1}\left(\frac{d}{dt}{|_{t=0}}\Big((\gamma(t),0)\Big)\right)+d\phi^{-1}_q\left(\frac{d}{dt}{|_{t=0}}\Big((p,tu)\Big)\right) 
=\frac{d}{dt}{|_{t=0}}\Big(\phi^{-1}(\gamma(t),0)\Big)+\frac{d}{dt}{|_{t=0}}\Big(\phi^{-1}(p,tu)\Big)\\
=\frac{d}{dt}{|_{t=0}}\Big((\gamma(t),0)\Big)+\frac{d}{dt}{|_{t=0}}\Big((p,tdf_p^{-1}(u))\Big)=(w,0)+(0,df_p^{-1}(u)) 
\end{gather*}
Obtemos uma identifica\c c\~ao can\^onica $T_{(p,0)}\nu{N}\simeq{T_pN}\oplus\nu_pN$. Finalmente, fazendo\footnote{Chamamos aqui a aplica\c c\~ao exponencial riemanniana de Exp, para diferenci\'a-la da exponencial no grupo, que chamaremos de exp.}, para $v\in{T_xP}$ , $\mbox{Exp}(x,v)=\mbox{Exp}_x(v)$ temos: 
\begin{gather*}
(d\mbox{Exp})_q(\xi)=\frac{d}{dt}{|_{t=o}}\Big(\mbox{Exp}(\gamma(t),0)\Big)+\frac{d}{dt}{|_{t=o}}\Big(\mbox{Exp}(p,tdf_p^{-1}(u))\Big)\\
=\frac{d}{dt}{|_{t=o}}\Big((\gamma(t),0)\Big)+\frac{d}{dt}{|_{t=o}}\Big(\mbox{Exp}_p(tdf_p^{-1}(u))\Big)=(w,0)+(0,df_p^{-1}(u))=\xi\\ 
\therefore~~ d(\mbox{Exp})_q=Id_{|T_q\nu{N}}\end{gather*}
 que \'e v\'alido para todos os pontos da forma $(p,0)$. Logo, pelo teorema da fun\c c\~ao inversa, ao redor de cada ponto $p\in{N}$, existe um aberto $V\subset{N}$ e vizinhan\c ca da se\c c\~ao nula de  $\nu{N}_{|\pi_N^{-1}V}$ que \'e levada difeomorficamente sobre a imagem. Lembremo-nos tamb\'em que todo vetor de $T_pP$ pode ser escrito de forma \'unica como soma de um vetor tangente \`a $N$ e um normal \`a $N$. Logo a aplica\c c\~ao exponencial leva um aberto de $\nu{N}_{|\pi_N^{-1}V}$ vizinhan\c ca da se\c c\~ao nula difeomorficamente sobre um aberto de $P$ que cont\'em $V$. Podemos tomar essa vizinhan\c ca da se\c c\~ao nula de $\nu{N}_{|\pi_N^{-1}V}$ como sendo da forma $\phi^{-1}(V\times{B}_\delta(0))$ para algum $\delta>0$.  

Pictoricamente, estamos levando uma vizinhan\c ca tubular ``reta'' (j\'a que tem seu di\^ametro em um fibrado vetorial), para uma vizinha\c ca tubular ``curva" ao redor de $N$ em $P$. N\'os temos que para  a pr\'opria se\c c\~ao nula,  $\mbox{Exp}:N\times\{0\}\rightarrow{N}$ \'e a identidade em $N$. \'E claro ent\~ao que para cada ponto $p\in{N}$  a exponencial de $B_\delta(0_p):=B_\delta(0)\cap\nu_pN\subset\nu_pN$ \'e uma subvariedade de $P$ de dimens\~ao $k$, transversal \`a $N$, que chamaremos de $\Sigma_p$.   

N\'os temos que a exponencial \'e ent\~ao um difeomorfismo local. Contudo, pode ainda ocorrer que a exponencial n\~ao leva vizinhan\c ca global da se\c c\~ao nula de $\nu{N}$ injetoramente sobre a imagem, ou seja, que para qualquer raio global do tubo que tomarmos teremos auto-intersec\c c\~ao ao mandarmos o tubo para $P$ atrav\'es da exponencial. Para completar a demonstra\c c\~ao do teorema, temos de provar que existe vizinhan\c ca de $N$ onde a Exp \'e de fato injetora.  
\begin{prop}
Sejam $X,Y$ variedades suaves, $f:X\rightarrow{Y}$ $C^\infty$, $f_{|N}$ injetora para uma dada $N$ subvariedade compacta de $X$ e $df_x:T_xX\rightarrow{T_yY}$ isomorfismo para $x\in{N}$. Ent\~ao existe vizinhan\c ca aberta $U$ de $N$ em $X$ tal que $f:U\rightarrow{f(U)}$ \'e difeomorfismo. \end{prop} 
{\bf{Dem:}}
Para provar o lema s\'o nos resta provar injetividade.  Seja ent\~ao o conjunto das vizinhan\c cas abertas de $N$:
$$\CC=\{U ~|~ U~ \mbox{\'e vizinhan\c ca aberta de} ~N\}$$
Constru\'imos a ordem parcial em $\CC$ pela inclus\~ao inversa, i.e.: $W\geq{U}$ se $W\subset{U}$, ent\~ao \'e f\'acil verificar que  $\CC$ \'e um conjunto dirigido. Seja $S$ um elemento de $\CC$, e $x_S$ um ponto em $S$, ent\~ao $\{x_S\}$ \'e uma rede  em $X$. Para qualquer $W\in\CC$, se $S\geq{W}$,  $x_S\in{W}$.   Logo como $N$ \'e compacto ( e portanto fechado) n\'os temos um lema de topologia geral que garante que uma rede $\{x_S\}$ converge para um ponto de $N$ (ver \cite{Engel} ). Al\'em disso, como $X$ \'e Hausdorff, este ponto \'e \'unico. Notamos ainda que para qualquer vizinhan\c ca aberta $V$ de $x$, existe $W\in\CC$, tal que para todo $S\geq{W}$ n\'os temos $x_S\in{V}$.  

Uma fun\c c\~ao $f:X\lra{Y}$ \'e cont\'inua se e somente se, para toda rede $\{x_S\}$, 
$$f(\lim_{S\in\CC}x_S)=\lim_{S\in\CC}f(x_S)$$ 
Claramente, $f$ \'e injetora em alguma vizinhan\c ca aberta de $N$ se e somente se $f_{|S}$ \'e injetora para algum $S\in\CC$. Logo, suponhamos por absurdo que para todo $S\in\CC$, $f_{|S}$ n\~ao \'e injetora. Ent\~ao qualquer que seja $S\in\CC$, existem $x_S\neq{y_S}\in{S}$  tal que $f(x_S)=f(y_S)$. Montamos duas dessas redes: $\{x_S\}$ e $\{y_S\}$, que j\'a sabemos convergir para pontos em $N$, $x$ e $y$ respectivamente. 
  Como $f$ \'e cont\'inua, 
$$\lim_{S\in\CC}f(y_S)=\lim_{S\in\CC}f(x_S)=f(x)=f(y)$$
 Mas $f$ \'e difeomorfismo de $N$ sobre sua imagem, em particular \'e injetora sobre $N$, ent\~ao essa rela\c c\~ao implica $x=y$. No entanto $f$ \'e difeomorfismno local ao redor de cada ponto de $N$, logo existe $V$ aberto $P$ tal que $x\in{V}$ onde $f$ \'e injetora. Portanto, como $y_S,x_S\lra{x}$, existe $W\in{\CC}$ tal que para $S\geq{W}$, $x_S\neq{y}_S$ implica $f(x_S)\neq{f}(y_S)$ o que contraria a nossa hip\'otese. Ou seja, existe $S\in\CC$ tal que se $x\neq{y}\in{S}, ~f(x)\neq{f}(y)$. $~~~~\blacksquare$\medskip 

Dado esse resultado, como $N$ \'e compacta n\'os temos que existe um raio m\'inimo $\delta_0$ para o qual podemos achar um difeomorfismo entre o tubo `` reto" dentro de $\nu{N}$ e o tubo ``curvo" em $P$ obtido pela exponencial. 

 Voltando ao nosso caso, $\Sigma_p=\Exp_pB_\delta(0_p)$, e as \'orbitas que passam por $x\in\Sigma_p$ ter\~ao a forma $g\cdot\Exp_p(w)$ para algum $w\in\nu_pN$ e algum $g\in{G}$. Lembremos que colocamos em $P$ uma m\'etrica $G$-invariante, portanto, como  $g$ age isometricamente, leva geod\'esica em geod\'esica. Mas se $g\neq{e}$ ent\~ao $g\cdot{p}\neq{p}$~ e $g\cdot\Exp(p,tu)$ \'e uma geod\'esica que passa por $g\cdot{p}$ com tangente $$\frac{d}{dt}{|_{t=o}}\Big(g\cdot\Exp(p,tu)\Big)=dg\circ\frac{d}{dt}{|_{t=o}}\Big(\Exp(p,tu)\Big)=dg(u)$$ assim como $\Exp(g\cdot{p},tdg(u))$, ent\~ao por unicidade,  
\beq\label{isom}g\cdot\Exp(p,w)=\Exp(g\cdot{p},dg_p(w))\eeq  

Al\'em disso, n\'os temos que, como  $G$ age por  isometrias, logo preserva a perpendicularidade do subespa\c co normal e o raio $\delta$, i.e.:
$dg(B_\delta(0_p))=B_\delta(0_{g\cdot{p}})\subset\nu_{g\cdot{p}}N$ ent\~ao $dg(u)\in\nu_{g\cdot{p}}N$. Mas como mostramos, para $\delta_0$ n\'os temos que: 
 $$\Exp(g\cdot{p},B_{\delta_0}(0_{g\cdot{p}}))\cap\Exp({p},B_{\delta_0}(0_p))=\emptyset$$
e portanto  teremos apenas uma intersec\c c\~ao entre cada \'orbita e  $\Sigma_p$.  

Finalmente, n\'os sabemos que $T_p\Sigma_p$ \'e transversal a $\I_p(\Lg)$. Como tanto $\Sigma$ quanto $\I(\Lg)$ s\~ao suaves, pela propriedade {\bf Cont.} de transversalidade que veremos no {\bf Teo.29}, eles se mant\'em abertos em uma vizinhan\c ca de $p$ em $\Sigma$ e provamos o teorema\footnote{Fica como exerc\'icio para o leitor descobrir como nossa constru\c c\~ao da faixa de Moebius ``torta" n\~ao se encaixa nas suposi\c c\~oes do teorema.}.  $~~~~~~~\blacksquare$\medskip 

Lembrando que temos um difeomorfismo entre uma vizinhan\c ca da se\c c\~ao nula de $\nu{N}$ e um aberto $U$ de $P$ dado pela aplica\c c\~ao exponencial, n\'os temos que para $y\in{U}$,  existe um \'unico $q\in{N}$~ e um \'unico $v\in{B_\delta(0_q)}\subset\nu_qN$ tal que $y=\Exp_q(v)$. Mas como acabamos de ver, existem  tamb\'em \'unicos $g\in{G}$ e $u\in{B_\delta(0_p)}\subset\nu_pN$, onde $dg(u)=v$ tal que 
$$y=\Exp_{g\cdot{p}}(dg(u))=g\cdot\Exp_p(u)=g\cdot{x}$$ onde $x\in\Sigma_p$, portanto, como o grupo $G$ age como grupo de difeomorfismos sobre $P$, n\'os temos um difeomorfismo
\begin{align*}
{{\psi}}_{p}:~ & U\lra{\Sigma_p\times{G}}\\ 
\phantom{add}~ & y\longmapsto(x,g) 
\end{align*}  
onde $(x,g)$ \'e o \'unico tal que $y=g\cdot{x}$~. 

Suponhamos que temos uma outra se\c c\~ao  $\Sigma$  de $P$ sobre $\theta$, i.e.: subvariedade de $P$ contida em $U$ que intercepta em um \'unico ponto as \'orbitas de $\theta$ e \'e transversal a elas, portanto $\Sigma$ \'e de mesma dimens\~ao que $\Sigma_p$, e intercepta unicamente tamb\'em as \'orbitas de $\Sigma_p$. Portanto, para cada $y\in\Sigma$ existe um \'unico elemento $g(y)\in{G}$ tal que $(g(y)\cdot{y})\in\Sigma_p$. Logo para $y\in\Sigma$ temos $y=g(y)^{-1}\cdot(g(y)\cdot{y})$ e portanto pela propriedade de $\psi_p$ n\'os temos uma \'unica decomposi\c c\~ao: 
\begin{align*}
\psi_{p|\Sigma}: & \Sigma\lra\Sigma_p\times{G}\\
\phantom{a} & y\longmapsto(g(y)\cdot{y},g(y)^{-1}) 
\end{align*}
Por $\Sigma$ ser subvariedade lisa e $\psi_p$ ser tamb\'em suave, as aplica\c c\~oes $g,g^{-1}:\Sigma\ra{G}$ s\~ao suaves. Portanto $y\mapsto{g(y)\cdot{y}}$ \'e um difeomorfismo entre $\Sigma$ e $\Sigma_p$, logo existe tamb\'em difeomorfismo $U\overset{\psi}{\simeq}\Sigma\times{G}$. 
Ou seja, n\'os temos, para  $(y,h)\in\Sigma\times{G}$, o mapa de transi\c c\~ao  $\psi_p\circ\psi^{-1}: \Sigma\times{G}\lra\Sigma_p\times{G}$, dado  por 
\beq
\psi_p\circ\psi^{-1}(y,h)=\psi_p(h\cdot{y})=\psi_p(hg^{-1}(y)g(y)\cdot{y})=(g(y)\cdot{y},hg^{-1}(y))\eeq 
que \'e uma composi\c c\~ao de difeomorfismos sobre a estrutura diferenci\'avel de $P$ e portanto podemos tomar um atlas para $P$ dado por cartas dessa forma. 

Como mencionamos, a topologia de $M$ \'e definida pela proje\c c\~ao ser aberta e cont\'inua. 
Como existe bije\c c\~ao entre $\Sigma$ e $U/G=\pi(U)=\theta$, e $\Sigma$ tem a topologia induzida, n\'os temos um homeomorfismo entre $\Sigma$ e $\theta$ dado por 
\begin{align*}
\pi_{|\Sigma} : ~& \Sigma\lra\theta\\ 
\pi_{|\Sigma}^{-1}=\mbox{pr}_\Sigma\circ\pi^{-1} : ~& \theta\lra\Sigma
\end{align*} 
onde $\mbox{pr}_\Sigma:\Sigma\times{G}\ra\Sigma$ \'e a proje\c c\~ao can\^nonica na primeira coordenada. 
Podemos induzir ent\~ao estrutura diferenci\'avel em $M$ pela estrutura diferenci\'avel dos $\Sigma$'s, isto \'e, tomando $\pi_{|\Sigma}$ como difeomorfismo sobre cada $\theta$.  Essa estrutura \'e compat\'ivel nas intersec\c c\~oes, isto \'e, se tivermos um outro aberto $\widetilde{U}\overset{\widetilde\psi}{\simeq}\widetilde\Sigma\times{G}$, mostramos que sobre a intersec\c c\~ao $U\cap\widetilde{U}$ existe um difeomorfismo entre os $\Sigma$'s, portanto est\'a bem definida a estrutura diferenci\'avel de $M$; existe e \'e \'unica. \'E claro que com essa exig\^encia $\pi$ \'e uma proje\c c\~ao diferenci\'avel e de posto m\'aximo.    

Reciprocamente, se exigirmos que a proje\c c\~ao $\pi:P\ra{M}$ seja diferenci\'avel e de posto m\'aximo, i.e.: se exigirmos que a estrutura diferenci\'avel de $M$ seja tal que $\pi$ \'e submers\~ao, obtemos que, dado um ponto qualquer $p\in{P}$ e uma se\c c\~ao $\Sigma$ que passa por $p$, 
$$d\pi_p:T_p\Sigma\oplus\I_p\Lg\lra{T_{\pi(p)}M}$$
mas dim$\Ker{d\pi_p}=\mbox{dim}G=\mbox{dim}\Lg=\mbox{dim}\I_p\Lg$ e para $X\in\Lg$ n\'os temos 
$$d\pi_p(\I_p(X))=\frac{d}{dt}{|_{t=o}}\left(\pi\circ({\exp}(tX)\cdot{p})\right)=\frac{d}{dt}{|_{t=o}}\pi(p)=0$$ 
i.e.:$\I_p\Lg\subset\Ker{d\pi_p}$ e por dimens\~ao  $\Ker{d\pi_p}=\I_p\Lg$. Portanto $T_p\Sigma\overset{d\pi_p}{\simeq}T_{\pi(p)}M$, e 
teremos difeomorfismos locais entre as se\c c\~oes e abertos de $M$.     

Provamos ent\~ao:     
\begin{theo} Existe uma \'unica estrutura diferenci\'avel em $M$ caracterizada por qualquer uma das condi\c c\~oes:
\begin{itemize}
\item{A proje\c c\~ao $\pi:P\lra{M}$ \'e uma submers\~ao suave.} 
\item{Se $\Sigma$ \'e uma se\c c\~ao   de $P$ sobre $\theta$  ent\~ao $\pi_{|\Sigma}$ \'e um difeomorfismo de $\Sigma$ sobre $\theta$.}\end{itemize}
\end{theo}
Fica claro ainda que  $P$ \'e localmente difeomorfo a conjuntos da forma $\theta\times{G}$, i.e.: podemos tomar um difemorfismo $\pi^{-1}(\theta)\overset{\phi}{\simeq}\theta\times{G}$ simplesmente aplicando a proje\c c\~ao $\pi_{|\Sigma}$ \`a primeira coordenada do difeomorfismo $\pi^{-1}(\theta)\overset{\psi}{\simeq}\Sigma\times{G}$ . Chamaremos tais difeomorfismos de triviliaza\c c\~oes locais. 

Ademais, podemos considerar uma se\c c\~ao $\Sigma$ sobre $\theta\subset{M}$ como uma imers\~ao suave que leva $\theta$ na subvariedade $\Sigma$,  $s:\theta\ra{P}$. Teremos que 
para $m\in\theta$, $\pi(s(m))=m$,  ou seja, para cada \'orbita $m\in\theta$, $s(m)$ est\'a na fibra sobre $m$.  Da mesmsa forma se $\widetilde{s}$ for outra se\c c\~ao sobre $\theta$, existe um \'unico mapa $g:\theta\ra{G}$ tal que $\widetilde{s}(m)=g(m)\cdot{s}(m)$. Chamamos $g$ de mapa de transi\c c\~ao entre as se\c c\~oes. 

\'E claro que se $\phi:\pi^{-1}(\theta)\ra\theta\times{G}$ for uma trivializa\c c\~ao local correspondente \`a subvariedade $\Sigma$, i.e.: \`a imers\~ao $s:\theta\ra{P}$, ent\~ao para todo $m\in\theta$,  $\phi(s(m))=(m,e)$ j\'a que $\psi(s(m))=(s(m),e)$ e $\pi_{|\Sigma}(s(m))=m$. Desta forma podemos ver que dada uma se\c c\~ao de $P$ sobre $\theta$, $s:\theta\ra{P}$,  existe uma \'unica trivializa\c c\~ao local $\phi^s:\pi^{-1}(\theta)\ra\theta\times{G}$ a ela  adaptada  de forma que para todo $m\in\theta$ n\'os tenhamos $\phi^s(s(m))=(m,e)$. Nos casos em que estiver subentendida \`a qual se\c c\~ao a trivializa\c c\~ao est\'a adaptada omitiremos o superscrito ``$s$''. 

N\'os vimos que para a decomposi\c c\~ao $\pi^{-1}(\theta)\overset{\psi}{\simeq}\Sigma\times{G}$, para todo $h\in{G}$, vale a propriedade 
\beq\psi^{-1}(p,h)=h\cdot\psi^{-1}(p,e)=h\cdot{p}\eeq 
Para $\pi^{-1}(\theta)\overset{\phi}{\simeq}\theta\times{G}$, para $m\in\theta$ tal que $\pi_{|\Sigma}^{-1}(m)=p$, n\'os temos $\psi^{-1}(p,h)=\phi^{-1}(m,h)$. Logo: 
\beq\label{cartaprop}\phi^{-1}(m,h)=h\cdot\phi^{-1}(m,e)\eeq
\'E ainda claro que temos um difeomorfismo 
$$\phi^{-1}(m,\cdot):G\lra{\pi^{-1}(m)}$$
ao qual chamaremos de $\phi_m:\pi^{-1}(m)\ra{G}$ e que, para qualquer $h\in{G}$ e $p\in\pi^{-1}(m)$, por (\ref{cartaprop}) acima obedece: 
$$\phi_m(h\cdot{p})=h\phi_m(p)$$ onde denotamos o produto no grupo $(h,g)\mapsto{hg}$. Agora, $\phi_m(h\cdot{p})\in{G}$ e $\left(\phi_m(h\cdot{p})\right)^{-1}=\left(\phi_m(p)\right)^{-1}h^{-1}$ , logo, se tivermos outro difeomorfismo $\widetilde\phi_m$ n\'os temos 
\begin{gather*}
\left(\phi_m(h\cdot{p})\right)^{-1}\left(\widetilde\phi_m(h\cdot{p})\right)=\left(\phi_m(p)\right)^{-1}h^{-1}h\widetilde\phi_m(p)=\left(\phi_m(p)\right)^{-1}\widetilde\phi_m(p)\\ \therefore~~~~~~~~~~~~\left(\phi_m(p)\right)^{-1}\widetilde\phi_m(p)=\left(\phi_m(q)\right)^{-1}\widetilde\phi_m(q)\end{gather*} 
 para todos $p,q\in\pi^{-1}(m)$. Isto \'e, para a transi\c c\~ao entre trivializa\c c\~oes locais existe um \'unico elemento de $G$ para cada fibra, ou seja, novamente a transi\c c\~ao \'e uma fun\c c\~ao suave $g_{\phi\widetilde\phi}:\theta\ra{G}$. 
Uma palavra de esclarecimento em rela\c c\~ao ao uso do termo ``transi\c c\~ao'' \'e aqui necess\'ario , j\'a que n\~ao \'e \ 'obvia a rela\c c\~ao entre $\widetilde\phi_m\circ\phi_m^{-1}\in{\mbox{Aut}(G)}$ e $g_{\phi\widetilde\phi}:\theta\ra{G}$. Seja ent\~ao $h\in{G}$ e $\phi_m^{-1}(h)=p$. \'E claro que $\left(\phi_m(p)\right)^{-1}=h^{-1}$. Ent\~ao obtemos 
$$h^{-1}\left(\widetilde\phi_m\circ\phi_m^{-1}(h)\right)=\left(\phi_m(p)\right)^{-1}\left(\widetilde\phi_m(p)\right)$$
que como vimos n\~ao depende de $p\in\pi^{-1}(m)$. Por isso a fun\c c\~ao de transi\c c\~ao de uma representa\c c\~ao \`a outra \'e r\'igida, um \'unico elemento de $G$ para cada ponto de $\theta$.

\section{Connections in Principal Bundles}
Sobre cada fibra, existe uma maneira can\^onica de identificar elementos de $TP$, a saber, pelo isomorfismo linear dado pela a\c c\~ao do grupo $G$ sobre a fibra; $dg:T_pP\ra{T_{g\cdot{p}}}P$. Incorporamos este princ\'ipio sempre que exigirmos que alguma estrutura seja $G$-invariante. Contudo sobre elementos de $TP$  que residem sobre  diferentes fibras, assim como em fibrados vetoriais, n\~ao h\'a identifica\c c\~ao can\^onica. Antecipando um pouco a nomenclatura que segue, se chamarmos o deslocamento sobre as fibras de `` vertical'', queremos uma forma de identifica\c c\~ao puramente `` horizontal'', uma forma de mantermos uma curva de $P$ ``\`a mesma altura".  Apesar deste `` deslocamento vertical'' estar canonicamente determinado, n\~ao existe na estrutura de $P$ algo que nos d\^e um  complemento, que especifique um `` deslocamento horizontal'' can\^onico. Para incorporar uma identifica\c c\~ao local em $TP$ ( i.e.: tanto horizontal quanto vertical) precisamos de uma decomposi\c c\~ao de $TP$ em subfibrados vertical e horizontal que sejam $G$-invariantes (precisamos manter a identifica\c c\~ao can\^onica sobre as fibras). Como veremos, a escolha de uma conex\~ao nos fornece tal decomposi\c c\~ao. 

Como $\pi:~P\ra{M}$ \'e uma submers\~ao, $\Ker(d\pi)$ \'e um sub-fibrado liso de $TP$, chamado de sub-fibrado vertical, cuja fibra em $p$ denotaremos por $\mbox{V}_p=\I_p\Lg$. Como $\I_p\Lg$  \'e inje\c c\~ao linear,  
$$\I_p:\Lg\lra{T_pP}$$ \'e um isomorfismo linear de $\Lg$ em $\mbox{V}_p$.  
\begin{defi} Para cada $p\in{P}$ definimos $\widetilde\omega_p:\mbox{V}_p\ra\Lg$ com sendo $\I_p^{-1}$. \end{defi}
Desse modo \'e claro que $\widetilde\omega$ \'e uma 1-forma suave em $\mbox{V}$ a valores em $\Lg$. 
 Para cada $g\in{G}$ n\'os obtemos uma 1-forma em $\mbox{V}$ a valores em $\Lg$, $g^*\widetilde\omega$ por 
$$(g^*\widetilde\omega)_p=\widetilde\omega_{g\cdot{p}}\circ{dg}$$
Utilizando \ref{Ip} obtemos 
\begin{gather*} 
\I_{g\cdot{p}}=dg\circ\I_p\circ\Ad{(g^{-1})}\Rightarrow(\I_{g\cdot{p}})^{-1}=\Ad{(g)}\I_p^{-1}dg^{-1}\\
~\therefore~\widetilde\omega_{g\cdot{p}}\circ{d}g=\Ad(g)\circ\widetilde\omega_p
\end{gather*}
e finalmente 
\beq{g}^*\widetilde\omega=\Ad(g)\circ\widetilde\omega\eeq

\begin{defi} Uma forma de conex\~ao em $P$ \'e definida como $\lambda\in\Gamma(\Lambda^1(TP)\otimes\Lg)$ que obede\c ca 
$g^*\lambda=\Ad(g)\lambda$
e tal que para $u\in{V}_p$ n\'os tenhamos :
$\lambda_p(u)=\I_p^{-1}(u)$.\end{defi} 
Em breve discutiremos o significado geom\'etrico de formas de conex\~ao. 
\begin{defi} Uma conex\~ao em $P$ \'e um subfibrado $\mbox{H}$ de $TP$ tal que $dg(\mbox{H}_p)=\mbox{H}_{g\cdot{p}}$ e $\mbox{H}_p\oplus\mbox{V}_p=T_pP$. Chamamos H de fibrado horizontal. \end{defi} 
Pela equa\c c\~ao (\ref{Ip}), como a a\c c\~ao de Ad$(g)$ \'e automorfismo de $\Lg$,  n\'os temos que $dg(\mbox{V}_p)=\mbox{V}_{g\cdot{p}}$. Portanto a decomposi\c c\~ao $T_pP=\mbox{H}_p\oplus\mbox{V}_p$ \'e invariante pela a\c  c\~ao de $G$. Denotamos as proje\c c\~oes suaves de $TP$ no subfibrado horizontal  de $\widehat{H}$ e no vertical de $\widehat{{V}}$ . Como a decomposi\c c\~ao do fibrado $TP$ \'e invariante pela a\c c\~ao de $G$, \'e claro que $\widehat{H}_{g\cdot{p}}\circ{d}g=dg\circ\widehat{H}_p$, o mesmo valendo para a proje\c c\~ao vertical. Explicitamente, se $w\in{T_pP}$ ent\~ao $w=w_h+w_v$ e $dg(w_h)\in\mbox{H}_{g\cdot{p}}~, ~dg(w_v)\in\mbox{V}_{g\cdot{p}}$,  portanto 
\begin{gather*}
dg(\widehat{H}_p(w))=dg(w_h)~~\mbox{mas}~~dg(w)=dg(w_h)+dg(w_v)\\
\therefore~~~\widehat{H}_{g\cdot{p}}dg(w)=\widehat{H}_{g\cdot{p}}dg(w_h)=dg(w_h)
\end{gather*}
\begin{theo}
Se H \'e  conex\~ao em $P$, ent\~ao para todo $p\in{P}$: 
$d\pi_p:T_pP\ra{T_{\pi(p)}M}$ se restringe a um isomorfismo linear $h_p:\mbox{H}_p\ra{T_{\pi(p)}M}$ para o qual vale $h_{g\cdot{p}}\circ{d}g{|_{\mbox{H}_p}}=h_p$.\end{theo}
{\bf{Dem:}}Que a proje\c c\~ao se restringe a um isomorfismo linear \'e claro, j\'a que o espa\c co vertical \'e o n\'ucleo da proje\c c\~ao, que tem posto m\'aximo,  e o horizontal \'e seu complemento. Agora n\'os temos que $h_{g\cdot{p}}\circ\widehat{H}_{g\cdot{p}}=d\pi_{g\cdot{p}}$, logo, aplicando os dois lados a $dg$ obtemos: 
\begin{gather*}
h_{g\cdot{p}}\circ\widehat{H}_{g\cdot{p}}\circ{dg}=h_{g\cdot{p}}\circ{dg}\circ\widehat{H}_{p}\\
=d\pi_{g\cdot{p}}\circ{dg}=d(\pi_{g\cdot{p}}\circ{g})=d\pi_p=h_p\circ\widehat{H}_p\\
\therefore~h_p\circ\widehat{H}_p=h_{g\cdot{p}}\circ{dg}\circ\widehat{H}_{p} 
\end{gather*}
e como $\widehat{H}_p$ \'e sobrejetor, obtemos o enunciado. ~~~~~~~~~~~~~$\blacksquare$

\begin{defi}
Se H \'e a conex\~ao em $P$ definimos a 1-forma de conex\~ao como 
$\omega:=\widetilde\omega\circ\widehat{V}$ . Isto \'e :
\begin{align*}
\omega_p:~& T_pP\lra\Lg\\
\phantom{\omega_p:}~& ~u\lra\I_p^{-1}\circ\widehat{V}_p(u) 
\end{align*}\end{defi}

\'E claro que se $v\in\mbox{V}_p$ ent\~ao $\omega(v)=\widetilde\omega(v)$. Al\'em disso 
\begin{gather*}
(g^*\omega)_p=\omega_{g\cdot{p}}\circ{dg}=\widetilde\omega_{g\cdot{p}}\circ\widehat{V}_{g\cdot{p}}\circ{dg}\\ 
=\widetilde\omega_{g\cdot{p}}\circ{dg}\circ\widehat{V}_p=(g^*\widetilde\omega)_p\circ{\widehat{V}}_p=\Ad(g)\circ\widetilde\omega_p\circ\widehat{V}_p=\Ad(g)\omega_p\\
\end{gather*} e portanto obtemos
\beq\label{foconex}~~~~g^*\omega=\Ad(g)\omega\eeq 
Portanto $\omega$ realmente \'e forma de conex\~ao. Como $\I_p$ \'e isomorfismo linear sobre $\V_p$~, \'e claro que $ \Ker\omega_p$ \'e complementar a $\mbox{V}_p$~. Al\'em disso, se $u\in\Ker\omega_p$ ent\~ao $dg(u)\in\Ker\omega_{g\cdot{p}}$ j\'a que 
$$ \omega_{g\cdot{p}}\circ{dg}(u)=\Ad(g)\omega_p(u)=0$$
logo, como $dg$ \'e isomorfismo linear, $\Ker\omega$ \'e um subfibrado $G$-invariante de $TP$ complementar a V. Ou seja, podemos definir uma conex\~ao em $P$ como $\HH_p=\Ker\omega_p$~. De fato temos: 
\begin{theo}\label{corresp}
O mapa $\HH\ra{\omega}$ \'e uma correspond\^encia bijetora. \end{theo}

{\bf{Dem}:}
Como $\I_p:\Lg\ra\V_p$ \'e isomorfismo linear, identificamos em cada ponto $\omega$ como a proje\c c\~ao no subespa\c co $\V_p$~. Como qualquer proje\c c\~ao, $\omega$ ser\'a caracterizada por seu n\'ucleo, $\HH$,  ou mais precisamente, por uma decomposi\c c\~ao $T_pP=\HH_p\oplus\V_p$. J\'a demonstramos a afirma\c c\~ao inversa. $\blacksquare$\medskip 

 Essa \'e uma forma geom\'etrica de encarar uma forma de conex\~ao; simplesmente como uma proje\c c\~ao em um subfibrado $G$-invariante de $TP$, complementar ao espa\c co tangente \`as \'orbitas.

Para cada se\c c\~ao $s:\theta\ra{P}$ e forma de conex\~ao $\omega$ em $P$, definimos a 1-forma em $\theta$ a valores em $\Lg$:
$$\omega^s:=s^* \omega$$ ou seja, se $u\in{T_mM}$ ent\~ao $\omega_m^s(u)=\omega_{s(m)}(ds_m(u))$. 
 
\begin{prop}
Se $s$ e $ \widetilde{s}$ s\~ao se\c c\~oes sobre $\theta$, $g:\theta\ra{G}$ \'e o mapa de transi\c c\~ao, e $\omega$ \'e uma 1-forma de conex\~ao em $P$, ent\~ao 
\beq\label{transconex} \omega^{\tilde{s}}=\Ad(g)\circ\omega^s+(dg)g^{-1}\eeq 
\end{prop}

{\bf{Dem:}}
 Por defini\c c\~ao n\'os temos, para $m\in\theta$, $u\in{T_m{M}}$,
 \beq\label{desconex1}\omega_m^{\tilde{s}}(u)=\omega_{\tilde{s}(m)}(d\tilde{s}(u))=\omega_{g(m)\cdot{p}}(d\tilde{s}(u))\eeq
  Como $g:\theta\ra{G}$, se $m\in\theta$ n\'os temos $dg_m:T_mM\ra{T_{g(m)}G}$, um sentido diferente do que est\'avamos usando:  $dg:TP\ra{TP}$, que leva um vetor tangente a uma curva $\alpha(t)$ passando por $p\in{P}$ \`a tangente \`a curva $g\cdot\alpha(t)$ passando por $g\cdot{p}$~. Portanto, denotaremos essa \'ultima aplica\c c\~ao, dado $m\in\theta$,  por $L_{{g(m)}_*}:TP\ra{TP}$, i.e.: utilizaremos a nota\c c\~ao comum  de $L_g$ como multiplica\c c\~ao \`a esquerda e $R_g$ como multiplica\c c\~ao \`a direita . Agora, por (\ref{foconex})
\beq\label{desconex2}\omega_{g(m)\cdot{p}}\circ{L_{g(m)}}_*=\left(L_{g(m)}^*\omega\right)_p=\Ad(g(m))\omega_p\eeq 
Novamente denotando a a\c c\~ao do grupo em $P$ por $\mu:{P}\times{G}\ra{P}$, temos $\tilde{s}=\mu(s,g)$ e ent\~ao $d\tilde{s}=d\mu(ds,dg)$. Aplicando (\ref{desconex2}) a ${L_{g(m)^{-1}}}_*$ em ambos os lados e colocando em (\ref{desconex1}) obtemos: 
\beq\label{desconex3}\omega_m^{\tilde{s}}(u)=\Ad(g(m))\omega_p\circ{L_{g(m)^{-1}}}_*d\mu_{(s(m),g(m))}\Big(ds_m(u),dg_m(u)\Big)\eeq
Agora utilizando a trivializa\c c\~ao local adaptada a $s$, dada por  $\pi^{-1}(\theta)\overset{\phi^{s}}{\simeq}\theta\times{G}$ n\'os temos, utilizando (\ref{cartaprop}) 
\begin{gather}\mu\left(s(m),g(m)\right)=({\phi^s})^{-1}(m,g(m))\Longrightarrow\mu(s,g)=({\phi^s})^{-1}(\mbox{Id}_\theta,g)\\
\therefore~~~~ d\mu(ds,dg)=(d{\phi^s})^{-1}(\mbox{Id}_{TM{|_\theta}},dg)\\
\label{desconex4}\therefore~~~~d\mu_{(s(m),g(m))}\Big(ds_m(u),dg_m(u)\Big)=(d{\phi^s})^{-1}_{(m,g(m))}(u,dg_m(u)) 
\end{gather}
Aplicando ${L_{g(m)^{-1}}}_*$ aos dois lados de (\ref{desconex4}) e utilizando novamente (\ref{cartaprop}) obtemos
\begin{gather}
{L_{g(m)^{-1}}}_*d\mu_{(s(m),g(m))}\Big(ds_m(u),dg_m(u)\Big)={L_{g(m)^{-1}}}_*(d{\phi^s})^{-1}_{(m,g(m))}(u,dg_m(u))\\ 
\label{desconex5}=(d{\phi^s})^{-1}_{(m,e)}\left(u,{L_{g(m)^{-1}}}_*(dg_m(u))\right)=(d{\phi^s})^{-1}_{(m,e)}(u,0)+(d{\phi^s})^{-1}_{(m,e)}\left(0,{L_{g(m)^{-1}}}_*(dg_m(u))\right)\end{gather}
Agora $(d{\phi^s})^{-1}_{(m,e)}(u,0)=ds_m(u)$ j\'a que se, para $t=0$, $\alpha:I\ra{\theta}$ \'e curva tangente a $u$ em $m\in\theta$ : 
$$(d{\phi^s})^{-1}_{(m,e)}(u,0)=\frac{d}{dt}{|_{t=o}}\left(({\phi^s})^{-1}(\alpha(t),e)\right)=\frac{d}{dt}{|_{t=o}}s(\alpha(t))=ds_m(u)$$
Al\'em disso, para qualquer $X\in\Lg$ n\'os temos
\begin{gather}
\label{desconex6}(d{\phi^s})^{-1}_{(m,e)}(0,X)=\frac{d}{dt}{|_{t=o}}\left(({\phi^s})^{-1}(m,\exp{tX})\right)=\frac{d}{dt}_{|_{t=o}}\left(\exp{tX}\cdot{s}(m)\right)=\I_{s(m)}X\\ 
\therefore~~~(d{\phi^s})^{-1}_{(m,e)}\left(0,{L_{g(m)^{-1}}}_*(dg_m(u))\right)=\I_{s(m)}\left({L_{g(m)^{-1}}}_*(dg_m(u))\right)
\end{gather}
Utilizando (\ref{desconex5}) e (\ref{desconex3}) obtemos finalmente:
\begin{gather}
\label{aiai}\omega_p\circ{L_{g(m)^{-1}}}_*d\mu_{(s(m),g(m))}\Big(ds_m(u),dg_m(u)\Big)=\omega_p(ds_m(u))+{L_{g(m)^{-1}}}_*(dg_m(u))\\
\therefore~~~\omega_m^{\tilde{s}}(u)=\Ad(g(m))\omega_p(ds_m(u))+{R_{g(m)^{-1}}}_*{L_{g(m)}}_*{L_{g(m)^{-1}}}_*(dg_m(u))\\ 
\label{reve8}\therefore~~~\omega_m^{\tilde{s}}(u)=\Ad(g(m))\omega_p^s(u)+{R_{g(m)^{-1}}}_*dg_m(u)
\end{gather}
$~~~~~~~~~~~~~~~~~~~~~~~~~~~~~~~~~~~~~~~~~~~~~~~~~~~~~~~~~~~~~~~~~~~~~~~\blacksquare$\medskip

\begin{prop}\label{conversamente}
Rec\'iprocamente, se para cada se\c c\~ao $s$ de $P$ sobre $\theta$ n\'os tivermos uma 1-forma sobre $\theta$ a valores em $\Lg$, $\omega^s$, e se essas 1-formas satisfizerem (\ref{transconex}), ent\~ao existe uma \'unica forma de conex\~ao $\omega$ em $P$ tal que $\omega^s=s^*\omega$. \end{prop}  

{\bf{Dem:}}
Seja $w\in{T_pP}$,  $p=s(m)\in{P}$, e $\phi:\pi^{-1}(\theta)\ra\theta\times{G}$ a trivializa\c c\~ao local adaptada\footnote{Abreviamos aqui $\phi^s$ por $\phi$.} \`a se\c c\~ao $s$. Ent\~ao, para cada $w\in{T_pP}$, existem \'unicos $u\in{T_mM}, X\in\Lg$ tal que 
 
\beq\label{reve}w=(d{\phi})^{-1}_{(m,e)}(u,X)=ds_m(u)+\I_{s(m)}X\eeq
Definimos ent\~ao $\omega\in\Gamma(\Lambda^1(TP_{|\theta})\otimes\Lg)$ como

\beq\label{reve1}\omega_p(w)=\omega_p\left((d{\phi})^{-1}_{(m,e)}(u,X)\right):=\omega^s_m(u)+X\eeq
Como sabemos, para todo $y\in\pi^{-1}(\theta)$ existem \'unicos $x\in\Sigma=s(\theta)$ e $h\in{G}$ tal que $y=h\cdot{x}$. Logo para $v\in{T_yP}$ estendemos a defini\c c\~ao de $\omega$: 

\beq\label{reve2}\omega_y(v)=\omega_{h\cdot{x}}(v):=\Ad(h)\omega_x({L_{h^{-1}}}_*(v))\eeq
Precisamos mostrar que se $\omega^s$ obedece (\ref{transconex}), essa defini\c c\~ao independe das se\c c\~oes que tomarmos; i.e.: a forma  $\omega$ em $P$ relativa \`a se\c c\~ao $s$ \'e igual a forma $\widetilde\omega$ obtida pela defini\c c\~ao a partir da se\c c\~ao $\tilde{s}=g\cdot{s}$. 
Claramente, se estiver bem definida, $h^*\omega=Ad(h)\omega$, o que pode ser facilmente visto colocando $\tilde{v}\in{T}_xP$ tal que ${L_h}_*(\tilde{v})=v$ em (\ref{reve2}). Al\'em disso, se $v\in\V_p$ , 
$$v=(d{\phi})^{-1}_{(m,e)}(0,X)=\I_{s(m)}X$$ e portanto pela nossa defini\c c\~ao:
$$\omega_p(v)=\omega_p(\I_p(X))=X=\I_p^{-1}(\I_p(X))=\I_p^{-1}(v)$$ 
portanto, se estiver bem definida, $\omega$ \'e forma de conex\~ao. 
Seja ent\~ao $\tilde\phi:\pi^{-1}(\theta)\ra\theta\times{G}$ uma trivializa\c c\~ao local adaptada \`a se\c c\~ao $\tilde{s}=g\cdot{s}$. Como ${L_{g(m)^{-1}}}_*$ \'e isomorfismo linear para todo $m\in\theta$, existem \'unicos $u\in{T_mM}, X\in\Lg$ tal que 
\beq\label{reve7}w={L_{g(m)^{-1}}}_*\left(d\tilde\phi^{-1}_{(m,e)}(u,X)\right)\eeq
onde como mencionamos $w\in{T_{p}}P$ e portanto $d\tilde\phi^{-1}_{(m,e)}(u,X)\in{T}_{g(m)\cdot{p}}P$. Mas
$$d\tilde\phi^{-1}_{(m,e)}(u,0)=d\tilde{s}_m(u)$$ e como vimos no exerc\'icio anterior 
\beq\label{reve3}{L_{g(m)^{-1}}}_*\left(d\tilde{s}_m(u)\right)=ds_m(u)+\I_p\left({L_{g(m)^{-1}}}_*(dg_m(u))\right)\eeq
e utilizando (\ref{desconex6}):
\beq\label{reve4}
{L_{g(m)^{-1}}}_*\left(d\tilde\phi^{-1}_{(m,e)}(0,X)\right)={L_{g(m)^{-1}}}_*\I_{g(m)\cdot{p}}(X)\eeq 
Mas por (\ref{Ip}) obtemos 
\beq\label{reve5}{L_{g(m)^{-1}}}_*\left(\I_{g(m)\cdot{p}}(X)\right)={L_{g(m)^{-1}}}_*\left({L_{g(m)}}_*\left(\I_p\left(\Ad\left(g(m)^{-1}\right)X\right)\right)\right)=\I_p\left(\Ad\left(g(m)^{-1}\right)X\right)\eeq 
Agora, por (\ref{reve3}) e (\ref{reve5}) obtemos
\begin{eqnarray}
\label{reve6}w&=&ds_m(u)+\I_p\left({L_{g(m)^{-1}}}_*(dg_m(u))+\Ad\left(g(m)^{-1}\right)X\right)\\
~&=&(d{\phi})^{-1}_{(m,e)}(u,{L_{g(m)^{-1}}}_*(dg_m(u))+\Ad\left(g(m)^{-1}\right)X)\\ 
~\therefore~~~\omega_p(w)&=&\omega^s_m(u)+{L_{g(m)^{-1}}}_*(dg_m(u))+\Ad\left(g(m)^{-1}\right)X
\end{eqnarray}
Por outro lado, utilizando as equa\c c\~oes (\ref{reve2}) e (\ref{reve7}):
\begin{eqnarray}
\widetilde\omega_p(w)&=&\widetilde\omega_{{g(m)}^{-1}\cdot{\tilde{s}(m)}}(w)\\
~&=&\Ad\left(g(m)^{-1}\right)\widetilde\omega_{\tilde{s}(m)}\left({L_{g(m)}}_*(w)\right)\\ 
~&=&\Ad\left(g(m)^{-1}\right)\widetilde\omega_{\tilde{s}(m)}\left({L_{g(m)}}_*\left({L_{g(m)^{-1}}}_*\left(d\tilde\phi^{-1}_{(m,e)}(u,X)\right)\right)\right)\\
~&=&\Ad\left(g(m)^{-1}\right)\widetilde\omega_{\tilde{s}(m)}\left(d\tilde\phi^{-1}_{(m,e)}(u,X)\right)\\
\label{reve9}~&=&\Ad\left(g(m)^{-1}\right)\left(\omega^{\tilde{s}}_m(u)+X\right) 
\end{eqnarray}
Finalmente aplicando a regra de transforma\c c\~ao (\ref{reve8}) a (\ref{reve9}) obtemos:
\begin{eqnarray*}
\widetilde\omega_p(w)&=&\Ad\left(g(m)^{-1}\right)\left(\Ad(g(m))\omega_p^s(u)+{R_{g(m)^{-1}}}_*dg_m(u)+X\right)\\ 
~&=&\omega_p^s(u)+{L_{g(m)^{-1}}}_*(dg_m(u))+\Ad\left(g(m)^{-1}\right)X
\end{eqnarray*}
 e portanto obtemos a igualdade almejada. Demonstrar unicidade da forma de conex\~ao \'e bem mais f\'acil. Se $\omega'$ \'e outra forma de conex\~ao, tal que para toda se\c c\~ao $s$, $s^*\omega=s^*\omega'$, elas claramente concordam sobre vetores da forma $ds_m(u)=d\phi^{-1}_{(m,e)}(u,0)$. Por outro lado sobre vetores verticais, $v=d\phi^{-1}_{(m,e)}(0,X)$ n\'os temos que $\omega(v)=\omega'(v)=\I_p^{-1}(v)=X$. Como $d\phi_p$ \'e isomorfismo, $\omega=\omega'$. $~~~~~~~~~~~~~~~~~~~~~~~~~~\blacksquare$\medskip

\subsubsection{Curvature in Principal Bundles}

\begin{defi} Se $\omega\in\Gamma\left(\Lambda^1(TP)\otimes\Lg\right)$ \'e forma de conex\~ao definimos a curvatura de $\omega$, $\Omega\in\Gamma\left(\Lambda^2(TP)\otimes\Lg\right)$ por $$\Omega:=d\omega+\omega\wedge\omega$$\end{defi} 
Lembramos que $d$ \'e a derivada exterior e que estamos utilizando a aplica\c c\~ao da \'algebra de Lie assim como em (\ref{omex}), i.e.: para $X_1,X_2\in\Gamma(TP)$, $$\omega\wedge\omega(X_1,X_2)=\frac{1}{2}\left(\omega(X_1)\omega(X_2)-\omega(X_2)\omega(X_1)\right)=[\omega(X_1),\omega(X_2)]$$ 

\begin{theo}\label{curvoutra} Se $\HH_p=\Ker\omega_p$ ent\~ao $d\omega\circ\widehat{H}=\Omega$. \end{theo}
{\bf{Dem}:} Queremos provar que $d\omega(\widehat{H}X,\widehat{H}Y)=\Omega(X,Y)$ para quaisquer $X,Y\in\Gamma(TP)$. Como j\'a foi demonstrado no {\bf{ Cap.I}} ambos os lados s\~ao bilineares e anti-sim\'etricos. Logo nos basta verificar a afirma\c c\~ao para tr\^es casos:
\begin{itemize}

\item{$X,Y\in\V$~:
N\'os temos, ainda no caso geral, 
\beq\label{omecurv}\Omega(X,Y)=d\omega(X,Y)+[\omega(X),\omega(Y)]\eeq
\beq\label{omecurv2}d\omega(X,Y)=Y[\omega(X)]-X[\omega(Y)]-\omega([X,Y])\eeq
Neste caso espec\'ifico claramente $d\omega(\widehat{H}X,\widehat{H}Y)=0$. 
Se $X$ \'e vertical, ent\~ao, $X_p=\I_p(\widetilde{X})$ para algum $\widetilde{X}\in\Lg$. Logo, como $d\omega$ \'e tensorial, $d\omega(X,Y)_p$ s\'o depende dos valores dos campos no ponto $p\in{P}$, portanto 
\beq{d\omega(X,Y)_p=d\omega\left(\I(\widetilde{X}),\I(\widetilde{Y})\right)_p}\eeq para $\widetilde{X},\widetilde{Y}\in\Lg$ apropriados. Portanto,  $\omega_p(X)=\I_p^{-1}(X)=\widetilde{X}$, \'e um elemento fixo da \'algebra de Lie, \'e um vetor constante, logo $Y_p[\omega(X)]=0$, o mesmo valendo trocando-se $X$ por $Y$,  e obtemos ent\~ao: 
\beq{d\omega(X,Y)_p=Y_p[\omega(X)]-X_p[\omega(Y)]-\omega_p([X,Y])=-\omega_p([X,Y])}\eeq
Agora, assumindo a identidade \cite{Spivak} $[\I(\widetilde{X}),\I(\widetilde{Y})]=\I([\widetilde{X},\widetilde{Y}])$, obtemos
\beq\omega_p([X,Y])=\omega_p([\I(\widetilde{X}),\I(\widetilde{Y})])=\omega_p(\I_p([\widetilde{X},\widetilde{Y}]))=[\widetilde{X},\widetilde{Y}]=[\omega_p(X),\omega_p(Y)]\eeq 
E pela equa\c c\~ao (\ref{omecurv}) obtemos $\Omega(X,Y)=0$. 
} 
\item{$X\in\V, Y\in\HH$: Novamente \'e claro que $d\omega(\widehat{H}X,\widehat{H}Y)=0$. Agora $\omega(Y)=0$ e portanto $\Omega(X,Y)=d\omega(X,Y)$. Utilizando o mesmo argumento do item anterior obtemos novamente de (\ref{omecurv2}): 
$$d\omega(X,Y)=-\omega([X,Y])$$ 
Mas se $X$ \'e vertical, $X$ \'e um campo tangente ao fluxo de $g(t)$ para algum $g:I\ra{G}$. Lembrando  que  
\beq[X,Y]=\lim_{t\ra0}\frac{1}{t}\left({L_{g(t)^{-1}}}_*Y-Y\right)\eeq  lembrando que $Y$ \'e horizontal, e o subfibrado horizontal \'e $G$-invariante n\'os temos que $[X,Y]\in\HH$. Logo $\omega([X,Y])=0$.  
}
\item{$X,Y\in\HH$ : Nesse caso $d\omega(\widehat{H}X,\widehat{H}Y)=d\omega(X,Y)$. Como $\omega(X)=0$,  de (\ref{omecurv}) obtemos tamb\'em $\Omega(X,Y)=d\omega(X,Y)=-\omega([X,Y])$.  
} 
\end{itemize}
$~~~~~~~~~~~~~~~~~~~~~~~~~~~~~~~~~~~~~~~~~~~~~~~~~~~~~~~~~~~~~~~~~~~~~~~~~~~~~\blacksquare$\medskip 

\'E interessante notar que a forma de curvatura  \'e nula se qualquer um dos campos for vertical, logo os campos relevantes em $TP$ ser\~ao aqueles levados de $M$ por alguma se\c c\~ao $s$, podemos ent\~ao considerar $\Omega$ como uma 2-forma em $TM$ a valores em $\Lg$.  A forma de curvatura de fibrados principais mede a falta de integrabilidade da distribui\c c\~ao horizontal em ser integr\'avel, em analogia com a forma de curvatura de fibrados vetoriais, que ``mede o quanto o mapa $X\longrightarrow\nabla_X$ falha em ser homomorfismo de \'algebras de Lie''. A analogia com o teorema de Frobenius \'e clara, e merece ser destrinchada. 

\subsubsection{Flat Connections}
Diremos que uma conex\~ao em um fibrado principal \'e {\it{plana}} ou {\it{flat}} se ao redor de cada ponto $p\in{P}$ existir  uma se\c c\~ao $s:\theta\ra\pi^{-1}(\theta)$ para a qual $p\in{s}(\theta)$ e tal que, para todo $q\in{s}(\theta)$ 
n\'os tenhamos um isomorfismo linear: 
$T_xM\overset{ds_x}{\simeq}\HH_q$, onde $\pi(q)=x$. Colocado de outra forma, uma conex\~ao \'e flat se existe uma trivializa\c c\~ao local $\phi:\pi^{-1}(\theta)\ra\theta\times{G}$ tal que para todo  $q\in{s}(\theta)$ 
$$\HH_q=d\phi^{-1}_{(x,e)}\left(T_xM\times\{0\}\right)$$
Como $\HH_{g\cdot{q}}={L_{g}}_*\HH_q$, por (\ref{cartaprop})
$$\HH_{g\cdot{q}}=d\phi^{-1}_{(x,g)}\left(T_xM\times\{0\}\right)$$
portanto podemos redefinir 
\begin{defi}Uma conex\~ao em um fibrado principal \'e {\it{plana}} ou {\it{flat}} se ao redor de cada ponto $p\in{P}$ existir
um aberto $\pi^{-1}(\theta)$ e trivializa\c c\~ao local $\phi:\pi^{-1}(\theta)\ra\theta\times{G}$ tal que para todo $q\in\pi^{-1}(\theta)$ existe $g\in{G}$ para o qual 
 \beq\label{plana}\HH_q=d\phi^{-1}_{(\pi(q),g)}\left(T_xM\times\{0\}\right)\eeq  \end{defi}

\begin{prop}Uma conex\~ao em $P$ \'e flat se e somente se a forma de curvatura \'e nula. 
\end{prop}
{\bf{Dem:}} ~Chamaremos a restri\c c\~ao de $\phi$ para um $g\in{G}$ fixo de $\phi^{-1}_g:\theta\ra\pi^{-1}(\theta)$. Suponhamos que a conex\~ao seja flat. N\'os temos que para todo $x\in\theta$, 
$$\HH_{\phi^{-1}{(x,g)}}=d\phi^{-1}_{(x,g)}\left(T_xM\times\{0\}\right)$$ ou seja
$$\HH_{\phi^{-1}_g{(x)}}=(d\phi_g^{-1})_{x}\left(T_xM\right)$$
o que por defini\c c\~ao significa que para cada $g\in{G}$,  $\phi^{-1}(\theta\times\{g\})=\phi^{-1}_g(\theta)$ \'e variedade integral da distribui\c c\~ao suave $\HH$. Mas se $X,Y\in\HH$, ent\~ao existem $\widetilde{X},\widetilde{Y}\in\Gamma(TM{|_\theta})$ tais que $d\phi^{-1}_g(\widetilde{X})=X\circ\phi^{-1}_g$ , o mesmo valendo para $\widetilde{Y}$, ou seja, estes campos s\~ao $\phi_g^{-1}$-relacionados. Portanto 
$$d\phi^{-1}_g([\widetilde{X},\widetilde{Y}])=[X,Y]\circ\phi^{-1}_g$$
ent\~ao, como para todo $q\in\pi^{-1}(\theta)$, existem $x\in\theta$ e $g\in{G}$ tais que $\phi^{-1}_g(x)=q$, n\'os temos 
$$d\phi^{-1}_g([\widetilde{X},\widetilde{Y}]_x)=[X,Y]_q$$ e portanto por hip\'otese: 
$$[X,Y]_q\in\HH_q$$
o que implica que mesmo se $X,Y\in\HH$, $\Omega(X,Y)=-\omega([X,Y])=0$, o que implica por sua vez que $\Omega=0$.

 Se por outro lado, supusermos que a forma de curvatura \'e nula, n\'os obtemos de cara que a distribui\c c\~ao horizontal \'e integr\'avel. Chamemos de $\Sigma'$ uma variedade integral de $\HH$ passando por $p\in{P}$. Como $\HH\oplus\V=TP$, para todo $q\in\Sigma$ obtemos que $T_q\Sigma'\oplus\V_q=T_qP$, ou seja, $\Sigma'$ \'e transversal ao subfibrado vertical.  Agora, n\'os sabemos que $d\pi_q:T_qP\ra{T_{\pi(q)}M}$ \'e submers\~ao, logo, como $\Ker{d\pi_p}=\V_p$,  n\'os temos que $d\pi_p:T_p\Sigma\ra{T_{\pi(p)}M}$ \'e isomorfismo linear. Portanto pelo teorema da fun\c c\~ao inversa, existe aberto  $\Sigma$ de $\Sigma'$ que \'e levado difeomorficamente por $\pi_{|\Sigma}$ em um aberto $\theta$ de $M$. Ent\~ao n\~ao podemos ter dois pontos distintos de $\Sigma$ sendo levados em um \'unico ponto de $M$, o que signifca que todos os pontos de $\Sigma$ intersectam as \'orbitas uma \'unica vez. Portanto n\'os obtemos que $\Sigma$ \'e uma se\c c\~ao de $P$ sobre $\theta$, e tamb\'em \'e uma variedade integral de H. Por ser se\c c\~ao, existe trivializa\c c\~ao local $\phi$ tal que para algum $g\in{G}$, $\phi^{-1}(\theta\times\{g\})=\Sigma$, como $\Sigma$ \'e variedade integral de $\HH$ obtemos o enunciado. Esse teorema significa que  a curvatura s\'o \'e  nula se existe algum referencial (ou sistema de coordenadas)  em que n\~ao se observa efeitos de curvatura (o que poderia ser chamado de referncial euclidiano, numa generaliza\c c\~ao da nomenclatura de $TM$), o que pode ainda ser considerada como outra faceta de seu car\'ater tensorial. $~~~~~~~~~~~\blacksquare$ 

\subsubsection{ Horizontal Liftings}
Uma constru\c c\~ao bastante utilizada em fibrados principais \'e a de {\bf{levantamento horizontal}} de campos e curvas. Dizemos que um campo $\widetilde{X}\in \Gamma(TP)$ \'e o levantamento horizontal de $X\in\Gamma(TM)$ se $\widetilde{X}\in\HH$ e para todo $p\in P$, $d\pi_p\widetilde{X}=X_{\pi(p)}$. A exist\^encia e unicidade de $\widetilde{X}$ \'e clara pela exist\^encia do isomorfismo linear $d\pi_p:\HH_p\ra T_{\pi(p)}M$. Que o levantamento \'e suave pode ser visto utilizando o fato que $\pi$ \'e submers\~ao, portanto existe  um campo suave $Y\in\Gamma(TP)$ que se projeta em $X\in\Gamma(TM)$, portanto sua componente horizontal \'e suave e tem a propriedade desejada. A invari\^ancia do campo $\widetilde{X}$ pela a\c c\~ao de $G$ \'e clara pela invari\^ancia de $\HH$ e unicidade de $\widetilde{X}$. 

Podemos tamb\'em tomar levantamento horizontal de curvas em $M$, o que pode ser demonstrado simplesmente tomando as curvas integrais do levantamento horizontal de $\gamma'$.

Outra forma de demonstra\c c\~ao \'e supor que $\alpha:I\ra\theta\subset M$ \'e uma curva suave tal que $\alpha(0)=x\in\theta$, ent\~ao, chamando uma dada trivializa\c c\~ao de $\phi:\pi^{-1}(\theta)\ra\theta\times G$, qualquer curva da forma $\gamma(t)=\phi^{-1}(\alpha(t),g(\alpha(t)))$ onde $g:\theta\ra G$ \'e curva suave tal que $g(\alpha(0))=e$, ser\'a um levantamento de $\alpha$ passando por $p$ em $t=0$. Agora,  
$$ \gamma'(t)=(d{\phi})^{-1}_{(\alpha(t),g(\alpha(t)))}(\alpha'(t),dg_{\alpha(t)}\alpha'(t))$$
e portanto utilizando (\ref{desconex5}), (\ref{desconex6}) e (\ref{aiai}), n\'os temos :
\beq \omega_{\gamma(t)} \Big({L_{g(\alpha (t))^{-1}}}_*(\gamma'(t))\Big)= \omega_{\gamma(t)}(ds_{\alpha(t)}(\alpha'(t)))+{L_{g(\alpha(t))^{-1}}}_*(dg_{\alpha(t)}(\alpha'(t)))\eeq 
Agora, n\'os sabemos que $\gamma'(t)$ \'e horizontal se e somente se ${L_{g(\alpha (t))^{-1}}}_*(\gamma'(t))$ for horizontal. Como $\omega_{\gamma(t)}(ds_{\alpha(t)}(\alpha'(t)))=\xi(t)$ \'e uma curva em $\Lg$, escrevendo $g(\alpha(t))=g(t)$, i.e.: ${L_{g(\alpha(t))^{-1}}}_*(dg_{\alpha(t)}(\alpha'(t)))=g(t)^{-1}g'(t)$, obtemos finalmente  que $\gamma(t)$ \'e horizontal se existe solu\c c\~ao, para a equa\c c\~ao
$$  g(t)^{-1}g'(t)=\xi(t)$$
onde $g(0)=e$. Ou seja, justamente a equa\c c\~ao (\ref{curvG}), cuja solu\c c\~ao existe e \'e \'unica pelo {\bf{Lema \ref{CurvaG}}}.  Portanto chegamos ao resultado de que se $\alpha(t)$ \'e uma curva suave em $M$ passando por $x=\alpha(0)$, dado um ponto $p\in P$ tal que $\pi(p)=x$, existe um \'unico levantamento horizontal de $\alpha(t)$ que passa por $p$ em $t=0$, que chamaremos de $\widetilde{\alpha}_p(t)$. Por unicidade, e novamente, como o subespa\c co horizontal \'e invariante por $G$, n\'os temos que $g\cdot (\widetilde{\alpha}_p)=\widetilde{\alpha}_{g\cdot p}$. 

\section{Frame Bundle}

Como mencionamos ao in\'icio do cap\'itulo, o conjunto de bases $G$-admiss\'iveis de um fibrado vetorial $E$ serviu de motiva\c c\~ao para a introdu\c c\~ao de fibrados principais. Verifiquemos ent\~ao que este conjunto \'e de fato um $G$-fibrado principal.  O conjunto das bases do $k$-fibrado vetorial $E$ $G$-admiss\'iveis, $P(E)$ \'e dado por $\underset{x\in{M}}{\bigcup}P(E)_x$ onde $P(E)_x$ \'e o conjunto de todas as bases admiss\'iveis de $E_x$. Ou seja $P(E)=\{(x,s(x)) ~|~x\in{M}, s(x)$ base admiss\'ivel de $E_x\}$ . Aqui a proje\c c\~ao 
leva simplesmente $(x,s(x))\mapsto{x}$. Como vimos ao in\'icio do cap\'itulo, existe bije\c c\~ao entre $G$ e $P(E)_x$ para todo $x\in{M}$, j\'a que dadas duas base admiss\'iveis existe um \'unico elemento de $G\subset{GL(k)}$ que leva uma na outra.   
Tomando uma se\c c\~ao lisa de bases sobre $E_{|\theta}$, i.e.: $s:x\mapsto{s(x)}\in{P(E)_x}$, definida  por $n$ se\c c\~oes lisas linearmente independentes de $E_{|\theta}$, e que chamaremos de agora em diante de referencial, temos a bije\c c\~ao 
\begin{gather*}
\phi^s:\theta\times{G}\rightarrow{P(E)_{|\theta}}\\
~~~~~(x,h)\longmapsto{h({s}(x))}
\end{gather*}  
Que \'e sobrejetor \'e claro, para mostrar que \'e tamb\'em injetora, basta notar que, j\'a que as fibras s\~ao disjuntas, se $x\neq{y}$  n\~ao existe $h\in{G}$ tal que $h({s}(x))=s(y)$, e se $x=y$ n\'os utilizamos o fato que  $G$ age livremente (injetoramente) sobre cada fibra. Com essa bije\c c\~ao induzimos uma estrutura diferenci\'avel em em $P(E)_{|\theta}$ pela estrutura diferenci\'avel em $\theta\times{G}$. 
 Consequentemente um mapa $f:{P(E)_{|\theta}}\ra{P(E)_{|\theta}}$ ser\'a suave se e somente se\footnote{Omitindo o super\'indice ``$s$".} $\widetilde{f}=\phi^{-1}\circ{f}\circ\phi$ for suave. 
\begin{eqnarray*} 
~{P(E)_{|\theta}} & \xrightarrow{f\phantom{1}} & {P(E)_{|\theta}}\\
\phi\uparrow & \phantom{\xrightarrow{\nabla\phantom{1}}} & \downarrow{\phi^{-1}}\\ 
~{P(E)_{|\theta}} & \xrightarrow[\widetilde{f}]{\phantom{\nabla{1}}} & {P(E)_{|\theta}}\\ 
\end{eqnarray*} 

Examinemos a a\c c\~ao do grupo $G$ sobre ${P(E)_{|\theta}}$. Seja $p\in{P(E)_{|\theta}}$, uma base sobre $x$ tal que $p=h_0({s(x))}=\phi(h_0,x)$, ent\~ao, fazendo $f=h\in{G}$, temos 
$$h(p)=h\left(h_0(s(x))\right)=\phi(x,hh_0)$$ 
mas ent\~ao $\tilde{h}:(x,h_0)\mapsto(x,hh_0)$, e como a a\c c\~ao $G\times{G}\ra{G}$ \'e suave, a a\c c\~ao de $\tilde{h}$ \'e suave e consequentemente a  a a\c c\~ao de $h$ \'e suave. Logo o grupo age suave e livremente sobre ${P(E)_{|\theta}}$. 

Se tivermos outro  referencial admiss\'ivel sobre $\theta$, $\tilde{s}:x\ra{P(E)_x}$, procedendo exatamente da mesma forma que fizemos na constru\c c\~ao de trivializa\c c\~oes locais adaptadas a se\c c\~oes de fibrados principais, por $\phi^s$ ser um difeomorfismo,  existe $g:\theta\ra{G}$ suave tal que $\tilde{s}(x)=g(x)(s(x))$, i.e.: $\phi^{-1}(\tilde{s}(x))=(x,g(x))$. Claramente $g^{-1}(x)=g(x)^{-1}$ que \'e suave $s(x)=g(x)^{-1}(\tilde{s}(x))$ portanto 
\begin{gather*}
\phi^{\tilde{s}}:\theta\times{G}\rightarrow{P(E)_{|\theta}}\\
~~~~~(x,h)\longmapsto{h({\tilde{s}}(x))} 
\end{gather*}   
induz estrutura difeom\'orfica \`aquela induzida por $\phi^s$ e portanto a estrutura diferenci\'avel de $P(E)$ est\'a bem definida, e portanto $P(E)$ \'e $G$-fibrado principal. Temos tamb\'em que, como as as se\c c\~oes de $P(E)$ s\~ao dadas localmente por $\phi(\theta\times\{g\})$ para $g$ fixo, elas s\~ao simplesmente referenciais admiss\'iveis, e $\phi^s$ \'e trivializa\c c\~ao local adaptada a $s$.   

Agora, dada uma conex\~ao admiss\'ivel $\nabla$ em $E$ (i.e.: tal que o transporte paralelo leva base admiss\'ivel em base admiss\'ivel), a maneira que pareceria \'obvia de obter uma conex\~ao em $P(E)$ seria aplicando essa conex\~ao em cada campo de um referencial. Em outras palavras, aplicando o transporte paralelo infinitesimal em cada elemento de um referencial de $E$ (que equivale a uma se\c c\~ao de $P(E)$).  Mais especificamente, dados um referencial $\{e_i\}_{i=1}^k$  sobre $\theta$ (que identificaremos com a se\c c\~ao $s:\theta\ra{P(E)}$), um vetor  $v\in{T_xM}$  tangente \`a curva $\sigma:I\ra{M}$ em $t=0$, e chamando sugestivamente a forma de conex\~ao correspondente a $\nabla$ nessa base de $\widetilde\omega^s$,  obtemos, utilizando (\ref{prinvet}) e  (\ref{con}): 
\begin{gather*}\prod_{i=i}^k\frac{d}{d{t}}_{|t=0}\Big(P_\sigma(0,t)e_i(\sigma(t))\Big)=:\frac{d}{d{t}}_{|t=0}\Big(P_\sigma(0,t)s(\sigma(t))\Big)=\widetilde\omega^s(v)s(x)\end{gather*} 
Pelo {\bf{Teorema {\ref{admiss}}}}, como $s$ \'e referencial admiss\'ivel, $\widetilde\omega^s(v)\in\Lg\equiv \End_\Lg(E_x)$, i.e.: temos uma representa\c c\~ao $d\rho:\Lg\ra\End(E_x)$ tal que $\widetilde\omega^s(v)=d\rho(\omega^s(v))$, onde $\omega^s(v)\in\Lg$. Al\'em disso, por (\ref{mudanca}) n\'os temos que dado outro referencial $\tilde{s}$, por $d\rho$ ser representa\c c\~ao linear vale 
$$ \omega^{\tilde{s}}=\Ad(g)\circ\omega^s+(dg)g^{-1}$$
portanto,  satisfizemos as hip\'oteses da {\bf{Proposi\c c\~ao \ref{conversamente}}}, e podemos definir unicamente a forma de conex\~ao $\omega=\I^{-1}\circ\widehat{V}$ no fibrado principal tal que tenhamos  $s^*\omega=\omega^s$. A saber, por (\ref{reve1}): 
$$\omega_p(w)=\omega_p\left((d{\phi^s})^{-1}_{(x,e)}(u,X)\right):=\omega^s_x(u)+X$$
onde $x\in\theta$, $s$ \'e um referencial tal que a base sobre $x$ \'e dada por $s(x)=p$ , $\phi^s$ \'e a trivializa\c c\~ao local adapatada a $s$ destrinchada acima,  $u\in{T_xM}$ , $X\in\Lg$  e $\omega^s$ \'e a forma de conex\~ao em $E$ relativa a $\nabla$ e ao referencial $s$. 
 
Por outro lado, se nos for dado $\omega\in\CC(P(E))$, dado um referencial $s$  basta definirmos a forma de conex\~ao em $E$ relativa a $s$ como uma representa\c c\~ao de $\omega^s=s^*\omega$, i.e.: $\widetilde\omega^s=d\rho(s^*\omega)$. 
Utilizando o resultado do  {\bf{Teo.\ref{corresp}}} e do {\bf{Teo.\ref{corresp2}}}, provamos ent\~ao:
\begin{theo}\label{correspond}
Existe correspond\^encia bijetora $\nabla\ra\HH$ entre conex\~oes em $E$ e em $P(E)$.\end{theo} 

\subsubsection{ Invariant M\'etrics Revisited}

Seja $\pi:P\ra{M}$ um $G$-fibrado principal, com $G$ ainda compacto. Seja $\langle\cdot,\cdot\rangle$ um produto interno Ad-invariante\footnote{Como $G$ \'e compacto, admite m\'etrica bi-invariante, se $G$ for conexo isto \'e equivalente a um produto interno Ad-invariante em $\Lg$.} em $\Lg$, i.e.: $\langle\Ad(g)u,\Ad(g)v\rangle=\langle u,v\rangle$ para quaisquer $u,v\in\Lg$ . Utilizando o isomorfismo dado por $\I_p:\Lg\simeq\V_p$, por (\ref{Ip}) n\'os temos uma m\'etrica riemanniana $G$-invariante sobre o fibrado vertical. Isto \'e, sejam $\I_p(u),\I_p(v)\in\V_p$ ent\~ao denotando da mesma forma o produto interno em $\V$ definimos:~ $\langle \I_p(v),\I_p(u)\rangle_p:=\langle u,v\rangle$. Logo, para $g\in G$ utilizando (\ref{Ip}) 
\begin{gather*}\langle L_{g_*}(\I_p(v)),L_{g_*}(\I_p(u))\rangle_{g\cdot p }=\langle \I_{g\cdot p}(\Ad(g)u),\I_{g\cdot p}(\Ad(g)v)\rangle_{g\cdot p}\\
=\langle\Ad(g)u,\Ad(g)v\rangle=\langle u,v\rangle
\end{gather*} 
Chamaremos uma m\'etrica $\gamma$  de invariante para $P$ se $\gamma$ for uma m\'etrica riemanniana em $P$ que \'e invariante em rela\c c\~ao \`a a\c c\~ao de $G$ e se restringe \`a m\'etrica definida acima sobre o fibrado vertical. 

Agora se $\V^\perp$ for o subfibrado de $TP$ ortogonal a $\V$ em rela\c c\~ao a $\gamma$, ent\~ao claramente $\V^\perp$  \'e transversal a $\V$. Como $\V$ e $\gamma$ s\~ao $G$-invariantes, tamb\'em o ser\'a $V^\perp$, portanto $\V^\perp=\HH$ \'e uma conex\~ao para $P$. Al\'em disso como vimos, h\'a um isomorfimo entre $\HH_p$ e $T_{\pi(p)}M$, e portanto induzimos uma \'unica m\'etrica sobre $M$ pela proje\c c\~ao $\pi$. Analogamente se tivermos uma m\'etrica $h$ em $M$ e um fibrado horizontal qualquer $\HH$, temos uma \'unica m\'etrica induzida em $\HH$ por $\pi^*h$. Portanto como $TP=\V\oplus\HH$ e a restri\c c\~ao da m\'etrica para $\V$ \'e can\^onica, temos uma \'unica m\'etrica $\gamma$ em $P$, tal que $\HH\perp\V$, definida por uma conex\~ao H, uma m\'etrica $h$ em $M$ e uma m\'etrica de $\Lg$ Ad-invariante $\alpha$. A saber, lembrando que dado $\HH$, h\'a uma \'unica forma de conex\~ao correspondente $\omega_p=\I_p^{-1}\circ\widehat{V}$, temos 
\beq\label{metricP}\gamma=\pi^*h+\alpha\circ\omega\eeq
Se $u,v\in\V_p$ ~ent\~ao $$\gamma_p(u,v)=\alpha(\I_p^{-1}(u),\I_p^{-1}(u))$$
j\'a que $d\pi_p(v)=d\pi_p(u)=0$. Se $u,v\in\HH_p$ temos 
$$\gamma_p(u,v)=\gamma_{\pi(p)}(d\pi_p(u),d\pi_p(v))$$ 
j\'a que $\omega_p(u)=\omega_p(v)=0$  e finalmente se $u\in\HH_p$ e $v\in\V_p$ ent\~ao $\gamma_p(u,v)=0$ j\'a que $\omega_p(u)=0$ e $d\pi_p(v)=0$. Como tanto $\pi^*h$ quanto $\alpha\circ\omega$ s\~ao $G$-invariantes, $\gamma$ \'e $G$-invariante. Chamamos tais m\'etricas de m\'etricas de fibrado (bundle metrics).

Como veremos mais tarde, existem rela\c c\~oes interessantes entre a geometria de $(M,h)$ e a de $(P,\gamma)$ envolvendo a conex\~ao. Estas rela\c c\~oes s\~ao centrais \`a unifica\c c\~ao da gravita\c c\~ao e campos de Yang-Mills. 

\section{Associated Bundles}
Dado um $G$-fibrado vetorial $E$ sobre $M$, com fibra t\'ipica isomorfa ao espa\c co vetorial $\R^k$, na {\bf{Sec.2.3}} n\'os constru\'imos um $G$-fibrado principal sobre $M$ correspondente. O que podemos dizer sobre o procedimento inverso, i.e.: dado um $G$-fibrado principal  sobre $M$ podemos associar a ele um fibrado vetorial com fibra t\'ipica isomorfa a $\R^k$ e variedade base $M$? Na verdade veremos que podemos atingir um resultado mais geral, com a fibra t\'ipica sendo difeomorfa a uma variedade suave qualquer onde o grupo age como grupo de transforma\c c\~oes. 

 Para ilustrar bem nosso objetivo, tomemos o fibrado dos referenciais $P(E)$, a partir do qual tentaremos reconstruir $E$. A id\'eia que surge naturalmente \'e tomar um vetor em $E_x$ como uma escolha de valores para os elementos de uma base em $p\in P(E)$, ou seja, como suas coordenadas $(a^1,\cdots, a^k)$ na base $p$. Suponhamos que $x\in \theta\subset M$ ,  $v\in E_x$ e $p=(e_1(x),\cdots,e_n(x))$, onde os $e_i$ s\~ao se\c c\~oes linearmente independentes de $E|_\theta$. Ent\~ao $v=a^1e_1(x)+\cdots+a^ke_k(x)$ \'e a a\c c\~ao natural de $P$ em $\R^k$. Chamando $(a^1,\cdots, a^k):=\tilde{v}$, esta a\c c\~ao \'e simplesmente dada por $\tilde{v}p$, a multiplica\c c\~ao de uma matriz linha por uma matriz coluna, e poder\'iamos pensar em identificar dessa forma,   $v\simeq(\tilde{v},p)\in \R^k\times P$. Obviamente essa identifica\c c\~ao \'e insuficiente, pois h\'a muitas bases e muitos elementos de $\R^k$ correspondentes que resultariam no vetor de $E_x$ em quest\~ao. Precisamos tomar o quociente pelos isomorfismos lineares das bases, i.e.: pela a\c c\~ao dos elementos da fibra $\pi^{-1}(x)$. Agora, identificando a matriz linha $\tilde{v}$ a uma 1-forma em $E_x^*$, n\'os temos que por um isomorfismo $g(x)$ de $E_x$, como vimos no primeiro cap\'itulo, temos as seguintes transforma\c c\~oes: 
$$\begin{array}{lll}
g(x):& \tilde{v} & \longmapsto \tilde{v}g(x)^{-1}\\
~& e_i(x)  & \longmapsto g(x)( e_i(x))\end{array}$$ 
Claramente temos ent\~ao que $\tilde{v} g(x)^{-1}g(x)(p)=\tilde{v}p=v$ e portanto tamb\'em identificar\'iamos $v\simeq(\tilde{v} g(x)^{-1},g(x)( e_i(x))$. A solu\c c\~ao para eliminarmos esta redund\^ancia \'e \'obvia, quocientarmos pela rela\c c\~ao de equival\^encia em  $P(E)\times \R^k$ dada por 
$(\rho(g)p,v\rho(g^{-1}))\sim(p,v)$, ou, substituindo $\rho(g)$ por $g$, onde $\rho:G\ra\Aut(E)$,  podemos escrever mais sucintamente:
$$(gp,vg^{-1})\sim(p,v)$$
o que \'e compat\'ivel com nosso uso de multiplica\c c\~ao \`a esquerda pelo grupo (a representa\c c\~ao do grupo em $\R^k$ vai \`a direita pela inversa, j\'a que $\R^k$ corresponde ao espa\c co dual). A nota\c c\~ao usual \'e feita para um fibrado $G$-principal $P$, no qual $G$ age \`a direita, e nos conformaremos a ela. Um breve adendo: se $G$ age \`a direita sobre dois espa\c cos, $A,B$ e existe uma a\c c\~ao de $\mu:A\times B\ra C$, para um outro espa\c co qualquer $C$, e se essa a\c c\~ao \'e invariante pela a\c c\~ao do mesmo elemento de $G$ nos dois espa\c cos,  i.e.: $\mu(a\cdot_Ag,b\cdot_Bg)=\mu(a,b)$ dizemos que ela \'e $G$-equivariante.

No nosso caso, a representa\c c\~ao do grupo \'e a mesma em $E_x\simeq\R^k$ e $P(E_x)$, portanto o quociente pela a\c c\~ao do grupo como acima est\'a inclusa na a\c c\~ao da base sobre as coordenadas, i.e.: $[p,v]=pv$. No caso geral, a a\c c\~ao do grupo pode ser distinta para os dois espa\c cos, por isso definimos:
\begin{defi}
Seja $(P,\pi,M)$ um $G$-fibrado principal sobre o qual $G$ age a direita por``$\cdot$'', e seja $F$ um espa\c co no qual $G$ age  \`a esquerda (cuja a\c c\~ao denotaremos por ``$\cdot$''), ent\~ao definimos o {\bf{fibrado associado a $P$ pela a\c c\~ao de $G$ em $F$}} como o fibrado 
$$P_F=P\times_GF:=P\times F/\sim$$ onde definimos a rela\c c\~ao de equival\^encia, juntamente com a a\c c\~ao do grupo em $P\times F$ (que tamb\'em denotaremos por ``$\cdot$''), como 
$$(p,v)\sim(p\cdot g,g^{-1}\cdot v):=(p,v)\cdot g$$ A proje\c c\~ao \'e dada por 
$$\begin{array}{ll}
\pi_F: & P_F \lra M\\
~ & [p,v]\mapsto \pi(p)
\end{array}$$

\end{defi}

 Notemos que que $\pi_F$ est\'a bem definida, j\'a que se $(p_1,v_1)\sim(p_2,v_2)$ ent\~ao $p_2=p_1\cdot g$ para algum $g\in G$ e portanto $\pi(p_2)=\pi(p_1)$. 
Precisamos agora voltar e mostrar que a motiva\c c\~ao do conceito de fibrado associado faz sentido, i.e.: que cada fibra de $P_F$ \'e difeomorfa a $F$.
\begin{theo}
Para cada $x\in{M}$, a fibra $\pi_F^{-1}(x)$ \'e difeomorfa a $F$.\end{theo}
{\bf{Dem:}}
Fixado um ponto $p\in P$ sobre $x\in M$ existe um mapa associado $\imath_p:F\ra P_F$ definido por $\imath_p(v):=[p,v]$, no nosso exemplo esse mapa corresponde a tomarmos $k$ coordenadas e associar-las \`a base $p$. Agora, $\pi_F[p,v]=\pi(p)=x$ logo $\imath_p(F)\subset\pi_F^{-1}(x)$. Agora, para $[q,u]\in\pi_F^{-1}(x)$ , definimos 
$$\begin{array}{ll}
\jmath_{p}: & \pi^{-1}_F(x)\ra F\\
~ & [q,v]\longmapsto g_p(q)\cdot v
\end{array}$$
onde $g_p:\pi^{-1}(x)\ra G$ \'e o mapa que fornece o elemento em $G$ que ``liga'' $p$ a $q$, i.e: $p\cdot g_p(q)=q$, como mostramos nas primeiras se\c c\~oes deste capi\'itulo, \'e um difeomorfismo. Provemos que o mapa $\jmath_p$ est\'a bem definido:  para $g\in G$, n\'os temos $ g_p(q\cdot g)=g_p(q)g$ j\'a que, utilizando a defini\c c\~ao de $g_p$: 
$$p\cdot (g_p(q\cdot g))=q\cdot g= (p\cdot g_p(q))\cdot g=p\cdot (g_p(q)g)$$
Logo  $$\jmath_p[q\cdot g,g^{-1}\cdot v]=g_p(q)gg^{-1}\cdot v=g_p(q)\cdot v$$
Ent\~ao obtemos, qualquer que seja $v\in F$, $$
\begin{array}{l} 
(\i)~~~\imath_p\circ\jmath_p[q,v]=\imath_p(g_p(q)\cdot v)=[p,g_p(q)\cdot v]=[p\cdot g_p(q),v]=[q,v]\\
(\i\i)~~\jmath_p\circ\imath_p(v)=\jmath_p[p,v]=g_p(p)\cdot v =v
\end{array}$$
 e portanto $\imath_p$ e $\jmath_p$ s\~ao inversas suaves. $~~~\blacksquare$\medskip 

N\'os temos ainda que se $\theta$ \'e um aberto de $M$, $$\pi_F^{-1}(\theta)\simeq(\pi^{-1}(\theta)\times F)/G\simeq(\theta\times G\times F)/G\simeq\theta\times F$$
Portanto \'e localmente trivial, logo podemos introduzir em $P_F$ uma estrutura diferenci\'avel requerendo que $\pi_F^{-1}(\theta) $ seja uma subvariedade aberta de $E$ difeomorfa a $\theta\times F$ pelo difeomorfismo induzido por uma  trivializa\c c\~ao qualquer $\pi^{-1}(\theta)\simeq\theta\times G$.  Dadas duas tais  trivializa\c c\~oes $\psi_1,\psi_2$ sobre $\theta$, i.e.: 
$$\begin{array}{ll}
\psi_i: & \theta\times F\ra  \pi_F^{-1}(\theta)\\
~ & (x,v)\longmapsto [s_i(x), v]
\end{array}$$
onde $s_i:\theta\ra \pi^{-1}(\theta)$ \'e se\c c\~ao suave de $P$ n\'os obtemos\footnote{Note que utilizamos a mesma constru\c c\~ao para obtermos a correspond\^encia entre diferentes trivializa\c c\~oes quaisquer no fibrado principal.}, utilizando o mapa de transi\c c\~ao  $g_{12}:\theta\ra G$ entre as se\c c\~oes $s_1,s_2$,  que 
$$\psi^{-1}_2\circ\psi_1(x,v)=\psi^{-1}_2[s_1(x),v]=\psi^{-1}_2[s_1(x)\cdot g_{12}(x),g_{12}^{-1}v]=[s_2(x),g_{12}^{-1}(x)\cdot v]=(x,g_{12}^{-1}(x)\cdot v)$$
N\'os obtemos ent\~ao que se come\c camos com um $E$ \'e um $G$-fibrado vetorial com fibra t\'ipica $V$, e tomamos $G\equiv\Aut_G(V)$ , ent\~ao $P(E)\times_G V=E$. 

Agora, n\'os j\'a provamos que existe correspond\^encia bijetora entre conex\~oes em $E$ e conex\~oes  em $P(E)$ ({\bf{Teo. \ref{correspond}}}), simplesmente definindo a forma de conex\~ao em $\pi_V^{-1}(\theta)$ como $\widetilde\omega^s_x=d\rho_x(s^*\omega)_x$ onde $\omega_p:T_pP\ra\Lg$ ,  $d\rho_x:\Lg\ra\End_\Lg(E_{\pi(p)})$ \'e uma representa\c c\~ao de $\Lg$ , $s$ \'e uma se\c c\~ao de $P$ sobre $\theta$ e $\omega$ \'e uma forma de conex\~ao em $P$. 

 Se tomarmos a descri\c c\~ao por transporte paralelo, isto \'e equivalente a, dado um ponto $p\in P(E)$ sobre $x\in M$, uma curva suave $\alpha:I\ra M$ tal que $\alpha(0)=x$ e um vetor de $E_x$, $v=[p,\tilde{v}]$ onde $\tilde{v}$ \'e a proje\c c\~ao de $v$ na base $p$,  definirmos 
$$P_\alpha(0,t)v:=[\widetilde{\alpha}_p(t),v]=\widetilde{\alpha}_p(t)v$$ 
onde $\widetilde\alpha_p(t)$ \'e o \'unico levantamento horizontal de $\alpha(t)$ em $P$ passando por $p$ em $t=0$, ou seja, mantemos as coordenadas do vetor fixo em termos de uma dada base paralela ao longo de $\alpha(t)$. 
Em outras palavras, derivamos como fun\c c\~oes em $\R^k$ as coordenadas de um campo (se\c c\~ao de $E$) em rela\c c\~ao a uma base paralela. Como mostramos que temos o conceito de levantamento horizontal de curvas em fibrados principais gerais, podemos tomar essa defini\c c\~ao para o caso geral de fibrado associado. 

 Relembrando, n\'os temos, como $\jmath_p:F\ra F_{\pi(p)}$ \'e difeomorfismo, que para cada $p\in \pi^{-1}(x)$ e $q\in\pi_F^{-1}(x)$ existe um \'unico $v(q,p)\in F$ tal que $(p,v(q,p)\in q$. Seja ent\~ao $s:M\ra P\times_G F$ uma se\c c\~ao, para todo $p\in P$  existe $v(s(\pi(p)),p)=:\varphi^s(p)$ tal que $[p,\varphi^s(p)]=s(\pi(p))$, onde denotamos a fun\c c\~ao que, dada essa se\c c\~ao, leva $p$ em $v(p)$ por
$\varphi^s:  P\ra F$. N\'os temos que $\varphi(p\cdot g)=g^{-1}\cdot \varphi^s(p)$ j\'a que temos, para uma dada se\c c\~ao $s$ e dados $x\in\theta$ , $p,q\in\pi^{-1}(x)$: 
$$[q,\varphi^s(q)]=[p\cdot g, \varphi^s(p\cdot g)]=s(x)=[p,\varphi^s(p)]=[p\cdot g,g^{-1}\cdot\varphi^s(p)]$$

Na nossa analogia do fibrado das bases, isso nos d\'a para cada campo $Y\in\Gamma(E)$ uma associa\c c\~ao entre  as bases sobre $\theta$ e a descri\c c\~ao de $Y$ sobre essas bases. O que nos resta agora para acharmos $\nabla^{P_F}_XY$, onde $\nabla^{P_F}$ \'e a conex\~ao em $P\times_G F$ e $X\in\Gamma(TM)$, \'e  tomarmos, na analogia do fibrado das bases, a derivada dos coeficientes sobre uma base paralela ao longo de $X$. Como o levantamento horizontal est\'a bem definido, \'e  exatamente isso que fazemos:
\beq \varphi^{\nabla^{P_F}_Xs}=\widetilde{X}[\varphi^s]\eeq
onde $\widetilde{X}$ \'e o levantamento horizontal de $X$, como descrevemos na se\c c\~ao anterior, e aqui o colchete designa deriva\c c\~ao, e n\~ao quociente. Ambos os lados pertencem a $C^\infty(P,F)$ e o que obtemos \'e que, para todo $p\in P$,
\beq \label{covassociado}[p,\widetilde{X}[\varphi^s](p)]= \nabla^{P_F}_Xs(\pi(p)) \eeq

Lembrando que ${R_g}_*\widetilde{X}=\widetilde X$, e portanto a equivari\^ancia da a\c c\~ao de $G$ \'e mantida e vemos que a equa\c c\~ao est\'a bem-definida. Explicitamente, n\'os utilizamos a seguinte proposi\c c\~ao:
\begin{prop}{Seja $X$ um campo $C^\infty$ em $M$, e $F:M\rightarrow{M}$ um difeomorfismo, e seja $\theta(t,p)$ o fluxo de $X$, ent\~ao $X$ \'e invariante por $F$ se e somente se ${F(\theta(t,p))}=\theta(t,F(p))$
}\end{prop} 

{\bf{Dem:}} Se $F(\theta(t,p))=\theta(t,F(p))$ ent\~ao n\'os temos que
$$dF_p\left(\frac{d}{dt}_{|_{t=o}}\theta(t,p))\right)=dF_p(X_p)$$
Mas $$\frac{d}{dt}_{|_{t=o}}\theta(t,F(p))=X_{F(p)}$$  
Da mesma forma $$\frac{d}{dt}_{|_{t=o}}F(\theta(t,p))=dF_p(X_p)=X_{F(p)}=\frac{d}{dt}_{|_{t=o}}(\theta(t,F(p)))$$ 
 portanto, pela unicidade de curvas integrais:
$$\theta(t,F(p))=F(\theta(t,F(p)))$$ 
{{$\blacksquare$}}\medskip

Logo temos, chamando ainda  $\theta(t,p)$ o fluxo de $X$ por $p$,
\begin{eqnarray*}
\varphi^{\nabla^{P_F}_Xs}(g\cdot p)&=&\widetilde{X}[\varphi^s](g\cdot p)\\
~&=&\frac{d}{dt}_{|_{t=o}}\left(\varphi^s(\theta(t,g\cdot p))\right)\\
~&=&\frac{d}{dt}_{|_{t=o}}\varphi^s\left(g\cdot(\theta(t,p))\right)\\
~&=&g^{-1}\cdot\varphi^{\nabla^{P_F}_Xs}( p)
\end{eqnarray*}
 Outra observa\c c\~ao importante \'e a de que, olhando bem para o lado esquerdo de (\ref{covassociado}), percebemos que na linguagem de fibrado das bases, ele representa a derivada dos coeficientes de um campo ao longo de uma curva horizontal de bases passando por $p$, justamente o que procur\'avamos. Notamos tamb\'em que a conex\~ao em $P_F$ ``desceu'' de uma $G$-conex\~ao de $P$, j\'a que utilizamos tanto levantamento horizontal quanto $G$-invari\^ancia.

Suponhamos agora que $V$ seja um espa\c co vetorial  riemanniano, com m\'etrica $\langle\cdot,\cdot\rangle$ sobre o qual age a representa\c c\~ao $\rho:G\ra V$. N\'os podemos inicialmente tentar induzir uma m\'etrica em $P_V=P\times_GV$  definindo, para qualquer $p\in\pi^{-1}(x)$ e $v,v'\in V$:
\beq \label{metassocia}\langle[p,v],[p,v']\rangle_{\pi^{-1}_V(x)}:=\langle v,v'\rangle\eeq
Lembramos que {\it{qualquer}} elemento de $\pi^{-1}_V(x)$ pode ser escrito em termos de {\it{qualquer}} $p\in\pi^{-1}(x)$. Mas para que este produto interno fa\c ca sentido precisamos tomar {\it{o mesmo}} $p$ (o que equivaleria, no fibrado das base, a tomar o produto interno das coordenadas escritas na mesma base). Verifiquemos se isto est\'a bem definido, i.e.: se o produto interno n\~ao depende do ponto $p$ que escolhemos: 
$$\langle[p,v],[p,v']\rangle_{\pi^{-1}_V(x)}=\langle[p\cdot g,\rho(g^{-1})v],[p\cdot g,\rho(g^{-1})v]\rangle_{\pi^{-1}_V(x)}=\langle \rho(g^{-1})v,\rho(g^{-1})v'\rangle$$
 Ent\~ao n\'os temos que s\'o podemos passar a m\'etrica consistentemente para o fibrado associado se a a\c c\~ao do grupo em $V$ for ortogonal em rela\c c\~ao a m\'etrica, i.e.: $\rho:G\ra\OG V$. Isso implica n\~ao s\'o que a m\'etrica induzida em $P_F$ \'e $G$-invariante, mas que \'e tamb\'em compat\'ivel com qualquer $G$-conex\~ao provinda de $P$. Vejamos como: denotando tamb\'em por $\rho_x$ a a\c c\~ao de $G$ em $\pi^{-1}_F(x)$, temos que $\rho_x:G\ra \OG{\pi^{-1}_F(x)}$. Logo 
\beq\label{represaadj}d(\rho_x)_e:\Lg\ra \mathfrak{s}\mathfrak{o}(\pi^{-1}_F(x))\eeq o que implica que a representa\c c\~ao de $\omega^s$ ser\'a uma  matriz anti-sim\'etrica, o que em seu turno implica que a conex\~ao \'e compat\'ivel com a m\'etrica.

\chapter{ Yang-Mills Fields and Characteristic Classes}

\begin{quote}
{\it{H\'a uma m\'ascara de teoria sobre toda a face da natureza.}} - William Whewell\smallskip

{\it{Como pode ser que a matem\'atica, sendo antes de tudo um produto do pensamento humano independente da experi\^encia, \'e t\~ao admiravelmente adaptada aos objetos da realidade?}} - Albert Einstein

\end{quote}
\section{Yang-Mills}
\subsubsection{Introduction}

A aplicabilidade da teoria da matem\'atica \'e um assunto filosoficamente interessante e profundo; ser\'a que a matem\'atica \'e t\~ao \'util para a descri\c c\~ao da Natureza porque a sele\c c\~ao natural de Darwin beneficiou um processamento de informa\c c\~oes adaptado \`a realidade, o que conhecemos por l\'ogica? Ou simplesmente \'e o que temos em m\~ao e procuramos a rela\c c\~ao inversa; adaptar a realidade \`a matem\'atica? A teoria de Gauge e a equa\c c\~ao de Yang-Mills constituem  exemplos dos mais formid\'aveis de uma converg\^encia n\~ao intencional de f\'isica e  matem\'atica, refor\c cando a primeira hip\'otese. \'E bem verdade que Yang e Mills procuravam exclusivamente uma generaliza\c c\~ao das equa\c c\~oes de Maxwell, sem nenhum conhecimento de sua rela\c c\~ao com uma interpreta\c c\~ao geom\'etrica por fibrados (interpreta\c c\~ao que procuramos esmiu\c car). 

Sua busca era mais que justificada: a teoria da eletrodin\^amica qu\^antica \'e uma das mais bem sucedidas da hist\'oria da f\'isica. O objetivo da f\'isica (ou o de uma grande parte dos f\'isicos) era (e talvez ainda seja) o de  colocar todas as part\'iculas no mesmo p\'e que o f\'oton. Apesar de n\~ao tratarmos aqui do aspecto qu\^antico das teorias de Gauge, segundo Atiyah \footnote{Ver \cite{Atiyah} que seguimos livremente nesta introdu\c c\~ao.}  ``podemos dizer que uma compreens\~ao profunda da teoria cl\'assica \'e provavelmente um pr\'e-requisito para o desenvolvimento da teoria qu\^antica''. 

 Faremos, antes de come\c carmos a exposi\c c\~ao matem\'atica mais pesada, uma breve introdu\c c\~ao f\'isica dos conceitos da teoria cl\'assica de Gauge, muitos deles j\'a explorados por n\'os nos cap\'itulos anteriores.  

Imaginemos uma part\'icula  em $M$, variedade semi-riemanniana quadridimensional. Suponhamos que essa part\'icula tenha alguma esp\'ecie de  estrutura interna  i.e.: ela tem uma posi\c c\~ao $x\in M$ e est\'a em um estado interno particular   neste ponto.  Suponhamos ainda que este espa\c co interno possua  simetrias suaves, modeladas pelo grupo de Lie $G$.   Consideraremos ent\~ao o espa\c co total de todos os estados de uma tal part\'icula, que chamaremos de $E$.

 A curvatura $\Omega$ pode ser tomada como a distor\c c\~ao das fibras provocada pelo campo\footnote{ Na \'area cercada por um paralelogramo infinitesimal, dado por duas dire\c c\~oes em $x$.}, se pensarmos no campo como dado por seus efeitos locais. 
Podemos identificar coerentemente nossos espa\c cos internos sobre $M$ se para quaisquer caminhos que tomarmos entre dois pontos o estado interno final da part\'icula for o mesmo. Se assumirmos que o estado interno da part\'icula \'e levado ao longo do trajeto por transporte paralelo, i.e.: de modo que ``conserve" seu estado interno, pelo {\bf Teo.\ref{variation}}  essa condi\c c\~ao \'e equivalente\footnote{\'E equivalente em um espa\c co simplesmente conexo, n\~ao o sendo temos o efeito intrigante de Aharonov-Bohm \cite {Baez}.} a n\~ao termos curvatura, ou campo externo, j\'a que n\~ao h\'a maneira de medi-lo por seu efeito se todos os caminhos entre dois pontos n\~ao induzem nenhuma diferen\c ca na estrutura interna da part\'icula.

 Qualquer  identifica\c c\~ao de espa\c cos internos \'e chamada de uma escolha de Gauge, e uma mudan\c ca de um gauge a outro \'e chamado de transforma\c c\~ao de gauge, que a cada ponto $x$  associa uma transforma\c c\~ao do espa\c co interno $G$. Sem curvatura, todas s\~ao equivalentes, e se ligarmos um campo externo todas detectar\~ao igualmente discrep\^ancias no estado interno final de part\'iculas tomando caminhos distintos.

Essa identifica\c c\~ao de campos com distor\c c\~oes geom\'etricas tamb\'em \'e central \`a teoria da relatividade geral. A diferen\c ca aqui \'e que essa distor\c c\~ao n\~ao ocorre {\it{no}} espa\c co-tempo, mas na geometria de um espa\c co de estrutura interna, superposto ao espa\c co-tempo. Como assumimos que o grupo de estados internos \'e bem mais simples do que o de transforma\c c\~oes de coordenadas (tem dimens\~ao finita), isto se traduz em uma relativa simplifica\c c\~ao da teoria em compara\c c\~ao com a geometria riemanniana da relatividade geral.

  Historicamente, potenciais foram introduzidos como um instrumento de simplifica\c c\~ao das equa\c c\~oes do campo, e a ambiguidade inerente em sua escolha (liberdade de Gauge) era tida como uma indica\c c\~ao de que n\~ao possu\'ia significado f\'isico. Pelo ponto de vista geom\'etrico, o potencial, identificado \`a conex\~ao,  t\^em exist\^encia pr\'opria e bem definida,  somente a escolha de um gauge (ou referencial) n\~ao tem significado f\'isico. Ou seja uma conex\~ao nos fornece um meio de quantificar a varia\c c\~ao de estados internos ao longo de trajet\'orias em $M$. 

Explicitamente, tomamos $P$ como o $G$-fibrado principal das bases $G$-admiss\'iveis sobre $M$ (ver {\bf Sec2.3}), por exemplo os eixos de isospin sobre $M$. A conex\~ao $\omega$ nos fornece uma identifica\c c\~ao intr\'inseca entre (bases de) estados internos sobre diferentes pontos, ela nos diz como manter uma base de estados internos (e.g.: de isospin) ``fixa" ao longo de uma trajet\'oria qualquer $\gamma$.  Em termos dessa base \'e f\'acil quantificar a varia\c c\~ao do estado interno da part\'icula (de seu isospin) ao longo de $\gamma$: simplesmente utilizamos a conex\~ao em $E$ como fibrado associado a $P$, explicitada\footnote{I.e.: simplesmente derivamos as coordenadas da decomposi\c c\~ao do estado interno em rela\c c\~ao a uma base ``fixa" ao longo de $\gamma$.} em (\ref{covassociado}).  

At\'e aqui consideramos uma part\'icula como no caso cl\'assico, como tendo trajet\'oria bem definida etc. Apesar da abordagem ortodoxa da mec\^anica qu\^antica ser feita atrav\'es de campos, essa nossa descri\c c\~ao pode ser associada a formula\c c\~ao por integrais de trajet\'oria de Feynman,  equivalente \`a formula\c c\~ao  usual. 

Mais explicita e  formalmente, a exemplo de \cite {Naber}, mencionamos e comentamos os ingredientes b\'asicos para a descri\c c\~ao cl\'assica da intera\c c\~ao de uma part\'icula com um campo de gauge:
\begin{enumerate}
\item{{\bf Uma variedade suave (semi) riemannana $M$.} - Este \'e simplesmente o espa\c co onde as part\'iculas vivem. }
\item{{\bf Um espa\c co vetorial de dimens\~ao finita $F$ equipado com um produto interno $\langle\cdot,\cdot \rangle$. } - Na interpreta\c c\~ao ortodoxa, este \'e o espa\c co onde as fun\c c\~oes de onda das part\'iculas tomam seus valores. Este espa\c co \'e determinado pela estrutura interna da part\'icula (fase, isospin, etc) e \'e chamado de espa\c co interno. Exemplos t\'ipicos s\~ao $\C,\C^2,\C^4$ ou  as \'algebras de Lie $\mathfrak{u}(1), \mathfrak{su}(2)$. Pelo produto interno se computa a norma de fun\c c\~oes de onda e portanto probablidades qu\^anticas.} 
\item{{\bf Um grupo de Lie $G$ e uma representa\c c\~ao 
$\rho:G\ra\GL F$ ortogonal  em rela\c c\~ao a $\langle\cdot,\cdot\rangle$. } - $G$ age ent\~ao sobre as bases dos estados internos sobre cada ponto.  Como vimos em (\ref{metassocia}), a ortogonalidade da representa\c c\~ao \'e necess\'aria para que o produto interno n\~ao dependa da base de estados internos que escolhemos. Como vimos em ({\bf{Sec.1.6}}), existe bije\c c\~ao entre $G$ e as bases ortonormais $G$ admiss\'iveis sobre um dado ponto $x\in M$.  }
\item{{\bf Um $G$-fibrado principal sobre $M$: $(P,\pi,M,G)$.} - Pela {\bf{Sec.2.3}} este fibrado pode ser identificado ao fibrado das bases $G$ admiss\'iveis sobre $M$. A fibra sobre cada ponto \'e uma c\'opia de $G$ vista como todas as bases ortonormais $G$-admiss\'iveis dos estados internos. Uma se\c c\~ao de $P$ \'e um  referencial $G$-admiss\'ivel relativo ao qual descrevemos nossa fun\c c\~ao de onda. }
\item{{\bf Uma conex\~ao $\omega$ em $P$, com curvatura $\Omega$.} -  Essa conex\~ao nos fornece a varia\c c\~ao intr\'inseca das bases. Aplicada sobre um referencial local $s$, n\'os obtemos o {\bf potencial de gauge local},~ $\mathcal{A}=s^*\omega$. Similarmente obtemos o {\bf campo local de gauge}~ $\mathcal{F}=s^*\Omega$. }
\item{{\bf Uma se\c c\~ao global  $\Phi$ do fibrado vetorial associado $P\times_\rho F$ } - Campos de mat\'eria ser\~ao associados a tais se\c c\~oes que satisfa\c cam as equa\c c\~oes de Euler-Lagrange de algum funcional de a\c c\~ao que envolva os potenciais locais $\mathcal A$. Como vimos em {\bf Sec.2.4}, dado um referencial podemos associar localmente estas se\c c\~oes a fun\c c\~oes $G$-equivariantes $\psi:\theta\ra F$, que nos fornecer\~ao as chamadas fun\c c\~oes de onda.  }
\item{{\bf Uma a\c c\~ao $S(\Phi,\omega)$ cujos pontos estacion\'arios s\~ao as solu\c c\~oes cl\'assicas. } - Tipicamente este funcional \'e da seguinte forma:$$S(\Phi,\omega)=c\int_M\|\Omega\|^2+c_1\|D\Phi\|^2$$ Onde $D$ \'e a derivada exterior covariante determinada por $\omega$ {\bf Teo.26}. A constante $c$ \'e chamada de {\bf constante de normaliza\c c\~ao} e $c_1$ \'e a {\bf constante de acoplamento}. Discutiremos as normas utilizadas na se\c c\~ao seguinte. N\~ao trataremos do caso dos campos com mat\'eria, ou seja, faremos $c_1=0$. Estamos interessados somente na din\^amica dos campos.}
\end{enumerate}

Ou seja, todo o trabalho que tivemos at\'e aqui foi o de apresentar boa parte do arcabou\c co te\'orico de uma teoria f\'isica baseada nestas suposi\c c\~oes. O que n\'os queremos agora, \'e, assim como na teoria da relatividade geral, minimizar um funcional escalar da curvatura, para que possamos achar uma configura\c c\~ao de campos que represente uma solu\c c\~ao cl\'assica do sistema. Para isso, ainda precisamos colocar alguns detalhes na teoria, como achar uma m\'etrica bem definida no espa\c co de estados internos. 

\subsubsection{Preliminaries}
Na teoria de Yang-Mills, como vimos, os campos f\'isicos de interesse s\~ao a curvatura e a conex\~ao, que no caso de fibrados vetoriais, s\~ao representa\c c\~oes da forma de conex\~ao $\omega^s$ do fibrado principal. Como  uma representa\c c\~ao de $G$ em um dado $x\in M$, \'e uma aplica\c c\~ao $\rho_x:G\ra \Aut(E_x)$,  onde $E$ \'e o $G$-fibrado vetorial em quest\~ao; a representa\c c\~ao correspondente da \'algebra de Lie \'e $d(\rho_x)_e:\Lg\ra \End(E_x)$, ou seja, s\~ao formas a valores em representa\c c\~oes da \`algebra de Lie $\Lg$ sobre os endomorfismos de cada fibra vetorial, $\End(E_x)$.  Estamos interessados em estudar se\c c\~oes de formas de conex\~ao em $E$, e portanto precisaremos considerar o fibrado $\End(E)\simeq E\otimes E^*$. Podemos proseguir de duas maneiras equivalentes: tomando diretamente o fibrado vetorial $\End(E)$ como o protuto tensorial de fibrados , ou pelo interm\'edio do fibrado associado  $P_F:=(P\times F)/G$ onde $G$ age pela conjuga\c c\~ao no espa\c co vetorial $F=\End(V)$, tentaremos oferecer uma compara\c c\~ao entre os dois.  
Para que possamos medir de alguma maneira a ``intensidade'' de formas de conex\~ao sobre um dado ponto da base, precisamos de alguma no\c c\~ao de produto interno, que, pelo que vimos na se\c c\~ao anterior, seja invariante pela a\c c\~ao do grupo em $F$ (\ref{represaadj}). 
 
Podemos induzir um mapa linear natural  
\begin{align*}
\mbox{tr}:\End(V)&\lra\R\\
v\otimes \lambda &\lra\lambda(v)
\end{align*} 
Pra vermos que esse mapa coincide com o que usualmente chamamos de tra\c co, escolhemos  bases duais $\{e_i\}$ e $\{e^i\}$ de  $V$ e de $V^*$ respectivamente, escrevendo $T\in\End(V)$ nessa base temos 
$$\tr(T)=T^i_je^j(e_i)=T^j_j$$ 
Claramente, se $T=v_1\otimes\lambda^1 ,S=v_2\otimes\lambda^2 \in\End(V)$, (lembramos que elementos dessa forma geram $\End(V)$) temos 
\beq\label{trace1}\tr(TS)=\tr\left(v_1(\lambda^1(v_2))\otimes\lambda^2\right)=\lambda^1(v_2)\tr(v_1\otimes\lambda^2)=\lambda^1(v_2)\lambda^2(v_1)=\tr(ST)\eeq  

Definimos o negativo da forma de Killing: 
$$\begin{array}{ll}
K: & \Lg\times\Lg\ra \R\\
~ & (h,j)\mapsto -\tr(\ad_h\circ\ad_j)
\end{array}$$
 que \'e assim claramente bilinear e sim\'etrica. \'E poss\'ivel ainda mostrar que a forma de Killing \'a invariante pela representa\c c\~ao adjunta do grupo na \'algebra  e que, se $G$ for compacto, ent\~ao $K$ \'e  positivo definido se e somente se $G$ \'e semi-simples \cite {Lie}. 
\begin{lem} Se $\Lg$ \'e a \'algebra de Lie de um grupo compacto semi-simples, ent\~ao a forma de Killing \'e positiva-definida em $\Lg$, i.e.: para todo $X\neq0~,~K(X,X)>0$.\end{lem}

 Usualmente essa \'e a m\'etrica  para $\Lg$, mas n\~ao a utilizaremos explicitamente, apenas como artif\'icio para induzir uma m\'etrica n\~ao -degenerada no fibrado associado e com isso um operador de Hodge. No caso de $\mathfrak{u}(n)$, como mostraremos a seguir, \'e poss\'ivel utilizar o operador tra\c co, que \'e o que faremos, pois, al\'em da simplicidade formal, fica mais direta a conex\~ao com classes de Chern e topologia de fibrados, que apresentamos no cap\'itulo seguinte.   

Agora, para $G$ compacto, \'e um fato que a \'algebra de Lie de $G$ admite a seguinte decomposi\c c\~ao 
$$\Lg=\mathfrak{z}\oplus[\Lg,\Lg]$$
onde $\mathfrak{z}=\Ker(\ad)$ \'e o centro da \'algebra, logo $K|_{\mathfrak{z}}\equiv 0$ e a parte $[\Lg,\Lg]$ \'e semi-simples. Logo, utilizando a m\'etrica produto natural (ortogonal) em $\Lg=\mathfrak{z}\oplus[\Lg,\Lg]$ temos que um outro produto interno em $\Lg$ que seja positivo definido sobre o centro e proporcional a $K$ na parte semi-simples ser\'a positivo definido em sua totalidade.  No caso de maior interesse para a f\'isica: $$\mathfrak{u} (n)=\mathfrak{u}(1)\oplus\mathfrak{s}\mathfrak{u} (n)$$ 
Nesse caso, o centro s\~ao os m\'ultilpos da matriz identidade e portanto \'e n\~ao degenerada nessa parte.  
Agora mostremos o tra\c co e a froma de Killing  s\~ao de fato proporcionais na parte semi-simples.

  Assumindo que um grupo $G$ \'e simples se e somente se ele admite representa\c c\~ao adjunta irredut\'ivel, suponhamos  que temos duas formas bilineares invariantes por uma representa\c c\~ao $\rho:G\ra \Aut(V)$ em um espa\c co vetorial $V$, $\langle\cdot,\cdot\rangle_1$ e  $\langle\cdot,\cdot\rangle_2$. Afirmamos que $B_1=\langle\cdot,\cdot\rangle_1$ e  $B_2=\langle\cdot,\cdot\rangle_2$ s\~ao proporcionais. Para provar que isto de fato se d\'a, consideremos a aplica\c c\~ao $B_1:V\ra V^*$. Como o n\'ucleo desta aplica\c c\~ao \'e  invariante pela a\c c\~ao de $\rho$, que \'e representa\c c\~ao irredut\'ivel de $G$ em $\Aut(V)$, n\'os temos que ou $\Ker B_1=0$ ou $\Ker B_1=V$. Portanto podemos assumir sem perda de generalidade que $\Ker B_1=0$, i.e.: \'e uma forma bilinear n\~ao degenerada. Podemos definir uma plica\c c\~ao $A:V\ra V$ como $B_1(A(u))=B_2(u)$ que est\'a bem definida por $B_1$ ser n\~ao degenerada. Agora n\'os temos, para $u,v\in V$, $g\in G$:
$$\langle A\rho(g)(u),\rho(g)v\rangle_1=\langle \rho(g)(u),\rho(g)v\rangle_2=\langle u,v\rangle_2=\langle A(u),v\rangle_1=\langle\rho(g) A(u),\rho(g)v\rangle_1$$ portanto $A$ comuta com todos os operadores $\rho(g)$ e portanto pelo Lema de Schur, \'e um m\'ultiplo da identidade. Portanto $\langle\cdot,\cdot\rangle_1=c\langle\cdot,\cdot\rangle_2$, e, fazendo $V\equiv\Lg$, $\rho=\Ad$ e $B_1=K$ e $B_2=\tr$ podemos assumir a m\'etrica mais trat\'avel em $\Lg$: 
$$ \langle h , j\rangle=-\tr(hj)$$ que, como j\'a mostramos,  tamb\'em \'e invariante pela a\c c\~ao adjunta de  representa\c c\~oes lineares de $G$, e portanto \'e um m\'ultiplo da forma de Killing. 

Voltemos agora  a considerar o fibrado associado $P_F$. \'E claro  que podemos (no caso $\mathfrak{u}(n)$) induzir uma forma bilinear sim\'etrica em $P_F$ pelo  tra\c co em $F=\End(V)$, j\'a que por (\ref{trace1}) o tra\c co \'e invariante pela a\c c\~ao de $\Aut_G(F)$. Como mostramos tamb\'em na \'ultima se\c c\~ao do cap\'itulo anterior (\ref{represaadj}), qualquer $G$-conex\~ao advinda de $P$ ser\'a automaticamente compat\'ivel com esse produto interno.  

Pela \'otica do fibrado vetorial $\End_G(E)$, definimos igualmente o tra\c co de uma se\c c\~ao de $\End(E)$, e para $\lambda\otimes T\in\Gamma\Big(\Lambda^p(TM^*{|_\theta})\otimes\End({E}){|_\theta}\Big)$ definimos: $\mbox{tr}(\lambda\otimes T)=\mbox{tr}(T)\lambda\in\Gamma(\Lambda^p(TM^*{|_\theta})$.  \'E claro que, como $\End(E_x)\simeq E_x\otimes E_x^*$,  uma se\c c\~ao $s\in\Gamma(\End(E))$ sob um isomorfismo de fibrados $g\in\Gamma(\Aut_G(E))$ se transforma pela conjuga\c c\~ao : $s(x)\ra g^{-1}(x)s(x)g(x)\in\End(E_x)$. Portanto a a\c c\~ao do  tra\c co  \'e invariante por uma tal transforma\c c\~ao (o que poderia ser visto de outra maneira, notando que o tra\c co \'e uma aplica\c c\~ao $C^\infty(M,\R)$-bilinear $E^*\times E\ra\R$, ou seja, \'e um tensor de ordem $(1,1)$, e portanto invariante por mudan\c cas de base, ou isomorfismos agindo pela conjuga\c c\~ao). 

Assim, o fibrado $\End_G(E)$ tem um produto interno can\^onico sobre cada fibra  que \'e preservado por transforma\c c\~oes de gauge, que como sabemos, agem pela conjuga\c c\~ao. 
Agora,  induzimos uma m\'etrica em $\End_G(E)$ pelas m\'etricas can\^onicas das trivializa\c c\~oes. Sem utilizar as formas de conex\~ao, veremos como este produto interno \'e compat\'ivel com a conex\~ao. 

Sejam $\phi,\psi$ duas trivializa\c c\~oes sobre $\theta$, i.e.: 
$$\pi_E^{-1}(\theta)\overset{\phi,\psi}{\simeq}\theta\times \End(V)$$
Lembramos que as bases $G$-admiss\'iveis de $\End(E)$ foram definidas como aquelas provindas das bases can\^onicas das trivializa\c c\~oes e que, para $x\in\theta,v\in \End(V)$ a transi\c c\~ao \'e dada por $$\phi\circ\psi^{-1}(x,v)=(x,g_{\phi \psi}(x)vg_{\phi \psi}(x)^{-1})$$ onde $\Ad(g_{\phi \psi}):\theta\ra \Aut(\End(V))$. Para que fa\c ca sentido induzirmos a m\'etrica em $\End(E)$ pelas m\'etricas das trivializa\c c\~oes, n\'os precisamos que $\Ad(g_{\phi \psi}):\theta\ra \OG{\End(V)}\subset\Aut(\End(V))$, agora 
$$\{\psi^{-1}(x,\tilde{e}^j_i)\}_{i,j=1}^k=\{e^j_i\}_{i,j=1}^k$$ 
onde $\tilde{e}^j_i=\tilde{e}^j\otimes \tilde e_i$ \'e a base can\^onica de $V\otimes V^*$ e portanto $\{e^j_i\}_{i,j=1}^k$ \'e base ortonormal em $\End(E_x)$; logo $\{\phi(e^j_i)\}_{i,j=1}^k=\{(x,g_{\phi \psi}(x)e_i^jg_{\phi \psi}(x)^{-1})\}_{i=1}^k$ \'e ortonormal. 
 
Sejam ent\~ao $x,y\in\theta$, e suponhamos que $\alpha:I\ra \theta$ \'e uma curva tal que $\alpha(0)=x$ e $\alpha(1)=y$. Como o transporte paralelo leva base admiss\'ivel am base admiss\'ivel, n\'os temos que existem trivializa\c c\~oes $\phi,\psi$, tais que se $\{\psi^{-1}(x,\tilde{e}_i^j)\}_{i,j=1}^k=\{e^j_i\}_{i,j=1}^k$ \'e base ortonormal em $\End(E_x)$, ent\~ao $$\{b_i^j\}_{i,j=1}^k=\{P_\alpha(e_i^j)\}_{i,j=1}^k=\{\phi^{-1}(y,\tilde{e}^j_i)\}_{i,j=1}^k=\{\psi^{-1}(y,g_{\phi \psi}(y)\tilde{e}^j_ig_{\phi \psi}(y))^{-1}\}_{i,j=1}^k$$
que \'e ortonormal. Portanto (ao menos dentro de uma trivializa\c c\~ao) o transporte paralelo leva base ortonormal em base ortonormal.
 Da mesma forma obtemos que  ser\'a invariante por qualquer m\'etrica em $\End(E)$ induzida por trivializa\c c\~oes locais.
Apesar de termos considerado curvas restritas a um domi\'inio trivial (precisar\'iamos ainda colar todas as partes da curva), esta descri\c c\~ao, como n\~ao poderia deixar de ser, \'e completamente equivalente a feita diretamente pelo fibrado associado.

Suponhamos agora que $\omega^s=s^*\omega$, onde $\omega$ \'e a forma de conex\~ao em $P$ e $s$ \'e uma se\c c\~ao de $P$,
como sabemos essa conex\~ao ter\'a uma representa\c c\~ao no fibrado associado, o que nos incita uma pergunta, ainda que perif\'erica ao estudo que estamos conduzindo, ser\'a que a representa\c c\~ao de uma forma de conex\~ao em P (que tem valores em $\Lg$) \'e compat\'ivel com a conex\~ao usual no fibrado vetorial $\End(E)$, dado por ((\ref{abcde}),(\ref{abcdef}))? Caso a constru\c c\~ao que fizemos at\'e agora esteja coerente, a resposta deve ser positiva. Chequemos pois que isso se d\'a.

{\bf Dem:}  Tomamos a representa\c c\~ao $\rho_x:G\ra \Aut(E_x)$ e\footnote{Na maioria dos casos, se usa a identifica\c c\~ao $\Aut(E_x)\cong \GL k$ e toma-se a representa\c c\~ao de $G$ simplesmente como a inclus\~ao, $G\hookrightarrow \GL k$, mas n\~ao o faremos aqui.}  ent\~ao, como sabemos,  a respectiva representa\c c\~ao do grupo que estamos tomando sobre  $\End(E)$ (ou, no contexto de fibrado associado, sobre $\End(V)$) \'e :
$$\begin{array}{ll}
\widetilde\rho_x: & G \lra \Aut(\End(E_x))\\
~ & g \longmapsto  \Ad_{\rho_x(g)}
\end{array}
$$
Consideremos a aplica\c c\~ao  $\Ad:G\ra\Aut(\Lg)$. Chamando $d(\Ad)_e=\ad:\Lg\ra\End(\Lg)$, temos que 
$$d(\widetilde\rho_x)_e=d(\Ad_{\rho_x(g)})_e=\ad_{ d(\rho_x)_e(h)}$$
sendo que consistentemente $(d\rho_x)_e:\Lg\ra\End(E_x)$. Portanto n\'os temos o mapa
$$\begin{array}{ll}
d(\widetilde\rho_x)_e: & \Lg\lra \End(\End(E_x))\\
~ & h\longmapsto \ad_{ d(\rho_x)_e(h)}
\end{array}
$$
 Agora, \'e poss\'ivel mostrar \cite{Lie} que 
$$\begin{array}{lll}
\ad_h: & \Lg & \lra  ~~~\Lg\\
& j & \longmapsto  [h,j]
\end{array}
$$
 Portanto para $X\in \Gamma(E)$ e $\lambda\in\Gamma(E^*)$, \beq\label{represag}d(\widetilde\rho_x)_e(\omega^s)(X\otimes\lambda)=\ad_{d(\rho_x)_e(\omega^s)}(X\otimes\lambda)=d(\rho_x)_e(\omega^s)X\otimes\lambda-X\otimes\lambda d(\rho_x)_e(\omega^s)\eeq e como 
$$\lambda d(\rho_x)_e(\omega^s)=\Big(d(\rho_x)_e(\omega^s)\Big)^T\lambda$$ 
reinserindo em (\ref{represag}) n\'os obtemos a forma usual da conex\~ao para o fibrado $E\otimes E^*$ ((\ref{abcde}),(\ref{abcdef})) (atrav\'es da forma de conex\~ao em $P$ pela representa\c c\~ao natural de $\Lg$ em $\End(E_x)$).$~~~~~~~~~~~~\blacksquare$\medskip
  
Ou seja, ao inv\'es de tomarmos o produto interno na fibra sobre $x\in M$ como $(h,j)\mapsto\tr(\ad_{d(\rho_x)_e(h)}\ad_{d(\rho_x)_e(j)})$, faremos $(h,j)\mapsto \tr(d(\rho_x)_e(h)d(\rho_x)_e(j))$.

\subsubsection{ Yang-Mills Equation}
 
Chamaremos de $\Ad(E)\subset\End(E)$ o espa\c co de endomorfismos cujas representa\c c\~oes locais em matrizes s\~ao anti-sim\'etricas em cada fibra $E_x$, i.e.: que s\~ao representa\c c\~oes em $\mathfrak{s}\mathfrak{o}(E_x)$.
 
 N\'os j\'a temos um produto interno para $p$-formas definido por, para 
 ${w_1,w_2}\in\Lambda^p{T^*_xM}$:
 
$$\langle{w_1,w_2}\rangle*1=w_1\wedge{*}w_2$$
 
Logo, sejam $$\mu_1\otimes{w_1},\mu_2\otimes{w_2}\in\Ad{(E_x)}\otimes\Lambda^p{T^*_x}M$$ 
 
Ent\~ao $$\langle\mu_1\otimes{w_1},\mu_2\otimes{w_2}\rangle{:=}\tr(\mu_1\mu_2)\langle{w_1,w_2}\rangle$$ 
 
ou seja, temos um produto escalar $L^2$ em $\Gamma(\Lambda^p(TM)\otimes\Ad{E})$:
\beq\label{integralYM}\langle\langle{\mu_1\otimes{w_1},\mu_2\otimes{w_2}}\rangle\rangle{:=\int_M}\langle\mu_1\otimes{w_1},\mu_2\otimes{w_2}\rangle{*1}\eeq
 
 Logo, como temos um produto interno bem definido, podemos definir o operador estrela de Hodge neste espa\c co tamb\'em. N\'os temos ainda que, a adjunta formal da derivada covariante \'e dada por $D^*=(-1)^{n(p+1)+1}*D*$. Este fato pode ser facilmente demonstr\'avel utilizando o Gauge can\^onico, j\'a que ali $$\tilde{D}\alpha(x_1, \cdots, x_{p+1})=d\alpha(x_1, \cdots, x_{p+1})$$ e n\'os j\'a demonstramos que $d^*=(-1)^{n(p+1)+1}*d*$.
 
Assumindo ainda que $D$ \'e uma derivada de gauge exterior compat\'ivel com a m\'etrica ($D$ \'e compat\'ivel com $\langle\cdot,\cdot\rangle$ se $D\langle\mu,\nu\rangle=d\langle\mu,\nu\rangle=\langle{D\mu,\nu\rangle}+\langle\mu,D\nu\rangle$ ou se a forma de conex\~ao tem representa\c c\~ao anti-sim\'etrica) com curvatura $\Omega_{D}\in\Gamma(\Lambda^2(TM)\otimes\Ad{E})$, utilizando (\ref{integralYM}) definimos o funcional de Yang-Mills aplicado a $D$: 
 
\beq\label{YMfunc}S_{YM}(D):=\langle\langle\Omega_D,\Omega_D\rangle\rangle=\int_M\langle\Omega_D,\Omega_D\rangle{*}1=\int_M\tr(\Omega_D\wedge*\Omega_D)\eeq
 
O integrando $\langle\Omega_D,\Omega_D\rangle=L_{YM}$ \'e chamado de Lagrangeana de Yang-Mills e $S_{YM}(D)$ \'e chamada de a\c c\~ao de Yang-Mills (ou funcional de Yang-Mills). Como vimos, uma transforma\c c\~ao de Gauge deixa $L_{YM}$ e portanto $S_{YM}$ invariantes. Para determinarmos as equa\c c\~oes de Euler-Lagrange para esse funcional utilizamos o fato que o espa\c co de todas as conex\~oes m\'etricas em $E$, $\mathcal{C}(E)$ \'e um espa\c co afim. Portanto, podemos considerar varia\c c\~ao da forma: 
$D+t\gamma$ onde $\gamma\in\Gamma(\Lambda^1(TM)\otimes{\Ad(E)})$ 
Agora por (\ref{kappa}):
\beq\label{variaomega} \Omega_{D+t\gamma}=\Omega_D+tD\gamma+t^2\gamma\wedge\gamma\eeq
logo $$\frac{d}{dt}_{|_{t=0}}S_{YM}(D+t\gamma)=\frac{d}{dt}_{|_{t=0}}\left(\int{\langle\Omega_{D+t\gamma}},\Omega_{D+t\gamma}\rangle*1\right)=2\int\langle\Omega_D,D\gamma\rangle*1=2\langle\langle{D^*\Omega_D,\gamma\rangle\rangle}$$ 
assim, $D$ extremiza $S_{YM}$ se $D^*\Omega_D=0$, onde $D^*=*D*$. Chamamos tal $D$ de conex\~ao de Yang-Mills.  Ent\~ao, levando em conta a identidade de Bianchi (\ref{Bianchi}), temos que a conex\~ao de Yang-Mills deve obedecer \`as seguintes equa\c c\~oes: 
\beq\label{eqsYang}
\left\{ \begin{array}{l}
 D*\Omega_D=0\\
 D\Omega_D=0\\
\end{array} \right.
\eeq

 Al\'em disso, denotando por $\delta\Omega$ a varia\c c\~ao de $\Omega$, como $\delta\omega=\gamma$ por (\ref{variaomega}),  emerge gratuitamente  que,  $$\delta\Omega=D\delta\omega$$ Portanto, como $\Omega=D\omega$ n\'os obtemos que a derivada funcional e a covariante comutam para a forma de conex\~ao.  

Al\'em da invari\^ancia de Gauge, as equa\c c\~oes de Yang-Mills t\^em outras propriedades interessantes advindas de simetrias em $L_{YM}$. Como mencionamos algumas vezes, o produto interno induzido sobre a fibra \'e invariante por automorfismos do fibrado, e se olharmos bem, notamos que a m\'etrica de $M$ entra de forma muito sutil nas equa\c c\~oes, somente atrav\'es do operador de Hodge (j\'a que $\Omega_D$ s\'o depende da forma de conex\~ao $\omega$), pela equa\c c\~ao (\ref{hodgeprod}).
 
 Seja $g$ uma m\'etrica em $M$, tomamos uma m\'etrica conformemente equivalente a $g$, $\tilde{g}=f^2g$ para $0\neq f\in C^\infty(M,\R)$. \'E f\'acil ver que se  $\{e_I\}$ \'e base ortonormal de $\Lambda^p(T^*M)$ em rela\c c\~ao a $g$, ent\~ao $\{\frac{e_I}{f}\}$ o \'e em rela\c c\~ao a $\tilde{g}$. 
Logo, como $\tilde{*}_p$ est\'a definido por sua aplica\c c\~ao a uma base de $\Lambda^p(T^*M)$, para cada elemento de $\{e_I\}$ temos 
\beq\label{11}
\begin{array}{l}
\tilde{*}\left(\tilde{e}_{i_1}\wedge\dots\wedge\tilde{e}_{i_p}\right)=\pm\tilde{e}_{i_{p+1}}\wedge\dots\wedge\tilde{e}_{i_n}\\
=\frac{1}{f^p}\tilde*\left(e_{i_1}\wedge\dots\wedge{e}_{i_p}\right)=\pm\frac{1}{f^{n-p}}e_{i_{p+1}}\wedge\dots\wedge{e_{i_n}}\\ 
\end{array}
\eeq
$$\therefore~~~~~~\tilde{*}\left(e_{i_1}\wedge\dots\wedge{e_{i_p}}\right)=\pm{f^{2p-n}}e_{i_{p+1}}\wedge\dots\wedge{e_{i_n}}=f^{2p-n}*\left(e_{i_1}\wedge\dots\wedge{e_{i_p}}\right)$$ 
E finalmente $$\tilde{*}=f^{2p-n}*$$ Logo se $n=2p$ obtemos  $\tilde{*}=*$.
 
Isso significa que se $\Psi:M^m\rightarrow{N^n}$ \'e um difeomorfismo entre $(M,\tilde{g})$ e $(N,g)$ tal que $m=n=2p$ e se para alguma fun\c c\~ao $f$, onde $0\neq f\in C^\infty(M,\R)$ n\'os tenhamos $\Psi^*g=f^2\tilde{g}$, ent\~ao  para todo ${w}\in\Lambda^p(T^*N)$, temos:
\beq\label{10}{\Psi^*}(*w)=\tilde{*}(\Psi^*w)\eeq
O que \'e facilmente obtido de (\ref{11}) fazendo a seguinte substitui\c c\~ao $\{\frac{e_I}{f}\}\ra\{\frac{\Psi^*e_I}{f}\}$.
 
Agora n\'os sabemos que dado $\nabla\in{C(E)}$, onde $\pi:E\rightarrow{N}$, n\'os formamos o pull-back $\phi^*(E)$ sobre $M$ e obtemos uma conex\~ao can\^onica em $\phi^*(E)$ dada pelas formas de conex\~ao 
$$\tilde{w}=\phi^*w$$ 
Logo, lembrando que para todo    
$\lambda\in\Gamma(\Lambda^p(T^*M)\otimes \End(E))$ por (\ref{Dlocal})
temos $$D\lambda=d\lambda+w\wedge\lambda$$ vale:
\beq\label{pullD}\phi^*(D\lambda)=\phi^*d\lambda+\phi^*w\wedge\phi^*\lambda=d(\phi^*\lambda)+\tilde{w}\wedge\phi^*\lambda=\tilde{D}(\phi^*\lambda)\eeq e portanto para o difeomorfismo $\Psi$, por (\ref{10}) e tomando uma conex\~ao de Yang-Mills em $E$:
$$ \Psi^*(D*\Omega)=\tilde{D}(\Psi^**\Omega)=\tilde{D}\tilde{*}(\Psi^*\Omega)=\tilde{D}\tilde{*}\tilde{\Omega}=0$$
 e a igualdade de Bianchi tamb\'em vale automaticamente, obtemos que se $D$ \'e conex\~ao de Yang-Mills, $\tilde{D}$ tamb\'em o \'e. Um caso particular 'e se um difeomorfismo $\Phi:M\ra M$ for uma isometria,  $\Phi^*$ preserva o operador de Hodge e portanto preserva $L_{YM}$ e as equa\c c\~oes de Yang-Mills. Por exemplo, as equa\c c\~oes de Yang-Mills para o fibrado trivial s\~ao invariantes pelo grupo de Poincar\'e.

Mas explorando essa maior liberdade conforme, como a m\'etrica can\^onica de $S^4-\{p\}$ \'e conforme a de $\R^4$ sob proje\c c\~ao estereogr\'afica existe uma correspond\^encia bijetora entre campos Yang-Mills ($\Omega_D$) em $\R^4$ e em $S^4-\{p\}$. 
 
\'E claro que qualquer campo de Yang-Mills em $S^4$ tem a\c c\~ao finita e portanto corresponde a um campo de Yang-Mills de a\c c\~ao finita em $\R^4$: 
$$\int_{s^4}\tr(\Omega\wedge*\Omega)=\int_{S^4-\{p\}}\tr( \Omega\wedge *\Omega) = \int_{\R^4}\tr( \Psi^*\Omega\wedge*\Psi^*\Omega)=\int_{\R^4}\tr(\tilde{\Omega}\wedge*\tilde{\Omega})$$ 
Um teorema de Karen Uhlenbeck n\'os fornece a rec\'iproca:  um campo $YM$ de a\c c\~ao finita em $\R^4$ se estende a um campo $YM$ suave em $S^4$. 
Portanto, os campos $YM$ de a\c c\~ao finita em $R^4$ podem ser identificados com todos os campos $YM$ de $S^4$. 

No caso mais geral de  $n=4$, como vimos na teoria de Hodge,  como  $(*_2)^2=1$ por (\ref{autodecomp}),  podemos tomar a soma direta
$$\Lambda^2(T^*M)\otimes \End(E)=\Lambda^2_+(T^*M)\otimes \End(E)\oplus\Lambda^2_+(T^*M)\oplus \End(E)$$ de subespa\c cos de autovetores de $*$. Em particular, $$\Omega=\Omega_++\Omega_-$$ 
e automaticamente satisfazemos a equa\c c\~ao de Yang-Mills: $D*\Omega=D\Omega=0$.  Chamamos uma forma de conex\~ao $\omega$ de auto-dual (anti-dual) se sua sua curvatura \'e $\Omega_+$ ($\Omega_-$). Normalmente essas conex\~oes s\~ao designadas por e $SD$ (self-dual) e $ASD$(anti-self-dual). \'E ainda f\'acil ver que essa decomposi\c c\~ao \'e ortogonal:
$$\pl\Omega_+,\Omega_-\pr=\int_M\Omega_+\wedge*\Omega_-=-\int_M*\Omega_+\wedge\Omega_-=-\int_M\Omega_-\wedge*\Omega_+=-\pl\Omega_-,\Omega_+\pr=0$$
Solu\c c\~oes duais da equa\c c\~ao de Yang-Mills s\~ao chamadas de instantons, e existe uma vasta quantidade de f\'isica e matem\'atica que revolvem ao seu redor. 
Um exemplo de uma solu\c c\~ao dual n\~ao trivial (i.e.: com curvatura n\~ao nula) de energia finita \'e
\beq \omega_x=\mbox{im}\left(\frac{\overline{x}dx}{1+|x|^2}\right)\eeq
onde $x$ \'e a vari\'avel quaterni\^onica em $\R^4\simeq \Ha$ e identificamos a \'algebra de Lie $\SU 2\simeq \SP 1$ ao conjunto de quaternions imagin\'arios \cite{Gil}. Em 1979, f\'ormulas similares foram descobertas para todas as conex\~oes duais e anti-duais em $\R^4$ euclidiano \cite{Atiyah}. 

Durante algum tempo, as \'unicas solu\c c\~oes de energia finita conhecidas para as equa\c c\~oes de Yang-Mills eram solu\c c\~oes duais e anti-duais. Em 1989, L.Sibner, R. Sibner e K. Uhlembeck \cite{Karen} publicaram uma prova da exist\^encia de solu\c c\~oes de algumas solu\c c\~oes n\~ao duais no $\R^4$  euclidiano. Uma outra boa refer\^encia para o assunto \'e \cite{Gil}. 

Exploraremos um pouco melhor  a rela\c c\~ao da equa\c c\~ao de Yang-Mills com a topologia e classifica\c c\~ao de fibrados na pr\'oxima se\c c\~ao.

\section{Classification of  Fiber Bundles}
Nesta se\c c\~ao estaremos preocupados em achar maneiras de dizer quando dois fibrados sobre a mesma base s\~ao ou n\~ao s\~ao isomorfos, quest\~ao  encontrada no Teorema de Classifica\c c\~ao de Fibrados Vetoriais, um resultado de ordem formal. Na verdade, o que procuramos \'e uma maneira de associar a cada fibrado vetorial $E$ sobre uma variedade base $M$, uma classe de cohomologia de $M$. A essas associa\c c\~oes damos o nome de {\bf classes caracter\'isticas}, e elas s\~ao invariantes globais e pode-se dizer que medem o desvio de uma estrutura produto global. Classes caracter\'isticas \'e um dos conceitos geom\'etricos que conecta topologia alg\'ebrica e geometria diferencial, e apesar de n\~ao fazermos tanto uso destas constru\c c\~oes neste trabalho, sua import\^ancia n\~ao pode ser exagerada para a f\'isica moderna. N\~ao faremos no entanto uma apresenta\c c\~ao formal, mesmo dos poucos t\'opico que abordamos, j\'a que exigiria uma bagagem alg\'ebrica que excederia nosso tempo dispon\'ivel  e conhecimento da \'area.  

\subsubsection{Fiber Bundle Topology}
Apresentaremos primeiramente uma introdu\c c\~ao aos resultados mais elementares sobre a topologia de fibrados vetoriais.

Denotamos por $\mbox{Vect}_G(M)$ o conjunto de classes de equival\^encia de $G$-fibrados sobre $M$, isto \'e, $E\sim F$ se e somente se existe difeomorfismo $\Psi:E\rightarrow F$  tal que $\pi\circ \Psi=\pi$ isto \'e, $\Psi$ \'e um isomorfismo de fibrados sobre $M$. \'E trivial verificar que isso realmente constitui uma rela\c c\~ao de equival\^encia. 
 
Dado $f:N\rightarrow M$, constru\'imos o mapa induzido pelo pull-back: 
$$f^*:\mbox{Vect}_G(M)\rightarrow \mbox{Vect}_G(N)$$ 
Lembramos que se $E$ \'e fibrado vetorial sobre $M$, ent\~ao 
$$f^*(E):=\{(x,v)~|~x\in N, v\in E_{f(x)}~\}$$
Para verificar que este mapa est\'a bem definido sobre as classes de equival\^encia, basta vermos que, se $E\overset{\Psi}{\simeq}F$, ambos fibrados vetoriais sobre $M$, ent\~ao o mapa 
$$\begin{array}{ll}
f^*(E)\lra & f^*(F)\\
(x,v)\longmapsto & (x,\Psi(v))
\end{array}$$ 
 \'e isomorfismo de fibrados. 
Dado $E$, denotamos $E\times I$ o fibrado sobre $N\times I$, onde a fibra sobre $(x,t)$ \'e a fibra de $E$ sobre $x$. Isto \'e, 
$$E\times I=\mbox{pr}^*_1(E)$$
onde $$\mbox{pr}_1:N\times I\rightarrow N$$
 
\begin{prop}\label{fibprod}{Todo $G$-fibrado $\widetilde{E}$ sobre $N\times I$ \'e isomorfo a um da forma $E\times I$ (ou seja $\mbox{pr}_1^*E$), para algum fibrado $E$ sobre $N$.}
\end{prop}

{\bf Dem:} Seja $\widetilde{E}$, tomamos
\begin{displaymath}
\begin{array}{ll}
\imath_o:&N\rightarrow N\times I\\ 
\phantom{a}&x\mapsto(x,0)\\ 
\end{array}
\end{displaymath}

definimos ent\~ao $E=\imath^*_o(\widetilde{E})$, ou seja $E_x=\widetilde{E}_{\imath_o(x)}=\widetilde{E}_{(x,0)}$. Agora, tomando o fibrado $E\times I$ definido acima temos:
$$(E\times I)_{(x,t)}=E_{\mbox{pr}_1(x,t)}=E_x=\widetilde{E}_{(x,0)}$$
Escolhida uma conex\~ao em $\widetilde{E}$, seja $P_{(x,t)}:\widetilde{E}_{(x,0)}\rightarrow\widetilde{E}_{(x,t)}$ o transporte paralelo ao longo da curva $t\rightarrow(x,t)$. 
 
Logo, fazendo $(E\times I)_{(x,t)}\equiv E_x\equiv \widetilde{E}_{(x,0)}$ basta tomarmos
$$\begin{array}{lll}
\phi_{(x,t)}: & (E\times I)_{(x,t)}\ & \rightarrow\widetilde{E}_{(x,t)}\\ 
~& ~~~~v & \mapsto P_{(x,t)}(v) \end{array}$$
como o transporte paralelo \'e um isomorfismo entre as fibras, e depende suavemente dos par\^ametros $(x,t)$,  n\'os temos que $\phi:E\times I\ra \widetilde{E}$ \'e um isomorfismo de fibrados. Sua a\c c\~ao explora a estrutura produto da base e equivale a transladar as fibras por transporte paralelo ao longo de cada reta $t\mapsto(x,t)$. $~~~~~~~~\blacksquare$

\begin{theo}\label{homotopicas}{Se $f_0,f_1:N\rightarrow M$ s\~ao homot\'opicas, ent\~ao $f_0^*:\mbox{Vect}_G(M)\rightarrow \mbox{Vect}_G(N)$ e $f^*_1:\mbox{Vect}_G(M)\rightarrow \mbox{Vect}_G(N)$ s\~ao iguais.}
\end{theo}

{\bf Dem:} Seja $E\in \mbox{Vect}_G(M),\mbox{~e~}F:N\times I\rightarrow M \mbox{~uma homotopia entre~}f_0\mbox{~e~}f_1.$
Ent\~ao $F^*(E)\mbox{~\'e fibrado sobre~}N\times I$.
Logo, para algum $\widetilde{E}$ sobre $N$, 
$$F^*(E)\simeq\widetilde{E}\times I$$
Portanto, definindo $f_t:=F\circ \imath_t:N\rightarrow M$, onde 
\begin{displaymath}
\begin{array}{ll}
\imath_t:&N\rightarrow N\times I\\
\phantom{a}&x\mapsto(x,t)\\
\end{array}
\end{displaymath}
n\'os temos que $f^*_t=\imath^*_t\circ F^*.$
Logo, como $$F^*E\simeq\widetilde{E}\times I\mbox{~~~e~~~} \imath^*_t:\mbox{Vect}_G(N\times I)\rightarrow \mbox{Vect}_G(N)$$ 
est\~ao bem definidas, 
$$f_t^*(E)\simeq \imath^*_t(\widetilde{E}\times I)=\widetilde{E}.$$ 
Ent\~ao, $f^*_0(E)\simeq f^*_1(E)$. ~~~~~~~~~~~$\blacksquare$
 
\begin{prop}\label{poincare}{Se $M$ \'e contr\'atil, ent\~ao todo $G$-fibrado sobre $M$ \'e trivial.} 
\end{prop}
 
{\bf Dem:} Seja $f_0=\mbox{Id}_M$ e $f_1:M\rightarrow\{p\}\in M$ mapa constante; seja $E$ um fibrado qualquer sobre $M$. Ent\~ao, como 
$f_0$ \'e homot\'opica a $ f_1\mbox{,~~e~~}f^*_0=\mbox{Id}_E$,
obtemos pelo teorema anterior que 
$$E=f^*_0(E)\simeq f^*_1(E)=M\times E_p$$
$~~~~~~~~~\blacksquare$ 
 
\begin{prop}\label{classifica} N\'os temos um isomorfismo $\mbox{Vect}_G(S^n)\ra\pi_{n-1}(G)$. \end{prop}

{\bf Dem.:}
Um $G$-fibrado $E$ sobre $S^n$ \'e equivalente a um fibrado formado pela colagem (como vimos na {\bf{Sec.1.1}}) dos fibrados  $E^+$ e $E^-$ sobre os hemisf\'erios norte $D^n_+$ e sul $D^n_-$ ao longo do equador $S^{n-1}=D^n_+\cap D^n_-$ por um mapa de transi\c c\~ao de gauge $g:S^{n-1}\ra G$. Pelo corol\'ario anteriror $E^+$ e $E^-$ s\~ao triviais.   Isso significa que  temos a liberdade de tomar as trivializa\c c\~oes 
$$\psi_1:E_{|D^n_+}\rightarrow D^n_+\times E_p~~,~~ \psi_2:E_{|D^n_-}\rightarrow D^n_-\times E_p$$
 
Portanto, como $\psi_1\circ\psi^{-1}_2:(D^n_+\times {E_p})_{|_{D^n_+\cap D^n_-}}=S^{n-1}\times E_p$ 
\begin{displaymath}
\begin{array}{ll}
\psi_1\circ\psi^{-1}_2:&S^{n-1}\times E_p\rightarrow S^{n-1}\times E_p\\
\phantom{a}&(x,q)\mapsto(x,g(x)\circ q)\\
\end{array}
\end{displaymath} 
 
Logo, a classe de equival\^encia (de isomorfismos de fibrados) de $E$ \'e determinada unicamente pela classe de equival\^encia de $g:S^{n-1}\rightarrow G$, ou seja, n\'os obtemos um isomorfismo:
$$\mbox{Vect}_G(S^n)\rightarrow\pi_{n-1}(G)$$ $~~~~~\blacksquare$
 
{\bf Exemplo:} Como para um grupo $G$ simples e compacto $\pi_3(G)=\Z$, ent\~ao $\mbox{Vect}_G(S^4)\simeq\pi_3(G)=\Z$.\smallskip

\subsubsection{Bundle Classification Theorem}

A proposi\c c\~ao {\bf \ref{classifica}} \'e na verdade um caso espec\'ifico de um fato muito mais geral, o {\bf Teorema de Classifica\c c\~ao de Fibrados}. Este teorema prova que  para qualquer grupo de Lie $G$ e inteiro positivo $N$, podemos construir um espa\c co $B_G=B_G^N$, chamado de espa\c co de classifica\c c\~ao de $G$ para variedades $M$ de dimens\~ao menor que $N$, e um $G$-fibrado vetorial sobre $B_G$, $\xi_G^N$, chamado de fibrado universal, tal que se $M$ for uma variedade suave de dimens\~ao menor que $N$ e $E$ for qualquer $G$-fibrado vetorial sobre $M$, existe um mapa $f:M\ra B_G$ tal que $E\simeq f^*B_G$.  Para o caso $G=\GL{k}$, provaremos este teorema juntamente com sua extens\~ao: se $E\simeq f_0^*\xi_G$ e $E\simeq f_1^*\xi_G$, ent\~ao $f_0$ e $f_1$ s\~ao homot\'opicas. 

Seja ent\~ao 
$$Q^r=\{T\in L(\R^k,\R^l)~|~\mbox{posto}(T)=r\}$$ come\c camos com a seguinte proposi\c c\~ao:
\begin{prop} $Q^r$ tem dimens\~ao $kl-(k-r)(l-r)$. \end{prop}

{\bf Dem:} Consideremos a a\c c\~ao 
$$\begin{array}{lll}
\varphi: & \GL l\times\GL k & \ra   \Aut(L(\R^k,\R^l))\\
~& (g,h) & \mapsto  L_gR_{h^{-1}}
\end{array}$$
Afirmamos que as \'orbitas de $\varphi$ s\~ao justamente as transforma\c c\~oes lineares de posto constante. Que elas mant\'em o posto constante \'e claro, vejamos porque todas as transforma\c c\~oes de posto $r$  est\~ao ligadas pela a\c c\~ao de $\varphi$ \`a matriz 
\beq\label{matrizid}\left(\begin{array}{c|c}
\mbox{Id}_{\R^r} & 0\\
\hline
0 & 0
\end{array}\right)\eeq
 Tomando $T$ com posto $r$, dim(Ker$T)=k-r$. Tomamos $\{e_i\}_{i=r+1}^k$ como base de $\Ker T$, e um completamento qualquer $\{e_i\}_{i=1}^r$ em $\R^k$. Em $\R^l$, tomamos uma base composta por $\{Te_i\}_{i=1}^r=\{\tilde{e}_i\}_{i=1}^r$ e um completamento $\{\tilde{e}_i\}_{i=r+1}^l$. \'E claro que nessas bases $[T]$ ser\'a uma matriz diagonal com um \'unico bloco $r\times r$ n\~ao nulo, justamente  a identidade em $\R^k$. Como transforma\c c\~oes de base em $\R^k$ e $\R^l$ s\~ao equivalentes a a\c c\~ao de $\varphi$, n\'os temos que $Q^r$ ser\'a dado pela \'orbita de T pela a\c c\~ao de $\varphi$.  Tomando as a\c c\~oes usuais de composi\c c\~ao nos grupos $\Aut(L(\R^k,\R^l))$ e $\GL l\times\GL k$, \'e trivial ver que $\varphi$ \'e um homomorfismo: 
$$\varphi(gg',hh')T=g(g'Th'^{-1})h^{-1}=\varphi(g,h)\varphi(g',h')T$$ 
logo, $Q^r\simeq(\GL l\times\GL k)/G_T$ onde $G_T$ \'e o grupo de isotropia de $T$ pela a\c c\~ao de $\varphi$. 
Suponhamos ent\~ao que $(\tilde g,\tilde h)\in G_T$. Temos ent\~ao , utilizando a base acima para representar $T$ por (\ref{matrizid}), a seguinte condi\c c\~ao 
\beq\label{matrizid2}
T\circ h=\left(\begin{array}{llll}
h^1_1 & \cdots & ~ & h^k_1\\
\vdots & ~ & ~ & \vdots\\
h^1_r & \cdots & ~ & h^k_r\\
\hline
~ & ~~0 & ~ & ~
\end{array}\right)=g\circ T=\left(
\begin{array}{lll|l}
g^1_1 & \cdots & g^r_1 & ~\\
\vdots & ~ & \vdots & 0\\
~ & ~ & ~ & ~ \\
g^1_l & \cdots & g^r_l & ~
\end{array}\right) \eeq
Para $h$, que \'e uma matriz quadrada $k\times k$, \'e f\'acil notar que s\'o temos de fixar os termos $\{h^i_j~| ~i=r+1,\cdots k, \mbox{ e ~} j=1,\cdots r~\}$ que t\^em de ser nulos pela igualdade \`a $g\circ T$ (temos liberdade para escolher os elementos que s\~ao levado em zero pela $T$ e o quadrado $r\times r$ esquerdo superior). Para a matriz $g\in\mbox{M}(l\times l)$, com j\'a fixamos $h$,  precisamos fixar todos os elementos que n\~ao  s\~ao levados a zero por $T$, ou seja todos os $r\times l$ elementos n\~ao nulos da matriz resultante. Isso nos d\'a uma dimens\~ao de $$kr+rl-r^2=kl-(kl-kr-rl+r^2)=kl-(l-r)(k-r)$$
   $~~~~~~ \blacksquare$\medskip

Suponhamos, sem perda de generalidade, que $l\geq k$. 
\begin{prop}Se $M$ \'e uma variedade suave com dimens\~ao $m\leq(l-k)$, ent\~ao para qualquer mapa $f:M\ra L(\R^k,\R^l)$ que seja transversal a todos os $Q^i$, para $i<k$, e para todo $x\in M$, $f(x)\in L(\R^k,\R^l)$ ter\'a posto $k$ constante. \end{prop}

{\bf Dem:} pelo {\bf Teo.\ref{transversalteo}}, n\'os temos que $f:M\ra L(\R^k,\R^l)$ \'e transversal a $Q^i$ se  $Q^i\cap f(M)=\emptyset$ ou  se
$$\I m(df_x)+T_{f(x)}Q^i=T_{f(x)}L(\R^k,\R^l)$$ Em termos de dimens\~ao, $(l-k)+\mbox{dim}(Q^i)\geq\mbox{dim}(M)+\mbox{dim}(Q^i)\geq \mbox{posto}(df_x)+\mbox{dim}(Q^i)\geq \mbox{dim}(L(\R^k,\R^l)$. Portanto temos:
\beq (l-k)+kl-(l-i)(k-i)\geq \mbox{posto}(df_x)+kl-(l-i)(k-i)\geq lk\eeq
Portanto  $$(l-k)\geq\mbox{posto}(df_x)\geq (l-i)(k-i)$$
 Mas $l\geq k> i$. Logo chegamos a  $(l-i)(k-i)\geq(l-i)>(l-k)$ e finalmente $$(l-k)\geq\mbox{posto}(df_x)>(l-k)$$ absurdo. Portanto, para dim$M\leq (l-k)$ e $f$ transversal a $Q^i$ n\'os temos $Q^i\cap f(M)=\emptyset$ logo se isso \'e v\'alido para todo $i<k$, chegamos a  $f(M)\in Q^k$. ~~~~~~~~~$\blacksquare$    

Antes de prosseguirmos, citamos um teorema sobre transversalidade \cite{Guillemin}

\begin{theo} Sejam $M,~N$ variedades suaves, $M$ variedade compacta,  e $f_0:M\ra N$ um mapa suave. Para qualquer $Z$ subvariedade mergulhada de $N$ temos os seguintes resultados:\begin{itemize}
\item[{\bf Gen.:}] {Existe homotopia $f_t:M\ra N$ de $f_0$ tal que para  $\epsilon>0$ suficientemente pequeno, $f_\epsilon$ \'e transversal a $Z$. Em outras palavras, podemos perturbar  a fun\c c\~ao para que se torne transversal. Essa propriedade \'e chamada de {\bf genericidade}.} 
\item[{\bf Est.:}]{Se $f_0$ \'e transversal a $Z$, para qualquer homotopia suave $f_t$, n\'os temos que $f_\epsilon$ \'e transversal a $Z$ para $\epsilon$ suficientemente pequeno. Ou seja, ela se mant\'em transversal por pequenas perturba\c c\~oes. Esta propriedade \'e chamada de {\bf estabilidade}.}
\item[{\bf Cont.:}] {Se $f:M\ra N$ \'e suave e transversal \`a subvariedade mergulhada $Z$ em $p\in M$, ent\~ao existe aberto em torno de $p$ , $U\subset M$ onde $f$ permanece transversal a $Z$.}

\end{itemize}\end{theo}

Agora podemos enunciar a proposi\c c\~ao chave: 
\begin{prop}Se $M$ \'e compacto  , ent\~ao qualquer fibrado vetorial com fibra t\'ipica isomorfa a $\R^k$ \'e isomorfo a algum subfibrado do fibrado produto $\R^l_M:=M\times \R^l$, onde e dim$(M)+ k \leq l$~.\end{prop}

{\bf Dem:} N\'os temos que $\pi_L:L(E,\R^l_M)\ra M$ \'e um fibrado vetorial.  Um isomorfismo de $E$ com um subfibrado de $\R^l_M$ pode ser visto ent\~ao como uma se\c c\~ao $\Psi:M\ra L(E,\R^l_M)$, no sentido que $\pi_L\circ\Psi=\I d$ e tal que $\Psi_x:E_x\ra \R^l$ \'e uma transforma\c c\~ao linear de posto m\'aximo, $k$, para todo $x\in M$. \'E uma tal se\c c\~ao que tentaremos construir.

 Consideremos o fibrado vetorial $L(E,\R^l_M)$. Em uma trivializa\c c\~ao $$U_\alpha\times L(\R^k,\R^l)\overset{\phi_\alpha}{\simeq}{L(E,\R^l_M)|_U}_\alpha $$  Agora tomamos a  se\c c\~ao nula, que \'e uma se\c c\~ao global suave $\Psi_0:M\ra L(E,\R^l_M)$. Sobre $U_\alpha$, n\'os tomamos o mergulho (das transforma\c c\~oes de posto $i$ sobre $\alpha$):
$$\phi_\alpha(U_\alpha\times Q^i):=Z^i_\alpha\subset L(E,\R^l_M) $$ 
Agora, por genericidade, existe homotopia de $\Psi_0$, que chamaremos de $\Psi_\alpha^0:M\times [0,1]\ra L(E,\R^l_M)$ que \'e transversal \`a subvariedade mergulhada $Z^0_\alpha$ em um dado $\epsilon_0\in]0,1]$~. 

Agora tomamos uma homotopia de $(\Psi_\alpha^0)_{\epsilon_0}$, que chamamos de $\Psi_\alpha^1:M\times [0,1]\ra L(E,\R^l_M)$. Por genericidade esta homotopia \'e transversal a $Z^1_\alpha$ para algum  $\epsilon_1\in]0,1]$ suficientemente pequeno para que, por estabilidade, continue sendo transversal a $Z^0_\alpha$. Procedendo dessa forma $k$ vezes obtemos uma se\c c\~ao global, transversal a todos os $Q^i$ sobre $U_\alpha$ ~(para $i<k$) . Logo pela proposi\c c\~ao anterior, esta se\c c\~ao tem  posto $k$ sobre $U_\alpha$. 

Ent\~ao, como $M$ foi assumido compacto, utilizamos exatamente o mesmo m\'etodo (atentando para mantermos as perturba\c c\~oes sobre o dom\'inio da estabilidade) um n\'umero finito de vezes, obtendo uma se\c c\~ao $\Psi$ de 
 $L(E,\R^l_M)$ de posto $k$, ou seja, n\'os temos uma imers\~ao de $E$ sobre uma $k$-distribui\c c\~ao em $M\times\R^l$.
 
Notemos que o \'unico caso em que este teorema permite que trivializemos $E$, \'e para  $l=k$ e portanto  dim$M=0$, ou seja, para $M=\{p\}$.
$~~~~~~\blacksquare$\medskip

Agora, suponhamos que $E$ seja um fibrado sobre  $N$, e $M$ seja subvariedade mergulhada fechada de $N$. Seja $\imath^*(E)=E|_M$  o fibrado sobre $M$, e suponha que j\'a temos uma se\c c\~ao $\Psi:M\ra L(E,\R^l_M)$ tal que $\Psi(x)$ tenha posto m\'aximo para todo $x\in M$.  
\begin{prop}\label{extension}
Existe se\c c\~ao $\zeta:N\ra L(E,\R^l_N)$ de posto m\'aximo tal que $\Psi=\zeta|_M$.\end{prop}

{\bf Dem:} Agora, ao inv\'es da se\c c\~ao nula que tomamos no in\'icio da demonstra\c c\~ao acima, tentaremos construir uma se\c c\~ao suave $\zeta$ tal que $\zeta|_M=\Psi$, i.e.: $\zeta$ \'e uma  extens\~ao suave qualquer de $\Psi$. 

Como $M$ \'e subvariedade mergulhada de $N$, temos ao redor de cada ponto $p\in M$ um aberto $U_p$ (pequeno suficiente para que seja dom\'inio de uma trivializa\c c\~ao de $E$), dom\'inio de coordenadas  c\'ubicas $\eta:U_p\ra \R^n$ centradas em $p$ tal que $M\cap U_p=S_p$ \'e uma \'unica fatia. Logo podemos tomar uma proje\c c\~ao suave $\pi_{U_p}:U_p\ra S_p$ (podemos definir $\pi_p=\eta^{-1}\circ\mbox{pr}_m\eta$) e assim definimos 
$$\bar\zeta_p:=\Psi\circ\pi_p:U_p\ra L(\R^k,\R^l)$$
Agora, a cole\c c\~ao $\{U_p~|~p\in M\}\cup (N-M)$ forma uma cobertura aberta de $N$, logo podemos tomar parti\c c\~ao da unidade $\{\sigma_i~|~i=1,2,\cdots\}$ subordinada a essa cobertura. Tomando a subsequ\^encia $\{\sigma_i~|~\mbox{supp}\sigma_i\cap M\neq \emptyset\}$, para cada tal $i$, como trivialmente $(N-M)\cap M=\emptyset$, existe ponto $p_i$ tal que supp$\sigma_i\subset U_{p_i}$. Ent\~ao constru\'imos a se\c c\~ao suave em $N$
\beq \zeta=\underset{i}{\sum}\sigma_i\bar\zeta_{p_i}\eeq 
tal que $\zeta|_M=\Psi$. 
   
 Utilizando o fato que $\zeta$ \'e suave, pela propriedade {\bf Cont.} acima, ao redor de cada ponto $p$ de $M$, onde sabemos que $\zeta$ \'e transversal a $Z$,  existir\'a um aberto de $N$, $V_p$, onde $\zeta$ permanece transversal a todos os $Q^j$. A cole\c c\~ao $\{U_p\cap V_p~|~p\in M\}\cup (N-M)$ ainda \'e cobertura aberta de $N$. Agora tomando qualquer 
cobertura de $N-M$ por dom\'inios de trivializa\c c\~oes podemos proceder exatamente como na constru\c c\~ao de $\Psi$, sendo que sobre $M$, i.e.: $\zeta|_M$ estar\'a a salvo das perturba\c c\~oes necess\'arias para deix\'a-la transversal a $Z$ em $N-M$. $~~~~~~~~\blacksquare$\medskip

Finalmente chegamos ao Teorema de Classifica\c c\~ao de Fibrados
\begin{theo}\label{TCF}
Qualquer que seja $E$, $G$-fibrado vetorial sobre $M$, $E\simeq f_0^*\xi_G$ para alguma fun\c c\~ao suave $f_0:M\ra B_G$. Se al\'em disso $f_1^*\xi_G\simeq E$, ent\~ao $f_0$ e $f_1$ s\~ao homot\'opicas 
. \end{theo}

{\bf Dem:} N\'os provaremos para o caso $G=\GL k$. Tomamos $B_G$ como sendo a veriedade Grassmaniana $G(k,l)$, i.e.: os pontos de $B_G$ s\~ao os subespa\c cos $k$-dimensionais de $\R^l$. Definimos ainda $$\xi_G=\{(p,v)\in G(k,l)\times \R^l ~|~v\in p\}$$ 
o que significa que a fibra sobre o ponto $p$ da variedade base $G(k,l)$ \'e justamente o subespa\c co de $\R^l$ correspondente a $p$.  Agora, pela proposi\c c\~ao anterior, n\'os sabemos que existe um isomorfismo $\varphi_0$ de $E$ com uma $k$-subfibrado de $M\times\R^l$. Portanto n\'os definimos  um mapa 
$$\begin{array}{lll}
f_0: & M\ra & G(k,l)\\
~& x\mapsto & \varphi_0(\pi_E^{-1}(x))
\end{array}$$
Lembramos que como  $\varphi_0$ \'e inje\c c\~ao linear (tem posto m\'aximo $k<l$), ou seja, \'e isomorfismo linear sobre a imagem.  Logo n\'os temos que \begin{gather*}
f_0^*\xi_G=\{(x,v)\in M\times\xi_G~|~v\in f_0(x)\}\simeq 
\{(x,v)\in M\times\xi_G~|~v\in\varphi_0(\pi_E^{-1}(x))\}\\
\therefore~~~~f_0^*\xi_G\simeq \{(x,v)~|~x\in M ~\mbox{~e~} v\in\pi_E^{-1}(x)\}=E\end{gather*}
 Agora suponhamos que $\varphi_1:E\ra M\times\R^l$ seja tal que temos 
$$f_1^*\xi_G=\{(x,v)\in M\times\xi_G~|~v\in\varphi_1(\pi_E^{-1}(x))\}\simeq E$$
Tomamos a subvariedade $(M\times\{0\})\cup (M\times\{1\})\subset M\times I$. Ent\~ao pela extens\~ao que constru\'imos na {\bf Prop.$\ref{extension}$}, e utilizando a {\bf Prop.\ref{fibprod}},  temos ent\~ao um isomorfismo $\varphi$ entre $E\times I$  e um subfibrado de $\R^l_{M\times I}$  tal que   $\varphi|_{M\times\{0\}}=\varphi_0$ e $\varphi|_{M\times\{1\}}=\varphi_1$. \'E claro ent\~ao que 
$$\begin{array}{ll}
F: & M\times I\ra G(l,k)\\
~& (x,t)\mapsto \varphi \left(\pi^{-1}_{E\times I}(x,t)\right)\end{array}$$
\'e uma homotopia entre $f_0$ e $f_1$. Ou seja, al\'em de exist\^encia estabelecemos unicidade (em classe de homotopia)$~~~\blacksquare$

\subsubsection{Characteristic Classes and Numbers}
Classes e n\'umeros caracter\'isticos s\~ao justamente os invariantes globais que associam \`a cada fibrado vetorial uma classe de cohomologia da variedade base. 

Dado $f:N\rightarrow M$, n\'os obtemos os mapas induzidos
\begin{displaymath}
\begin{array}{ll}
f^*_{\mbox{vect}}:&\mbox{Vect}_G(M)\rightarrow \mbox{Vect}_G(N)~~~~~\mbox{e}\\
f^*:&H^*(M)\rightarrow H^*(N)\\
\end{array}
\end{displaymath}
onde $H^*(M)$ \'e o anel de cohomologia de Rham de $M$.

Assumindo que $f^*$ tamb\'em s\'o depende da classe de homotopia, n\'os temos que  ambos os mapas s\'o dependem da classe de homotopia de $f$.
Uma classe caracter\'istica $c$ (para G-fibrados) associa a $E$ uma classe de cohomologia $c(E)$ em $H^*(M)$ de uma forma natural.
\begin{defi} Uma classe caracter\'istica $C$ para G-fibrados \'e um mapa\footnote{Na linguagem de teoria das categorias, $C$ \'e uma transforma\c c\~ao natural do funtor Vect$_G$ ao funtor $H^*$, ambos considerados como funtores contravariantes na categoria de variedades suaves.}:
$$C:\mbox{Vect}_G\rightarrow H^*$$
tal que dado um $G$-fibrado vetorial sobre a variedade suave $M$ e um mapa suave $f:N\rightarrow M$, onde $N$ \'e tamb\'em   suave, ent\~ao
$$f^*(c(E))=c(f^*_{\mbox{vect}}(E))$$

\end{defi}

O sub-\'indice  ``vect" \'e na verdade desnecess\'ario, e o suprimiremos de agora em diante. Denotaremos por $\mbox{Char}(G)$ o conjunto das classes caracter\'isticas de G-fibrados.

\begin{prop} Seja $\xi_G$ o G-fibrado universal sobre o espa\c co $B_G$ , ent\~ao
$$c\mapsto c(\xi_G)$$
\'e isomorfismo entre $\mbox{Char}(G)$ e $H^*(B_G)$.\end{prop}

{\bf Dem:}  Seja $\sigma\in H^*(B_G)$, 
 definimos
\begin{displaymath}
\begin{array}{llll}
c:&H^*(B_G) & \rightarrow & \mbox{Char}(G)\\
~ &~~~~\sigma & \mapsto &c_{\sigma}\\
\end{array}
\end{displaymath}

onde
\begin{displaymath}
\begin{array}{l}
c_{\sigma}:\mbox{Vect}_G\rightarrow H^*\\
E=  f_E^*\xi_G\mapsto f_E^*(\sigma)\\
\end{array}
\end{displaymath}
onde agora $c_{\sigma}$ \'e um funtorial de $\mbox{Vect}_G$ em $H^*$ que leva $E$ em $f_E^*(\sigma)$, onde $f_E:M\rightarrow B_G$ \'e tal que $f_E^*(\xi_G)=E\in \mbox{Vect}_GM$, logo existe um \'unico (a menos de classes de homotopia) tal $f_E$ pelo {\bf Teo.\ref{TCF}}.

Resumidamente poder\'iamos definir

\begin{displaymath}
\begin{array}{lll}
c_E: & H^*B_G & \rightarrow H^*M\\
~ &\sigma &\mapsto f_E^*\sigma\\
\end{array}
\end{displaymath}
Isto \'e, $c_E=f_E^*$, ou seja $c_E(\sigma)=c_\sigma(E)$. N\'os temos ent\~ao que   $c_E=c_F$ se $f_E$ \'e homot\'opica a $f_F$, ou seja, se  $F\simeq E$.  Logo, $c_{\sigma}$ est\'a bem definido.

Antes de mais nada precisamos mostrar que $c_{\sigma}$ \'e uma transforma\c c\~ao natural de $\mbox{Vect}_G$ em $H^*$. Isto \'e, se $g:N\rightarrow M$ \'e liso, ent\~ao temos:
$$c_{\sigma}(g^*(E))=c_{\sigma}(g^* f_E^* \xi_G)=c_{\sigma}((f_E\circ g)^*\xi_G):=(f_E\circ g)^*(\sigma)=g^*f_E^*(\sigma)=g^*c_{\sigma}(E)$$
Colocando de outra forma $f_{g^*(E)}(\sigma)=g^*f_E$ . Provamos, ent\~ao, que para todo $\sigma\in B_G^*$ existe $c_{\sigma}\in \mbox{Char}G$, tal que $c_{\sigma}(\xi_G)=\sigma$, ou seja, $c\mapsto\xi_G$ \'e sobreje\c c\~ao de $\mbox{Char}(G)$ em $H^*B_G$. 

Al\'em disso, sejam $c_1, c_2 \in \mbox{Char}(G)$ tal que $c_1(\xi_G)=c_2(\xi_G)\in H^*B_G$
Ent\~ao
$$f^*c_1(\xi_G)=f^*c_2(\xi_G)=c_1(f^*\xi_G)=c_2(f^*\xi_G)$$
Portanto, como
qualquer que seja $ E \in \mbox{Vect}_GM$ existe $f:M\rightarrow B_q ~\mbox{tal que}~ f^*\xi_G=E$
obtemos que 
$c_1(E)=c_2(E)$ para todo $ E\in \mbox{Vect}_G$. Ou seja, $c_1=c_2$. $~~~~~~~\blacksquare$\bigskip

Apesar de ter grande import\^ancia conceitual e de ser indiscutivelmente uma bela constru\c c\~ao te\'orica,  esta descri\c c\~ao de classes caracter\'isticas n\~ao nos fornece uma maneira pr\'atica de calcular classes caracter\'isticas.

 N\'umeros caracter\'isticos s\~ao definidos da seguinte forma: se $\lambda\in H^m(M)$, com $M$ fechado, ent\~ao os n\'umeros 
$$\int_Mc(E)~~~~~\mbox{onde ~~~} c\in \mbox{Char}(G)$$
s\~ao chamados de {\bf n\'umeros caracter\'isticos} de um $G$-fibrado sobre $M$. Eles s\~ao claramente invariantes (por classes de isomorfismo de fibrados), portanto, como um fibrado trivial \'e induzido por um mapa constante, todas suas classes caracter\'isticas e n\'umeros caracter\'isticos s\~ao nulos, o que nos fornece uma boa maneira de testar n\~ao trivialidade. Estudemos agora um caso particular de classes caracter\'isticas.

\subsubsection{ Chern Classes} 
As classes de Chern fornecem um teste siples de verificar se dois fibrados vetoriais sobre a mesma base n\~ao s\~ao isomorfos (nada garante que se tiverem a mesma forma de Chern ser\~ao isomorfos). \'E muitas vezes importante `` contar" quantas se\c c\~oes linearmente independentes um fibrado vetorial possu\'i, as classes de Chern fornecem informa\c c\~oes sobre isso atrav\'es do Teorema de Riemann-Roch e o do Teorema do \'Indice de Atiyah-Singer. 

A classe de Chern de um fibrado em $M$, pelo {\bf Teorema de Classifica\c c\~ao de Fibrados Vetoriais}, pode ser dada pelo pull-back das classes de Chern no fibrado universal (que podem por sua vez serem descritas explicitamente por ciclos de Schubert \cite{chern}), mas n\~ao ser\'a essa a nossa abordagem. 

De qualquer forma, o significado intuitivo das classes de Chern (e outras classes caracter\'isticas) concerne os ``zeros necess\'arios" de uma se\c c\~ao (n\~ao nula) de um fibrado vetorial, ou obstru\c c\~oes para a constru\c c\~ao de certas se\c c\~oes\footnote{Um exemplo de tais obstru\c c\~oes que podem surgir \'e dado pelo teorema da ``bola peluda", que nos diz ser imposs\'ivel construir um campo vetorial global n\~ao nul o sobre $S^2$.}. Daremos a seguir um tratamento informal e pouco alg\'ebrico de formas de Chern, em acordo com o resto da se\c c\~ao. Para um tratamento mais formalizado indicamos \cite{Morita}.

Uma  propriedade interessante do operador tr que necessitaremos agora,  \'e que se $\lambda$ \'e uma p-forma a valores em $\End(E)$, utilizando a derivada covariante exterior $D$, por (\ref{Dtio}):
$$D\lambda=d\lambda+\omega\wedge\lambda-(-1)^p\lambda\wedge\omega$$
portanto em coordenadas temos $\lambda=\lambda^i_je_i\otimes e^j$ onde $\lambda^i_j$ \'e uma $p$-forma,
$$D\lambda=d\lambda^i_j e^j\otimes e_i+\omega^k_i\wedge\lambda^i_j e^j\otimes e_k-(-1)^p\lambda^j_i\wedge\omega^i_ke^k\otimes e_j$$ e portanto
\beq\label{tracoD}\tr(D\lambda)=d(\tr\lambda)+\omega^j_i\wedge\lambda^i_j-(-1)^p\lambda^j_i\wedge\omega^i_j=d(\tr\lambda)\eeq  

Voltamos ao fato que o funcional de Yang-Mills 
$$YM(D):=\langle\langle\Omega_D,\Omega_D\rangle\rangle=\frac{1}{2}\int_M \tr(\Omega_D\wedge*\Omega_D)$$
s\'o depende da m\'etrica atrav\'es de $*$. N\'os poder\'iamos tentar escrever uma a\c c\~ao que n\~ao envolve a m\'etrica, por exemplo em 4D: 
$$S_4(D)=\int_M \tr(\Omega\wedge\Omega)$$

 \'E claro que se $\Omega=\Omega_+$ ent\~ao $*\Omega=\Omega$ e n\'os temos que ent\~ao $YM(D)=S_4(D)$. \'E interessante notar que nossa nova a\c c\~ao est\'a intimamente relacionada com  dualidade e com a teoria de Yang-Mills. 
 
Comecemos calculando os pontos cr\'iticos de $S_4$. Seja 
$\tilde{D}=D+t\gamma$
onde 
$\gamma~\in~\Gamma(\Lambda^1(TM)\otimes \Ad(E))$. 
Ent\~ao, como sabemos $\Omega_{D+t\gamma}=\Omega_D+tD\gamma+t^2\gamma\wedge\gamma$, portanto
\begin{gather*}
\frac{d}{dt}_{|_{t=0}}\Omega_{D+t\gamma}\wedge\Omega_{D+t\gamma}=2(\Omega_D\wedge D\gamma)\\
\therefore~~\delta S_4(D)=2\int_M \tr(\Omega_D\wedge D\gamma)=2\int_M \tr(D\Omega_D\wedge\gamma)=0 
\end{gather*} 
Onde usamos a identidade de Bianchi na \'ultima igualdade e na pen\'ultima $$D\gamma\wedge\Omega_D=D\gamma\wedge*(*\Omega_D)=\langle D\gamma,*\Omega_D\rangle$$ 
\beq\label{minha}\therefore~~~~\langle\langle D\gamma,*\Omega_D\rangle\rangle=\langle\langle\gamma,*D*(*\Omega_D)\rangle\rangle=\langle\langle\gamma,*D\Omega_D\rangle\rangle=\int_M \tr(D\Omega_D\wedge\gamma)\eeq Outra forma de obter este resultado\footnote{Lembramos que $M$ \'e uma variedade sem  bordo. } \'e utilizar  Bianchi  \beq D\gamma\wedge\Omega_D=D(\gamma\wedge\Omega_D)+\gamma\wedge D\Omega_D=D(\gamma\wedge\Omega_D)\eeq e (\ref{tracoD}) juntamente com Stokes:
\beq 2\int_M \tr(\Omega_D\wedge D\gamma)=2\int_M \tr(D(\gamma\wedge\Omega_D))=2\int_M d\tr(\gamma\wedge\Omega_D)=0\eeq
 Ora, mas n\'os obtemos ent\~ao que qualquer conex\~ao $D$ \'e ponto cr\'itico de $S_4$! 
 
Ainda assim, vamos prosseguir mais um pouco nessa linha e generalizar:
$$S_{2n}(D)=\int_M \tr\Omega^n$$
onde $M^{2n}$ e $\Omega^n=\underbrace{\Omega\wedge\dots\wedge\Omega}_n$ . Tomando a varia\c c\~ao, obtemos, empregando (\ref{tracoD}):
\begin{eqnarray}
\frac{d}{dt}_{|_{t=0}}\Omega^n_{D+t\gamma}&=&n(\Omega^{n-1}_D\wedge D\gamma)=nD(\Omega^{n-1}\wedge \gamma)\\
\label{chern}\therefore~~ \tr\frac{d}{dt}_{|_{t=0}}\Omega^n_{D+t\gamma}&=&nd(\tr(\Omega_D^{n-1}\wedge \gamma))\end{eqnarray}
Logo:  
\beq\label{chern2}\delta S_{2n}(D)=\int_M \tr(\frac{d}{dt}_{|_{t=0}}\Omega^n_{D+t\gamma})=n\int_M \tr(\Omega^{n-1}_D\wedge D\gamma)=n\int_M d\tr(\Omega^{n-1}_D\wedge \gamma)=0\eeq 
 Chamamos $\tr\Omega^n$ a $n$-\'esima forma de Chern.
 
Ainda por (\ref{tracoD}), obtivemos que $ d(\tr\Omega^n_D)=\tr(D\Omega^n_D)=0$. Portanto  as formas de Chern s\~ao fechadas. Isso significa que a $k$-\'esima forma de Chern define uma classe de cohomologia, em $H^{2k}(M)$. Agora, a forma de Chern realmente depende da conex\~ao $\omega$, mas sua classe de cohomologia n\~ao, isto \'e, se mudarmos a conex\~ao $\omega$, a forma de Chern se desloca por uma forma exata. 
 Para ver esse resultado, tomemos $\omega'=\omega+\gamma$ e $\omega_s=\omega+s\gamma$, ent\~ao n\'os temos (utilizando (\ref{chern})): 
$$\tr(\Omega'^{k})-\tr(\Omega^k)=\int^1_0\frac{d}{ds}\tr(\Omega^k_s)ds=k\int^1_0 d\tr(\gamma\wedge\Omega^{k-1}_s)ds=kd\int^1_0 \tr(\gamma\wedge\Omega^{k-1}_s)ds$$ 
que \'e, portanto, exata. 

Utilizando o {\bf Teo. \ref{homotopicas}},   outra forma de percebermos que a forma de Chern s\'o depende da classe de homotopia de $M$, isto \'e, que duas formas de Chern diferem por uma forma exata (est\~ao na mesma classe de cohomologia)  (Milnor): 
 
Dadas duas conex\~oes $\nabla^0$ e $\nabla^1$ em $E$, n\'os usamos $\mbox{pr}_1:M\times I\rightarrow M$ para traz\^e-las para $E\times I$, chamando $\mbox{pr}_1^*{\nabla}^1=\widetilde\nabla^1$ e $\mbox{pr}_1^*{\nabla}^0=\widetilde\nabla^0$. Agora n\'os constru\'imos em $E\times I$ a conex\~ao $$\widetilde{\nabla}=(\mbox{pr}_2)(\widetilde\nabla^1)+(1-\mbox{pr}_2)(\widetilde\nabla^0)$$ onde a fun\c c\~ao $\mbox{pr}_2:M\times I\rightarrow \R$ \'e a proje\c c\~ao $M\times I\ra I\subset\R$. 
Agora seja
 \begin{displaymath}
\begin{array}{ll}
\varepsilon_t: & M\rightarrow M\times I\\ \phantom{a}&x\mapsto(x,t)\end{array}\end{displaymath} 
Ent\~ao $\varepsilon^*_t(E\times I)=E$ e portanto $\varepsilon^*_0\widetilde{\nabla}=\nabla^0\mbox{~~e~~}\varepsilon^*_1\widetilde{\nabla}=\nabla^1$. \begin{gather*} 
\therefore~~~~~~\tr(\Omega_{t})=\tr(\varepsilon^*_t\Omega_{\widetilde{D}})=\varepsilon^*_t \tr(\Omega_{\widetilde{D}})\mbox{~~ ent\~ao~~~}\\
 \tr(\Omega_0)=\varepsilon^*_0( \tr(\Omega_{\widetilde{D}}))\mbox{~~~e~~~}\tr(\Omega_1)=\varepsilon^*_1 \tr(\Omega_{\widetilde{D}})
\end{gather*} 
Onde utilizamos que $f^*\tr(\lambda)=\tr(f^*\lambda)$, resultado v\'alido para qualquer $\lambda$ $p$-forma a valores em $\End(E)$. Isto pode ser visto lembrando que se $\{e_i\}$ \'e base de $E_{f(x)}$, ent\~ao $\{f^*e_i\}=\{e_i\}$ \'e base de $(f^*E)x$, agora escrevendo  escrevendo em coordenadas o resultado \'e trivial.

 Logo,  obtemos  $\tr(\Omega_1)$ e $\tr(\Omega_0)$ s\~ao pull-backs da mesma forma fechada $\tr(\Omega_{\widetilde{D}})$ por dois mapas diferentes mas homot\'opicos, logo  
$\varepsilon^*_0(\Omega_{\widetilde{D}})\simeq \varepsilon^*_1(\Omega_{\widetilde{D}})$ 
isto \'e, s\~ao co-hom\'ologas.  

Logo podemos definir a {\bf{$k$-\'esima classe de Chern}} $c_k(E)$ do fibrado vetorial $E$ sobre $M$ como a classe de cohomologia de $\tr(\Omega^k)$, onde $\Omega$ \'e a forma de curvatura de {\it{qualquer}} forma de curvatura em $E$. Estes invariantes s\~ao ferramentas importantes na classifica\c c\~ao de fibrados vetoriais.  Utilizando uma defini\c c\~ao de car\'ater mais topol\'ogico para as formas de Chern \'e poss\'ivel mostrar que, quando apropriadamente normalizados, suas integrais sobre uma variedade compacta orientada $M$ s\~ao n\'umeros inteiros.  A normaliza\c c\~ao requerida \'e 
$$c_k(E)=\frac{(i/2\pi)^k}{n!}\int_M\tr(\Omega^k)$$
A integralidade destas classes s\~ao de extrema import\^ancia na Teoria de Chern-Simons. Mostraremos uma aplica\c c\~ao destas classes aos monop\'olos magn\'eticos, quando estudarmos o eletromagentismo como teoria de gauge. Estritamente falando, classes de Chern se aplicam \`a fibrados vetoriais complexos, sendo seu an\'alogo para fibrados reais as classes de Pontryagin, mas n\~ao adentraremos nesse ponto aqui. Na verdade os fibrados que realmente nos interessam, devido \`a quantiza\c c\~ao (a fun\c c\~ao de onda ter valores complexos), s\~ao os complexos.

\chapter{Applications}
\begin{quote}{\it Os maiores matem\'aticos, como Archimedes, Newton, e Gauss, sempre uniram teoria e aplica\c c\~oes em igual medida} - Felix Klein. \end{quote}

\section{Eletromagnetism}
\begin{quote}

{\it As equa\c c\~oes de Maxwell tiveram maior impacto sobre a hist\'oria humana do que quaisquer dez presidentes americanos.} - Carl Sagan\end{quote}

{\it E a continuidade de nossa ci\^encia n\~ao foi afetada por todos estes turbulentos acontecimentos, assim como as teorias antigas sempre s\~ao inclusas como casos lim\'itrofes nas novas.} - Max Born, referindo-se \`a relatividade restrita e \`a mec\^anica qu\^antica. 
\subsubsection{ Maxwell Equations}
Poder\'iamos adicionar \`a cita\c c\~ao de Born que n\~ao s\'o s\~ao as teorias antigas inclusas nas novas teorias, sendo a ci\^encia uma forma progressiva de conhecimento, mas que s\~ao elas muitas vezes a pedra fundamental na compreens\~ao da teoria mais evolu\'ida, os ombros dos gigantes sobre os quais todos subimos\footnote{``Se eu enxerguei mais longe, foi porque me apoiei em ombros de gigantes." Isaac Newton.}. A teoria eletromagn\'etica de Maxwell por exemplo serviu como inspira\c c\~ao tanto para a relatividade geral quanto para a teoria de gauge. Nesta se\c c\~ao, partindo das equa\c c\~oes de Maxwell, constru\'iremos uma teoria de gauge em um $\U1$-fibrado sobre a variedade 4-dimensional $M$. 

As equa\c c\~oes de Maxwell descrevem o comportamento de dois campos vetoriais, o campo el\'etrico $\vec E$ e o magn\'etico $\vec B$. Estes campos s\~ao fun\c c\~oes do tempo, que \'e um par\^ametro real $t$, e s\~ao definidos sobre todo o espa\c co que (em sua forma usual) \'e tido como $\R^3$. Os campos dependem ainda da densidade de carga el\'etrica $\rho$ e da densidade de corrente $\vec\jmath$, que \'e um campo vetorial dependente do tempo em $R^3$. 

  As quatro equa\c c\~oes, na nota\c c\~ao vetorial devida a Hertz (forma em que \'e mais conhecida), em unidades apropriadas para que a velocidade da luz seja $c=1$ s\~ao:
\begin{eqnarray}
\label{nomonopolo}\nabla\cdot\vec{B}=0\\
\label{conservcurr}\nabla\times\vec E+\frac{\partial \vec B}{\partial t}=0\\
\label {gauss}\nabla\cdot\vec E=\rho\\
\label{fluxomag}\nabla\times\vec B-\frac{\partial \vec E}{\partial t}=\vec \jmath
\end{eqnarray}
A semelhan\c ca entre as equa\c c\~oes (\ref{nomonopolo}), (\ref{conservcurr}) \`as equa\c c\~oes (\ref{gauss}), (\ref{fluxomag}) respectivamente, \'e intrigante. No entanto h\'a uma disparidade: o operador divergente em $\R^3$ \'e associado a um operador diferencial de primeira ordem sobre 2-formas, enquanto o operador rotacional atua sobre 1-formas, mas queremos aplicar ambos sobre $\vec E$ e $\vec B$. Com um aux\'ilio do operador estrela de Hodge podemos transformar um no outro, quem sabe estabelecendo uma rela\c c\~ao entre o primeiro par de equa\c c\~oes  e o segundo.

 Na relatividade, $M$ \'e uma variedade suave lorentziana, i.e.: equipada com um produto bilinear sim\'etrico n\~ao-degenerado, mas que n\~ao \'e positivo definido, tem assinatura $(3,1)$.  Para facilitar a nota\c c\~ao, vamos considerar o caso de uma m\'etrica de Minkowski\footnote{Na verdade, o Princ\'ipio de Equival\^encia de Einstein afirma que localmente um referencial inercial \'e equivelente ao espa\c co de Minkowski, o que seria o an\'alogo de tomarmos um gauge quase-can\^onico em $TM$ {\bf Sec.1.6}.}, que chameremos de $\eta $.

 Comecemos analisando as equa\c c\~oes (\ref{nomonopolo}) e (\ref{conservcurr}). Para escrever o campo el\'etrico como uma 1-forma, precisamos compatibilizar  grandezas vetoriais, usualmente utilizadas no eletromagnetismo, com a descri\c c\~ao em formas que estamos perseguindo. Para isso basta usar o operador $\sharp$ (definido na {\bf Sec.1.3}). No entanto, como estamos na m\'etrica de Minkowski, isso simplesmente significa fazermos $\partial_i^\sharp=dx^i$. E portanto tomamos $E=\vec E^\sharp$. Para o campo magn\'etico, que queremos encarar como uma 2-forma\footnote{Poder\'iamos colocar os dois campos como 1-formas, deixando o opererador $*$ cuidar da ordem necess\'aria \`a aplica\c c\~ao da derivada exterior adequada, ao prosseguir por essa linha no entanto n\~ao encontramos equa\c c\~oes de Yang-Mills, e as equa\c c\~oes se tornam menos elegantes. } , faremos $B=*_S(\vec B^\sharp)$, onde $*_S$ \'e  o operador de Hodge restrito ao subespa\c co puramente espacial (com a m\'etrica euclidiana). \'E f\'acil ver que 
\beq\label{rotdiv}(\nabla\times \vec v)^\sharp=-*_Sd_S(\vec v)^\sharp \mbox{~~~e~~~}(\nabla\cdot\vec v)^\sharp=*_S(d_S(*_S(\vec v)^\sharp))\eeq 
 A forma natural de produzirmos uma 2-forma a partir de $E$ (sem utilizar $*$) seria tomar $E\wedge dt$, e agora unimos os dois em uma \'unica 2-forma chamada de {\bf campo eletromagn\'etico}:
\beq\label{campoeletro1}F=B+E\wedge dt=*_S(\vec B^\sharp)+(\vec E)^\sharp\wedge dt\eeq

 E agora,  separando a diferencial exterior em sua parte espacial (que opera com diferencia\c c\~ao s\'o nos \'indices espaciais) e sua parte temporal, i.e.: $d=d_S+d_t$ temos
\beq\label{gradEB}\begin{array}{ll}
 dF & =dB+dE\wedge dt\\
~&=d_SB+\partial_tB \wedge dt+(d_SE+\partial_tE\wedge dt)\wedge dt\\
~& = d_SB+(\partial_tB+d_SE)\wedge dt
\end{array}\eeq
Notemos que o primeiro termo da \'ultima express\~ao n\~ao cont\'em $dt$, logo os dois termos s\~ao linearmente independentes, portanto a equa\c c\~ao $dF=0$ equivale a:
\begin{eqnarray}\label{eletro1} d_SB=0\\
\label{eletro2}\partial_tB+d_SE=0
\end{eqnarray}
 E obtemos de (\ref{eletro1}) e \ref{eletro2}), em uma forma compar\'avel \`as equa\c c\~oes (\ref{nomonopolo}) e (\ref{conservcurr}): 
\beq\label{eletro
3}\begin{array}{l}
(\nabla\cdot \vec B)\mbox{vol}_S=0\\
*_S(\partial_t\vec B)^\sharp+*_S(\nabla\times \vec E)^\sharp=*_S\left((\partial_t\vec B+\nabla\times \vec E)^\sharp\right)=0\end{array}\eeq
Portanto claramente equivalem a  (\ref{nomonopolo}) e (\ref{conservcurr}). 
Uma vantagem da linguagem de formas diferenciais \'e sua generalidade. Podemos tomar nosso espa\c co-tempo como sendo qualquer variedade $M$, definindo o campo eletromagn\'etico como uma 2-forma $F$ em $M$, as primeiras equa\c  c\~oes de Maxwell dizem simplesmente que 
$$dF=0$$
A nossa divis\~ao de espa\c co-tempo em espa\c co {\it e} tempo \'e que foi de alguma forma arbitr\'aria, j\'a que raramente podemos tomar $M=S\times\R$, onde $S$ \'e uma variedade riemanniana de tr\^es dimens\~oes. Somente quando fizermos uma tal separa\c c\~ao (localmente) \'e que podemos falar de campo el\'aetrico e campo magn\'etico, separadamente. 

 Utilizando a n\~ao degeneresc\^encia da m\'etrica de Minkowski podemos definir o operador $*$ de Hodge normalmente (lembramos que definimos por (\ref{priformas}) uma m\'etrica em $\Lambda^p(V)$ dada uma m\'etrica em $V$). N\~ao \'e dif\'icil mostrar que se $M$ for uma variedade semi-riemanniana de dimens\~ao $n$ e assinatura $(s,n-s)$ n\'os temos $*_{n-p}*_p=(-1)^{p(n-p)+s}$.   \'E relativamente f\'acil perceber que $*(*_S(\vec v^\sharp))=\vec v^\sharp\wedge dt$, j\'a que, se o $*_S$ leva $\vec v^\sharp$ em formas a ele ortogonais  espacialmente, o operador $*$ seleciona as duas dire\c c\~oes ortogonais a esses dois vetores, uma \'e a dire\c c\~ao temporal, logo a outra \'e no sentido inicial do vetor, restando checar os sinais apropriados (orienta\c c\~ao), o que deixaremos a cargo do leitor. Da mesma maneira \'e f\'acil verificar que $*(\vec v^\sharp\wedge dt)=*_S(\vec v^\sharp)$. Obtemos ent\~ao:  
\beq\begin{array}{ll}
~& *(E\wedge dt)= *_S(\vec E^\sharp)\\
~& *B=*(*_S(\vec B^\sharp))=(\vec B)^\sharp\wedge dt \\
\therefore~~&*F=-*_S(\vec E^\sharp)+(\vec B)^\sharp\wedge dt\end{array}\eeq
Ou seja, fizemos a mudan\c ca $$\vec E\ra-\vec B\mbox{~~~e~~~}\vec B\ra\vec E$$
Que \'e uma (de duas) das diferen\c cas entre as equa\c c\~oes (\ref{nomonopolo}) , (\ref{conservcurr}) \`as (\ref{gauss}) e (\ref{fluxomag}). Escrevemos  (\ref{gradEB}) na forma
 $$dF=d_SB+(\partial_tB+d_SE)\wedge dt=d_S*_S(\vec B^\sharp)+(*_S\partial_t(\vec B^\sharp)+d_s(\vec E^\sharp))\wedge dt$$ Logo podemos escrever (utilizando  que $*(A\wedge dt)=*_SA$,  e $(*_S)_2\circ(*_S)_1=-1$ na passagem da segunda para a terceira linha e $*_1*_3=1$ na passagem da primeira para a segunda): 
\beq\begin{array}{ll}
 d*F & =d_S*_S(\vec E^\sharp)+(*_S\partial_t(\vec E^\sharp)-d_s(\vec B^\sharp))\wedge dt\\
~& =(\nabla\cdot \vec E)\mbox{vol}_S+*\left(*(*_S\partial_t(\vec E^\sharp)-d_s(\vec B^\sharp))\wedge dt)\right)\\
~& =*\left((\nabla\cdot \vec E)dt-\partial_t(\vec E^\sharp)-*_Sd_s(\vec B^\sharp))\right)
\end{array}
\eeq
E obtemos ent\~ao que:
\beq\label{eletro6} *d*F=(\nabla\cdot \vec E)dt-\partial_t(\vec E^\sharp)-*_Sd_s(\vec B^\sharp))\eeq 
A outra diferen\c ca entre os dois pares de equa\c c\~oes, \'e que as \'ultimas cont\'em  termos n\~ao homog\^eneos, a carga e a corrente. Mas carga \'e simplesmente a corrente que est\'a, em rela\c c\~ao ao nosso referencial, parada, i.e.: \'e um vetor sem proje\c c\~ao espacial, cuja \'unica componente n\~ao nula \'e a temporal. Escrevemos pois, tomando coordenadas $\{x_\mu\}_{\mu=0}^3$: 
$$\vec J=\rho\partial_0+j^1\partial_1+j^2\partial_2+j^3\partial_3$$
  E assim unimos a carga e a corrente em uma \'unica 1-forma 
$$J=-\rho dx^0 +j_ idx^i$$
Agora, simplesmente aplicando $\flat$ em (\ref{eletro6}), e lembrando que $(dx^0)^\flat=-\partial_0$, chegamos ao outro par das equa\c c\~oes: 
\beq\label{Mills} *d*F=J\eeq 
Historicamente, foi Faraday que descobriu, em 1831, que um campo magn\'etico variando no tempo induziria um rotacional n\~ao-nulo no campo ele\'trico (\ref{conservcurr}). Em 1861 Maxwell percebeu que a lei de conserva\c c\~ao de cargas (el\'etricas) poderia ser introduzida automaticamente na Lei de Biot-Savart:  
$$\nabla\times\vec B=\vec \jmath$$  Simplesmente adicionando-se a ela um termo $-\frac{\partial \vec E}{\partial t}=\vec \jmath $ e tomando o divergente chegava-se \`a equa\c c\~ao de continuidade
\beq\label{continuidade}\frac{d \rho}{d t}=\nabla\cdot\vec\jmath\eeq
\'E interessante notar que Einstein fez uma generaliza\c c\~ao muito parecida para chegar \`a sua equa\c c\~ao:
$$R_{\mu\nu}=\kappa{T}_{\mu\nu}\ra R_{\mu\nu}-\frac{1}{2}g_{\mu\nu}R=\kappa{T}_{\mu\nu}$$ 
que dessa forma automaticamente incorporava a conserva\c c\~ao infinitesimal\footnote{Ver \cite{Penrose} Cap.19, ou \cite{Wald} Cap.4 para uma explica\c c\~ao das dificuldades de passarmos de uma equa\c c\~ao de  conserva\c c\~ao infinitesimal para uma lei de conserva\c c\~ao de carga e momento.}  do tensor de energia-momento atrav\'es da identidade de Bianchi contra\'ida\footnote{Havia ainda o agravante de que sem esse termo, como se pode notar pela identidade abaixo, que a curvatura escalar \'e uma constante .} que na nota\c c\~ao de `` \'indices abstratos'' \cite{Wald}\'e :
$$\nabla^c(R_{cd}-\frac{1}{2}g_{cd}R)=0$$ 
Na verdade essa similaridade t\^em uma ra\'iz  comum, a identidade de Bianchi (\ref{Bianchi}). Isso pode ser visto no nosso caso escrevendo-se a (\ref{continuidade})  na forma $$d*J=0$$
agora, simplesmente aplicando $d*$ em ambos os lados da equa\c c\~ao (\ref{Mills}) obtemos o resultado. A disparidade entre as duas equa\c c\~oes para $F$ pode ser atribu\'ida a inexist\^encia (pelo que at\'e hoje se sabe) de monop\'olos magn\'eticos. 
Qualquer  semelhan\c ca com as equa\c c\~oes de Yang-Mills n\~ao \'e mera coincid\^encia.
 
Utilizando o Lema de Poincar\'e, pela equa\c c\~ao $dF=0$, se $U$ for uma regi\~ao contr\'atil\footnote{Lembramos ainda que pela {\bf Prop \ref{poincare}} todo fibrado vetorial sobre uma variedade contr\'atil \'e isomorfo a uma variedade trivial, ou seja, todas as suas se\c c\~oes podem ser vistas como fun\c c\~oes, elementos de $C^\infty(U\ra V)$ onde $V$ \'e a fibra t\'ipica.} existe  uma 1-forma $A$ sobre $U$ tal que $F|_U=dA$.  Logo a segunda equa\c c\~ao fica 
$$*d*dA=0$$
Claramente, qualquer transforma\c c\~ao $A\ra A+df$ onde $f:U\ra \R$ n\~ao afeta as equa\c c\~oes. Por outro lado, suponhamos que n\'os tenhamos um $\U1$-fibrado vetorial, $E$. Para simplificar, assumimos que $E=M\times\C$. Pelo  {\bf Teo.\ref{corresp2}} uma conex\~ao $\nabla$ em $E$ pode ser ent\~ao descrita por uma forma de conex\~ao, que aqui chamaremos de potencial vetor, que \'e uma 1-forma a valores em $\End(E)$. Mas como $\End(\C)\simeq\C$ canonicamente, isto equivale a dizer que temos uma forma a valores complexos. Tomamos a representa\c c\~ao fundamental de $\U1$ sobre $\C$. A \'algebra de Lie de $\U1$ \'e dada por 
$$\mathfrak{u}(1)=\{ix~|~x\in\R\}$$
 O que significa que as componentes de $A$ s\~ao fun\c c\~oes puramente imagin\'arias, o que pode causar certo disconforto, mas que em todo caso pode ser consertado meramente estipulando que essa conex\~ao seja $i$ vezes a conex\~ao real. 
Agora suponhamos que apliquemos uma transforma\c c\~ao de gauge  a este potencial vetor. Como $E$ \'e trivial n\'os podemos pensar na transforma\c c\~ao de gauge como  uma fun\c c\~ao $g:M\ra\U1$. E obtemos ent\~ao, lembrando que $\U1$ \'e abeliano: 
$$\widetilde A=gAg^{-1}+gdg^{-1}=A+gdg^{-1}$$
logo se pudermos escrever $g=e^{-f}$ para alguma fun\c c\~ao a valores imagin\'arios $f$, n\'os recuperamos 
$$\widetilde A=A+df$$
Claramente agora a curvatura \'e simplesmente o campo eletromagn\'etico e as equa\c c\~oes de Maxwell no v\'acuo s\~ao as equa\c c\~oes de Yang-Mills para $G=\U1$. Lembramos que se o grupo n\~ao for abeliano, a pr\'opria derivada exterior covariante envolve termos do potencial , deixando as equa\c c\~oes de Yang-Mills n\~ao lineares. 

Ainda n\~ao comentamos um assunto important\'issimo, na verdade, o que d\'a sentido a toda a discuss\~ao precedente. Precisamos discutir a  forma como o campo eletromagn\'etico interage com a mat\'eria, as equa\c c\~oes de movimento, i.e.: a {\bf for\c ca de Lorentz}. \'E claro que uma carga situada em um campo n\~ao s\'o \'e sujeita \`a a\c c\~ao do campo como tamb\'em age sobre o campo, transformando-o. Claramente, consideraremos aqui uma carga pequena suficiente para que sua retroa\c c\~ao seja insignificante. A f\'ormula usual para a for\c ca de Lorentz $\vec F$ sobre uma part\'icula com carga $q$ com velocidade $v$ \'e dada por
 \beq\label{lorentz1}\vec F=q(\vec E+\vec v\times \vec B)\eeq
no entanto, essa forma n\~ao \'e abrangente o suficiente para descrever o caso relativ\'istico. Por exemplo, levando o tempo em considera\c c\~ao da equa\c c\~oes de Euler-Lagrange deve emergir tamb\'em  a energia da part\'icula no campo.  

Como definimos o campo eletromagn\'etico, n\~ao seria a for\c c\a resultante simplesmente a a\c c\~ao de $F$ sobre uma dada part\'icula, i.e.: se  $\mathbf{v}$ \'e a velocidade da part\'icula de carga $q$, na variedade que estamos considerando, 
dever\'iamos ter $q\left(F(\mathbf{v})\right)^\flat$ para a for\c ca exercida na part\'icula. 

Isto de fato se d\'a,\footnote{ Tomando coordenadas $\{x_\mu\}_{\mu=0}^3$, onde como sempre, \'indices romanos significam somente os campos espaciais.} \'e poss\'ivel mostrar de uma forma extraordinariamente simples \cite{Landau},  que no caso relativ\'istico, em coordenadas, denotando a quadri-velocidade da part\'icula de carga $q$, por $\mathbf{v^\sigma}$, a equa\c c\~ao de movimento (em unidades tal que $c=1$) toma a forma:
 $$m\frac{dv^\sigma}{ds}=F^{\sigma\beta}qv_\beta=qF^\sigma_\beta v^\beta=q\left(F(\mathbf{v})\right)^\flat$$
onde chamamos a aten\c c\~ao que $s$ representa o par\^ametro afim da curva da part\'icula.

Agora escrevendo os (co-)campos em coordenadas temos
\begin{eqnarray}
B & = &B_xdy\wedge dz+B_ydz\wedge dx+B_zdx\wedge dy\\
E & = & E_xdx+E_ydy+E_zdz\end{eqnarray}
 Agora, por  (\ref{campoeletro1}), temos:
\begin{eqnarray*}
F(\mathbf{v})& =& (B_xdy\wedge dz+B_ydz\wedge dx+B_zdx\wedge dy+E_xdx\wedge dt+E_ydy\wedge dt+E_zdz\wedge dt)(\mathbf{v})
\end{eqnarray*}
 Escrevendo $\mathbf v$ coordenadas, n\~ao \'e dif\'icil  calcular que a componente espacial de $q\left(F(\mathbf{v})\right)^\flat$ nos d\'a exatamente a equa\c c\~ao de Lorentz, e que sua parte temporal nos fornece $qE_iv^i=qE(\vec v)$ que \'e produto da velocidade pela for\c ca que o campo el\'etrico exerce na part\'icula, ou seja, a pot\^encia inserida na part\'icula pelo campo. Terminamos essa se\c c\~ao com uma defini\c c\~ao que ser\'a ampliada para Kaluza-Klein na pr\'oxima se\c c\~ao.

\begin{defi} A {\bf co-for\c ca de Lorentz} sobre uma part\'icula com quadri-velocidade $\mathbf{v}$ \'e dada por 
$ F(\mathbf{v})$.\end{defi}

\section{Kaluza-Klein}
\begin{quote}
{\it De agora em diante, espa\c co por si s\'o e tempo por si s\'o, descender\~ao a meras sombras, e somente uma uni\~ao dos dois deve sobreviver.} - Hermann Minkowski.\end{quote}

\subsubsection{History}
Essa frase, dita por Minkowski em seu discurso na 80$\mbox{th}$ Assembl\'eia de Ci\^encia Natural Alem\~a, \'e uma das mais famosas cita\c c\~oes da f\'isica. Ela representa a mudan\c ca de paradigma advinda da relatividade: o tempo deixou de ter seu car\'ater absoluto, mesclou-se \`as outras dimens\~oes espaciais, e juntas perderam seu car\'ater absoluto, rig\'ido, para fornecer uma bela explica\c c\~ao   da gravita\c c\~ao\footnote{Segundo o legend\'ario Lev Landau \cite{Landau},``[a teoria da relatividade geral] representa provavelmente a mais bela de todas as teorias f\'isicas existentes.''}.

Em 1919, a teoria eletromagn\'etica de Maxwell estava bem estabelecida, Einstein havia h\'a pouco formulado sua teoria generalizada, enquanto  as for\c cas nucleares ainda n\~ao eram compreendidas. Era natural portanto, procurar unificar as for\c cas conhecidas pelo aparato te\'orico da relatividade geral, ou seja atrav\'es da geometria do espa\c co-tempo. Theodor Kaluza  
alcan\c cou essa unifica\c c\~ao atrav\'es de um  surpreendente artefato:  postular uma dimens\~ao extra. Apesar de atraente, a id\'eia de Kaluza tinha dois graves defeitos: a depend\^encia da quinta coordenada era suprimida por nenhum motivo aparente, e uma quinta dimens\~ao jamais havia sido observada.  Essas duas cr\'iticas foram sanadas por Oscar Klein, que postulou uma topologi circular para a quinta dimens\~ao. Ele  mostrou que se o raio fosse pequeno o suficiente era poss\'ivel manter a depend\^encia na quinta coordenada, justificar sua n\~ao observabilidade e preservar os resultados de Kaluza. A interpreta\c c\~ao deste formalismo seria um pouco distinta da interpreta\c c\~ao das teorias de gauge com espa\c co interno,  as {\it part\'iculas} realmente percorreriam todas essas dimens\~oes, sendo que o eletromagnetismo emergiria da proje\c c\~ao em nosso espa\c co-tempo quadri-dimensional dessa din\^amica em 5 dimens\~oes. 

Muito anos mais tarde, generaliza\c c\~oes de Kaluza-Klein para grupos de dimens\~eos  maiores deram luz \`a supergravidade, e ainda depois, como argumenta M.J. Duff \cite{duff}, a revitaliza\c c\~ao da teoria de supercordas deveu mais a supergravidade de Kaluza-Klein do que ao modelo de resson\^ancia dual, ao contr\'ario do que revis\~oes hist\'oricas normalmente comentam.  Faremos um tratamento matem\'atico de um caso mais geral do que Kaluza-Klein original, considerando um fibrado principal com grupo estrutural arbitr\'ario . 

\subsubsection{Mathematical Foundations}
Ao longo desta se\c c\~ao, $P$ ser\'a um $G$-fibrado principal sobre  $M$, variedade (pseudo)riemanniana $m$-dimensional.  $\{X_i\}_{i=1}^m$ ser\'a um referencial ortonormal sobre um aberto $\theta\subset M$ e $\{\lambda^i\}_{i=1}^m$ seu co-referencial. Novamente $G$ denota um grupo de Lie compacto $k$ dimensional, $\Lg $ sua \'algebra de Lie dotada de produto interno $\Ad$-invariante ($K$), como mencionado na   {\bf Sec 2.3} e explicitado na {\bf Sec. 3.1}. O dotaremos de base $\{e_\sigma\}_{\sigma=m+1}^n$ (onde $m+k=n$) e base dual para $\Lg^*$ , $\{e^\sigma\}_{\sigma=m+1}^n$. De uma forma geral os \'indices latinos variam de $1$ a $m$, e os gregos de $m+1$ a $n$. Denotaremos ainda ${C^\gamma_\sigma}_\beta$ as constantes estruturais de $\Lg$:
$$[e_\sigma,e_\beta]={C^\gamma_\sigma}_\beta e_\gamma$$
\'E claro que ${C^\gamma_\sigma}_\beta=-{C^\gamma_\beta}_\sigma $. Ademais, lembrando que como $\Ad:G\ra\SO\Lg$ implica $ad:\Lg\ra\mathfrak{s}\mathfrak{o}(\Lg)$, que s\~ao as matrizes anti-sim\'etricas, escrevendo $\ad(e_\sigma)e_\beta=[e_\sigma,e_\beta]$ n\'os obtemos tamb\'em que  ${C^\gamma_\sigma}_\beta =-{C^\beta_\sigma}_\gamma $. Dada uma m\'etrica $h$ em $M$, e uma conex\~ao $H$ (associada a forma $\omega$) em $P$, lembramos que, por (\ref{metricP}) podemos induzir uma \'unica m\'etrica $\alpha$ em $P$ $G$-invariante, tal que a decomposi\c c\~ao $TP=\HH\oplus\V$ seja ortogonal, que $d\pi|_{\HH}$ seja isometria, e que  o produto interno de vetores verticais seja $K$:
$$ \alpha=\pi^*h+K\circ\omega$$
Para facilitar a nota\c c\~ao, denotaremos  como $e^i$ a 1-forma sobre $TP$ dada por $\pi^*\lambda^i$ (note que os \'indices s\~ao latinos, n\~ao gregos) i.e.: $\pi^*\lambda^i\ra e^i$, e identificaremos tamb\'em $e^\beta\circ\omega\ra e^\beta$, obtendo uma 1-forma em V. 
Portanto, deixando as letras  latinas {\it ma\'iusculos} correrem sobre {\it todos} os \'indices, temos que $\{e^A\}_{A=1}^n$ \'e um co-referencial em $P$. Denotamos o referencial dual a $\{e^A\}_{A=1}^n$ pela m\'etrica $\gamma$ de $P$ por $\{e_A\}_{A=1}^n$. \'E f\'acil ver que $d\pi (e_i)=X_i$. N\'os temos tamb\'em  que $\{e^A\}_{A=1}^n$ \'e de fato um co-referencial, j\'a  que $e_\sigma\in\V$ por defini\c c\~ao, e portanto 
$$e^i(e_\sigma)=\lambda^i(d\pi(e_\sigma))=0=\gamma(e_i,e_\sigma)=e^\sigma(e_i)$$
Portanto obtemos um pouco mais, que \'e um referencial ortonormal, e como H \'e o subfibrado ortogonal a V,   
 n\'os temos que $$\mbox{span}\left[\{e_\sigma\}_{\sigma=m+1}^n\right]=\V|_\theta\mbox{~~~e~~~span}\left[\{e_i\}_{i=1}^m\right]=\HH|_\theta$$

Terminada essa parte preliminar, o nosso objetivo ser\'a calcular a conex\~ao de Levi-Civita em $P$, a \'unica  compat\'ivel com a nossa m\'etrica e sem tors\~ao. A partir da\'i, descobriremos fatos surpreendentes ligados \`a $P$ e sua rela\c c\~ao com $M$. Devemos, em alguns momentos, tentar pensar em $P$ mais como uma variedade riemanniana do que como um fibrado principal.  

Seja $\Omega\in\Gamma(\Lambda^2(TP^*)\otimes\Lg)$ a forma de curvatura em $P$, que pelo {\bf Teo.\ref{curvoutra}} \'e dada por 
\beq\label{a2}\Omega=d\omega\circ\widehat H\eeq
onde a forma de conex\~ao $\omega$  \'e nada mais que a proje\c c\~ao no subespa\c co vertical, seguida do isomorfismo entre este e $\Lg$. 
Como temos um referencial local em $\V$, podemos definir as 2-formas reais $\Omega^\sigma\in\Gamma(\Lambda^2(\HH^*))$:  
$$\Omega^\sigma\otimes e_\sigma:=\Omega$$ 
onde $\Omega^\sigma=e^\sigma([e^i\otimes e_i,e^j\otimes e_j])$. Ou podemos tomar a descri\c c\~ao ainda mais desmembrada, definindo fun\c c\~oes reais $F_{ij}^\sigma$: 
\beq\label{fiji} (\frac{1}{2}F_{ij}^\sigma e^j\wedge e^i)\otimes e_\sigma:=(\Omega^\sigma)\otimes e_\sigma\eeq
Estamos ficando sem letras para denotar todos os objetos que queremos, ent\~ao vamos denotar a forma de conex\~ao {\it riemanniana, de Levi-Civita}\footnote{Que n\~ao deve ser confundida com {\it a forma de conex\~ao $\omega$ de $P$ como fibrado principal}, por exemplo, aqui $\w\in\Gamma(\Lambda^1(TP^*)\otimes \End(TP))$ e portanto n\~ao tem valor na \'algebra de Lie. } para $P$, relativa ao referencial que constru\'imos $e_A$, como $\w^A_B$. Ou seja
$$\nabla e_B=\w_B^Ae_A\mbox{~~~~ou equivalentemente~~~~}d(e^A)=\w^A_B\wedge e^B $$
Sendo de Levi-Civita, temos que $\w_B^A=-\w_A^B$. Como $\omega=e^\sigma\otimes e_\sigma$. n\'os temos que\footnote{Lembremo-nos que aqui $e_\sigma$ \'eum elemento fixo da \'algebra de Lie (ver {\bf Teo.\ref{curvoutra}}).} 
\beq\label{a3} d\omega=d(e^\sigma)\otimes e_\sigma=\w^\sigma_B\wedge e^B\otimes e_\sigma\eeq
E portanto, por (\ref{a2}): 
\beq\label{a1}\Omega=\w^\sigma_i\wedge e^i\otimes e_\sigma\eeq
O que \'e um fato curioso, sendo que a priori a conex\~ao de Levi-Civita n\~ao precisaria estar relacionada a curvatura da conex\~ao de $P$.  A rela\c c\~ao existe porque a curvatura da conex\~ao, (sendo a derivada covariante exterior da pr\'opria conex\~ao) mede o quanto estes campos variam, o que \'e claramente relacionado \`a conex\~ao Levi-Civita.

 Comparando (\ref{a1}) com (\ref{fiji}) obtemos
\beq\label{a4} \w^\sigma_i\wedge e^i=  \frac{1}{2}F^\sigma_{ij}e^j\wedge e^i\eeq

Denotando por $\Gamma^i_j$ a forma de conex\~ao em $M$ relativas ao referencial $X_i=d\pi (e_i)$. Chamaremos o pull-back de $\Gamma^i_j$ de $\bar\Gamma^i_j:=\pi^*\Gamma^i_j$. N\'os temos que 
\beq d\lambda^i=\Gamma^i_j\wedge\lambda^j\eeq
Aplicando $\pi^*$ em ambos os lados temos:
\beq\label{a7}\begin{array}{ll}
d(e^i)& =d(\pi^*\lambda^i)=\pi^*d(\lambda^i)=\pi^*(\Gamma^i_j\wedge\lambda^j)\\
~& =\bar\Gamma^i_j\wedge e^j\\
~& =\w^i_B\wedge e^B=\w^i_j\wedge e^j+\w^i_\sigma\wedge e^\sigma
\end{array}\eeq

N\'os temos ainda, escrevendo explicitamente $d\omega=\Omega-\omega\wedge\omega$:
\beq\label{a5}\begin{array}{ll}
(\w^\sigma_B\wedge e^B)\otimes e_\sigma & =(\frac{1}{2}F_{ij}^\sigma e^j\wedge e^i)\otimes e_\sigma-\frac{1}{2}e^\beta\wedge e^\nu[e_\beta,e_\nu]\\\\
\therefore~~\w^\sigma_B\wedge e^B & =\frac{1}{2}F_{ij}^\sigma e^j\wedge e^i-\frac{1}{2}C^\sigma_{\beta\nu}e^\beta\wedge e^\nu\\\\
~& =\w^\sigma_\beta\wedge e^\beta + \w^\sigma_i\wedge e^i\mbox{~~~~mas por (\ref{a4}):}\\\\
\w^\sigma_i\wedge e^i= \frac{1}{2}F_{ij}^\sigma e^j\wedge e^i & \Rightarrow~ \w^\sigma_\beta\wedge e^\beta=-\frac{1}{2}C^\sigma_{\beta\nu}e^\beta\wedge e^\nu
\end{array}\eeq
Agora, as contas j\'a est\~ao maduras o suficiente para introduzirmos os an\'alogos dos s\'imbolos de Christoffel, definindo:
$$\w^A_{BC}:=\w^A_B(e_C)\Longrightarrow \w^A_B=\w^A_{BC}e^C$$
Introduzindo nas duas \'ultimas equa\c c\~oes de (\ref{a5}), como temos uma base completa de 2-formas, obtemos sem maiores esfor\c cos que  
$$\w^\sigma_{ij}e^j\wedge e^i= \frac{1}{2}F_{ij}^\sigma e^j\wedge e^i$$ portanto $\w^\sigma_{i\beta}=0=\w^i_{\sigma\beta}$ e 
finalmente
\beq\label{a6} \w^\sigma_i=\frac{1}{2}F_{ij}^\sigma e^j \eeq 
Da mesma forma obtemos que $\w^\sigma_{\beta i}=0$ e 
\beq \w^\sigma_\beta=  -\frac{1}{2}C^\sigma_{\nu\beta}e^\nu\eeq
Substituindo (\ref{a6}) em (\ref{a7}) obtemos:
\beq \w^i_j\wedge e^j=\bar\Gamma^i_j\wedge e^j +\frac{1}{2}\underset{\sigma}{\sum} F^\sigma_{ij}e^j\wedge e^\sigma=\bar\Gamma^i_j\wedge e^j -\frac{1}{2}\underset{\sigma}{\sum} F^\sigma_{ij}e^\sigma\wedge e^j \eeq
e estamos prontos para enunciar o primeiro resultado obtido:
\begin{prop} 
\beq\label{continhas}
\left\{\begin{array}{ll}
\w^i_j= & \bar\Gamma^i_j-\frac{1}{2}\underset{\sigma}{\sum} F^\sigma_{ij}e^\sigma\\\\ 
\w^\sigma_i= & \frac{1}{2}F^\sigma_{ij}e^j\\\\
\w^\sigma_\beta= & -\frac{1}{2}C^\sigma_{\nu\beta}e^\nu
\end{array}\right.\eeq\end{prop} $\blacksquare$\medskip

Agora, suponhamos que $\gamma:I\ra P$ seja uma geod\'esica e $\bar\gamma$ sua proje\c c\~ao em $M$. Definimos uma fun\c c\~ao $q:I\ra \Lg$ chamada de {\bf carga espec\'ifica} por $q(t)=\omega(\gamma'(t))$. Note que esse nome garboso nada mais significa que a proje\c c\~ao da velocidade em V (ou $\Lg$), e tentaremos provar que a proje\c c\~ao da velocidade da part\'icula nas dimens\~oes extras \'e sua carga, e que portanto deve ser mantida constante para geod\'esicas. Lembremos tamb\'em que no modelo considerado, as part\'iculas sempre percorrem geod\'esicas, pois n\~ao estamos considerando for\c cas n\~ao inclusas em nossa geometria (i.e.: for\c cas externas).

J\'a que  $\omega=e_\sigma\otimes e^\sigma$ ent\~ao para qualquer vetor $v$ em $P$ n\'os temos que   $$\omega(v)=\langle v, e_\sigma\rangle e^\sigma$$ Para provar que para uma geod\'esica $\gamma(t)$ em $P$ a carga espec\'ifica $q(t)=\omega(\gamma'(t))$ \'e constante, basta mostrar ent\~ao que $$\langle \gamma'(t), e_\sigma\rangle= \mbox{cte}$$
Mas a m\'etrica que escolhemos para $P$ \'e invariante pela a\c c\~ao de $G$, ent\~ao como $e^\sigma$ s\~ao os campos tangentes \`a \'orbita do grupo, eles s\~ao {\bf campos de Killing}. Logo a const\^ancia de $\langle \gamma'(t), e_\sigma\rangle$ \'e um caso especial do seguinte teorema:
\begin{theo}
Se $X$ \'e um campo de  Killing em uma variedade riemanniana $N$ e $\gamma$ \'e uma geod\'esica em $N$ ent\~ao o produto interno de $X$ com $\gamma'(t)$ independe de $t$.\end{theo}

{\bf Dem:}
Necessitaremos da primeira f\'ormula da varia\c c\~ao:

Seja $\gamma:I\ra N$ uma curva suave e 
$$\begin{array}{ll}
F:[-\epsilon,\epsilon]\times I\rightarrow{N}\\
(s,t)\longmapsto{F(s,t)}
\end{array}$$ uma varia\c c\~ao de $\gamma$, i.e.: tal que $F(0,t)=\gamma(t)$ exatamente como constru\'imos para a prova do {\bf Teo. \ref{variation}}. tal que, para $s$ fixo: $F(s,t)=F_s(t)=\gamma_s(t)$ Agora sejam 
\beq\frac{\partial}{\partial{s}}_{\vert(s_0,t_0)}:=F_*\frac{d}{d{s}}_{\vert(s_0,t_0)}~~~;~~~~ \frac{\partial}{\partial{t}}_{\vert(s_0,t_0)}:=F_*\frac{d}{d{t}}_{\vert(s_0,t_0)}\eeq 
Como n\~ao temos tors\~ao, e $s,t$ s\~ao coordenadas, 
$$\frac{D}{ds}\frac{\partial}{\partial t}=\frac{D}{dt}\frac{\partial}{\partial s}$$

Para um dado $s$ temos:
$$E(F(s,t))=\int_a^b\langle \frac{\partial}{\partial{t}}_{\vert(s,t)},\frac{\partial}{\partial{t}}_{\vert(s,t)}\rangle dt$$
Ent\~ao,  temos 

\begin{eqnarray*}
\frac{1}{2}\frac{d}{ds}E(F(s,t)) & = &\frac{1}{2}\frac{d}{ds}\int_a^b\langle \frac{\partial}{\partial{t}}_{\vert(s,t)},\frac{\partial}{\partial{t}}_{\vert(s,t)}\rangle dt  =\int_a^b\langle\frac{D}{ds} \frac{\partial}{\partial{t}}_{\vert(s,t)},\frac{\partial}{\partial{t}}_{\vert(s,t)}\rangle dt\\
\phantom{\frac{d}{ds}E(\gamma_s)} & =    & \int_a^b\langle \frac{D}{dt}\frac{\partial}{\partial{s}}_{\vert(s,t)},\frac{\partial}{\partial{t}}_{\vert(s,t)}\rangle dt=\left.\langle \frac{\partial}{\partial s}_{\vert(s,t)},\frac{\partial}{\partial{t}}_{\vert(s,t)}\rangle \right|_a^b-\int_a^b\langle \frac{\partial}{\partial s}_{\vert(s,t)},\frac{D}{dt}\frac{\partial}{\partial{t}}_{\vert(s,t)}\rangle dt
\end{eqnarray*}

Em particular, 
$$\frac{1}{2}\frac{d}{ds}E(F(s,t))|_{s=0}=\left.\langle \frac{\partial}{\partial s}_{\vert(0,t)},\gamma'(t)\rangle\right|^b_a-\int_a^b\langle \frac{\partial}{\partial s}_{\vert(0,t)},\frac{D}{dt}\gamma'(t)\rangle dt$$

Se $\gamma$ \'e geod\'esica ent\~ao: 
$$\frac{1}{2}\frac{d}{ds}E(F(s,t))|_{s=0}=\left.\langle \frac{\partial}{\partial s}_{\vert(0,t)},\gamma'(t)\rangle\right|^b_a$$
Agora suponhamos que $\Phi_s$ seja o subgrupo a 1-par\^ametro de isometrias gerado por $X$, fazemos ent\~ao $$F(s,t)=\Phi_s(\gamma(t))$$
Agora
$$F_*\frac{d}{d{t}}_{\vert(s,t)}={(\Phi_s)}_*(\gamma'(t))=\frac{\partial}{\partial{t}}_{\vert(s,t)}$$ e como $\Phi_s$ \'e isometria:
$$\langle {(\Phi_s)}_*(\gamma'(t)),{(\Phi_s)}_*(\gamma'(t))\rangle \mbox{~~~n\~ao depende de~~~} s$$ portanto  
$$\frac{1}{2}\frac{d}{ds}E(F(s,t))|_{s=0}=0$$ e finalmente
$$\langle \frac{\partial}{\partial s}_{\vert(0,t)},\gamma'(t)\rangle|_b=\langle \frac{\partial}{\partial s}_{\vert(0,t)},\gamma'(t)\rangle|_a$$ para todo $[a,b]\subset I$. Ou seja, o produto interno de $X$ com $\gamma'(t)$ independe de $t$. 

Uma forma mais direta de obtermos este resultado seria assumindo que um campo de Killing $X$ obedece, para quaisquer campos $Y$ e $Z$ :
$$\langle \nabla_YX,Z\rangle+\langle \nabla_ZX,Y\rangle=0$$ e portanto como $\gamma $ \'e geod\'esica
$$\frac{d}{dt}\langle X,\gamma'(t)\rangle=\langle \nabla_{\gamma'(t)}X,\gamma'(t)\rangle=0$$
n\'os fizemos a primeira prova porque a consideramos mais geom\'etrica, menos alg\'ebrica (al\'em do qu\^e, simult\^aneamente mostra que $X_\gamma$ \'e um campo de Jacobi,  i.e.: $e^\sigma|\gamma$ s\~ao campos de Jacobi, o que pode vir a ser \'util). 

$~~~~~~\blacksquare$\medskip

 Agora definimos um funcional linear sobre a proje\c c\~ao da nossa curva $\bar\gamma(t)=\pi(\gamma(t)$, ou seja, sobre $T_{\bar\gamma(t)}M$, da seguinte forma
\begin{defi} A {\bf{co-for\c ca de Lorentz}}, $\eta(\gamma(t))$ \'e definida por: 
$$ \begin{array}{ll}
\eta(\gamma(t)) : & T_{\bar\gamma(t)}M\ra\R\\
~& v\mapsto K\left(\Omega_{\gamma(t)}\left(\widetilde v, \widehat H( \gamma'(t))\right),\omega(\gamma'(t))\right)
\end{array}$$
 onde $K$ \'e o produto interno em $\Lg$, $\widetilde v$ \'e o levantamento horizontal de $v$ e $\widehat H$ \'e a proje\c c\~ao no subfibrado horizontal. \end{defi}  
Notemos que 
$$\Omega_{\gamma(t)}\left(\widetilde v, \widehat H(\gamma'(t))\right)=\Omega_{\gamma(t)}\left(\widetilde v, \gamma'(t)\right)$$ Notemos ainda que esta express\~ao fica um pouco mais complicada devido ao uso do levantamento horizontal do vetor $v$, que \'e necess\'ario j\'a que queremos ver os efeitos em $M$ e n\~ao em $P$. Se defin\'issemos $v$ como proje\c c\~ao de um vetor qualquer $X$, ter\'iamos simplesmente $\Omega_{\gamma(t)}\left( X, \gamma'(t)\right)$. 

Agora lembramos que provamos que  a carga espec\'ifica \'e uma constante, e portanto a forma de Killing $K$ aqui \'e s\'o uma maneira de multiplicar pela carga.
 Finalmente, {\bf associando a curvatura da conex\~ao com o campo}, assim como fizemos na se\c c\~ao anterior, temos que a for\c ca de Lorentz aqui \'e a proje\c c\~ao da for\c ca eletromagn\'etica na part\'icula sobre a dire\c c\~ao $X$, ou melhor sobre a dire\c c\~ao $d\pi(X)$. Se retirarmos a dire\c c\~ao $X$ (ou $v$), n\'os temos que (substituindo a nota\c c\~ao $K$ por ``$\cdot$'')$$\eta(\gamma'(t))=\Omega(\gamma'(t))\cdot q$$ 
\'e uma 1-forma, equivalente a {\bf co-for\c ca de Lorentz}, que nos fornece tamb\'em  a dire\c c\~ao com que o campo eletromagn\'etico atua sobre a part\'icula. 
Escrevendo em termos de nossa base $\{e_A\}$, temos,
$$\gamma'(t)=u^ie_i+q^\sigma e_\sigma$$ onde claramente\footnote{Agora que nossa compreens\~ao f\'isica do assunto j\'a se aprofundou minimamente, vale a pena apontar que $\gamma(t)$ tem $n-m$ ``cargas",  distintas, que n\~ao se misturam e n\~ao se alteram. Humm....} 
$$q^\sigma e_\sigma=\omega(\gamma'(t))=q\mbox{~~~~~e~~~~}\bar\gamma(t)=u^iX_i$$
Agora, por (\ref{fiji}), $\Omega=\frac{1}{2}F^\sigma_{ij}e^i\wedge e^j\otimes e_\sigma$, portanto, como $\{e_\sigma\}$ \'e base ortonormal (em rela\c c\~ao a $K$) para $\Lg$, temos: 

\begin{eqnarray}
\label{forcaeletro}\Omega(\gamma'(t))\cdot q & = & \frac{1}{2}\underset{\sigma}{\sum}F^\sigma_{ij}u^iq^\sigma e^j\\
\therefore~~(\Omega(\gamma'(t))\cdot q)^\flat & = &  \frac{1}{2}\underset{\sigma,j}{\sum}F^\sigma_{ij}u^iq^\sigma e_j\end{eqnarray}
E cuja proje\c c\~ao corresponde a $$ F(\bar\gamma'(t))\cdot q=\frac{1}{2}\underset{\sigma,j}{\sum}F^\sigma_{ij}u^iq^\sigma X_j$$
De fato, qualquer que seja $v\in T_{\bar\gamma(t)}M$, $v=v^iX_i$, e portanto $\widetilde v=v^ie_i$ e n\'os temos que tanto faz aplicarmos a for\c ca de Lorentz como definida em $M$, o que chamamos de $F$,  ou no levantamento horizontal, em $P$ (al\'em disso \'e claro que, como as bases s\~ao duais pela m\'etrica, as ccomponentes da descri\c c\~ao como formas ou como campos s\~ao as mesmas).

\'E claro que ainda n\~ao provamos o principal, i.e.: 
\begin{theo}
H\'a uma discrep\^ancia entre a equa\c c\~ao da geod\'esica em $M$ e a proje\c c\~ao  da geod\'esica em $P$ que  \'e exatamente a  for\c ca de Lorentz:
  \end{theo}   

{\bf Dem:} 
A demonstra\c c\~ao consiste em calcular a proje\c c\~ao de $$\frac{D\gamma'(t)}{dt}=\nabla^P_{\gamma'(t)}\gamma'(t) \mbox{~~e compar\'a-la a~~ }\frac{D\bar\gamma'(t)}{dt}=\nabla^M_{\bar\gamma'(t)}\bar\gamma'(t)$$
Escrevendo $\gamma'(t)=u^ie_i+q^\sigma e_\sigma$ , temos calculando e em seguida projetando no referencial  espacial\footnote{Note que n\~ao s\~ao necessariamente campos coordenados, ent\~ao deve-se resistir ao h\'abito de usar este termo ao longo dessa se\c c\~ao. Utilizaremos fortemente a nota\c c\~ao de soma de Einstein, muitas vezes mudando \'indices repetidos (i.e.: que s\~ao mudos) de nome.} :
\begin{eqnarray*}
{\nabla^P}_{\gamma'(t)}\gamma'(t) &=& u^j\nabla^P_{e_j}(u^ie_i+q^\sigma e_\sigma)+q^\beta\nabla^P_{e_\beta}(u^ie_i+q^\sigma  e_\sigma)\\\\
~ &=& u^j\left(du^i(e_j)e_i+u^i\w^A_i(e_j)e_A\right)+u^jq^\sigma\w^A_\sigma(e_j)e_A+q^\beta  u^i\w^A_i(e_\beta)e_A+q^\beta q^\sigma\w^A_\sigma(e_\beta)e_A\\\\
e^i({\nabla^P}_{\gamma'(t)}\gamma'(t))&=&u^jdu^i(e_j)+u^ju^k\w^i_k(e_j)+u^jq^\sigma\w^i_\sigma(e_j)+u^jq^\beta\w^i_j(e_\beta)+q^\beta q^\sigma\w^i_\sigma(e_\beta)
\end{eqnarray*}
Agora utilizando (\ref{continhas}), e a anti-simetria de $\w_A^B$ temos que 
\begin{eqnarray*}
u^ju^k\w^i_k(e_j)& =& u^ju^k\bar\Gamma^i_k(e_j)\\
u^jq^\sigma\w^i_\sigma(e_j)&=&-\frac{1}{2}\underset{\sigma}{\sum} u^jq^\sigma F^\sigma_{ij}\\
u^jq^\beta\w^i_j(e_\beta)&=&-\frac{1}{2}\underset{\beta}{\sum} u^jq^\beta F^\beta_{ij}\\
q^\beta q^\sigma\w^i_\sigma(e_\beta)&=&0
\end{eqnarray*}
E finalmente, substituindo 
\begin{eqnarray*}
 e^i(\nabla^P_{\gamma'(t)}\gamma'(t))& =& \left(u^jdu^i(e_j)+u^ju^k\bar\Gamma^i_k(e_j)\right)-\underset{\sigma}{\sum} u^jq^\sigma F^\sigma_{ij}\\
=\pi^*\lambda^i(\nabla^P_{\gamma'(t)}\gamma'(t))&=& \lambda^i\left(u^jdu^i(X_j)+u^ju^k\Gamma^i_k(X_j)\right)-\underset{\sigma}{\sum}u^jq^\sigma F^\sigma_{ij}\\\\
~&=&\lambda^i(u^j\nabla^M_{X_j}(u^kX_k))-\underset{\sigma}{\sum}u^jq^\sigma F^\sigma_{ij}\\\\
~&=& \lambda^i(\nabla^M_{\bar\gamma'(t)}\bar\gamma'(t))-\underset{\sigma}{\sum} u^jq^\sigma F^\sigma_{ij}\end{eqnarray*}$\blacksquare$\medskip

Este resultado significa que habitantes de $M$ que n\~ao soubessem da exist\^encia de outras dimens\~oes, n\~ao saberiam que  na verdade as geod\'esicas que percorrem est\~ao em um espa\c co com maior dimensionalidade. Encontrariam certas caracter\'isticas das part\'iculas que perceberiam permanecer constantes, e as chamariam de carga, cor, charme, ou qualquer outro nome estranho que escolhessem. Perceberiam tamb\'em que a discrep\^ancia entre as trajet\'orias que as part\'iculas deveriam percorrer segundo suas in\'ercias (geod\'esicas), era proporcional a essas cargas de uma maneira espec\'ifica, de acordo com uma lei que n\~ao envolvia explicitamente outras dimens\~oes. 

No entanto, o objetivo da Teoria de kaluza-Klein, \'e a de derivar as equa\c c\~oes de Einstein e de Yang-Mills a partir das equa\c c\~oes de Einstein em uma dimensionalidade maior. Logo  n\'os precisamos derivar estas equa\c c\~oes de campo  a partir da extremaliza\c c\~ao da integral do escalar de curvatura $^PR$. 

Agora, $P$ tem uma curvatura escalar $^{P}{R}$ que pela invari\^ancia da m\'etrica \'e constante em \'orbitas de $G$, e portanto pode ser calculada como fun\c c\~ao real sobre $M$. Al\'em disso, a exist\^encia de uma m\'etrica Ad-invariante em $\Lg$ \'e equivalente \`a exist\^encia de uma m\'etrica bi-invariante em $G$ (por transla\c c\~ao), logo $G$ tem uma curvatura escalar que \'e claramente constante \cite{Riem}, e tamb\'em bem definida sobre cada ponto de $M$. Finalmente, como tanto $M$ quanto $\Lg$ tem m\'etrica, e $\Omega$ \'e uma 2-forma em $M$ a valores em $\Lg$, $\Omega$ tem uma norma bem definida 
$$\|\Omega\|^2=\underset{\sigma,i,j}{\sum}(F^\sigma_{ij})^2$$
que tamb\'em \'e fun\c c\~ao escalar em $M$.   Acharemos uma rela\c c\~ao entre a curvatura do espa\c co total, $^PR$, a do espa\c co base, $^MR$, a do grupo, $^GR$, e a do campo eletromagn\'etico $\|\Omega\|$. 

\begin{theo}
A rela\c c\~ao entre as curvaturas escalares \'e dada por\footnote{Notemos que mudamos livremente \'indices para cima e para baixo, ou seja, n\~ao mantemos a lei de conserva\c c\~ao dos \'indices. Isso ocorre porque estamos em um referencial ortonormal, cujo dual \'e feito tamb\'em pela m\'etrica, ou seja $\langle\cdot,e_A\rangle=e^A$, onde obviamente violamos a conserva\c c\a~ao. }
:
$$^PR=~^MR-\frac{1}{2}\|\Omega\|^2+~^GR$$
\end{theo}

{\bf Dem:} 
N\'os sabemos que 
\begin{eqnarray*}
^P\Omega(e_A,e_B)e_C& =& ~^PR^D_{ABC}e_D\mbox{~~~ou ainda~~~}^PR_{ABCD}=\langle~^P\Omega(e_A,e_B)e_C,e_D\rangle
\end{eqnarray*}
$\mbox{onde~} ^P\Omega=d\w+\w\wedge\w\in \Gamma(\Lambda^2(TP)\otimes \End(TP))$ , e que, a exemplo das conex\~oes $\omega$ e  $\w$, n\~ao deve ser confundida com $\Omega$. Agora vejamos:
\begin{eqnarray*}
^PR:&=&\underset{A,B}{\sum}~^PR_{ABAB}=\underset{A,B}{\sum}\langle~^P\Omega(e_A,e_B)e_A,e_B\rangle\\\\
~&=&\underset{A,B}{\sum} \langle d\w(e_A,e_B)e_A, e_B\rangle +\langle (\w\wedge\w(e_A,e_B))e_A, e_B\rangle
\end{eqnarray*}
Primeiramente\footnote{A partir de agora n\~ao colocaremos mais explicitamente o sinal da somat\'oria quando \'indices se repetirem \`a mesma altura.}, escrevemos $\w=\w^C_De^C\otimes e_C$, portanto $d\w=(d\w^C_D)e^D\otimes e_C$ assim\footnote{Note que a derivada exterior n\~ao se aplica em $e^D$, isso ocorre porque a derivada exterior nesse caso, tem um significado espec\'ifico, como mostramos na equa\c c\~ao (\ref{Oomega}), onde t\'inhamos  uma forma em $M$ a valores no fibrado vetorial, e a derivada exterior s\'o atuava na forma em $M$. Isso n\~ao muda se o fibrado for $TM$(ou , no nosso caso, $TP$).}  temos
\beq\label{kalkle2}\langle d\w(e_A,e_B)e_A,e_B\rangle=(d\w^C_D)(e_A,e_B)e^D(e_A) \langle e_C,e_B\rangle= (d\w^B_A)(e_A,e_B)\eeq
Similarmente 
$\w\wedge\w=\w^C_D\wedge\w^E_Fe_C\otimes e^D(e_E)\otimes e^F=\w^C_D\wedge\w^D_Fe_C\otimes e^F$, assim como explicitado em (\ref{contrac}). Claramente
\beq\label{kalkle1}\langle (\w\wedge\w(e_A,e_B))e_A, e_B\rangle=\w^B_D\wedge\w^D_A(e_A,e_B)\eeq
Antes de mais nada, afirmamos que a curvatura de $G$, equipado com m\'etrica bi-invariante \'e $\frac{1}{4}C^\sigma_{\alpha\beta}C^\sigma_{\alpha\beta}$. Um caminho das pedras para a demonstra\c c\~ao, \'e, assumindo  que quando temos uma m\'etrica bi-invariante em $G$ vale $^G\nabla_XY=\frac{1}{2}[X,Y]$, encontrar $^GR(X,Y)Z=\frac{1}{4}[[X.Y],Z]$ e da\'i segue que $^GR=\frac{1}{4}\langle [e_\alpha,e_\beta],[e_\alpha, e_\beta]\rangle$ de onde segue nossa proposi\c c\~ao. 

Continuando, n\'os temos, a partir de (\ref{kalkle1}) e (\ref{kalkle2}):
\begin{eqnarray}
\label{kal3}\w^B_D\wedge\w^D_A(e_A,e_B)&=&\w^B_D(e_A)\w^D_A(e_B)-\w^B_D(e_B)\w^D_A(e_A)\\
\label{kal4}d\w^B_A(e_A,e_B)&=&e_B[\w^B_A(e_A)]-e_A[\w^B_A(e_B)]-\w^B_A([e_A,e_B])
\end{eqnarray} 
Agora, a curvatura de $M$ ser\'a dada por uma express\~ao id\^entica a de $P$, substituindo $\w$ por $\Gamma$. Para  (\ref{kal3}), pela anti-simetria em $A$ e $B$, precisamos considerar tr\^es casos, $A,B$ espaciais, $A,B$ ``gregos" e $A$ espacial, $B$ grego. Substiuindo $A$, $B$ e $D$ em (\ref{kal3}) por \'indices espaciais, n\'os obtemos a equa\c c\~ao  an\'aloga para $M$:
\begin{eqnarray}\label{kal5}
\w^j_k\wedge\w^k_i(e_i,e_j)&=&\bar\Gamma^j_k\wedge\bar{\Gamma}^k_i(e_i,e_j)=\Gamma^j_k(X_i)\Gamma^k_i(X_j)-\Gamma^j_k(X_j)\Gamma^k_i(X_i)
\end{eqnarray}
Nos restaram da somat\'oria os termos para $D$ grego, por (\ref{continhas}), lembrando que $F^\sigma_{ij}$ \'e anti-sim\'etrico em $i,j$:
\begin{eqnarray}
\label{110}\w^j_\sigma\wedge\w^\sigma_i(e_i,e_j)&=&\w^j_\sigma(e_i)\w^\sigma_i(e_j)-\w^j_\sigma(e_j)\w^\sigma_i(e_i)\\
~&=&\w^\sigma_j(e_i)\w^\sigma_i(e_j)\\
\label{111}~& =&-\frac{1}{4}F^\sigma_{ij}F^\sigma_{ij}\end{eqnarray}

Faltam dois casos para terminarmos de analisar (\ref{kal3}). Para $A,B$ e $D$ gregos n\'os temos que  
\begin{eqnarray} 
\w^\sigma_\beta\wedge\w^\beta_\alpha(e_\alpha,e_\sigma)&=&\w^\sigma_\beta(e_\alpha)\w^\beta_\alpha(e_\sigma)-\w^\sigma_\beta(e_\sigma)\w^\beta_\alpha(e_\alpha)\\
~&=&\w^\sigma_\beta(e_\alpha)\w^\beta_\alpha(e_\sigma)=\frac{1}{4}C^\sigma_{\alpha\beta}C^\beta_{\sigma\alpha}\\
\label{kal7}~&=&-\frac{1}{4}C^\sigma_{\beta\alpha}C^\sigma_{\beta\alpha}
\end{eqnarray}
Para $D$ espacial, \'e f\'acil verificar que :

\beq \w^\sigma_i\wedge\w^i_\alpha(e_\alpha,e_\sigma)=0\eeq

Para $A$ espacial $B,D$ gregos temos
\begin{eqnarray}\w^\beta_\sigma\wedge\w^\sigma_i(e_i,e_\beta)&=&\w^\beta_\sigma(e_i)\w^\sigma_i(e_\beta)-\w^\beta_\sigma(e_\beta)\w^\sigma_i(e_i)\\
~&=&-\w^\beta_\sigma(e_\beta)\w^\sigma_i(e_i)=0
\end{eqnarray}
 Pela anti-simetria de $C^\sigma_{\alpha\beta}$. 
Para $A,D$ espaciais e $B$ grego obtemos assim como em  (\ref{110})-(\ref{111}):
\beq \label{kal10}\w^\sigma_j\wedge\w^j_i(e_i,e_\sigma)=\w^\sigma_j(e_i)\w^j_i(e_\sigma)-\w^\sigma_j(e_\sigma)\w^j_i(e_i)=-\frac{1}{4}F^\sigma_{ij}F^\sigma_{ij}\eeq

Agora, seguindo para a equa\c c\~ao (\ref{kal4}).  
Para $A,B$ gregos temos:
\begin{eqnarray} d\w^\beta_\alpha(e_\alpha,e_\beta)&=&e_\beta[\w^\beta_\alpha(e_\alpha)]-e_\alpha[\w^\beta_\alpha(e_\beta)]-\w^\beta_\alpha([e_\alpha,e_\beta])\\
~&=&-\w^\beta_\alpha([e_\alpha,e_\beta])=
-C^\sigma_{\beta\alpha}\w^\beta_\alpha(e_\sigma)\\
\label{kal8}~&=&\frac{1}{2}C^\sigma_{\beta\alpha}C^\sigma_{\beta\alpha}
\end{eqnarray} 
Onde utilizamos a const\^ancia e a anti-simetria de $C^\sigma_{\alpha\beta}$ e (\ref{continhas}). 
Para $B$ grego e $A$ espacial temos
\begin{eqnarray}
 d\w^\beta_i(e_i,e_\beta)&=&d\left(\frac{1}{2}F^\beta_{ij} e^j\right)(e_i,e_\beta)=\frac{1}{2}\left(dF^\beta_{ij}\wedge e^j+F^\beta_{ij}de^j\right)(e_i,e_\beta)\\
~&=&-\frac{1}{2}dF^\beta_{ii}(e_\beta)+\frac{1}{2}F^\beta_{ij}\left(\w^j_A\wedge e^A(e_i,e_\beta)\right)\\
~&=&\frac{1}{2}F^\beta_{ij}\left(\w^j_\beta(e_i)-\w^j_i(e_\beta)\right)\\
\label{ultimate}~&=&-\frac{1}{2}F^\beta_{ij}F^\beta_{ij}
\end{eqnarray}  
Onde utilizamos anti-simetria para vermos que  $dF^\beta_{ii}=0$.
 
Finalmente, para os bravos que conseguiram suportar essa infinitude de contas, chegamos ao caso final: $A, B$ espacial.   
\begin{eqnarray}
d\w^i_j(e_j,e_i)&=& d\left(\bar\Gamma^i_j-\frac{1}{2} F^\sigma_{ij}e^\sigma\right)(e_j,e_i)\\
~&=&\left(d\bar\Gamma^i_j-\frac{1}{2} dF^\sigma_{ij}\wedge e^\sigma-\frac{1}{2} F^\sigma_{ij}de^\sigma\right)(e_i,e_j)\\
~&=&d\bar\Gamma^i_j(e_i,e_j)-\frac{1}{2} F^\sigma_{ij}\left(\w^\sigma_A(e_i)e^A(e_j)-\w^\sigma_A(e_j)e^A(e_i)\right)\\
~&=&d\bar\Gamma^i_j(e_i,e_j)-\frac{1}{2}F^\sigma_{ij}\left(\w^\sigma_j(e_i)-\w^\sigma_i(e_j)\right)\\
\label{ultimate2}~&=&d\bar\Gamma^i_j(e_i,e_j)+\frac{1}{2}F^\sigma_{ij}F^\sigma_{ij}
\end{eqnarray}
Agora, somando (\ref{ultimate}) a (\ref{ultimate2}) obtemos $d\bar\Gamma^i_j(e_i,e_j)$ que somado a (\ref{kal5}) nos fornece justamente $~^MR$. Agora, somando (\ref{kal7}) a  (\ref{kal8}) obtemos $^GR$, e finalmente, somando (\ref{111}) a (\ref{kal10}) obtemos $-\frac{1}{2}\|\Omega\|^2$. 

$~~~~\blacksquare$\medskip

Agora, como vimos na deriva\c c\~ao da equa\c c\~ao (\ref{metricP}), h\'a uma bije\c c\~ao entre as m\'etricas de fibrado (bundle metrics) para $P$ e  $(h,\omega)$ onde $h$ \'e m\'etrica de $M$ e $\omega$ \'e forma de conex\~ao em $P$. Logo, ao inv\'es de variarmos o funcional de a\c c\~ao  em rela\c c\~ao a  m\'etrica, podemos tomar a varia\c c\~ao independente de $h$ e $\omega$. Ent\~ao temos :
$$S(\alpha)=S(h,\omega)=\int_P ~^PR\mbox{vol}_P=\int_M \left(~^MR-\frac{1}{2}\langle\Omega,\Omega\rangle+~^GR\right)\mbox{vol}_M$$
Lembramos que m\'etrica $H$ entra sorrateiramente em $\langle\Omega,\Omega\rangle$ atrav\'es do operador $*$ de Hodge. \'E poss\'ivel mostrar \cite{bleecker} que,  ao variarmos independentemente $h$ e $\omega$, i.e.: tomando $h_t=h+t\mu$ e $\omega_s=\omega+s\gamma$ onde $\mu$ \'e um $(0,2)$-tensor sim\'etrico em $TM$ e $\gamma\in\Gamma(\Lambda^1(TM)\otimes{\Lg})$, obtemos, para a varia\c c\~ao de $h$:
\beq\label{Einstein} \frac{d}{dt}_{|_{t=0}}S(h_t,\omega)=\int_M~\left(^MR_{ij}-\frac{1}{2}~^MRh_{ij}-T_{ij}\right)\mu^{ij}\mbox{vol}_M\eeq
onde $T_{ij}$ \'e o tensor de energia momento relativo ao campo de Yang-Mills. Claramente (\ref{Einstein}) resulta na equa\c c\~ao de Einstein tendo o campo de Yang-Mills como fonte. A demonstra\c c\~ao de (\ref{Einstein}) em si envolve longos e tediosos c\'alculos e pode ser encontrada em qualquer bom livro de relatividade geral, por exemplo \cite{Wald} e \cite{Baez}, ou como mencionamos, em \cite{bleecker} de forma mais completa. A varia\c c\~ao em rela\c c\~ao a $\omega$, como somente o termo $\langle\Omega,\Omega\rangle$ depende de $\omega$, j\'a  foi calculada em (\ref{eqsYang}) e fornece justamente a equa\c c\~ao de Yang-Mills:
$$D^*\Omega=0$$
Portanto obtivemos que os campos obedecem as equa\c c\~oes corretas para ambos a relatividade geral e campos de Yang-Mills, ou seja, obtivemos toda a din\^amica resultante das duas teorias a partir de uma m\'etrica de fibrado em $P$. Dessa forma, \'e poss\'ivel n\~ao s\'o unificar as duas teorias, mas  ao alcan\c carmos esta unifica\c c\~ao, os campos de Yang-Mills, assim como o gravitacional, desaparecem como for\c cas e s\~ao descritos por pura geometria, no sentido que part\'iculas percorrem geod\'esicas de uma geometria riemanniana apropriada.

\end{document}